\documentclass[11pt, double, phd]{osudiss-2} 

\usepackage{graphicx} 
\usepackage{lipsum} 
\usepackage{verbatimbox}

\usepackage{bm} 
\usepackage{booktabs} 
\usepackage[titletoc]{appendix}

\usepackage{bookmark} 
\hypersetup{colorlinks=true,linkcolor=yaleblue, citecolor=yaleblue, urlcolor=yaleblue, breaklinks} 
\usepackage[all]{hypcap}

\usepackage{hyperref}

\usepackage[numbers,square,sort&compress]{natbib}

\usepackage{subfig}
\usepackage[export]{adjustbox}
\PassOptionsToPackage{obeyspaces}{url}

\usepackage[utf8]{inputenc}
\usepackage[T1]{fontenc}
\usepackage{lmodern}
\usepackage{roboto}

\usepackage{mathtools}
\usepackage{amssymb}
\usepackage{amsmath}
\usepackage{amsthm}

\usepackage[dvipsnames]{xcolor}

\definecolor{offblack}{RGB}{35,31,32}
\definecolor{yaleblue}{rgb}{0.06, 0.3, 0.57}
\titleformat{\chapter}[display]{\Huge\sffamily\color{offblack}\filcenter
}{\thechapter}{1ex}{}
\titleformat*{\section}{\Large\bfseries\color{yaleblue}}
\titleformat*{\subsection}{\large\bfseries\color{yaleblue}}
\titleformat*{\subsubsection}{\bfseries\color{yaleblue}}
\titleformat*{\paragraph}{\bfseries\color{offblack}}

\usepackage[acronym, section=chapter]{glossaries}
\makeglossaries 

\newacronym{lhc}{LHC}{Large Hadron Collider}
\newacronym{rhic}{RHIC}{Relativistic Heavy Ion Collider}
\newacronym{pca}{PCA}{Principal Component Analysis}
\newacronym{qcd}{QCD}{Quantum Chromodynamics}
\newacronym{eos}{EoS}{Equation of State} 

\newcommand{\be}{\begin{equation}}
\newcommand{\ee}{\end{equation}}
\newcommand{\bs}{\begin{subequations}}
\newcommand{\es}{\end{subequations}}
\newcommand{\beal}{\begin{align}}

\newcommand{\trento}{\texttt{T$_\mathrm{R}$ENTo}}
\newcommand{\URQMD}{\texttt{UrQMD}}
\newcommand{\SMASH}{\texttt{SMASH}}
\newcommand{\vishnew}{\texttt{VISHNew}}
\newcommand{\music}{\texttt{MUSIC}}

\newcommand{\sqrts}{\sqrt{s_\textrm{NN}}}
\newcommand{\Tsw}{T_{\text{sw}}}
\newcommand{\taufs}{\tau_{\text{fs}}}
\newcommand{\bigCI}{\mathrel{\text{\scalebox{1.07}{$\perp\mkern-10mu\perp$}}}}

\DeclareMathOperator{\sign}{sign}

\newcommand{\eq}[1]{Eq.~(\ref{#1})}

\newcommand{\fig}[1]{Fig.~\ref{#1}}
\newcommand{\figs}[1]{Figs.~\ref{#1}}

\newcommand{\Appendix}[1]{Appendix~\ref{#1}}

\newcommand{\Table}[1]{Table~\ref{#1}}

\title{Quantifying the \\Quark-Gluon Plasma}
\author{Derek Everett}
\advisorname{Ulrich Heinz}
\degree{Doctor of Philosophy} 
\member{Richard Furnstahl}
\member{Michael Lisa}
\member{Jay Gupta}
\authordegrees{M.Sc.}
\graduationyear{2021}
\unit{Graduate Program in Physics} 

\begin{document}

\frontmatter

\begin{abstract}

The study of heavy-ion collisions presents a challenge to both theoretical and experimental nuclear physics. Due to the extremely short ($10^{-23}$ s) lifetime and small size ($10^{-14}$ m) of the collision system, disentangling information provided by experimental observables and progress in physical insight requires the careful application of plausible reasoning. 

I apply a program of statistical methodologies, primarily Bayesian, to quantify properties of the medium in specific models, as well as compare and criticize differing models of the system. 
Of particular interest are estimations of the specific shear and bulk viscosities, where we find that information carried by the experimental data is still limited. In particular we find a large sensitivity to prior assumptions at high temperatures. Moreover, sensitivities to model assumptions are present at low temperatures, and this source of model uncertainty is propagated with model averaging and model mixing.

\end{abstract}

\dedication{
To my wife, Katherine, for her abiding love and support.} 
\begin{acknowledgments}

Allocation of supercomputing resources (Project: PHY180035) were obtained in part through the Extreme Science and Engineering Discovery Environment (XSEDE), which is supported by National Science Foundation grant number ACI-1548562. Calculations were performed in part on Stampede2 compute nodes, generously funded by the National Science Foundation (NSF) through award ACI-1134872, within the Texas Advanced Computing Center (TACC) at the University of Texas at Austin \cite{TACC}, and in part on the Ohio Supercomputer \cite{OhioSupercomputerCenter1987} (Project PAS0254).

I am very grateful to my advisor, Ulrich Heinz, for being a patient mentor and teacher since my very first semester at The Ohio State University. He has invested countless hours of his time instructing me in lectures, in group meetings, and in private conversations. I have been lucky to have such a well-rounded advisor who, despite a very successful career spanning more than forty years, always confronts each new problem or methodology with an open and inquisitive mind. I am hopeful that this scientific humility has rubbed off on me and that I can approach each new problem with a healthy dose of uncertainty regarding what \textit{I think} I know. 

This work would not have been possible without the close collaboration of many individuals, particularly many members of the JETSCAPE/XSCAPE collaboration. In particular, much of this work was produced in daily collaboration with Weiyao Ke, Jean-Francois Paquet, and Gojko Vuyanovich. Moreover, these results were shaped by many fruitful discussions with members of the JETSCAPE SIMS working group, including Lipei Du, Matthew Heffernan, Dan Liyanage, Matthew Luzum, Michael McNelis, Chun Shen, and Yingru Xu. 
They each have my sincerest thanks and respect. Although I have only included in this manuscript results and plots I have personally generated, I have included the JETSCAPE logo on those which were generated with their collaboration. 
Many results shown in Ch.\ref{ch5} and Ch.\ref{ch6} can also be found in Refs.~\cite{Everett:2020yty, Everett:2021ulz}, which should be considered the primary sources in such cases. 
Only when referring to a result found in this thesis which is not in those references should this thesis be cited. 

In addition, this work would not have been possible without pioneering work of the MADAI collaboration and the Duke Heavy-Ion Group. The MADAI collaboration publications \cite{Novak:2013bqa, Sangaline:2015isa} pioneered Bayesian parameter estimation and sensitivity analyses (using a principal component reduced model emulator) in the field of heavy-ions, which opened the door to a new era of heavy-ion modeling with \textit{quantified uncertainties}.  
Additionally, the high degree of quality and readability of the Bayesian parameter estimation code packages developed by J. Bernhard, J. Scott Moreland and Weiyao Ke often allowed for straightforward reuse with minimal modification~\cite{Moreland:2019szz, Bernhard:2018hnz}.

I would also like to thank Prof. Hannah Elfner, Dmytro Oliinychenko, and the entire \SMASH{} group for their collaboration and guidance in using \SMASH{} as an afterburner. The comparisons made between \SMASH{} and \URQMD{} were aided by a fruitful visit to Frankfurt, which was supported by Prof. Elfner's group. 

I would like to thank Prof. Dick Furnstahl for teaching an excellent course in Bayesian inference, which helped shape how I think about the scientific method and introduced many tools used throughout this manuscript. The course materials~\cite{furnstahl} are \textit{highly recommended} to all researchers whose interests align with this thesis.

Finally, I thank Lipei Du, Michael McNelis, Chandrodoy Chattopadhyay, and Dan Liyanage for being excellent peers in learning and collaboration, and equally excellent friends.

\end{acknowledgments}
\begin{vita}
\dateitem{January 3, 1992}{Born---Detroit, U.S.}
\dateitem{May, 2015}{B.S. Physics, Wayne State University, Detroit, MI}
\dateitem{May, 2015}{B.A. Mathematics, Wayne State University, Detroit, MI}
\dateitem{May, 2017}{M.S. Physics, The Ohio State University, Columbus, OH}

\begin{publist}

\pubitem{``Maximum entropy kinetic matching conditions for heavy-ion collisions'', 
D. Everett,  Chandrodoy Chattopadhyay, Ulrich Heinz.
\href{https://arxiv.org/abs/2101.01130}{https://arxiv.org/abs/2101.01130}}

\pubitem{``Multi-system Bayesian constraints on the transport coefficients of QCD matter'', 
D. Everett, W. Ke. J.-F. Paquet, Gojko Vujanovich {\it et al.}
\href{https://arxiv.org/abs/2011.01430}{https://arxiv.org/abs/2011.01430}}

\pubitem{``Phenomenological constraints on the transport properties of QCD matter with data-driven model averaging'', 
D. Everett, W. Ke. J.-F. Paquet, Gojko Vujanovich {\it et al.}
\href{https://arxiv.org/abs/2010.03928}{https://arxiv.org/abs/2010.03928}}

\pubitem{``Particlization in fluid dynamical simulations of heavy-ion collisions: The iS3D module'', 
Mike McNelis, D. Everett, Ulrich Heinz.
\href{https://arxiv.org/abs/1912.08271}{https://arxiv.org/abs/1912.08271}}

\end{publist}

\begin{fieldsstudy}
\majorfield{Physics}
\onestudy{High Energy Nuclear Physics}{Heinz group} 
\end{fieldsstudy}

\end{vita}

\tableofcontents 

\clearpage 
\listoffigures 



\mainmatter
\chapter{Introduction}
\label{ch1}

\section{Understanding the origins of our universe}
\label{ch1:bigbang}

\begin{figure*}[!htb]
\centering
\includegraphics[width=0.8\textwidth]{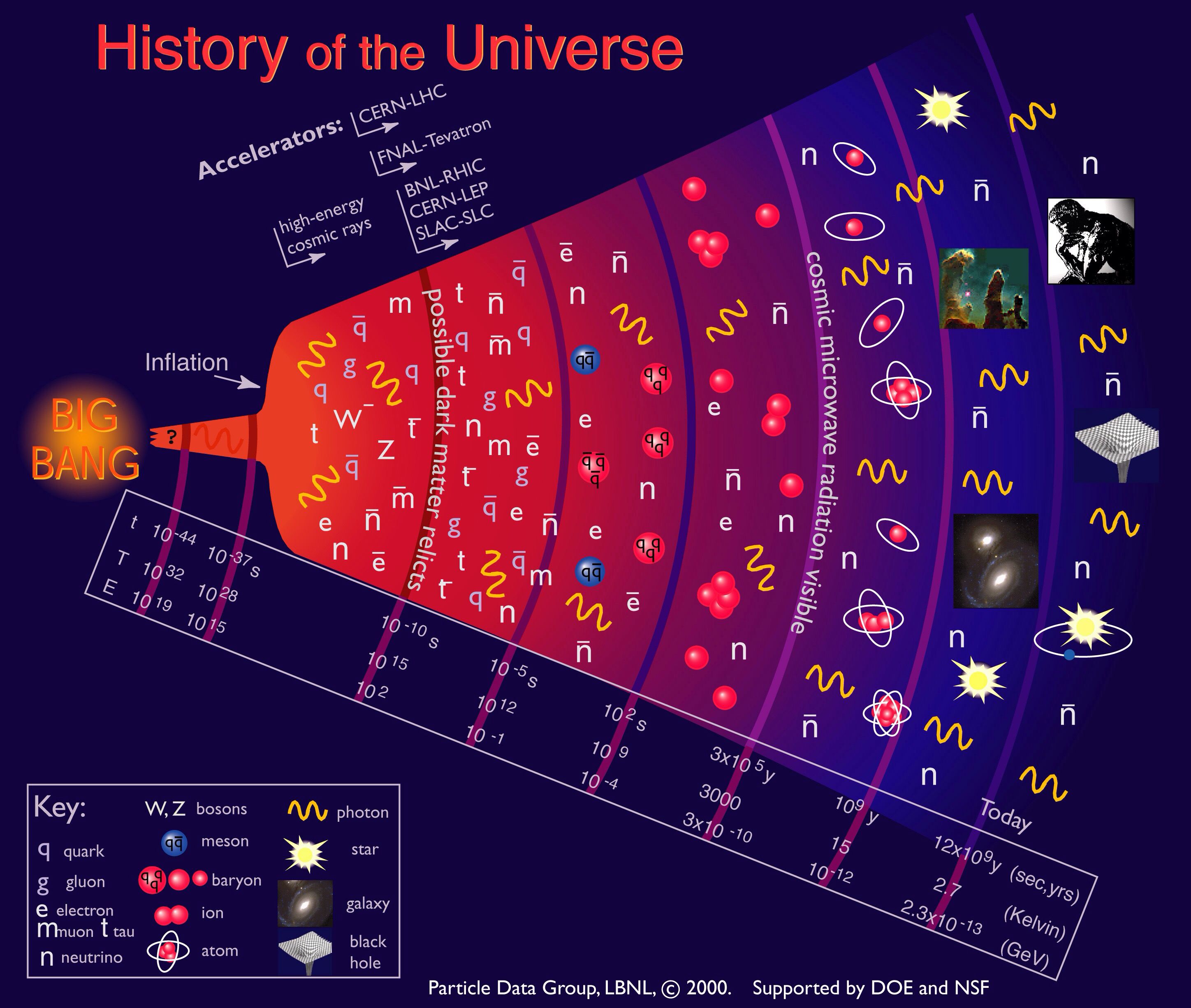}
    \caption{A schematic timeline of the universe ~\cite{PDG_history_universe}. The first microseconds were occupied by a state of Quark-Gluon plasma until the universe had sufficiently cooled to require the formation of stable hadrons.}
\label{fig:universe_history}
\end{figure*}

According to the current understanding of cosmology, our universe began with a ``Big Bang'' roughly fourteen billion years ago. A schematic history of the timeline of the universe is shown in Fig. \ref{fig:universe_history}, where we see that the universe passed through several distinct phases. In its first moments, the temperature of the expanding universe was sufficiently high that quarks and gluons and other elementary particles could propagate and scatter. After about ten microseconds, the universe had cooled down to a temperature of $0.15$ GeV, below which quarks and gluons are confined inside hadrons. Stable hadrons, like protons and neutrons, became the building blocks for the matter that now occupies the visible universe and of which we ourselves are composed~\cite{PDG_history_universe}. 

Understanding the first microseconds of the universe requires a model to describe the interactions of quarks and gluons; the state-of-the-art microscopic model is Quantum Chromodynamics (QCD). QCD posits that all partons, consisting of quarks and gluons, carry a `color' charge and interact via a force mediated by gluons. For distances larger than roughly the proton radius, this inter-parton force is constant -- independent of the distance between them. This strong force is profoundly dissimilar from the forces beween electrically charged particles, for example. Quantum Electrodynamics (QED) predicts that electrically charged particles interact with a force that is inversely proportional to the squared distance between the two particles. 

This feature of QED, that electrically charged particles and force-carrying photons can propagate in the vacuum, makes the theory accessible to both the scientific method and the canonical methods physicists have for performing predictive calculations. An experiment can be performed in which two electrically charged particles are fired at each other from asymptotically large distances and their deflection measured by a detector. Meanwhile, a physicist can employ the methods of perturbation theory to describe the scattering of the two weakly interacting charged particles and predict the outcome as a function of the free parameters in the theory. In the case of QED,\footnote{%
I've restricted this discussion to scales below the electroweak boson masses for simplicity.} 
the only necessary parameters are the electron mass and the dimensionless coupling constant $\alpha$, which encodes the strength of the interaction between charged particles (mediated by photons). Finally, the predictions and measurements are carefully compared, and the experimental data have sufficient information to constrain the model parameters to an impressive precision~\cite{Achard:2005it}. 

\begin{figure*}[!htb]
\centering
\includegraphics[width=0.5\textwidth]{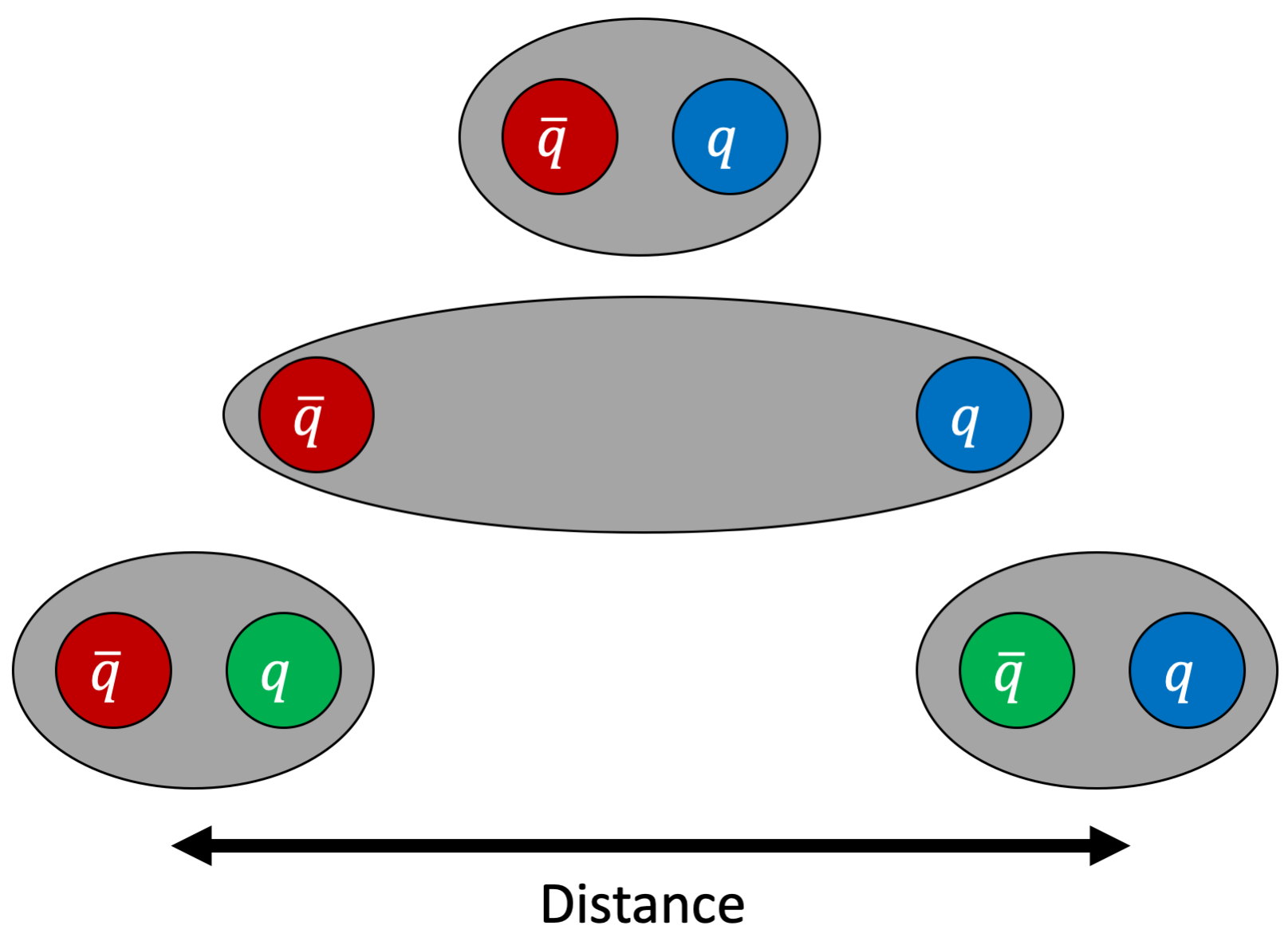}
    \caption{A schematic visualization of quark confinement within hadrons. Here a force tries to separate a quark and antiquark from within a meson (gray). A new quark anti-quark pair is created, resulting in a pair of mesons rather than freely propagating quarks. }
\label{fig:quark_confinement}
\end{figure*}

\begin{figure*}[!htb]
\centering
\includegraphics[width=\textwidth]{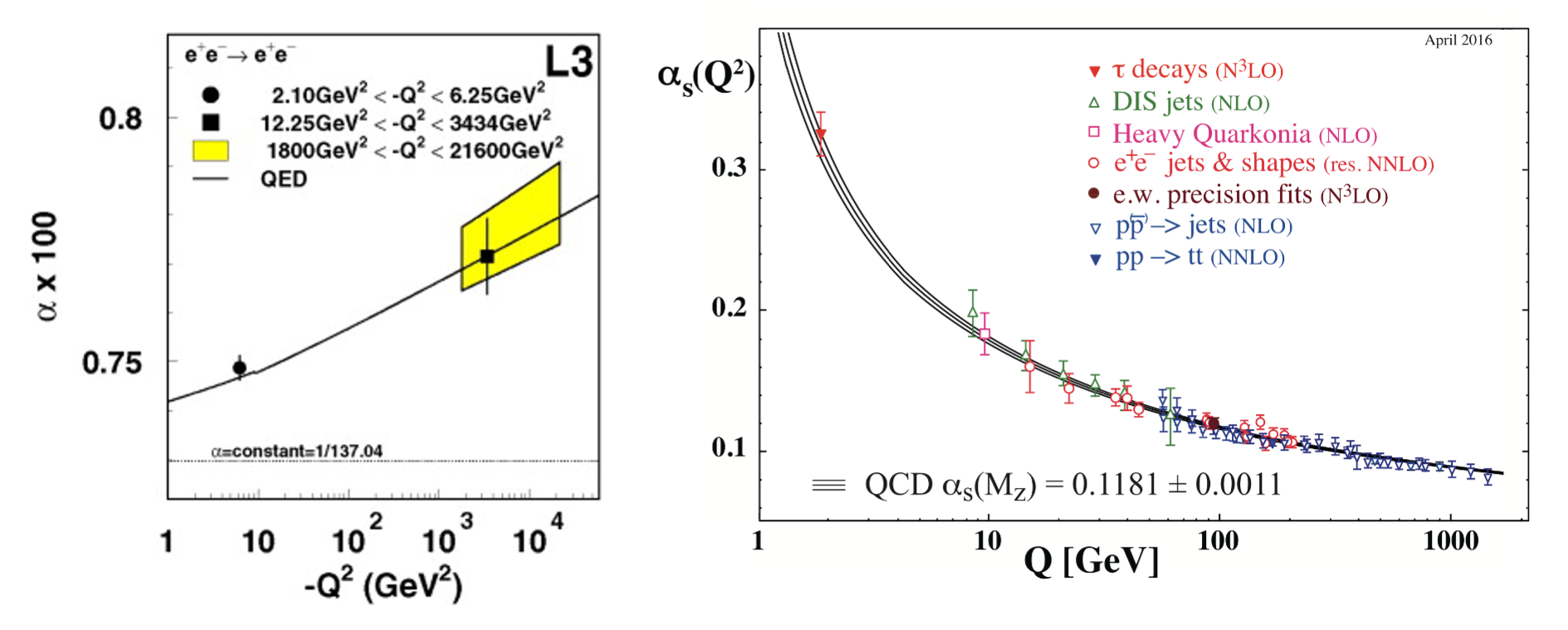}
    \caption{The running of the QED interaction strength~\cite{Achard:2005it} (left) and QCD interaction strength~\cite{Tanabashi:2018oca} (right) with momentum scale. The two theories have opposite behaviors: in QED the coupling increases with momentum scale (`screening'), while in QCD the coupling strength decreases (`anti-screening'). }
\label{fig:running_coupling}
\end{figure*}

Now we consider a bound state of two quarks, which are confined together inside a meson by a strong field of gluons, and consider trying to measure the properties of these quarks in vacuum. As we pull the quarks further apart we have to perform a huge amount of work. At a distance of roughly the proton radius the potential energy in their interacting field is so large that a new quark anti-quark pair is spontaneously created from the gluons. The strong `flux-tubes' of gluons break into two shorter flux-tubes, and the vacuum yields two meson bound-states, visualized in Fig. \ref{fig:quark_confinement}. 
Because quarks and gluons are bound together inside hadrons at everyday temperatures (below 0.15 GeV), it is very difficult for physicists to measure their \textit{dynamical} properties.
An experimental physicist can not fire two quarks at each other from asymptotically far away and measure their deflection. `Bare' quarks simply cannot exist in the vacuum; the vacuum would rather `dress' them into a hadronic bound-state. 
For the same reasons, at lower energy scales (larger distance scales) the theoretical physicist is unable to apply the methods of perturbation theory to make predictions for the hypothetical scattering experiment because those methods require the interaction strength to be sufficiently weak. 
Plots demonstrating the evolution of the interaction strength (characterized by dimensionless coupling parameter $\alpha$) are shown in Fig. \ref{fig:running_coupling} for QED and QCD. 
We see that QED is weakly coupled for small energy scales $\alpha_{\rm QED} (Q \rightarrow 0) \ll 1$, while QCD is strongly coupled $\alpha_{\rm QCD} (Q \rightarrow 0) \gg 1$.
QCD is only `asymptotically free'; at asymptotically large momentum scales the partons become non-interacting $\lim_{Q^2 \rightarrow \infty} \alpha_{\rm QCD}(Q^2) = 0$. 

\section{Creating the ``Little Bang'' in the laboratory}
\label{ch1:littlebang}

To try and measure the properties of interacting quarks and gluons, scientific collaborations perform collisions of heavy ions at ultra-relativistic speeds. The idea is straightforward -- try to recreate in the laboratory an extremely hot and dense medium resembling the early universe by colliding heavy nuclei with enormous kinetic energies~\cite{Heinz:2000ba}. As such, collisions of heavy nuclei are performed at the Relativistic Heavy Ion Collider (RHIC)\cite{rhic}, Large Hadron Collider (LHC)\cite{lhc}, and other facilities, where a wealth of observables have been measured.

\begin{figure*}[!htb]
\centering
\includegraphics[width=0.5\textwidth]{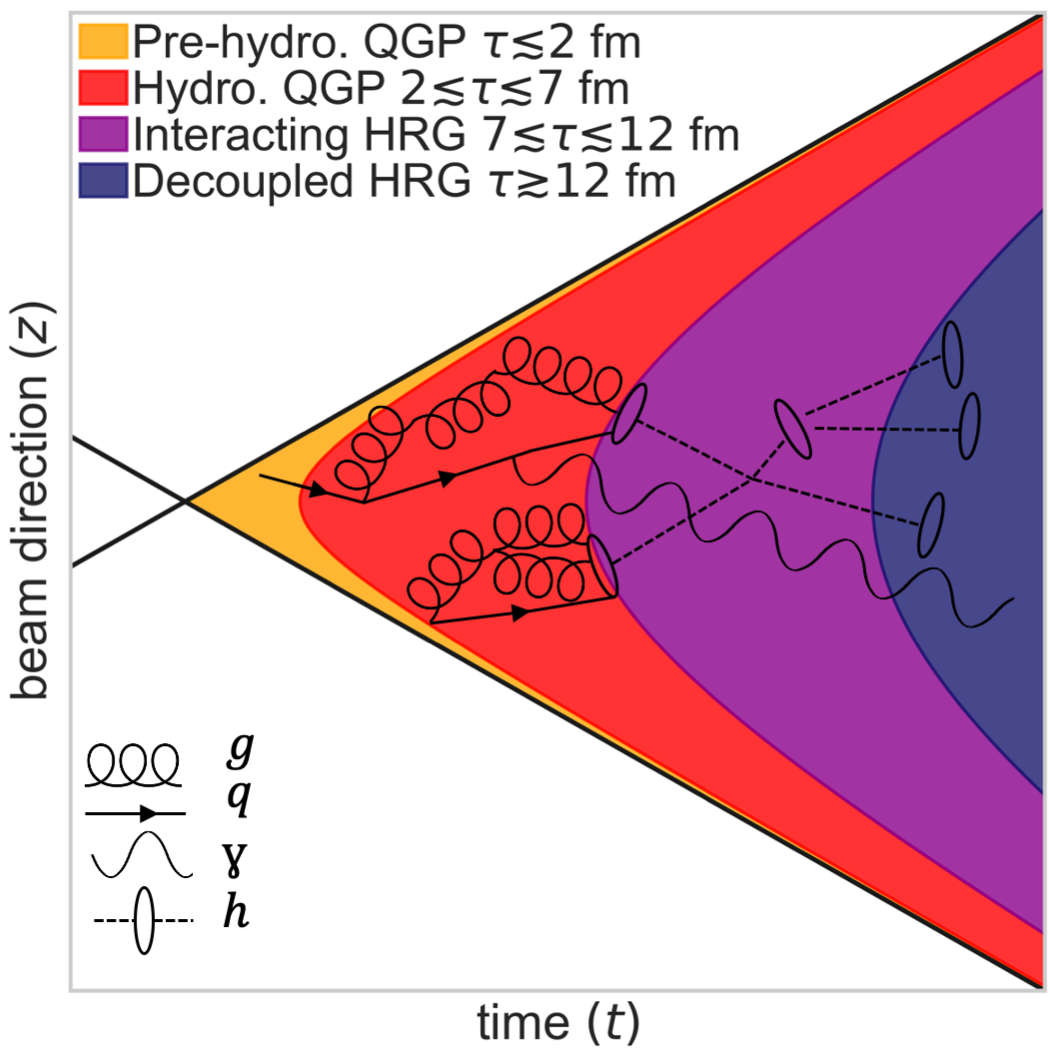}
    \caption{A timeline of a heavy-ion collision according to current modeling and phenomenological understanding. Gluons, quarks, photons, and hadrons are labeled by $g$, $q$, $\gamma$ and $h$ respectively.}
\label{fig:little_bang}
\end{figure*}

A spacetime diagram of the timeline of a heavy-ion collision is depicted in Fig. \ref{fig:little_bang}. For concreteness, we have labeled various phases of the collision and assigned them windows of proper time $\tau$. These labels should be understood to represent coarse pictures which omit many important details in the modeling of heavy-ion collisions. Similarly, the quoted proper times should be understood to be rules-of-thumb, not hard-and-fast numbers.\footnote{One femtometer, $10^{-15}$ meters, is abbreviated fm from here on.} There are no \textit{external} experimental probes\footnote{%
If you want to measure the temperature of your cup of coffee, you can place a thermometer in contact with it. The QGP fireball is far too small and short-lived to permit any external experimental probe.
} which can measure the evolution of the collision; the dynamics must be \textit{reconstructed} from the final state. This is only possible by comparing final state particles with physical models describing all stages of the collision.\footnote{Actually, electromagnetic radiation, including photons and dileptons, is emitted throughout the entire collision. Those emitted at early times usually escape the fireball `unscathed', making them important gauges of the collision at early times. Certainly electromagnetic probes will continue to be important measurements for understanding heavy-ion collisions, but they are omitted from the studies in this thesis. }

A very large energy density is deposited in the laboratory frame following the collision of the two nuclei. In its first ioctoseconds, the system begins to rapidly expand. The first $2$ fm/c of evolution are perhaps the phase of the collision whose dynamics are most uncertain. There are currently a variety of models to describe the matter at these earliest times, ranging from models with strongly coupled fields (AdS/CFT) to models of weakly coupled partons. Despite large differences in physical assumptions, many of these models can qualitatively describe the data well. Due to this large theoretical uncertainty, we have assigned this phase of the collision the name `Pre-hydrodynamic QGP'. This label applies trivially in the context of the models used in this thesis -- it is the phase of the collision \textit{before} the hydrodynamic phase.\footnote{%
This of course only labels what this phase is \textit{not}.} 

After only $1-2$ fm/c, there is strong evidence that the system can be described by hydrodynamic transport, so we have labeled the phase from then until about $7$ fm/c the `Hydrodynamic QGP'. Again, this label is intentionally ambiguous as to the relevant microscopic degrees of freedom. The success of hydrodynamic models with strong coupling (small viscosities) may preclude a picture of interacting partonic quasi-particles,\footnote{Quasi-particles are propagating degrees of freedom with wavelengths much smaller than the typical distance between scattering.} and a \textit{microscopic} description of this phase may require a fully dynamical theory of strong QCD quantum fields.\footnote{Propagating and infrequently interacting particles are only particular solutions to quantum field equations in the limit of weak coupling.} Therefore the pseudo-Feynman diagrams of interacting quarks and gluons in Fig. \ref{fig:little_bang} should not be understood to evoke a weakly coupled or perturbative system of quarks and gluons; they are simply a visual aid. 

After about $7$ fm/c the temperature has dropped sufficiently for hadrons to form. In the vicinity of this time, the hadronic density has also decreased sufficiently that the hadronic collision rate is outpaced by the rapid expansion. When the interaction rate among microscopic degrees of freedom is slower than the expansion rate, hydrodynamic descriptions break down. The continued scattering and decays of the weakly interacting hadron resonance gas can instead be described by kinetic theory. Finally, the gas of hadrons and other particles `freezes out' -- the momenta of the hadrons are no longer changed by interactions, and the stable particles stream freely to the detector.  

\begin{figure*}
\centering
\includegraphics[width=0.8\textwidth]{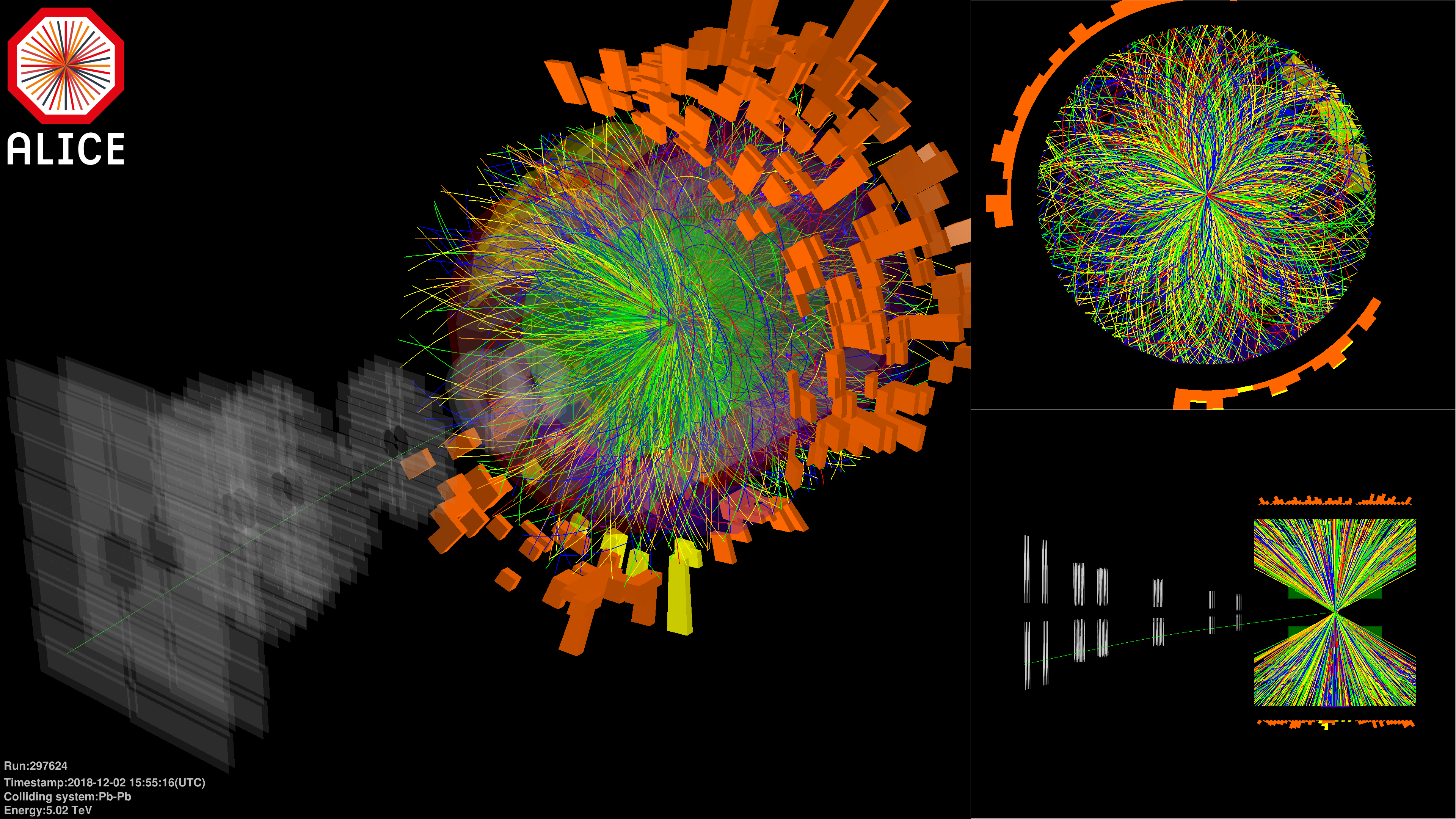}
    \caption{An event display for a Pb-Pb event measured by the ALICE detector~\cite{ALICE_event}. Each line represents the trajectory of a charged particle created in the collision, and each block represents its energy. }
\label{fig:alice_event}
\end{figure*}

Thousands of hadrons, including mesons like pions and kaons, and baryons like protons, neutrons, and heavier exotic species, are emitted from the collision. From the finally emitted particles that hit the detector -- usually the stable charged hadrons -- scientists try to reconstruct the entire dynamical timeline of the collision. 
An event display of a Pb-Pb event at $\sqrts{} = 5.02$ TeV center-of-mass energy measured by the LHC ALICE collaboration is shown in Fig. \ref{fig:alice_event}.
Nearly all of the thousands of hadrons which are detected were created only at late times, i.e. during the final stages at $\tau \gtrsim 7$ fm/c.
Therefore, progress of understanding in the field of heavy-ion collisions requires a continuous loop of building and tearing down physics models based on their ability or failure to describe observed data.

\section{Overview of the phenomenology of heavy-ion collisions}
\label{ch1:phenom}

\begin{figure*}[!htb]
\centering
\includegraphics[width=0.8\textwidth]{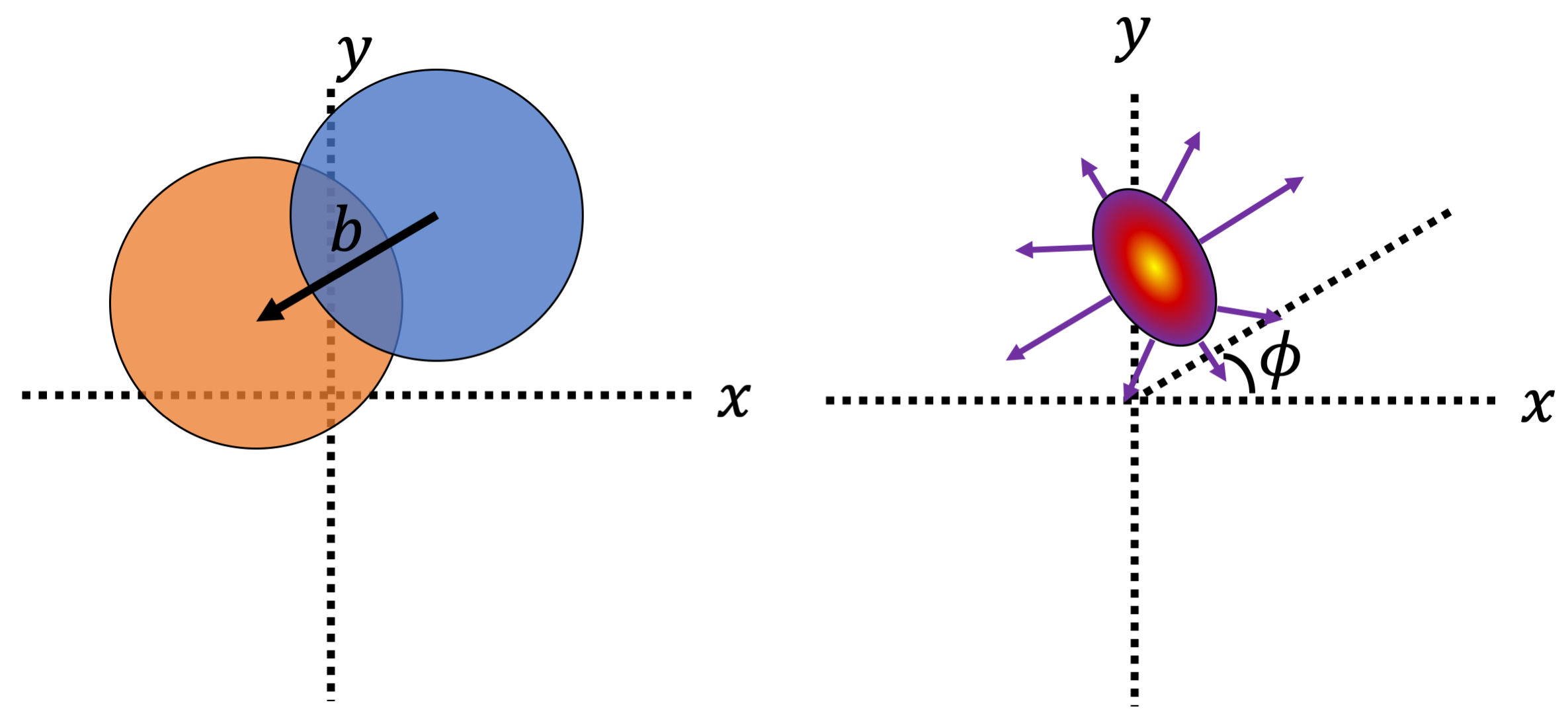}
    \caption{(Left) A projectile (blue) and target (orange) nucleus collide at an impact parameter $\bm{b}$. The $x, y$ axes are fixed according to the detector, and therefore the impact parameter in each collision points at a random angle in azimuthal coordinate $\phi$ which cannot be directly measured. (Right) An almond-shaped blob of energy density is deposited by the collision, which begins to expand anisotropically.}
\label{fig:impact_param}
\end{figure*}

In this section, we briefly introduce some of the key concepts and measurements relevant in driving the phenomenology of heavy-ion collisions. 
When collisions of heavy ions are performed at experimental facilities, there are only certain properties of the collision system that are under the control of the experimenter. The collision projectile and target, composed of specific isotopes of atomic nuclei, are controlled by injection of a pure beam of ions into the collider. Additionally, the center-of-mass energy of the collision system is controlled by the strength of the electromagnetic fields. Beyond these two controls, the specific configurations of each collision event are subject to random chance. The projectiles and targets consist of many randomly oriented nuclei colliding continuously. The impact-parameter $\bm{b}$, which is the vector that points from the center of the projectile nucleus to the center of the target nucleus, cannot be controlled nor measured directly. This situation is visualized in Fig. \ref{fig:impact_param}. Also depicted in the right frame of this figure is the deposited energy density in the transverse plane immediately following the collision. The nucleons in the two nuclei that scatter in the collision are called \textit{participants}, while the \textit{spectators} are the nucleons which do not. Therefore the deposited energy has a geometry controlled by the positions of the participant nuclei. The magnitude of the deposited energy density can vary depending on how many nucleons in that transverse proximity are scattered. Near the edges of the overlap region there are fewer nucleons and less matter is deposited, while more nucleons are scattered near the center of the overlap region. This gradient in local energy density is illustrated by the color gradient in right of Fig. \ref{fig:impact_param}.\footnote{This smooth almond-shaped energy density should be understood as an event-averaged approximation, to simplify exposition. In any given collision event, the positions and magnitudes of energy deposited by the participant nucleons can have strong fluctuations.} 

\begin{figure*}[!htb]
\centering
\includegraphics[width=0.8\textwidth]{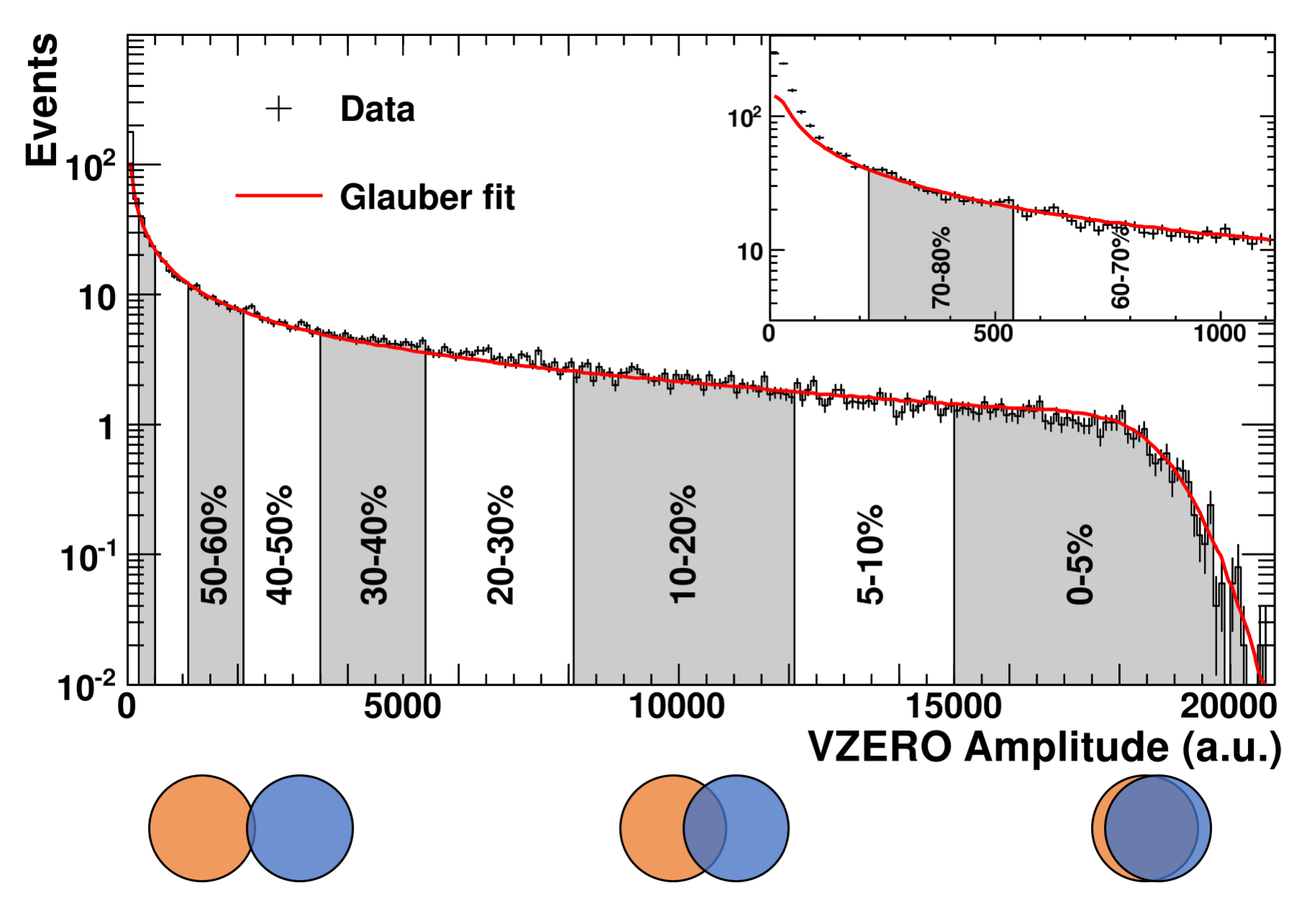}
\caption{Determination of centrality bins in the ALICE experiment~\cite{Aamodt:2010cz}. Below the figure are cartoons depicting the expected impact parameter when the projectile and target are identical spherical nuclei.} 
\label{fig:centrality_alice}
\end{figure*}
Because neither the angle nor magnitude of the impact parameter can be measured, we must use an experimental proxy to characterize different collisions. Experiments can typically detect with good efficiency the electrically charged particles which are produced in a given collision event. Each particle's momentum can be measured to within a finite resolution fixed by the design of the detection pads. This momentum can be decomposed into three orthogonal directions, and a set of cylindrical coordinates are typically employed. The particles' momentum in the plane transverse to the beamline, denoted by $\{x, y\}$, is decomposed into a magnitude $p_T \equiv \sqrt{p_x^2+p_y^2}$ and azimuthal angle $\phi_p \equiv \text{atan2}(p_x,p_y)$. The longitudinal momenta are typically characterized by the rapidity 
\be
    y \equiv \frac{1}{2}\ln \Bigl( \frac{E+p_z}{E-p_z} \Bigr),
\ee
where $E$ denotes the particles energy $E = \sqrt{\bm{p}^2 + m^2}$.
Another convenient momentum-space variable is the pseudo-rapidity $\eta$, and is defined
\be
    \eta \equiv \frac{1}{2}\ln \Bigl(\frac{|\bm{p}|+p_z}{|\bm{p}|-p_z}\Bigr).
\ee
For massless particles $E=|\bm{p}|$, and the rapidity and pseudo-rapidity are the same. Additionally, for particles emitted in the transverse plane $p^z = 0$, the rapidity and pseudorapidity are both zero.   
One of the coarsest observables is simply to count all the charged particles whose momenta lie within a certain acceptance. When we count all charged particles with pseudorapidity  $|\eta| \leq 0.5$, the observable is denoted $dN_{\rm ch}/d\eta$ and often called the \textit{charged particle multiplicity} or \textit{charged particle yield}. 

The number of charged particles fluctuates largely event-by-event. Some events may indeed produce thousands of charged particles, while other events produce only a handful. The amount of matter produced in an event is typically the proxy used to describe the \textit{centrality} of a collision event, shown in Fig. \ref{fig:centrality_alice}. Event-by-event, the experiment records the amount of matter produced using an experimental signature; in the case of the ALICE detector, this signature is the amplitude of the VZERO detector. Then, this distribution is binned in a histogram. The width of each bin corresponds to the probability (mass) that a randomly sampled event would be found with charged particle yield occupying that bin. The $5\%$ of events which produce the largest yields are denoted by the $0-5\%$ centrality class, and so on. 
Below the plot are cartoons depicting the expected impact parameter corresponding to different centrality bins. In collisions of nuclei which are head-on, we expect the largest amount of matter to be produced. On the other hand, for collisions in which the nuclei barely glance each other, we expect the smallest yields. Therefore, typically in collisions of heavy spherical nuclei, the event centrality is interpreted to be a control of the impact parameter and thus the size and average geometry of the deposited matter.\footnote{This introduction is restricted to spherical nuclei for simplicity.}


A discriminating set of observables are the identified particle yields. Usually particle identification is possible for the charged species. The number of species inside an acceptance of rapidity $|y| \leq 0.5$, denoted $dN_{i}/dy$, gives additional information about the \textit{chemistry} of the produced final state. Historically, hydrodynamic phenomenology has paid great attention towards describing the yields of charged pions, charged kaons and (anti)protons $dN_{i}/dy$, $i \in \{\pi, K, p\}$, in different centrality bins.

More differentially, experiments construct observables which characterize the momentum-dependence of the produced particles in each collision. The invariant $p_T$-spectrum of particles $dN_i/dy p_T dp_T$ emitted at midrapidity $|y| \leq 0.5$ provides information about the particle production of each species as a function of energy/momentum scale, and describing this spectrum has been a proving ground for many models of heavy-ion collisions. In particular, it is empirically observed that the $p_T$
spectrum of particles pions, kaons and protons with momenta $p_T \lesssim 2$ GeV can be reasonably fit with geometrically motivated models for the expansion. 
Simple metrics to quantify the $p_T$ dependence of hadrons include the mean momenta of each species $ \langle p_T \rangle_i$ in each centrality bin.

Additional information is provided by measuring the azimuthal $\phi_p$ momentum dependence of produced particles. For example, the azimuthal dependence of charged particles integrated over transverse momentum is provided by $dN_{\rm ch}/dy d\phi_p$. The decomposition of this spectrum in Fourier series gave rise to the so called `anisotropic flow' observables $v_n$:
\be
    \frac{dN_{\rm ch}}{dy d\phi_p} = \frac{1}{2\pi} \frac{dN_{\rm ch}}{dy} \left\{ 1 + 2 \sum_{n=1}^{\infty} v_n \cos[n(\phi_p - \Psi_n)] \right\}.
\ee
Measurements of the \textit{event-planes} $\Psi_n$ in an individual event are plagued by large uncertainties, and
therefore anisotropic flow coefficients are typically defined using only the directly observable correlations among pairs, triplets, quadruplets, etc. of particles comprising each event.

\section{Comparing physics models with observed heavy-ion data}
\label{ch1:comparing}

\begin{figure*}[!htb]
\centering
\includegraphics[width=0.3\textwidth]{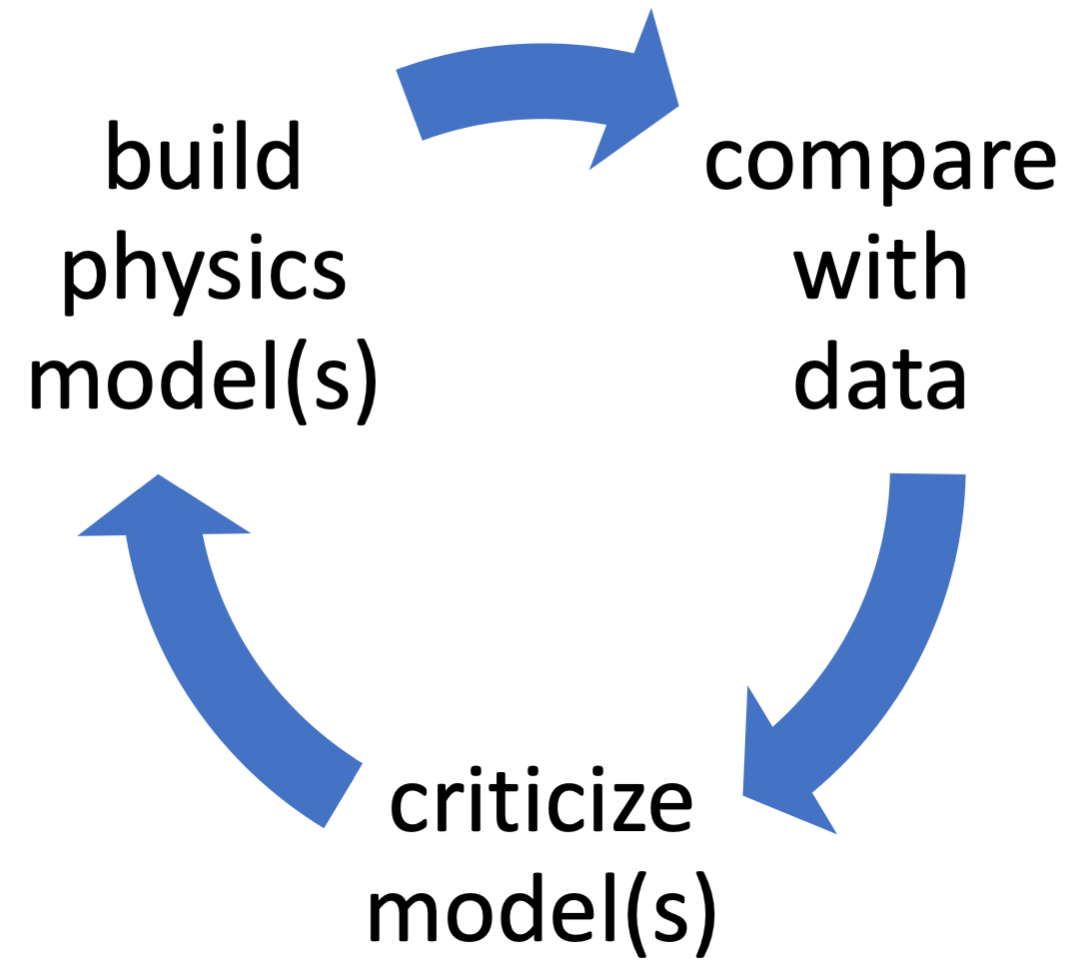}
    \caption{A visualization of the iterative loop of model building by which the phenomenology of heavy-ion collisions, and physical science more broadly, make progress. }
\label{fig:model_building}
\end{figure*}

Advances in the understanding of heavy-ion collisions are largely driven by the iterative loop visualized in Fig.~\ref{fig:model_building}. This simple graphic hides all of the important details of each analysis. Nevertheless, this graphic is intended to represent a suitable frame of mind, especially when we wish to investigate questions which are \textit{statistically well-posed}.\footnote{`Well-posed' is not intended to refer to questions which can be addressed with any particular statistical metric. Rather, it denotes questions which can be posed and calculated as conditional probabilities in a consistent formalism of plausible reasoning. This will be the subject of Ch. \ref{ch2}.} The community of researchers investigating heavy-ion physics, and more broadly low- and high-energy nuclear physics, has now been engaged in this model-building loop for many years. As such, the physics models which are now compared with data have become quite mature. Models to describe the evolution of the heavy ion medium now typically involve multiple stages, each being described by different physical dynamics and potentially requiring a large set of uncertain parameters.\footnote{%
The specific models which will be used in comparison to data in this thesis could be said to be composed of five sub-models and total nearly twenty model parameters.}

To introduce some of these models in a simple (albeit apocryphal) way, we consider the model paradigm that existed roughly fifteen years ago. Early in the $21^{\rm st}$ century, collaborations published reports that suggested the expansion of the QGP fireball could be well-described by the expansion of a nearly ideal fluid~\cite{Arsene:2004fa, Back:2004je, Adams:2005dq, Adcox:2004mh, Kolb:2003dz}. An \textit{ideal} fluid is one which flows with nearly zero specific shear-viscosity. These reports were based on the success of ideal hydrodynamic models in describing the momentum-dependence and azimuthal momentum anisotropy of particles produced in heavy-ion collisions, as well as other experimental signatures including the quenching of jets. That hydrodynamic models suceeded in qualitatively describing the soft-particle data, while many models of weakly-interacting partons or hadrons failed, was interpreted as a signal of hydrodynamic behavior in heavy-ion collisions.

\begin{figure*}[!htb]
\centering
\includegraphics[width=0.5\textwidth]{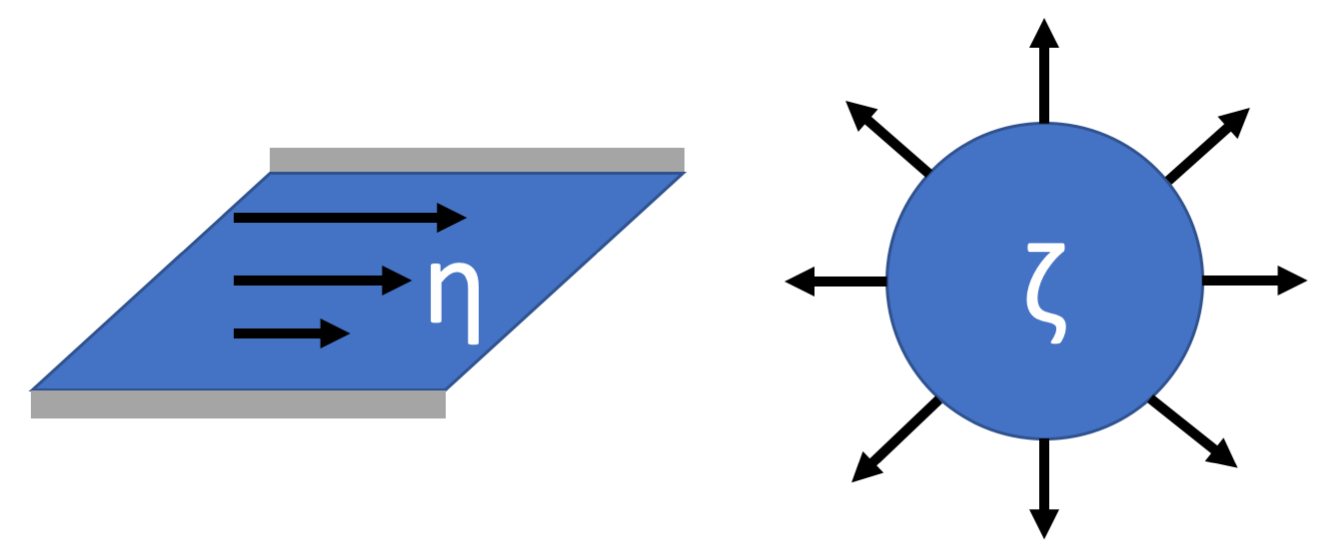}
    \caption{Visualizations of the shear viscosity $\eta$ (left) as a resistance to shearing flows, and bulk viscosity $\zeta$ (right) as a resistance to isotropic compressions and expansions. }
\label{fig:viscosity_visuals}
\end{figure*}

Ideal hydrodynamics can be considered as a subspace of a more general theory of viscous hydrodynamics. A natural step in the model-building loop was the introduction of non-zero viscosities to describe the heavy-ion collision expansion; a new parameter, the specific shear-viscosity $\eta/s$, was introduced into hydrodynamic models, and many comparisons with data were performed~\cite{Romatschke:2007mq, Luzum:2008cw, Song:2007fn}. Indeed, it was found that inclusion of a non-zero specific shear-viscosity $1/4\pi \lesssim \eta/s \lesssim 3/4 \pi$ could improve the model description of the observed data.  
Because QCD is a non-conformal theory, practitioners of hydrodynamic phenomenology soon introduced a non-zero specific bulk-viscosity~\cite{Denicol:2009am, Monnai:2009ad}, which controls the resistance of the fluid to isotropic compressions and expansions. The shear and bulk viscosities, $\eta$ and $\zeta$ respectively, are illustrated in Fig. \ref{fig:viscosity_visuals}. 

During the same period, the models employed to describe the initial conditions of the deposited matter were undergoing their own iterative changes. Simpler models based on smooth and event-averaged geometries gave way to models incorporating fluctuations of the deposited energy at smaller scales~\cite{Miller:2007ri, Hirano:2009ah}. 
These generalizations were necessary to describe the observed nonzero momentum anisotropy of particles produced in central collisions, for example, which were predicted to have isotropic geometries in models ignoring the fluctuations of nucleons inside the nuclei. 
These two different stages of the collision, initial conditions and hydrodynamic expansion, could not be developed in isolation, however. 
As we've observed, the final state observables predicted by such models depend both on the initial conditions, as well as the hydrodynamic evolution. More compact initial conditions yield larger hydrodynamic expansion rates, and more anisotropic initial geometries yield larger momentum anisotropies in the final state. But bulk viscosity acts to resist the expansion rate, and shear viscosity resists the build-up of momentum anisotropy. Because the final state observables are a combination of all of these effects, as well as many others, disentangling them systematically requires the global analysis of all stages of the collision with flexible models.

\section{The need for plausible reasoning}
\label{ch1:inference}

When models had only a handful of parameters, practitioners of heavy-ion phenomenology were able to gain meaningful insights by comparing model predictions against experimental measurements qualitatively. This practice, often referred to as doing `chi-by-eye', is a process by which a scientist compares model predictions at a single point (or handful of points) in parameter space by `hand-tuning' the parameters until the predictions `best' agree visually with the data. This method, and its playful name, are posed in contrast to quantitative prescriptions for measuring statistical consistency, e.g. the reduced $\chi^2$ metric~\cite{andrae2010dos}.  
These qualitative methods have been historically successful in discriminating models in heavy-ion phenomenology. However, as the dimension of the parameter space and space of observables increases, performing such investigations exhaustively becomes untenable. Moreover, it is not clear that any meaningful conclusions can be drawn when the model has approximate degeneracies -- different points in parameter space which yield approximately the same outputs.\footnote{%
The reader is strongly encouraged to visit \url{http://jetscape.org/sims-widget/} and try adjusting each of the parameters. Judged only by visual consistency with the data, there are numerous well-separated points in parameter space that could be judged to be in `good-agreement' (especially when fewer observables are compared). 
} 

A fundamental problem in comparing model predictions to data at only a point in parameter space is the complete neglect of uncertainty. Firstly, the experimental data are always measured with finite and quantified uncertainties. These measurements become more valuable to us when we compare our model's ability or failure to describe them with quantitative rather than qualitative metrics, and when honestly accounting for measurement uncertainties we are less prone to overfit our model, or over-interpret potential model-data (dis)agreements.    
Similarly, computing the outputs of a model at a single point in parameter space neglects the uncertainty in those parameters, and can quickly lead to the same pitfalls. 

The need for systematic and quantitative methods of plausible reasoning to understand heavy-ion collisions was realized, and several works were published in the last ten years~\cite{Novak:2013bqa, Sangaline:2015isa, Bernhard:2015hxa, Ke:2016jrd, Xu:2017hgt, Xu:2018cvp, Moreland:2018gsh, Auvinen:2020mpc, Ke:2020clc, Cao:2021keo}, including analyses of both soft hadron production (the topic of this thesis) and hard and electromagnetic probes. 
This thesis is developed towards a similar pursuit -- the robust inference of the properties of quark-gluon plasma. 
However, there remain sources of theoretical model biases that we are aware of at this time and that, for practical reasons, we will not be able to quantify in this manuscript. Moreover, there are likely sources of theoretical bias in our models that we are unaware of at this time, which could nevertheless be important.\footnote{I don't necessarily mean my own ignorance, although it is certainly at play. Rather, I mean that there may be deficiencies and biases in state-of-the-art models that the entire heavy-ion community is currently unaware of. This is why comparing models with the observated data and criticizing their failures is an important component of progress.} Therefore, we reiterate that this thesis is in the \textit{pursuit} of robust modeling and understanding of quark-gluon plasma, and the author hopes that the ideas, methods, and approach presented herein 
have made contributions towards these goals and
will be useful to others with the same goals. The Bayesian formalism of probability provides a way to manipulate quantified uncertainties very straightforwardly, and will be used throughout this thesis. We provide a description of these statistical methods in Ch. \ref{ch2}.

\chapter{Probability and Bayesian Inference}
\label{ch2}

%
\section{Introduction}
\label{ch2:intro}
%

How should we \textit{quantify} our uncertainty about some proposition, given related but incomplete information? 
In the same vein, how should one make an optimal decision in light of uncertainty? These questions have nearly universal relevance because everyday we make hundreds or thousands of decisions in light of incomplete information, using `common sense', our prior knowledge, and the relevant information presented to us. Consider going to a new restaurant and ordering from the menu. You have prior information about the kinds of foods you like, and you have short descriptions containing information about the ingredients and preparation of each dish. Which dish you choose will likely be informed by both sources of information. If you leave the choice up to chance by closing your eyes and randomly pointing to the menu, the odds of selecting a dish you will enjoy are much worse. If you only consider the data provided by the descriptions on the menu while ignoring all your prior culinary experiences thus far, you are also in danger of making a less-than-optimal decision.\footnote{The menu may read `Our anchovie pizza is rated best in the world', but you have found anchovies consistently disagreeable in the past. There is a small chance you will be pleasantly surprised, but its more likely you wouldn't enjoy this anchovie pizza.} 

There are many other problems of at least as much importance as what you will order for lunch.  For example, consider investigating the chance that some medication will cause a fatal reaction in a given individual. This is not a question to be addressed lightly given the consequences and requires sound reasoning which can be elucidated, checked, and quantified. 
For such important problems, we require a systematic and rigorous theory that allows us to quantify our uncertainty. The formal system of mathematics that was developed to systematically tackle such questions is the theory of \textit{probability}. Laplace wrote ``la th\'eorie des probabilit\'es n'est, au fond, que le bon sens r\'eduit au calcul'' -- in English, that `the theory of probability is nothing but common sense reduced to calculations'. We will elucidate some of the formalism and mathematics of probability in the sections that follow, focusing especially on the Bayesian methodology. A more complete exposition of probability can be found in Refs.~\cite{Sivia2006, jaynes03}, which are recommended as very readable texts.

%
\section{The axioms of probability, Bayes' theorem, and \\ quantifying information}
\label{ch2:prob}
%

Consider a proposition $\theta$, a set of relevant data $D$, and an additional set of information or assumptions $I$. The probability that $\theta$ is realized, conditional on the information $D$ and $I$ will be denoted $\mathcal{P}(\theta|D, I)$. The laws of probability can be stated with a few axioms, which we briefly review. 
Given a complete set of $N$ possible and disjoint outcomes $\{\theta_i\}_{i=1}^{N}$, the probabilities of all outcomes must sum to unity,
\begin{equation}
    \sum_{i=1}^N  \mathcal{P}(\theta_i| I) = 1.
\end{equation}
The analogous constraint for a random variable $\theta$ which is continuously distributed with probability density function $\mathcal{P}(\theta|I)$ is given by 
\begin{equation}
    \int d\theta  \mathcal{P}(\theta| I) = 1.
\end{equation}

The \textit{joint} probability $\mathcal{P}(\theta_1, \theta_2| I)$ of two propositions $\theta_1$ and $\theta_2$, (the probability that both $\theta_1$ and $\theta_2$ are realized, conditional on $I$) cannot be simply related to the probabilities $\mathcal{P}(\theta_1| I)$ and $\mathcal{P}(\theta_2| I)$ in general.
However, when two propositions $\theta_1$ and $\theta_2$ are logically \textit{independent} given $I$,\footnote{%
If we are being careful, we say that $\theta_1$ and $\theta_2$ are \textit{conditionally independent} given $I$. In many texts this is denoted $\theta_1 \bigCI \theta_2 | I$. 
Conditional independence does \textit{not} imply absolute (unconditional) independence.
} then the probability of both being realized is given by the simple product
\begin{equation}
    \mathcal{P}(\theta_1, \theta_2| I) = \mathcal{P}(\theta_1| I) \mathcal{P}(\theta_2| I) \textrm{      (if $\theta_1$ and $\theta_2$ indep. given $I$ )}.
\end{equation}
Finally, the law for joint and conditional probabilities can be stated as
\begin{equation}
    \mathcal{P}(\theta,D| I) = \mathcal{P}(\theta| D, I)  \mathcal{P}(D|I) = \mathcal{P}(D| \theta, I)  \mathcal{P}(\theta| I),
\end{equation}
which is simply rearranged into \textit{Bayes' Theorem},
\begin{equation}
    \mathcal{P}(\theta| D, I) = \mathcal{P}(D| \theta, I)  \mathcal{P}(\theta| I) / \mathcal{P}(D| I).
\end{equation}
Bayes' theorem allows us to invert the order of conditioning on the known information. This will be the essential tool throughout this thesis for making inferences or predictions in light of known and incomplete information. 

We have been careful to this point to explicitly indicate that the probabilities we are manipulating are conditional on certain information/assumptions $I$. This practice is important as we introduce these ideas in reminding us that in \textit{every} problem of inference, we start with a non-empty set of assumptions and information. As we introduce other formulae throughout this manuscript, we may neglect to explicitly write that the probabilities are conditional on $I$; the reader should understand that it is always there implicitly. 

There is a long standing debate within the statistics community about the definition and interpretation of probability. For the purposes of this thesis, we will loosely follow a {\it subjective Bayesian} philosophy. 
This asserts that probability, the quantity defined mathematically to follow the rules given above, quantifies our \textit{degree of belief} about the proposition. On the other hand, the \textit{Frequentist} formalism defines probability by the fraction of times that the proposition would be realized in the limit of infinitely many identical realizations. For the most part, the philosophical differences between these two interpretations will be of no concern to us, nor to the results quantified in the body of this thesis. However, methodological differences between these two formalisms can sometimes yield quantitatively different answers, which we will note when we find it fruitful to do so. We caution that the term `probability' will be abused throughout this thesis, applying the word to mean either a probability mass function when possible outcomes belong to a discrete set or a probability density function when the outcomes are continuously distributed. The precise meaning should be clear from the context. 

It is also useful to define a quantity to measure the information (or lack thereof) contained in a probabilistic statement.
Typically, the amount of information $\mathcal{I}$ we would obtain by \textit{observing} an event $\theta_i$ which has probability $\mathcal{P}(\theta_i)$ is defined by
\begin{equation}
    \mathcal{I} \equiv - \log(\mathcal{P}(\theta_i)).
\end{equation}
Then, the Shannon entropy is defined as the expected value of information to be gained by making an observation, $\mathcal{S} \equiv \langle \mathcal{I} \rangle $. These definitions satisfy all of the common sense expectations that we would require~\cite{Jaynes:1957zza}. Therefore, for a probability mass function $\mathcal{P}(\theta_i)$ with disjoint outcomes $\theta_i$, the Shannon entropy $\mathcal{S}$ is given by
\begin{equation}
    \mathcal{S} \equiv -\sum_{i} \mathcal{P}(\theta_i) \log(\mathcal{P}(\theta_i)).
\end{equation}
This formula cannot be trivially extended to the continuous case without encountering some difficulty.\footnote{
A generalization without the introduction of a measure yields the `differential entropy', a quantity which actually lacks the properties we would desire to measure information.
} 
For a continuous random variable $\theta$ with probability density $\mathcal{P}(\theta)$, the commonly employed generalization of the Shannon entropy is called the `limiting density of discrete points'~\cite{Jaynes1968}, which we also denote by $S$:
\begin{equation}
    \mathcal{S} \equiv -\int d\theta \mathcal{P}(\theta) \log \Bigl( \frac{\mathcal{P}(\theta)}{m(\theta)} \Bigr),
\end{equation}
where $m(\theta)$ defines an invariant measure.\footnote{We don't want to get sidetracked by these technical issues, but hopefully it is clear that we require a quantity such as $m(x)$ inside the logarithm to cancel the dimensions of $[\mathcal{P}(\theta)] = [\theta]^{-1}$. }

It is also useful to define a real number which measures the amount of information contained in one probability distribution $\mathcal{P}_1(\theta | I)$ relative to another $\mathcal{P}_2(\theta| I)$. This quantity goes by many names, including the \textit{Kullback}-\textit{Leibler (KL) Divergence}, \textit{mutual information}, and \textit{information gain} and is defined by
\begin{equation}
    \mathcal{D}_{\rm KL} \equiv \sum_{i} \mathcal{P}_1(\theta_i| I) \log \left[ \frac{\mathcal{P}_1(\theta_i| I)}{\mathcal{P}_2(\theta_i| I)}  \right]
\end{equation}
for discrete distributions,
and 
\begin{equation}
    \mathcal{D}_{\rm KL} \equiv \int d\theta \mathcal{P}_1(\theta| I) \log \left[ \frac{\mathcal{P}_1(\theta| I)}{\mathcal{P}_2(\theta| I)}  \right]
\end{equation}
for continuous distributions. 
Using the base-$2$ logarithm in these expressions yields information in units of \textit{bits}, while if we use instead the natural logarithm the units are \textit{nats}. Both the entropy and Kullback-Leibler divergence will be useful in quantifying uncertainty and changes in uncertainty, respectively. 
%
\section{Bayesian parameter estimation}
\label{ch2:bayes_param_est}
%

Suppose we have a set of observed data, a parametrized model which we believe can describe these data, and we want to quantify what are the most likely values of the model's parameters. Consider addressing this problem by using Bayes' theorem, where $\bm{\theta}$ represents our model parameters, and $\bm{y}_{\rm exp}$ the set of observed data: 
\begin{equation}
\label{eq:bayes_thm}
    \mathcal{P}(\bm{\theta}| \bm{y}_{\rm exp}, I) = \mathcal{P}(\bm{y}_{\rm exp}| \bm{\theta}, I)  \mathcal{P}(\bm{\theta}| I) / \mathcal{P}(\bm{y}_{\rm exp}| I),
\end{equation}
where $\mathcal{P}(\bm{y}_{\rm exp}| \theta, I)$ is the \textit{likelihood} that we would observe these data $\bm{y}_{\rm exp}$ given that the parameters had values $\bm{\theta}$, $\mathcal{P}(\bm{\theta}| I)$ is our \textit{prior} belief about the values of the parameters $\bm{\theta}$ before we have observed the data $\bm{y}_{\rm exp}$, and $\mathcal{P}(\bm{\theta}| \bm{y}_{\rm exp}, I)$ is our \textit{posterior}, which encodes our belief about the values of the parameters given all sources of information. The Bayes evidence $\mathcal{P}(\bm{y}_{\rm exp}| I)$ is irrelevant to the problem of estimating model parameters because it is independent of the parameters $\bm{\theta}$. Therefore, for the purposes of performing parameter estimation with a given model, it is sufficient to consider the proportionality
\begin{equation}
    \mathcal{P}(\bm{\theta}| \bm{y}_{\rm exp}, I) \propto \mathcal{P}(\bm{y}_{\rm exp}| \bm{\theta}, I)  \mathcal{P}(\bm{\theta}| I). 
\end{equation}
Thus, we can estimate our posterior belief about the parameters once we have quantified both our prior belief about the parameters and the likelihood function.

\subsection{The likelihood function}
\label{ch2:likelihood_func}

The Likelihood function encodes the conditional probability of observing the data $\bm{y}_{\rm exp}$, given that the parameters $\bm{\theta}$ are realized. This function is a probability (density) function in the space of observables, \textit{not} in the space of parameters. In particular, it is not normalized to unity by integrating over the space of parameters in general
\begin{equation}
    \int d \bm{\theta} P(\bm{y}_{\rm exp}| \bm{\theta}, I) \neq 1. 
\end{equation}
Rather, the likelihood is normalized in the space of observables
\begin{equation}
    \int d \bm{y}_{\rm exp} P(\bm{y}_{\rm exp}| \bm{\theta}, I) = 1. 
\end{equation}
Roughly speaking, the likelihood function quantifies the degree to which we penalize a model when it predicts a misfit with the observed data; this penalty depends on our uncertainty in the observed data. 

We rarely know the exact distribution of errors in a given experimental situation. 
There are some simple cases in which we can derive a likelihood distribution which maximizes our ignorance (entropy) subject to the given information at hand~\cite{Jaynes:1957zza}. For example, consider a problem in which you know the mean value $\mu$ and variance $\sigma^2$ of a distribution $\mathcal{P}(y| I)$. In this case, the probability distribution which maximizes the Shannon entropy functional, subject to the stated constraints, is the normal distribution $\mathcal{N}(\mu; \sigma^2)$.\footnote{This is readily derived by setting to zero the functional derivative of the entropy functional $\delta S/\delta \mathcal{P} = 0$ with constraints imposed by Lagrange multipliers.} This generalizes to the multivariate normal distribution when we know the mean value vector $\bm{\mu}$ and covariance matrix $\bm{\Sigma}$ of random variables $\bm{y}$,
\begin{equation}
    \mathcal{N}(\bm{\mu}; \bm{\Sigma}) = \frac{1}{(2\pi |\bm{\Sigma}|)^{N/2}}\exp\left[-\frac{1}{2} \Delta \bm{y}^T \bm{\Sigma}^{-1} \Delta \bm{y}  \right],
\end{equation}
where we have defined the \textit{discrepancy} $\Delta \bm{y} \equiv \bm{y} - \bm{\mu}$.

The multivariate normal distribution will be used as a likelihood function throughout this thesis. In heavy-ion collisions, individual measurements are often reported by a mean and variance. Less often reported are the covariances among pairs of measurements at the same experiment, although they are nonzero in principle. Therefore this multivariate normal likelihood function is probably the most agnostic model for the true distributions of errors given the information we currently have. However, relaxing or changing the likelihood function will affect the results of Bayesian inference in a non-trivial manner; the reader should keep this awareness in the back of their mind as we proceed.\footnote{This is another assumption occupying the set of all our assumptions/information $I$.} There exist many other likelihood functions that have not yet been explored in Bayesian analyses of heavy-ion collisions.\footnote{%
Some notable examples of likelihood functions with heavier tails than the normal distribution are the Huber and Cauchy distributions.} 

In general, a model can take a set of input parameters $\bm{\theta}$ and yield probability distributions for each output. 
We will denote by $\bm{y}_m(\bm{\theta})$ the vector of the mean of the predicted observable distributions, given a fixed set of input parameters $\bm{\theta}$. The model may have sources of uncertainty in these predictions, and we will denote the covariance matrix of these uncertainties by $\bm{\Sigma}_m(\bm{\theta})$. As an example, the model may have an intrinsic randomness in its prediction of a some output given fixed parameters. Many computer models that include a Monte Carlo component have statistical uncertainty in their predictions when averaged over a finite number of samples. There can be other sources of predictive uncertainty as well. In particular, the physics models which will be employed in this thesis are computationally expensive, and statistical emulators provide a much faster surrogate at the expense of additional predictive uncertainty (more on this in Ch. \ref{ch4:emu}). 
We will denote the mean values of the experimentally observed data by $\bm{y}_e$ and their error covariance matrix by $\bm{\Sigma}_e$. 

By defining the discrepancy between the model and observed data $\Delta \bm{y}(\bm{\theta}) \equiv \bm{y}_m(\bm{\theta}) - \bm{y}_e$ to now be a function of parameters, and assuming that the model predictive uncertainty and experimental uncertainty are independent $\bm{\Sigma}(\theta) = \bm{\Sigma}_m(\theta) + \bm{\Sigma}_e$,  we can express the multinormal likelihood function of observing the data given our model and its input parameters by
\begin{equation}
    \mathcal{P}(\bm{y}_e| \bm{\theta}, I) = \frac{1}{(2\pi |\bm{\Sigma}(\theta)|)^{N/2}}\exp\left[-\frac{1}{2} \Delta \bm{y}^T(\bm{\theta}) \bm{\Sigma}^{-1}(\bm{\theta}) \Delta \bm{y}(\bm{\theta})  \right].
\end{equation}
%
The sampling of the likelihood will typically require numerical methods; we return to this in Ch. \ref{ch4:mcmc}. 

\subsection{The Bayesian prior}
\label{ch2:priors}

The prior encodes our belief about the probabilities, conditional on our assumptions, before\footnote{The word `before' here does not mean chronologically before the data are measured, but logically \textit{without} the information in those data. 
Similarly, the name `prior' does not indicate a position in time. The separation of what constitutes the `prior' and what constitutes the data in the `likelihood' is a \textit{logical} separation, and is useful in grouping different sources of information~\cite{jaynes03}.
} we have observed the  relevant data. 
Herein lies a significant source of confusion and mistrust in the Bayesian methodology. In this manuscript we will abide by Laplace's original definition, and say that the prior distribution should be a quantification of our `common sense' on the matter, \textit{before} we have utilized the data. Here `common sense' does not mean the expectations of a sensible person picked at random from a crowd, but rather the aggregate of domain-expert knowledge, concerns of model self-consistency, and hard and fast scientific constraints.

The first inclination in posing problems of Bayesian inference with complex models is to choose a joint prior which is a product of uniform or `flat' priors for each model parameter
\begin{equation}
    \mathcal{P}(\bm{\theta}| I) = \prod_{i} \mathcal{P}(\theta_i| I)
\end{equation}
where 
\begin{equation}
  \mathcal{P}(\theta_i| I) =
  \begin{cases}
    \frac{1}{(\theta_{i, \rm max} - \theta_{i, \rm min})} & \text{ if } \theta_{i, \rm min} < \theta_{i} < \theta_{i, \rm max} \\
    0 & \text{else}.  \\
  \end{cases}
\end{equation}
This prior has the appearance of being agnostic or `uninformed' regarding the possible values of the parameters $\theta_i$, but this appearance is deceiving. Our definition of a probability function requires that the probability mass be invariant under a transformation of variables. Given a probability density function $\mathcal{P}(\bm{\theta})$, and any variable transformation $\bm{\omega} = \bm{\omega}(\bm{\theta})$, we must have that 
\begin{equation}
    d\mathcal{P} = \mathcal{P}(\bm{\theta}) d\bm{\theta} = \mathcal{P}(\bm{\omega}) d\bm{\omega}.
\end{equation}
It follows that 
\begin{equation}
    \mathcal{P}(\bm{\omega})  = \mathcal{P}(\bm{\theta}) \Bigl| \frac{\partial \bm{\omega} }{\partial \bm{\theta} }\Bigr|, 
\end{equation}
where $\Bigl| \frac{\partial \bm{\omega} }{\partial \bm{\theta} }\Bigr|$ denotes the Jacobian of the variable transformation. In particular, for any non-linear transformation $\bm{\omega} = \bm{\omega}(\bm{\theta})$ the Jacobian will not be constant, but a function of the parameters. It often happens that a model's predictions depend non-linearly upon the parameters $\bm{\theta}$, in which case it becomes clear that the uniform prior $\mathcal{P}(\bm{\theta})$ may contain a lot of information in the relevant input space. Maximum entropy methods are often employed in the selection of priors, which may help to reduce the impact of potential biases. This however is not a universal outcome, and depends on the context of the problem.

\begin{figure*}[!htb]
\centering
\includegraphics[width=0.5\textwidth]{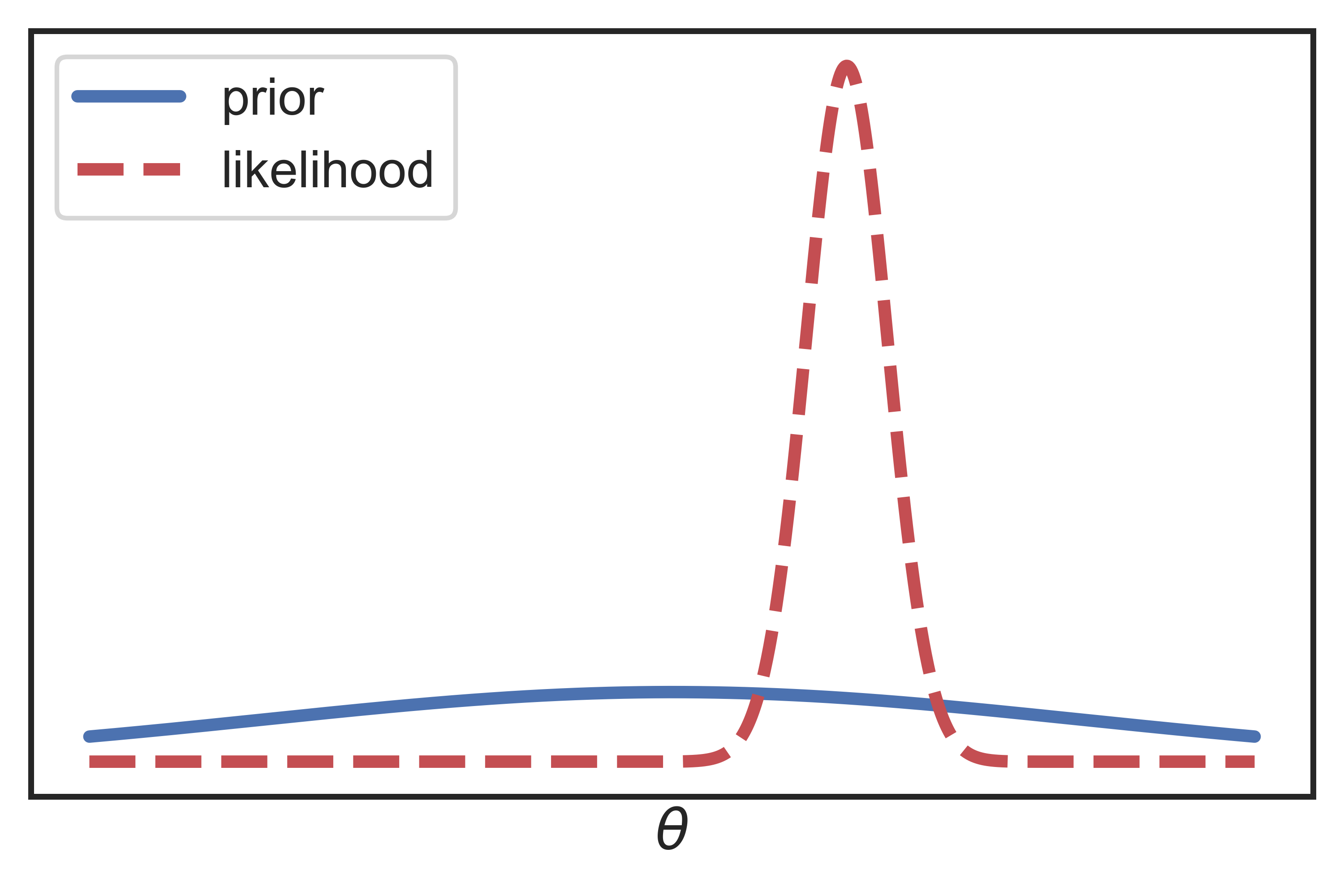}
\caption{A prior for a parameter $\theta$ which is locally uniform with respect to the likelihood function. }
\label{fig:locally_uniform_prior}
\end{figure*}

It is difficult at this time for practitioners of Bayesian inference in heavy-ion phenomenology to employ non-uniform priors due to the following practical issues. One problem is the required usage of statistical model surrogates; heavy-ion collision models require a lot of computing power. In principle, the formulation of the parameter prior is completely independent of the design of the model surrogate. However, it is dangerous in practice to trust \textit{extrapolations} made by model surrogates (see Ch. \ref{ch4:gp} for an example). Consequently, the parameter priors in such Bayesian analyses often have a finite domain of support. There are probably methods to side-step this problem, for instance by first performing a non-linear transformation which maps the real line of a model parameter to a compact interval, before training the surrogate to \textit{interpolate} between design points in the transformed space.\footnote{%
The function $\theta' = \tanh(\theta)$ comes to mind.} This may not always be a solution, however, because many of our models can break down in the limits of certain parameters becoming very large or very small. 
Following Ref.~\cite{box2011bayesian}, ideally the priors for parameters of most interest would ``let the data speak for themselves''. This motivates the desire for priors which are locally uniform with respect to the likelihood function, a situation depicted in Fig. \ref{fig:locally_uniform_prior}. Again, this situation is preferable but not always possible due to the theoretical limitations of our models. We will return to these concerns by example when we introduce the priors selected for our hydrodynamic models in Ch. \ref{ch5:priors}. Briefly, our models can break down in rather spectacular ways in the limit that certain parameters become very small or very large, and in practice we simply cannot employ those models in such regions of parameter space.

To avoid potential risk of sounding cavalier to this point, we emphasize that one should take great care in eliciting the prior distribution. Frequently throughout this manuscript we will distinguish which of our conclusions are driven by the likelihood (when the `data speak'), and which were already fixed by an informed (narrow) prior. In addition, we will explore methods to address and quantify the extent to which our conclusions are sensitive to our priors. It is difficult to provide a simple recipe for formulating the `best' prior for all circumstances. Rather, the best prior will depend entirely on the context of the problem, especially the model.
As we explore Bayesian inference in the context of particular problems, we will elucidate the priors that were chosen for that particular problem. The interested reader should see Refs.~\cite{box2011bayesian, Jaynes1968} for further relevant discussion on Bayesian priors, especially as they relate to scientific investigation.

\subsection{Maximizing the posterior}
Besides exploring the posterior distribution of a Bayesian parameter estimation, often the maximum of the posterior or `maximum a posteriori (MAP)' is quoted as well. This defines the point in the $n$-dimensional parameter posterior which has the largest posterior probability density. In the specific case when the priors are uniform functions, this point is also the maximum of the likelihood function, usually abbreviated `Max. Likelihood'. Whether we should assign any special meaning to this point in parameter space depends on the problem. For instance, in the case that the posterior is not sharply peaked and is diffuse, the MAP may not have any significance at all. In the opposite case, when the posterior is tightly peaked around its maximum value, then the model evaluated at its MAP provides the `best' predictions of the model. In any situation, the MAP alone does not provide a measure of \textit{uncertainty}, and we must be careful not to over-interpret the predictions of a model when they are unaccompanied by any measure of uncertainty.

%
\section{A simple example of Bayesian parameter estimation}
\label{ch2:ex_bayes_param_est}
%

\begin{figure*}[!htb]
\noindent\makebox[\textwidth]{%
  \centering
  \begin{minipage}{0.5\textwidth}
    \includegraphics[width=\textwidth]{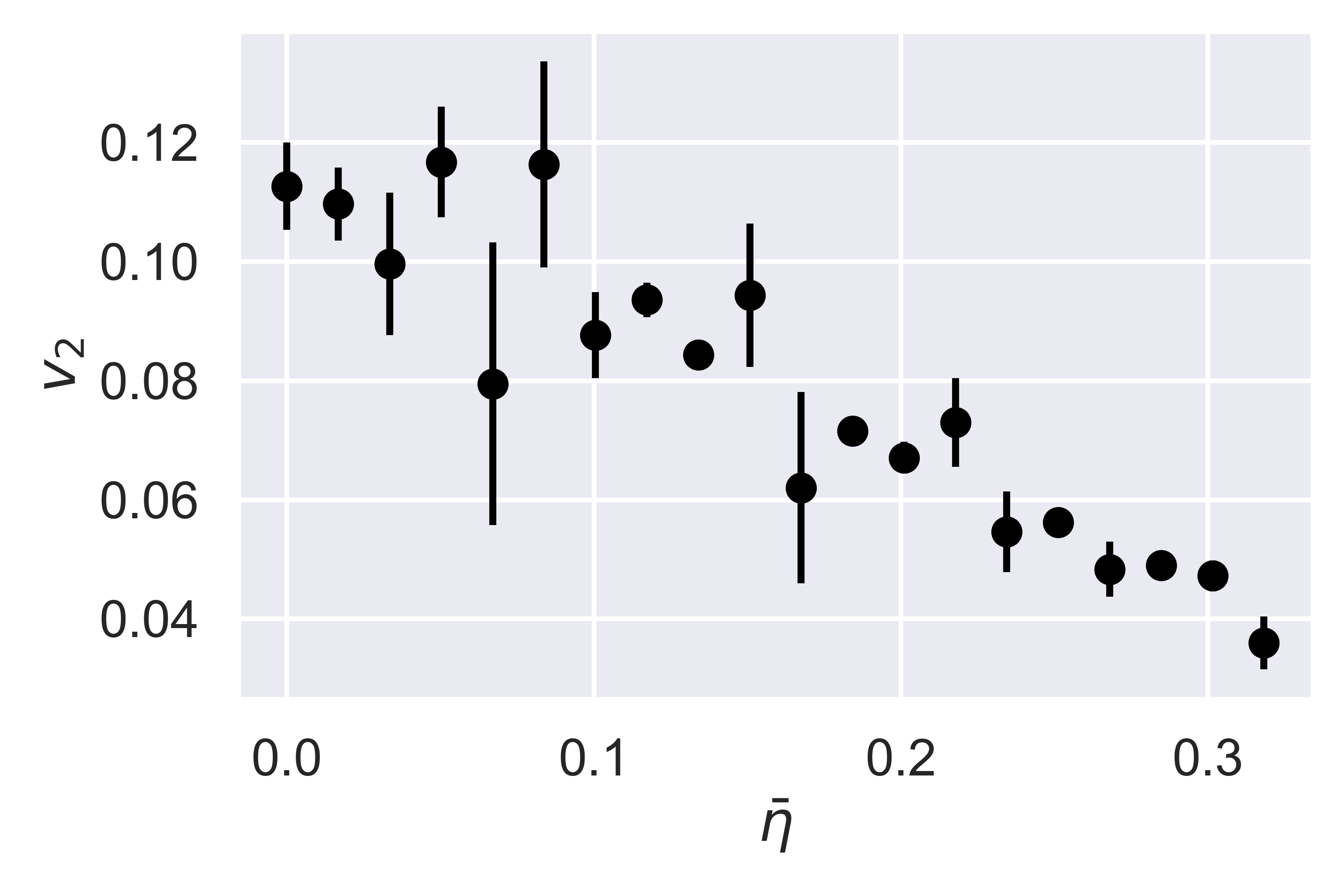}
  \end{minipage}
  \begin{minipage}{0.5\textwidth}
    \includegraphics[width=\textwidth]{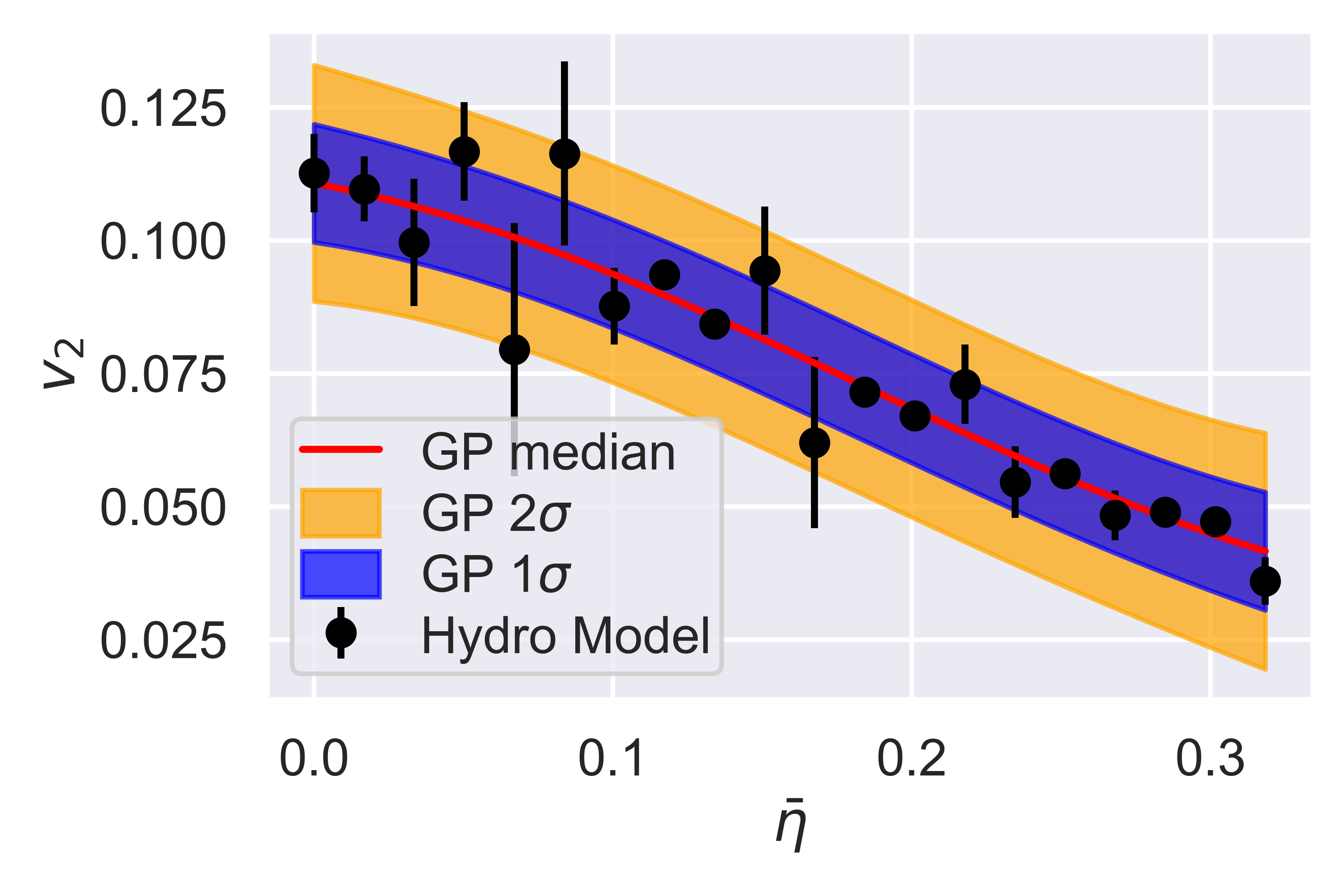}
  \end{minipage}
  }
  \caption{(Left) The `hydro' model prediction $v_2$ evaluated at twenty different values of the model parameter $\bar{\eta}$. (Right) The `surrogate' or interpolator for the model, given by Gaussian process regression, shown against its training points.}
  \label{fig:design_and_gp}
\end{figure*}

This section is provided as an illustration of some concepts in Bayesian inference which have been discussed to this point. The context is motivated by the heavy-ion phenomenology discussed in Ch. \ref{ch1:comparing}. Specifically, we consider a very simplified one parameter `hydrodynamic' model. According to our model, the elliptic flow $v_2$ for a particular peripheral centrality bin is inversely related to the specific shear-viscosity  $\bar{\eta} \equiv \eta/s$, 
\be
    v_2(\bar{\eta}) = c_0 + c_1 \bar{\eta} + \epsilon
\ee
with a fixed intercept $c_0$ and negative slope $c_1$. Added to the linear model is a random noise $\epsilon$, intended to mimic the type of stochastic uncertainties that manifest in more realistic analyses with fluctuating events. The predictions of the model are shown in Fig. \ref{fig:design_and_gp}.

These plots can be generated in a notebook intended for pedagogy~\cite{bayes_notebook}. The notebook is also intended to introduce tools for implementing Bayesian analyses with computationally intensive models, including Gaussian Process regression and numerical sampling via Markov Chain Monte Carlo. Rather than complicate the discussion in this section with these practical concerns, we will return to them in Ch. \ref{ch4}. For the purposes of this section, consider the Gaussian process regressor, whose predictions are shown in the right of Fig. \ref{fig:design_and_gp}, to be a `surrogate' for our physics model that we trust. Moreover, this surrogate provides a probability distribution for the predicted outputs given a \textit{fixed} set of inputs.

\begin{figure*}[!htb]
\noindent\makebox[\textwidth]{%
  \centering
  \begin{minipage}{0.5\textwidth}
    \includegraphics[width=\textwidth]{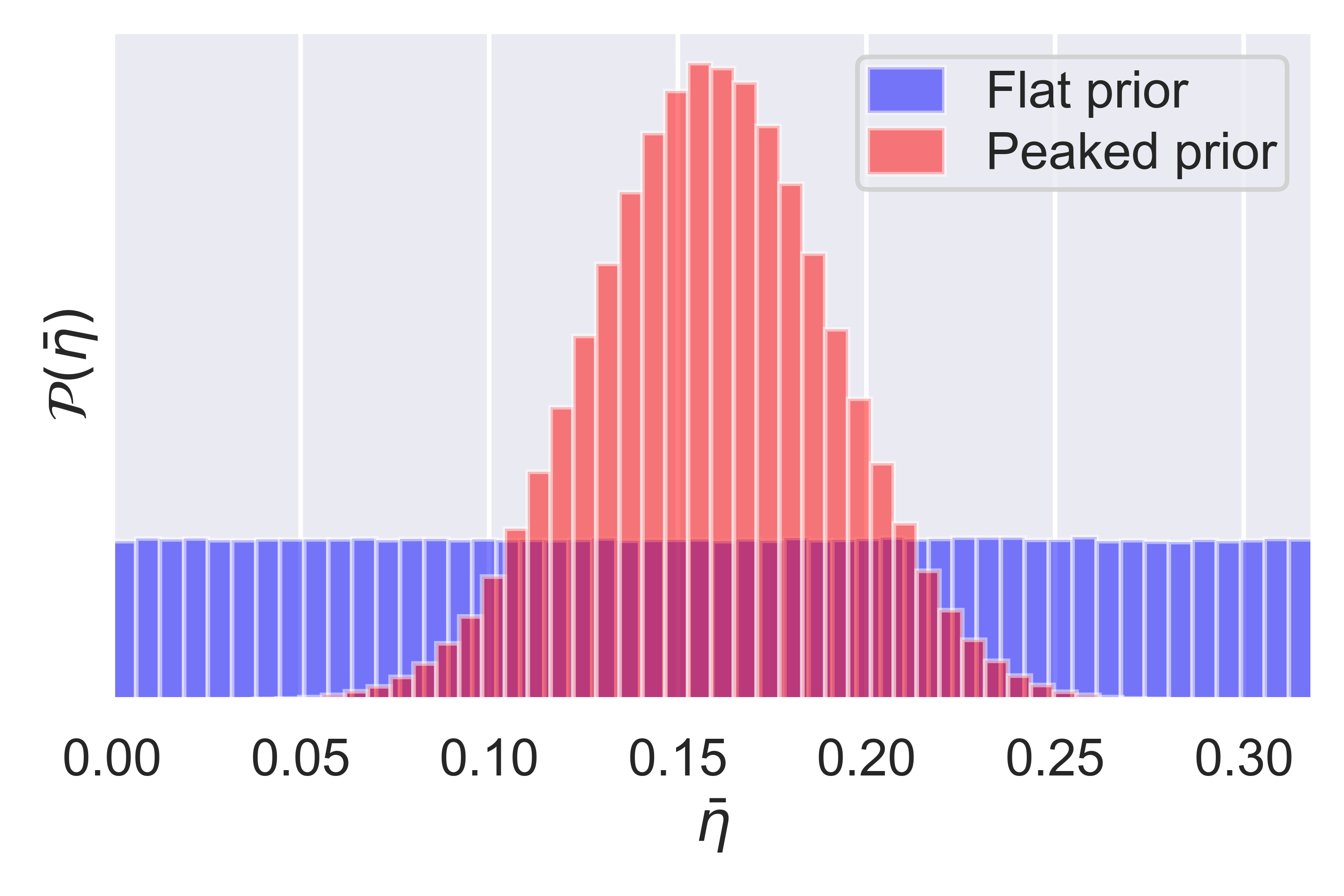}
  \end{minipage}
  \begin{minipage}{0.5\textwidth}
    \includegraphics[width=\textwidth]{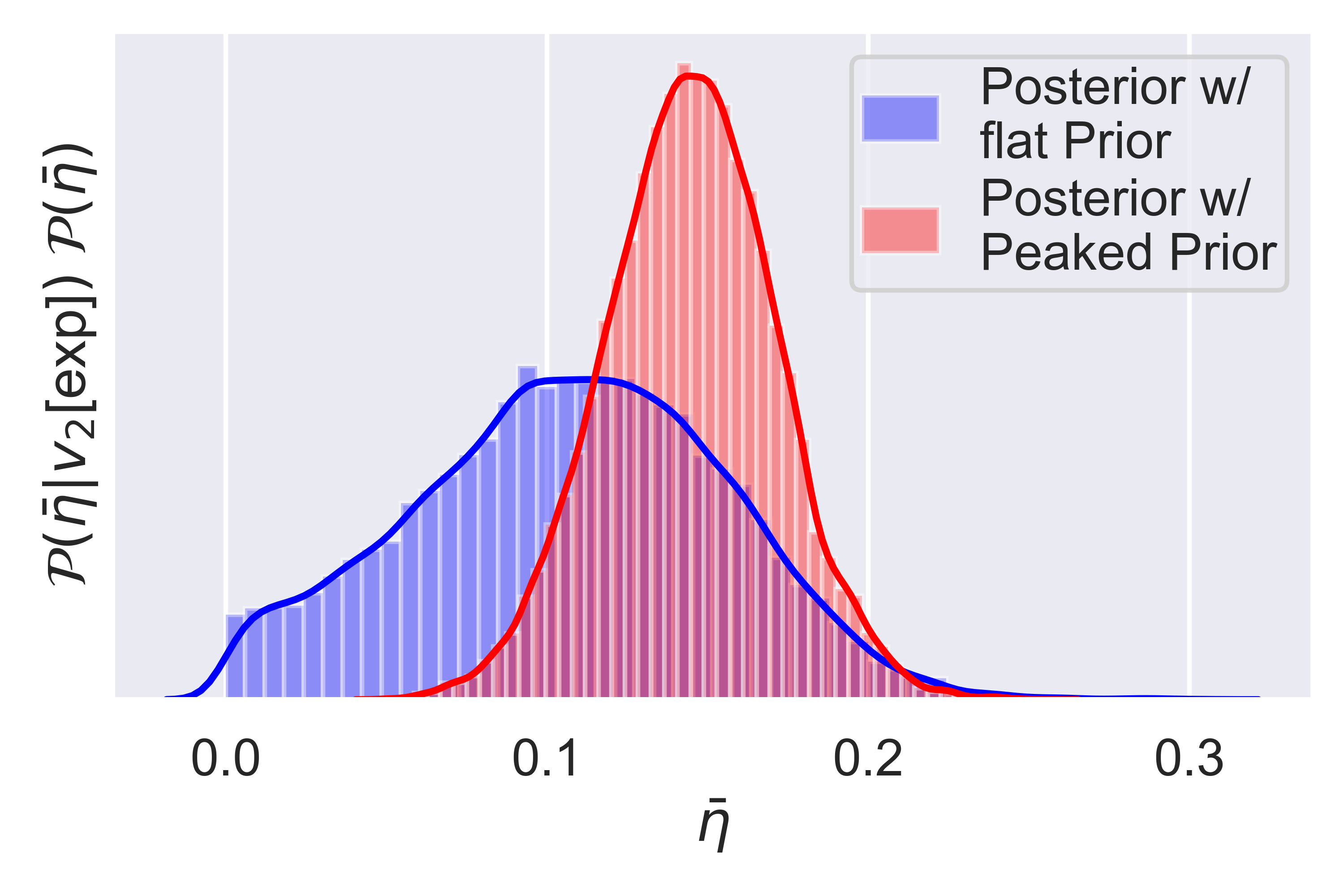}
  \end{minipage}
  }
  \caption{The priors (left) for the specific shear-viscosity $\bar{\eta}$ and the posteriors (right) after comparing to the pseudo-data $v_2$. The more informed prior (red) leads to a sharper posterior (red). }
  \label{fig:priors_and_posteriors}
\end{figure*}

Now, suppose that we want to compare our linear hydrodynamic model against some measured value of the elliptic flow. This problem is naturally suited to Bayesian parameter estimation introduced in the previous section. We will begin by eliciting priors; as an illustration of the importance of prior specification, we will do the analysis twice with two different priors. The first prior, which we will call the flat prior, will be a uniform distribution of probability for the specific shear viscosity between an upper and lower limit,
\begin{equation}
  \mathcal{P}_1(\bar{\eta}) =
  \begin{cases}
    \frac{1}{(\bar{\eta}_{\rm max} - \bar{\eta}_{\rm min})} & \text{ if } \bar{\eta}_{\rm min} < \bar{\eta} < \bar{\eta}_{\rm max} \\
    0 & \text{else}.  \\
  \end{cases}
\end{equation}

The second prior, which we call the peaked prior, will be more strongly informed by a belief that the shear viscosity should be rather close to a specific value $\bar{\eta}_0$. This will be implemented by fixing the prior to be a Gaussian distribution with mean $\bar{\eta}_0$ and a small but finite uncertainty $\sigma$:
\be
    \mathcal{P}_2(\bar{\eta}) = \frac{1}{\sqrt{2\pi}\sigma} \exp \left[ -\frac{(\bar{\eta} - \bar{\eta}_0)^2}{2\sigma^2} \right]
\ee
Both of these priors are displayed in the left panel of Fig. \ref{fig:priors_and_posteriors}. 

Next, we require a model for the likelihood of observing some value of elliptic flow $v_2$, given a value of specific shear-viscosity $\bar{\eta}$. Assuming that these pseudo-data are reported as a most-likely value together with a symmetric standard deviation, i.e. $v_{2, 0} \pm \delta v_{2, \rm 0}$, we will model the likelihood with the normal distribution:
\be
    \mathcal{P}(v_2|  \bar{\eta}, I) = \frac{1}{\sqrt{2\pi}\delta v_{2, \rm 0}} \exp \left[-\frac{(v_{2} - v_{2, \rm 0})^2}{2(\delta v_{2, \rm 0})^2} \right].
\ee

Our posteriors follow from Bayes' theorem, 
\be
    \mathcal{P}_i(\bar{\eta} |  v_{2, \rm 0}) \propto \mathcal{P}_i(\bar{\eta})\mathcal{P}(v_2|  \bar{\eta}, I),
\ee
where $i=1, 2$ labels the two different priors and posteriors. 
The two posteriors are shown in the right panel of Fig. \ref{fig:priors_and_posteriors}, and they are different (hopefully this doesn't come as a surprise). The posterior corresponding to the flat prior (shown in blue) is proportional to the likelihood function. This is a consequence of the flat prior being independent of $\bar{\eta}$ in the allowed region
\be
    \mathcal{P}_1(\bar{\eta}| v_{2, \rm 0}) \propto \mathcal{P}_1 (\bar{\eta})\mathcal{P}(v_{2, \rm 0}|  \bar{\eta}, I) \propto \mathcal{P}(v_{2, \rm 0}|  \bar{\eta}, I).
\ee
The posterior that included a more informed prior for the specific shear-viscosity $\bar{\eta}$ is shown in red. It is clearly visible that the posterior with the more informed prior is much tighter that the posterior with the flat prior. But this was not an effect of the data we used; the smaller uncertainty was already present prior to utilizing these data. 
As a result, the blue posterior reflects more strongly the new information contained in the experimental data. On the other hand, the red posterior is so strongly influenced by prior information or prejudice that new data barely have sway.

\begin{figure*}[!htb]
\noindent\makebox[\textwidth]{%
  \centering
  \begin{minipage}{0.5\textwidth}
    \includegraphics[width=\textwidth]{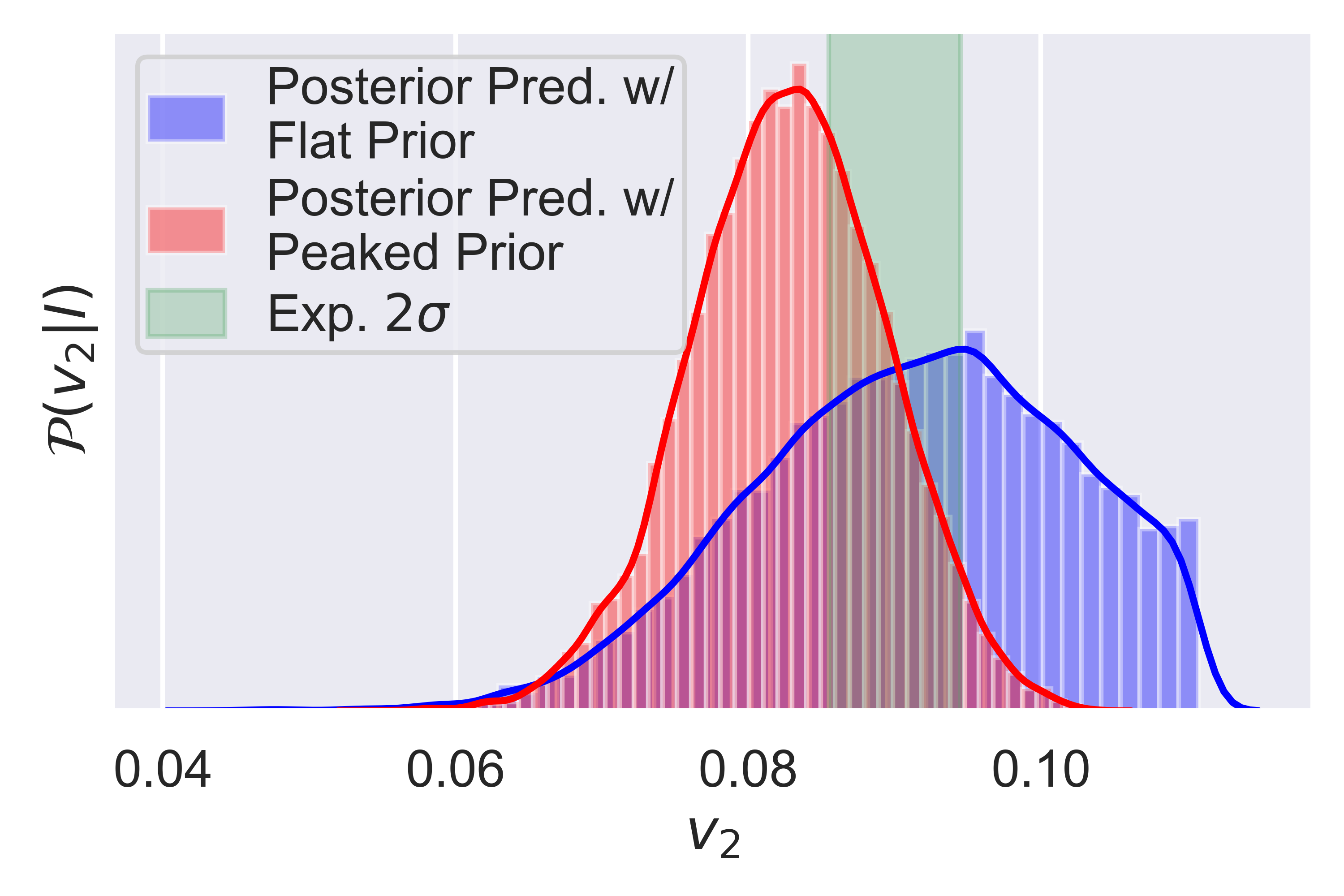}
  \end{minipage}
  \begin{minipage}{0.5\textwidth}
    \includegraphics[width=\textwidth]{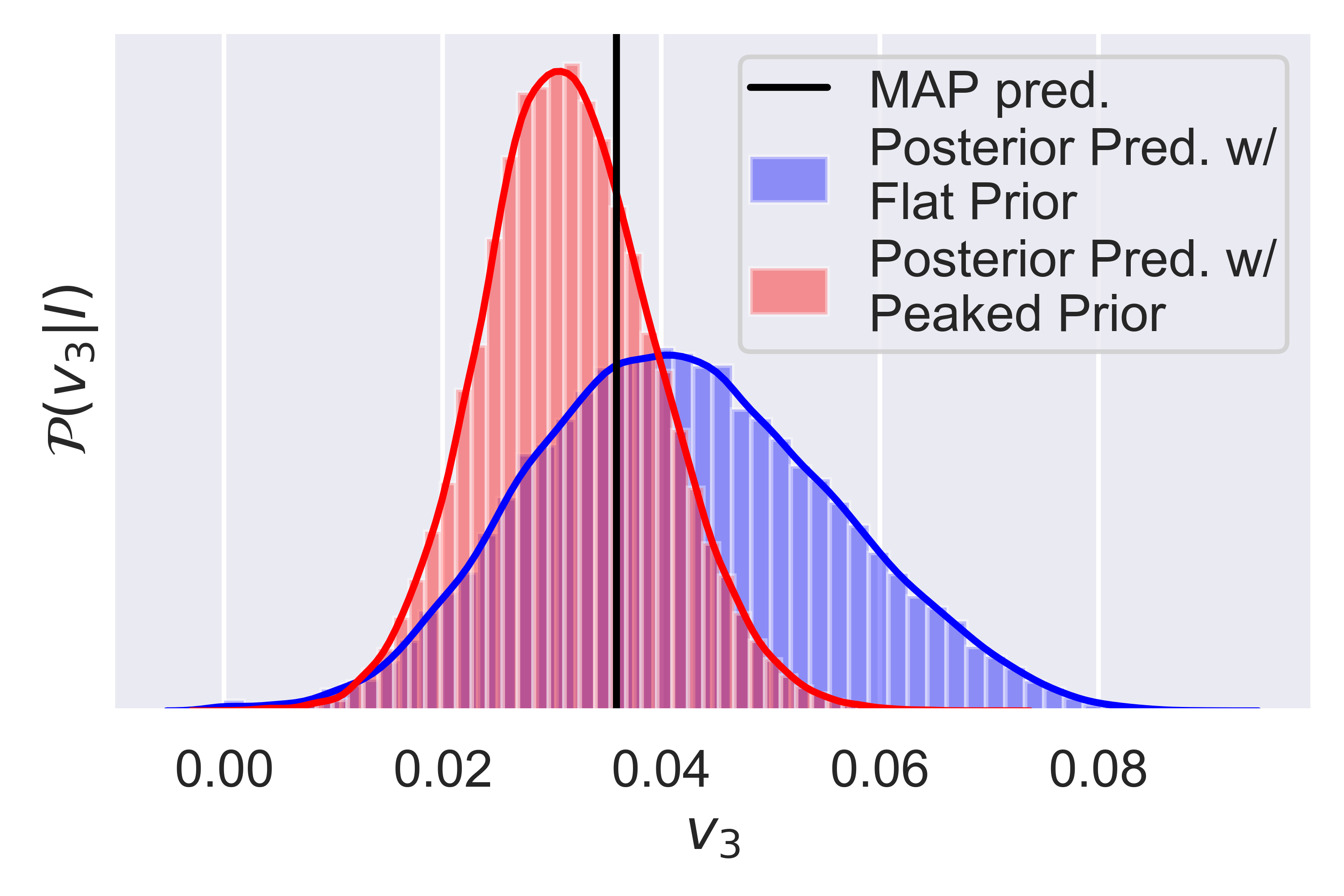}
  \end{minipage}
  }
  \caption{Posterior predictive distributions of $v_2$ (left) and $v_3$ (right) given by the model, which has been calibrated against only the $v_2$ pseudo-data. }
  \label{fig:posterior_pred_dists}
\end{figure*}

An invaluable tool in model-building is checking whether models make reasonable predictions for observables to which they were not calibrated. We discuss this again by example, by introducing a new `observable' $v_3$ which our hypothetical hydrodynamic model predicts:
\be
    v_3(\bar{\eta}) = c'_0 + c'_1 \bar{\eta} + c'_2 \bar{\eta}^2 + \epsilon'.
\ee
A common practice is a call for theoretical model predictions in advance of the publishing of a new experimental measurement. We suppose that is the case here, and our experimental collaborator asks us to predict the observable $v_3$ using our model. Consider the possible insights that can be drawn depending on whether we provide the prediction given the model evaluated at the MAP parameters (without any quantified uncertainty), or if we provide the posterior predictive distribution. These are both shown on the right panel of Fig.  \ref{fig:posterior_pred_dists}. 

The blue distribution is our posterior predictive distribution for $v_3$, which marginalizes over uncertainty regarding the value of our model parameter $\bar{\eta}$. On the other hand, our MAP prediction is shown by the vertical black bar, and does not include any uncertainty. Suppose that the measurement is found to be $v_3 = v_{3, 0} \pm \delta v_{3, 0}$. How should we judge whether our model's prediction is actually consistent with this measurement? This question is well-defined when we include our theoretical uncertainties, as they are in the posterior predictive distribution. This question is not well-defined if we only have the MAP prediction. In the case that there is significant tension between our posterior predictive distribution and the measured $v_3$ distribution, we learn that our model is not capable of \textit{simultaneously} describing the observables $v_2$ and $v_3$. This phenomena is usually called `tension', and it may teach us a great deal. It may be that our priors were actually too constraining, and/or that our model lacks some important component/physics/process to simultaneously describe both observables. Tension may even inform us that there is an unquantified error in one or both of the measurements, but this is a possibility that theoretical modelers would usually care not to entertain unless supported by additional evidence.

Notwithstanding this criticism of making model predictions using point-estimates (not probability distributions), it is a standard practice in the field. Indeed, many models of heavy-ion collision dynamics can be so computationally expensive that marginalizing over many points in parameter-space could require a huge computing allocation. Nevertheless, we have to be careful not to over-interpret point-estimate predictions when comparing models with data, and if its feasible predictive distributions should always be preferred.

%
\section{Model sensitivity analysis}
\label{ch2:model_sensitivity_analysis}
%

The ideas discussed in this section are not necessarily Bayesian and are also employed often in Frequentist analyses. However, it is useful to introduce the concept of `model sensitivity' at this point, as it will aid in our understanding and interpretation of Bayesian parameter estimation and model selection. Broadly speaking, a model sensitivity analysis explores the map between uncertainty/change in the model parameter space and uncertainty/change in the model output space. As usual, we proceed by example. Consider again two `hydrodynamic' models $\mathcal{M}_A$ and $\mathcal{M}_B$ which can both predict a single output $v_2$ given a single parameter, the `specific shear-viscosity' $\bar{\eta}$. To simplify our exposition and magnify the insights, we again take both models to be linear models of $\bar{\eta}$: model $\mathcal{M}_A$ predicts $v_{2, A} = c_A \bar{\eta}$, while model $\mathcal{M}_B$ predicts $v_{2, B} = c_B \bar{\eta}$, with both $c_A < 0$ and $c_B < 0$. In this example, a simple \textit{sensitivity index} can be defined by the gradient (slope) of the output with respect to the input parameter. In this case, model $\mathcal{M}_A$ has a sensitivity index of $c_A$, and model $\mathcal{M}_B$ has sensitivity index $c_B$.\footnote{%
Note this sensitivity index is a property pertaining to a pair of a specific output and specific input.}

\begin{figure*}[!htb] 
\centering
\includegraphics[width=0.6\textwidth]{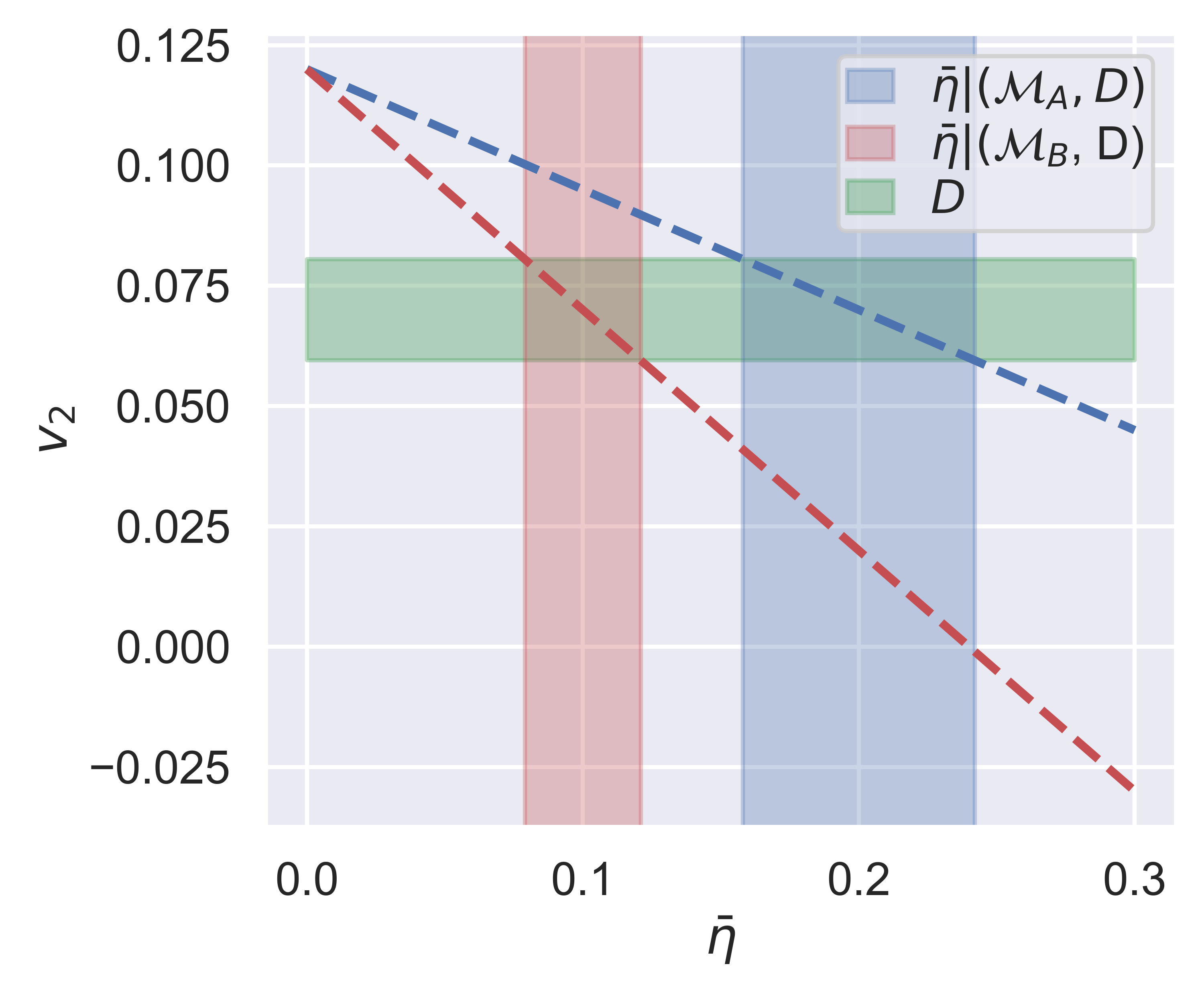}
    \caption{A visualization of the interplay between model sensitivity and posterior inference, as depicted with two linear models $\mathcal{M}_A$ and $\mathcal{M}_B$, and observed data $D$. The more sensitive model is consistent with the observed data in a smaller region of parameter space.}
\label{fig:sensitivity_schematic}
\end{figure*}

Suppose that we also compare our models against an experimentally measured value $v_{2, 0} $ which has an uncertainty $\delta v_{2, 0}$. This situation is depicted in Fig. \ref{fig:sensitivity_schematic}, and we see that model $\mathcal{M}_B$ is much more sensitive to the model parameter $\bar{\eta}$ than model $\mathcal{M}_A$.  
In this figure, the green band denotes the possible outputs $v_2$ which are consistent with the observed data. The two models have been chosen such that $\mathcal{M}_B$ is twice as sensitive as $\mathcal{M}_A$, $|c_B| = 2 |c_A|$. For each model separately, the regions in parameter space which are consistent with the observed data are shown as vertical (blue or red) bands. We see graphically that model $\mathcal{M}_A$ is consistent with the observed data within a range of values $\delta\bar{\eta}_A$ which is twice as large as the possible range for model $\mathcal{M}_B$, $\delta\bar{\eta}_A = 2 \delta\bar{\eta}_B$. 

This insight was motivated by inspection, but it is fully consistent with the result of Bayesian parameter estimation. Suppose that both models are assigned the same uniform prior on the possible values of $\bar{\eta}$, $\mathcal{P}(\bar{\eta} |  \mathcal{M}_A, I) = \mathcal{P}(\bar{\eta} |  \mathcal{M}_B, I)$. 
Suppose we assume the normal likelihood function for both models. 
The result is made more transparent if we first redefine our observables by subtracting the measured value: $v_2' = v_2 - v_{2, 0}$. It follows that $v'_{2, 0} = 0$ and $\delta v'_{2,0} = \delta v_{2,0}$. 
In this case, the likelihood function for model $\mathcal{M}_A$ is given by
\be
    \mathcal{P}(v'_2|  \bar{\eta}, \mathcal{M}_A, I) = \frac{1}{\sqrt{2\pi}\delta v'_{2, \rm 0}} \exp \left[-\frac{(v'_{2})^2}{2(\delta v'_{2, \rm 0})^2} \right] = \frac{1}{\sqrt{2\pi}\delta v'_{2, \rm 0}} \exp \left[-\frac{(\bar{\eta})^2}{2(\delta v_{2, \rm 0} / c_A)^2}\right].
\ee
Similarly, the likelihood function for model $\mathcal{M}_B$ is given
\be
    \mathcal{P}(v'_2|  \bar{\eta}, \mathcal{M}_B, I) = \frac{1}{\sqrt{2\pi}\delta v'_{2, \rm 0}} \exp \left[-\frac{(\bar{\eta})^2}{2(\delta v_{2, \rm 0} / c_B)^2}\right],
\ee
and, because $c_B = 2 c_A$, the likelihood function for model $\mathcal{M}_B$ has half the width of the likelihood of model $\mathcal{M}_A$. 
These conclusions may seem somewhat elementary given the simple examples and linear models presented here, but this discussion will prepare us to confront more complicated models in Chapters \ref{ch5} and \ref{ch6}. 

%
\section{Bayesian model selection}
\label{ch2:ex_bayes_model_selection}
%

Another ubiquitous question can be posed as follows: ``Which model, among several competing models, is more likely to be consistent with the observed data $\bm{y}_{\rm exp}$ and assumptions $I$?'' That is, suppose we have $n$ models $\mathcal{M}_i$, $\{i=1, \cdots, n\}$, and we want to compare the performance of our models and \textit{quantify} our belief in each of the given models. We denote the probability that model $\mathcal{M}_i$ is correct given the information as $\mathcal{P}(\mathcal{M}_i| \bm{y}_{\rm exp}, I)$, and use Bayes' theorem:
\begin{equation}
    \mathcal{P}(\mathcal{M}_i| \bm{y}_{\rm exp}, I) = \mathcal{P}(\bm{y}_{\rm exp}| \mathcal{M}_i, I) \mathcal{P}(\mathcal{M}_i|  I) / \mathcal{P}(\bm{y}_{\rm exp}|  I).
\end{equation}
We notice that the denominator $\mathcal{P}(\bm{y}_{\rm exp}|  I)$ is the probability that we would observe the data $\bm{y}_{\rm exp}$, independent of the model(s) -- a quantity which we usually have no means to compute in practice. But if we define the \textit{odds} between two competing models by 
\begin{equation}
\label{eq:bayes_factor}
    B_{ij} \equiv \frac{\mathcal{P}(\mathcal{M}_i| \bm{y}_{\rm exp}, I)}{\mathcal{P}(\mathcal{M}_j| \bm{y}_{\rm exp}, I)} = \frac{\mathcal{P}(\bm{y}_{\rm exp}| \mathcal{M}_i, I)}{\mathcal{P}(\bm{y}_{\rm exp}| \mathcal{M}_j, I)} \frac{\mathcal{P}(\mathcal{M}_i|  I)}{\mathcal{P}(\mathcal{M}_j|  I)},
\end{equation}
the common factor $\mathcal{P}(\bm{y}_{\rm exp}|  I)$ cancels! 
The odds $B_{ij}$ is often called the \textit{Bayes factor} between the two models $\mathcal{M}_i$ and $\mathcal{M}_j$. When $B_{ij} >> 1$, we prefer model $\mathcal{M}_i$ over model $\mathcal{M}_j$, and vice versa. The prior odds $\mathcal{P}(\mathcal{M}_i|  I)/\mathcal{P}(\mathcal{M}_j|  I)$ are simply the odds we assign between the two models based on our `common sense'. If one model is much less likely based solidly on prior considerations that aren't informed by the data $\bm{y}_{\rm exp}$, then it may be best to choose this ratio to be different from unity. However, in the applications of model selection in this thesis, we will always assume $\mathcal{P}(\mathcal{M}_i|  I)/\mathcal{P}(\mathcal{M}_j|  I) = 1$.

Now, we consider the computation of $\mathcal{P}(\mathcal{M}_i| \bm{y}_{\rm exp}, I)$, which we recognize as the Bayesian evidence appearing in Eqn.~(\ref{eq:bayes_thm}). If model $\mathcal{M}_i$ has a set of parameters $\bm{\theta}_i$, then according to our rules for marginalization and conditional probabilities, 
\begin{equation}
    \mathcal{P}(\bm{y}_{\rm exp}| \mathcal{M}_i, I) = \int d\bm{\theta}_i \mathcal{P}(\bm{y}_{\rm exp}, \bm{\theta}_i| \mathcal{M}_i, I) = 
    \int d\bm{\theta}_i \mathcal{P}(\bm{y}_{\rm exp}| \mathcal{M}_i, \bm{\theta}_i, I) \mathcal{P}(\bm{\theta}_i| \mathcal{M}_i, I).
\end{equation}
In this equation we recognize the very same prior $\mathcal{P}(\bm{\theta}_i| \mathcal{M}_i, I)$ and likelihood $\mathcal{P}(\bm{y}_{\rm exp}| \mathcal{M}_i, \bm{\theta}_i, I)$ which were necessary to perform parameter estimation in Ch.~\ref{ch2:bayes_param_est}, except our notation is now more explicit.\footnote{%
When we wrote Eqn.~(\ref{eq:bayes_thm}), it was implied that our parameter estimates were \textit{conditional} on the model we were using. Therefore we should interpret $I$ in that equation to include the assumptions of our model.} 
The Bayesian evidence is just the average of the model likelihood over the parameter space, with respect to the probability density given by the parameter prior (remember, the likelihood is not normalized to unity over the parameter-space).

In the context of model selection, a `model' refers to a specific set of parameters together with their prior, and a unique map from the parameters to a set of observables. As an example, a polynomial of second degree is a different model than a third-degree polynomial. In this case, however, the second degree polynomial model is `nested' inside the polynomial of third degree; when the coefficient of the cubic term is fixed to zero, we recover the second-order polynomial model. The polynomial of third degree has additional model complexity given by the additional parameter and its prior. On the other hand, we can also compare a model given by a second degree polynomial with a model given by a sinusoid having an uncertain amplitude and phase velocity. These models give altogether different predictions, and their coefficients have different meanings. Whether the models share similar features or not, we will refer to them as different models when comparing them with the Bayes factor (which is the standard terminology). If one of the models happens to be nested inside the other, we will make a note.

We note that when using uniform priors, this expression can be simplified, yielding
\begin{equation}
\label{eqn:bayes_evidence_vol}
    \mathcal{P}(\bm{y}_{\rm exp}| \mathcal{M}_i, I) = \frac{1}{\mathcal{V}_i} \int_{\mathcal{D}_i} d\bm{\theta}_i \mathcal{P}( \bm{y}_{\exp} |  \bm{\theta}_i, \mathcal{M}_i, I ),
\end{equation}
where we have defined the total volume of the prior $\mathcal{V}_i$ for model $\mathcal{M}_i$. This is the volume of the hypercube $\mathcal{D}_i$ inside which the prior for model $\mathcal{M}_i$ is nonzero. 
All of the models which we will compare via Bayes factors in this thesis have uniform priors. The Bayesian evidence of each model is the integral over the model likelihood inside the prior bounds, then divided by the volume of the prior. Belief in a model is increased by ability to fit the data (larger likelihood), averaged\footnote{%
Note the difference: The Frequentist likelihood ratio test is the ratio of the model likelihoods at their peak values in parameter space, while the Bayes factor \textit{marginalizes} over the uncertainty in parameter space. 
} inside of the prior bounds. But belief in the model is decreased by its complexity, called the `Occam penalty'. The complexity penalty is proportional to the subvolume of the model's prior excluded by the data (excluded by the likelihood function). In a situation where the likelihood $\mathcal{P}( \mathbf{y}_{\exp} |  \bm{\theta}_i, \mathcal{M}_i, I )$ does not actually depend on a particular model parameter $\theta_{i, m}$, we see from \eq{eqn:bayes_evidence_vol} that there is no Occam penalty: The volume of the prior cancels in the numerator and denominator. Therefore, the Bayes factor does not penalize a model for having parameters unconstrained by the data. 

The Bayes factor thus rewards models which are more \textit{predictive} (require less fine-tuning). We return to the example posed in Ch. \ref{ch2:model_sensitivity_analysis} as an illustration. Both models $\mathcal{M}_A$ and $\mathcal{M}_B$ were found to be equally capable of describing the measured data, however model $\mathcal{M}_B$ could only do so in a narrower region of parameter space: The likelihood function for model $\mathcal{M}_B$ was half as wide as that of model $\mathcal{M}_A$. Consequently, having chosen the same uniform prior for both models, model $\mathcal{M}_B$ incurs a larger Occam penalty and $B_{AB} > 1$.\footnote{If we instead compare models $\mathcal{M}_{A}$ and $\mathcal{M}_{B}$ via the Frequentist likelihood ratio test, which is the ratio of the model likelihood values at their peaks, we find odds of one and neither model is preferred. Similarly, a point estimate comparison given by the Akaike Information Criterion (AIC), a quantity designed to include a penalty for overfitting, would also be inconclusive: $\text{AIC} \equiv 2k - 2\ln \mathcal{L}_{\rm max}$, with $k$ the number of model parameters and $\mathcal{L}_{\rm max}$ the value of the likelihood at its maximum.} This is equivalent to the statement that model $\mathcal{M}_B$ is less predictive: Correctly predicting the output $v_2$ with model $\mathcal{M}_B$ requires more precise knowledge about the parameter $\bar{\eta}$ (there is less room for error). Understanding the influence of model sensitivity is important for interpreting posterior inferences, including the Bayes factor.

There has been doubt cast on the usefulness of the Bayes factor in discriminating between models due to its sensitivity to the specification of the prior. If the prior $\mathcal{P}( \bm{\theta}_i|  \mathcal{M}_i, I )$ for any of the models is changed, the resulting Bayes factor $B_{ij}$ changes. This sensitivity leads some statisticians and scientists to give up entirely on the notion of model comparison via Bayes factors and devise alternative metrics. We will instead explore certain methods which can explore and quantify the sensitivity of our model comparisons to prior specification. This will be demonstrated with specific examples in Ch. \ref{ch6:bf_prior_sens}.

Finally, we note that the definition of the Bayes factor arises naturally from our Bayesian probability axioms; 
we have no need for any ``\textit{ad hoc} devices''~\cite{jaynes03} of model comparison.
Moreover, this statistic only allows us to quantify which models are \textit{better} or \textit{worse} at describing the data, but never to address whether a model is \textit{right} or \textit{wrong}. Strictly speaking, the question `is my model right or wrong, given the data?' is usually ill-posed. Rather, meaningful insights will always be defined in the context of the model-building and criticism loop shown in Fig.~\ref{fig:model_building}.

%
\section{Bayesian model averaging}
\label{ch2:ex_bayes_model_avg}
%

Suppose that we have several models $\mathcal{M}_i$ which are more-or-less capable of describing a set of data $\bm{y}_{\rm exp}$ by inspection, we are uncertain which model to employ. 
Now, we consider that we want to either (i) make a prediction of some observable to which our models are not calibrated, or (ii) infer the likely values of parameters which are shared among the competing models, while marginalizing over uncertainty regarding the models. Bayesian model averaging (BMA) provides a means to address this problem quantitatively, and we outline it below.

\subsection{Predicting an observable}
\label{ch2:ex_bayes_model_avg_pred}

Suppose that we have models $\mathcal{M}_i$, each of which has a set of parameters $\bm{\theta}_i$ and is capable of predicting outputs $\bm{y}_i^{*}(\bm{\theta}_i)$. Model averaging provides a means to estimate the probability distribution $\mathcal{P}(\bm{y}^{*})$ of the outputs $\bm{y}^{*}$ in a manner which marginalizes over the \textit{model-space} uncertainty. Recall that the rule for marginalization requires
\begin{equation}
    \mathcal{P}(\bm{y}^{*}| I) = \sum_i \mathcal{P}(\bm{y}^{*}|  \mathcal{M}_i, I) \mathcal{P}(\mathcal{M}_i| I)
\end{equation}
where the sum over $i$ must include \textit{all possible disjoint} models. Let's address each these requirements in turn.

For the sum over $i$ to include all possible models, we would need the means to elicit every possible model which can describe the data, as well as the means to compute predictions with every model. Merely enumerating all possible models, when our model space is usually infinite-dimensional, is hopeless. In practice, we usually have the means to elicit and compute the predictions of finitely many models -- perhaps just a handful (as in the examples in this thesis). This situation is referred to as being \textit{$\mathcal{M}$-open} and is the reality we often face in scientific modeling of phenomena. This will require us to make the approximation 
\begin{equation}
    \mathcal{P}(\bm{y}^{*}| I) \approx \sum_{i=1}^{m} \mathcal{P}(\bm{y}^{*}|  \mathcal{M}_i, I) \mathcal{P}(\mathcal{M}_i| I)
\end{equation}
where the sum now includes only the $m$ models that we are capable of eliciting and calculating. 

The condition that the models be \textit{disjoint} becomes increasingly important when averaging over more than two or three models. 
We can consider a simple thought experiment in which we have $m=10$ `different models' and want to use these models to make a prediction about a particular outcome. Unbeknownst to us, nine out of the ten models are actually extremely similar, for instance, because all the physics assumptions in those models are nearly identical. Suppose that model $\mathcal{M}_1$ which predicts $\bm{y}_1^{*}$ and has model evidence $\mathcal{P}(\mathcal{M}_1| I)$  is the unique model, while models $\mathcal{M}_j$, $j=2, \dots, 10$ are nearly identical, all predicting $\bm{y}^{*}_j \approx \bm{y}^{*}_s \neq \bm{y}_1^{*}$ and all having similar Bayesian evidences $\mathcal{P}(\mathcal{M}_j| I) \approx \mathcal{P}(\mathcal{M}_s| I) $.
It follows that the model-averaged prediction 
\begin{equation}
    \mathcal{P}(\bm{y}^{*}| I) 
    \approx 
    \mathcal{P}(\bm{y}^{*}|  \mathcal{M}_1, I) \mathcal{P}(\mathcal{M}_1| I) + 9 \mathcal{P}(\bm{y}^{*}|  \mathcal{M}_2, I) \mathcal{P}(\mathcal{M}_2| I). 
\end{equation}
Therefore the predictions are more heavily weighted towards those of the correlated models, unconditional on their performance, simply because they are more numerous. 
The reader should see Ref.~\cite{huang2020combining} for an interesting study confronting this problem in climate science predictive modeling. In their case, their model space included approximately $m \approx 40$ models, and quantifying the extent to which `different' climate models were correlated was essential in defining the averaging weights. 

\subsection{Inferring a common set of parameters}
\label{ch2:ex_bayes_model_avg_infer}

Suppose that we have models $\mathcal{M}_i$, where each has a set of parameters $\bm{\theta}_i$ and is capable of predicting outputs $\bm{y}_i(\bm{\theta}_i)$ used to calibrate each of the models.
Further, suppose that all of the models share a certain common set of parameters $\bm{\omega}$ which exist as a subspace of each individual model's parameter space. 
We can infer the likely values of the shared parameters $\bm{\omega}$ while marginalizing over uncertainty in the model-space by model averaging. Therefore we want to compute 
\begin{equation}
    \mathcal{P}(\bm{\omega}| \bm{y}_{\rm exp}, I) \approx \sum_{i=1}^{m} \mathcal{P}(\bm{\omega}| \bm{y}_{\rm exp}, \mathcal{M}_i, I)
    \mathcal{P}(\mathcal{M}_i|  I). 
\end{equation}
This expression can be written in a more convenient form using the Bayes factor $B_{ij}$, defined in Eqn.~(\ref{eq:bayes_factor}): 
\begin{equation}
    \mathcal{P}(\bm{\omega}| \bm{y}_{\rm exp}, I) \approx  \mathcal{P}(\mathcal{M}_1|  I) \sum_{i=1}^{m} \mathcal{P}(\bm{\omega}| \bm{y}_{\rm exp}, \mathcal{M}_i, I) B_{i1}.
\end{equation}

Because the prefactor $\mathcal{P}(\mathcal{M}_1|  I)$ is a normalization independent of the parameters $\bm{\omega}$, we can simply sample the relative probability distribution of $\bm{\omega}$ according to the proportionality
\begin{equation}
\label{eq:bma_param_posterior}
    \mathcal{P}(\bm{\omega}| \bm{y}_{\rm exp}, I) \propto  \sum_{i=1}^{m} B_{i1} \mathcal{P}(\bm{\omega}| \bm{y}_{\rm exp}, \mathcal{M}_i, I). 
\end{equation}
To evaluate this expression, we need only the posteriors for the parameters $\bm{\omega}$ for each of the models $\mathcal{M}_i$, as well as the Bayes factors between each model and a single reference model, which, without loss of generality, we've denoted by $\mathcal{M}_1$. This will be the method by which we compute Bayes model averages of heavy-ion parameters in section \ref{ch6:bma_df}. We note that the same subtleties regarding our model space discussed in the previous section apply equally well to this problem. 

We have now discussed many of the most essential ideas and methods in Bayesian inference and model building and exploration. In the next chapter we will focus on the physics models which we will compare with experimental data observed in heavy-ion collisions. 
\chapter{Multistage Models for Heavy-Ion Collisions}
\label{ch3}

\section{Units and physical scales}
\label{ch3:units_scales}

At ultra-relativistic energies a set of natural units become useful and are given by fixing the speed of light $c = 3 \cdot 10^8$ m/s $\equiv 1$, reduced Planck's constant $\hbar = 1.054 \cdot 10^{-34}$ J $\cdot $ s $\equiv 1$, and Boltzmann constant $k_B = 1.380 \cdot 10^{-23} \rm{J} \cdot \rm{K}^{-1} \equiv 1$. 
The Heisenberg uncertainty principle
\begin{equation}
    \Delta x \cdot \Delta p \gtrsim \hbar / 2 \approx 1
\end{equation}
in these units provides a useful `back-of-the-envelope' correspondence between length and momentum scales. Furthermore, by expressing the relativistic free-particle dispersion of a particle with energy $E$, mass $m$, and spatial momentum $\bm{p}$,  
\be
E^2 = \bm{p}^2c^2 + m^2c^4 = \bm{p}^2 + m^2 
\ee
in these units, we find that energy, momentum and mass all have the same natural units $[E] = [\rm{length}]^{-1}$.
Similarly, when expressing Boltzmann's equipartition theorem
\be
E = (k_B T/2)_{\rm d.o.f.} = (T/2)_{\rm d.o.f.}
\ee
in these units we find temperature $T$ has the same units as energy/momentum/mass $[T] = [\rm{length}]^{-1}$. 
Electron-volts are a commonly used unit for energy in high-energy particle physics, and a useful relationship between natural units and experimental units is given by $\hbar \cdot c = 0.197$ GeV$\cdot$fm $\equiv$ 1.

Typical energy scales in low-energy nuclear physics are roughly estimated by $\Lambda_{QCD} \approx 0.2$ GeV, which corresponds to a length scale of $\sim 1$ fm -- the proton radius. Quantum Chromodynamics (QCD) is a non-conformal theory, meaning that the theory is not invariant under re-scaling transformations of dimensions. The up and down quarks have a small but nonzero rest mass, $m_u \approx 0.002$ GeV and $m_d \approx 0.005$ GeV, but even were they massless, conformal symmetry is still {\it anomalously}\footnote{If we write down a classical Lagrangian for QCD with zero bare quark and gluon masses it is scale-invariant. But once we consider a quantized theory of QCD, re-normalization introduces a non-perturbative energy scale $\Lambda_{\rm QCD} \approx 0.2$ GeV.} broken by the interactions in QCD.  

The natural length and timescales governing a heavy-ion collision's dynamics can be motivated by considering the spacetime-geometry of the collision, including Au-Au collisions at the Relativistic Heavy Ion Collider (RHIC) and Pb-Pb and Xe-Xe collisions at the Large Hadron Collider (LHC), among others.
Both Pb and Au nuclei have radii $R_{\rm Pb} \approx R_{\rm Au} \approx 6.5 $ fm. 
Moreover, nuclei are not smooth densities of matter, but are quantum mechanical and dynamical bound-states of neutrons and protons, each with a radius $R_{\rm p} \approx 1 $ fm. 
Therefore, in the plane transverse to the beamline before the nuclei collide, we have small length scales of roughly one fm and have larger length scales of the nuclear radius $\approx 7 $ fm. 
From effective theories of QCD at high energy emerge even smaller length scales which are fixed by the physics of saturation. The Color Glass Condensate (CGC) effective field theory~\cite{McLerran:2008es} predicts an energy-dependent scale, roughly $R_{\rm sat} \approx 0.2$ fm at top RHIC energies and $R_{\rm sat} \approx 0.1$ fm at LHC energies~\cite{Gelis:2014qga}. 
Because the nuclei travel with very nearly the speed of light $v_z \approx 1$, the natural geometric time scales transverse to the beam can be as small as $\tau \approx R_p / v_z \approx 1$ fm/c for low-energy descriptions of nuclei and $\tau \approx R_{\rm sat} \lesssim 0.2$ fm for high-energy nuclear descriptions. 

If the medium is to be described by a microscopic dynamical equation, such as the Boltzmann equation, then additional relevant scales are given by the microscopic time scales of particle interactions (see section \ref{ch3:pre_hydro}). Hydrodynamic theories are constructed assuming a hierarchy of microscopic and macroscopic scales. If $\tau_{\rm micro}$ is a proxy for the relevant microscopic time scales, while $\tau_{\rm macro}$ is a proxy for the macroscopic scales, we trust hydrodynamic predictions when the Knudsen-number ${\rm Kn} \equiv (\tau_{\rm micro}/ \tau_{\rm macro}) < 1$. When the Knudsen number increases, the relative importance of the viscous corrections grows and the applicability of hydrodynamic theory decreases. We will return to this point when we discuss hydrodynamics in section \ref{ch3:hydro}. In the latest stages of a heavy-ion collision, which we propagate according to hadronic transport models, there is not a single timescale relevant for describing particle interactions; rather, there are many timescales corresponding to different hadron scattering and resonance decay channels.

\section{Boost invariance and Milne coordinates}
\label{ch3:boost_invar_milne}
Due to the ultra-relativistic nature of heavy-ion collisions, an approximate boost-invariant symmetry manifests which is reviewed below. This introduces a more convenient set of coordinates than the usual Cartesian coordinates $(t, x, y, z)$ on four-dimensional spacetime. Milne coordinates $(\tau, x, y, \eta_s)$ are defined by the following non-linear transformations: 
the longitudinal proper-time $\tau$ is defined
\begin{equation}
    \tau \equiv \sqrt{t^2 - z^2}
\end{equation}
and the spacetime-rapidity $\eta_s$ is defined
\begin{equation}
    \eta_s \equiv \frac{1}{2} \ln \left( \frac{t+z}{t-z}\right).
\end{equation}
Under a longitudinal Lorentz boost with velocity $V_z$, the Cartesian time $t$ and longitudinal $z$ coordinates are mixed according to the Lorentz transformations $t' = \gamma(t - z V_z)$, $z' = \gamma(z - t V_z)$, where $\gamma \equiv (1-V_z^2)^{-1/2}$. However, the longitudinal proper time $\tau$ is manifestly invariant under a boost $V_z$.
The spacetime rapidity $\eta_s$ is not invariant under such a boost; however, its transformation is particularly simple. Defining the boost-rapidity $y_V \equiv \tanh^{-1}(V_z)$, an analog to the usual Cartesian velocity, in the boosted frame the spacetime rapidity is given by $\eta_s' = \eta_s - y_V$. 

\begin{figure*}[!htb]
\centering
\includegraphics[width=0.7\textwidth]{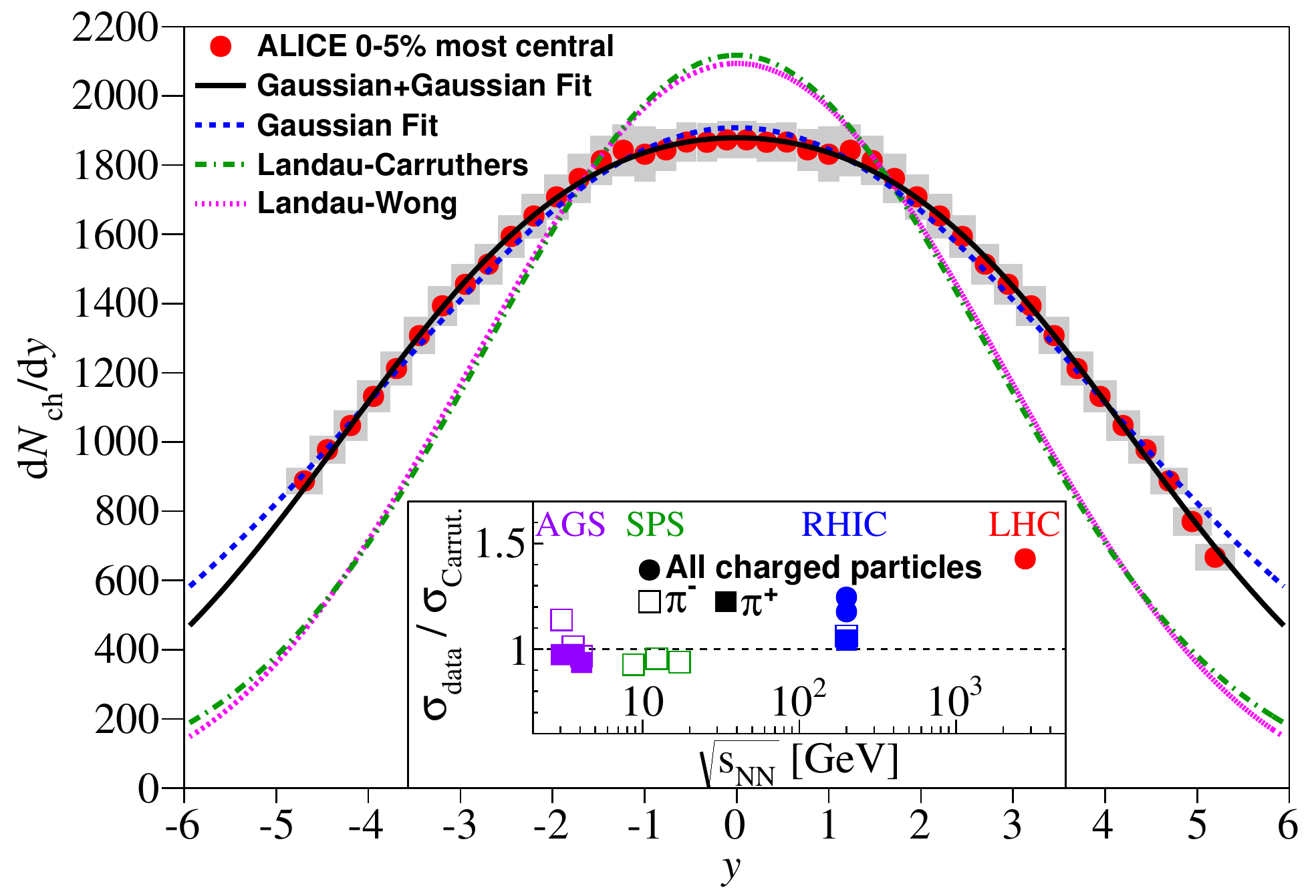}
    \caption{The rapidity distribution of charged particles measured by ALICE in Pb-Pb collisions at $\sqrts{} = 2.76$ TeV ~\cite{Abbas:2013bpa}; please see Ref. for explanations of the various curves.}
\label{fig:ALICE_dNdy}
\end{figure*}

The momentum-space rapidity $y$, called simply rapidity throughout this thesis, is defined by 
\begin{equation}
y \equiv \frac{1}{2} \ln \left( \frac{p^0+p_z}{p^0-p_z}\right),
\end{equation}
where $p^0$ is the energy of a particle and $p^z$ the component of its momentum in the $z$-direction. 
Bjorken argued that ultra-relativistic energies, the processes which describe particle production should be approximately invariant under longitudinal boosts~\cite{Bjorken:1982qr}. This symmetry is a good approximation for the particles near mid-rapidity ($y \approx 0$) in heavy-ion collisions. In a small window of rapidity $|y| \lesssim 0.5$, the distribution of observed particles $dN/dy$ is approximately flat; this is shown in Fig. \ref{fig:ALICE_dNdy} for central Pb-Pb events at $\sqrts{}=2.76$ TeV. 
Consequently, if we restrict our observables to include only particles near midrapidity, then boost-invariance is approximately satisfied. Under these conditions, we can assume the physics describing midrapidity is independent of spacetime rapidity $\eta_s$, and then we only need to propagate spacetime fields in \textit{two} transverse spatial dimensions and the longitudinal proper time. This approximation will allow us to model the evolution of heavy-ion collisions in a manner which would be prohibitively computationally expensive otherwise. All models which will be compared to data in this thesis employ this boost-invariant approximation. 

Being curvilinear, the Milne coordinate system has a non-trivial metric and metric connection. The metric $g$, whose components are $g_{\mu\nu}$, defines the invariant product between two vectors, $A \cdot B \equiv g(A, B) =  g_{\mu\nu} A^{\mu} B^{\nu}$, and in Milne coordinates is given by
\begin{equation}
    g_{\mu\nu} = \text{diag} (1, -1 ,-1, -\tau^2).
\end{equation}
The connection (also called Christoffel symbols), with components $\Gamma^{\mu}_{\alpha\beta}$, is necessary to define a covariant derivative -- a derivative operation which applied to a tensor yields a new tensor. 
The metric connection has components defined by 
\begin{equation}
\label{eq:metric_connection}
    \Gamma^{\alpha}_{\mu\nu} \equiv \frac{1}{2}g^{\beta \alpha}(\partial_\nu g_{\mu \beta} + \partial_\mu g_{\nu \beta} - \partial_\beta g_{\mu\nu}).
\end{equation}
Specifically in Milne coordinates we have
\begin{equation}
    \Gamma^{\eta}_{\tau\eta} = \Gamma^{\eta}_{\eta\tau} = \frac{1}{\tau}, \Gamma^{\tau}_{\eta\eta} = \tau,
\end{equation}
while all other components are zero. Despite being written with the appearance of a 3-index tensor, the connection $\Gamma^{\mu}_{\alpha\beta}$ is \textit{not} a tensor.\footnote{%
A \textit{tensor} is a quantity which under a coordinate transformation transforms in a particular way. A general coordinate transformation from coordinates $x$ to coordinates $x'$ is given by the matrix of derivatives $\mathcal{L}^{\mu}_{\nu} \equiv \frac{\partial x'^{\mu}}{\partial x^{\nu}}$. A tensor $\bm{A}$ with $m$ contravariant components $\mu_1, \dots, \mu_m$ and $n$ covariant components $\nu_1, \dots, \nu_n$ must transform under such a coordinate transformation according to $A^{\mu'_1 \cdots \mu'_n}_{\nu'_1 \cdots \nu'_n} = L^{\mu_1}_{\alpha_1} \cdots L^{\mu_m}_{\alpha_m} (L^{-1})^{\nu_1}_{\beta_1} \cdots (L^{-1})^{\nu_n}_{\beta_n} A^{\alpha_1 \cdots \alpha_m}_{\beta_1 \cdots \beta_n}$.  
}
The covariant derivative of a vector $A^{\mu}$ is given by
\begin{equation}
    D_{\alpha} A^{\mu} \equiv \partial_{\alpha}A^{\mu} + \Gamma^{\mu}_{\alpha\beta}A^{\beta},
\end{equation}
the covariant derivative of a rank-two tensor $A^{\mu\nu}$ is given by
\begin{equation}
    D_{\alpha} A^{\mu\nu} \equiv \partial_{\alpha}A^{\mu\nu} + \Gamma^{\mu}_{\alpha\beta}A^{\beta\nu} + \Gamma^{\nu}_{\alpha\beta}A^{\mu\beta},
\end{equation}
and the covariant derivative of a scalar $\theta$ is equivalent to the ordinary partial derivative, 
\begin{equation}
    D_{\alpha} \theta \equiv \partial_{\alpha}\theta.
\end{equation}
%

\section{Conservation laws}
\label{ch3:conserv_laws}

The physics describing heavy-ion collisions respects certain conservation laws. The local conservation of energy and momentum requires the energy-momentum tensor $T^{\mu\nu}$ be divergence-free,
\begin{equation}
\label{eqn:zero_div_T}
    D_{\mu} T^{\mu\nu} = 0.
\end{equation}
In Milne coordinates, this becomes 
\begin{eqnarray}
\label{eq:conservation_laws_milne}
{\partial }_{\mu }T^{\mu\tau}& =&-\frac{1}{\tau}(T^{\tau\tau}+\tau^{2}T^{\eta\eta})\ , \label{dT001}\\
{\partial }_{\mu }T^{\mu x}& =&-\frac{1}{\tau}T^{\tau x}\ , \quad 
{\partial }_{\mu }T^{\mu y}=-\frac{1}{\tau}T^{\tau y}\ , \quad 
{\partial }_{\mu }T^{\mu \eta }=-\frac{3}{\tau}T^{\tau \eta}\ . \qquad
\label{dT0i1}
\end{eqnarray}
The stress-tensor, which we will require to be symmetric $T^{\mu\nu}=T^{\nu\mu}$, has ten independent components in $3+1$ dimensions\footnote{%
``$s+t$ dimensional'' is a shorthand denoting $s$ spatial dimensions and $t$ temporal dimensions.}
in the absence of symmetries. The conservation laws Eq.~(\ref{eqn:zero_div_T}) provide only four equations, which are not sufficient to propagate the stress tensor dynamically. At early proper times, we will describe the medium by a Boltzmann equation in the weakly-coupled limit with partonic degrees of freedom. A significant duration of the collision will be described using viscous hydrodynamics. Finally, the last stage of the collision will be described using a Boltzmann transport of hadronic degrees of freedom. 
Although all of these dynamical models describe very different microscopic and macroscopic evolution, they all satisfy the local \textit{conservation} of energy-momentum. 

In principle, there are other conserved charges in heavy-ion collisions. For example, the local conservation of baryon charge is expressed by zero divergence of the baryon four-current $N^{\mu}_B$, %
\begin{equation}
\label{eqn:zero_div_N}
    D_{\mu} N^{\mu}_{B} = 0.
\end{equation}
However, this thesis will focus on describing heavy-ion collisions at the highest energies, where near mid-rapidity the conservation of baryon charge can be approximately neglected. 

Besides the conservation of energy-momentum during each stage of the hybrid-model, we also usually require inter-stage matching conditions to leave the stress tensor continuous. 
Across any spacetime surface $\Omega$ where we change our physical description/theory, e.g. changing from description $A$ to description $B$, we will typically require $T^{\mu\nu}_A(\Omega) = T^{\mu\nu}_B(\Omega)$.

\section{Initial conditions for deposited matter}
\label{ch3:initial_conditions}

Because the nuclei travel down the beam-line at ultra-relativistic speeds, they are Lorentz-contracted in the direction parallel to the beam, denoted by the $z$-axis:
\begin{equation}
    L_z = L_{0, z} / \gamma,
\end{equation}
where $L_z$ denotes the extent of each nucleus in the $z$-direction in the lab frame and $L_{0, z}$ denotes the nucleus' proper length in the same direction. 
The boost factor $\gamma \equiv (1-v_z^2)^{-1/2}$ is extreme, for RHIC Au-Au collisions at $\sqrts = 0.2$ TeV we find $\gamma \approx 106$, while for LHC Pb-Pb collisions at $\sqrts = 2.76$ TeV we have $\gamma  \approx 1470$. 
Therefore the amount of time required for the two-nuclei to pass through each other $\tau_{\rm cross}$ in the laboratory frame is extremely small. If $R$ denotes the proper radius of each nucleus, the diameter is given by $2R$. In the lab frame, this crossing time is approximated by 
\begin{equation}
    \tau_{\rm cross} \approx 2R / \gamma v_z  \approx 2R / \gamma
\end{equation}
where we have used that $v_z \approx 1$ at top RHIC and LHC energies. 
This yields $\tau_{\rm cross} \approx 0.1$ fm for Au-Au collisions at $\sqrts = 0.2$ TeV and
$\tau_{\rm cross} \approx 0.01$ fm for Pb-Pb collisions at $\sqrts = 2.76$ TeV. 

Consider that we start a clock $\tau_0=0$ in the lab frame at the moment of the very first nucleon-nucleon (or parton-parton) collision between the two nuclei, which happens at some location $(x_0,y_0)$ in the transverse plane. The transverse distance which a physical signal can propagate until the time of the very last collision (the time when the nuclei have passed each other) is given by the light-cone boundary $(x-x_0)^2 + (y-y_0)^2 \equiv r_T^2 \leq \tau^2 \leq (0.1$fm$)^2$. Therefore, during the time in which the nuclei overlap, every region in the transverse plane separated by more than $0.1$ fm is causally disconnected. As a result, one can treat each collision in the transverse plane \textit{independently} in this high-energy approximation. This is the approximation used in the initial condition model \trento ~\cite{Moreland:2014oya}, which we will employ to parametrize the initial conditions of heavy-ion collisions throughout this thesis. 

\trento{} uses Woods-Saxon distributions to describe the density of nucleons in each colliding heavy nucleus before the collision. The Woods-Saxon parameters are fixed with radius $R_{\rm Au}=6.38$\,fm 
and surface thickness $a_{\rm Au}=0.535$\,fm while those for Pb are $R_{\rm Pb}=6.62$\,fm and $a_{\rm Pb}=0.546$\,fm.
%
%
The model parametrizes the transverse energy deposition at $\tau = \tau_0$ via a reduced thickness function $T_R$:
\begin{eqnarray}
    \bar\epsilon(\mathbf{x}_\perp) 
    &\equiv& \frac{dE}{d\eta_s d^2\mathbf{x}_\perp} 
    = N T_R(\mathbf{x}_\perp; p),  
\label{eq:TrentoEd}
\\
\label{eq:TR}
    T_R(\mathbf{x}_\perp; p) &=& \left(\frac{T^p_A(\mathbf{x}_\perp) + T^p_B(\mathbf{x}_\perp)}{2}\right)^{1/p},
\end{eqnarray}
where $N$ is a free parameter which controls the normalization of energy density, and $T_A$ and $T_B$ represent the participant nucleon densities.

\begin{figure*}[htb]
\noindent\makebox[\textwidth]{%
  \centering
  \begin{minipage}{0.33\textwidth}
    \includegraphics[width=\textwidth]{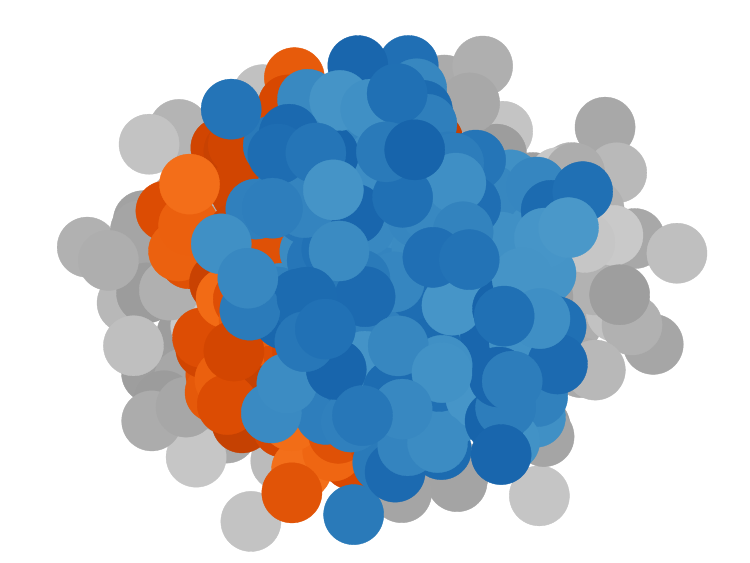}
  \end{minipage}
  \begin{minipage}{0.33\textwidth}
    \includegraphics[width=\textwidth]{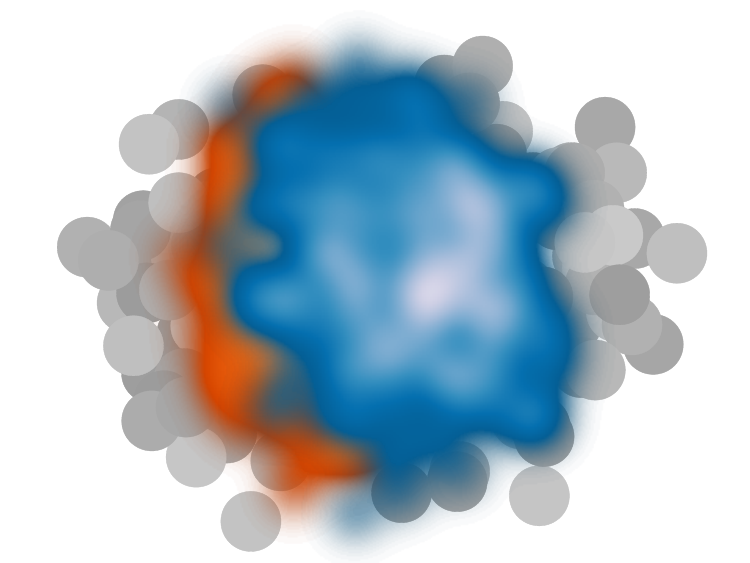}
  \end{minipage}
  \begin{minipage}{0.33\textwidth}
    \includegraphics[width=\textwidth]{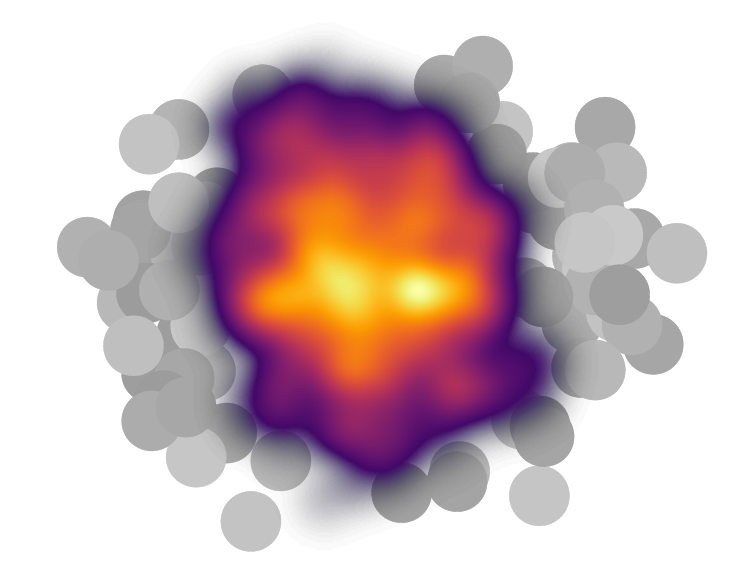}
  \end{minipage}
  }
  \caption{Figure from Ref.~\cite{Moreland:2019szz}. (Left) Spectator nucleons are shown as gray circles, while participant nucleons are shown as blue and orange circles. (Middle) The thickness functions of the participating nucleons are shown in blue and orange. (Right) The reduced thickness function of all participants, defining the initial energy density.  }
  \label{fig:trento_red_thickness}
\end{figure*}

In Eq.~(\ref{eq:TrentoEd}) we define $\bar\epsilon(\mathbf{x}_\perp)\equiv\lim_{\tau\to0^+}\tau\epsilon(\tau,\mathbf{x}_\perp,\eta_s{=}0)$. The free parameter $p$ defines a continuous family of mappings from the participant thickness functions $T_{A}$ and $T_B$ to the energy density deposition and interpolates between other widely-used models for energy density deposition. For example,  
when $p\to 0$, $T_R \to \sqrt{T_A T_B}$ and the model shares similar geometric features, the relations between eccentricies and centrality~\cite{Moreland:2014oya} 
to those given by the IP-Glasma initial condition model~\cite{Schenke:2012wb}. 
For $p=1$, the energy deposition is equivalent to the wounded nucleon Glauber model \cite{Miller:2007ri}. The positions of nucleons are sampled from the Woods-Saxon distributions of the nuclei; a free parameter $d_{\min}$ fixes the minimum allowed distance between any pair of sampled nucleons (to model the short range repulsive forces of the inter-nucleon potential).
Nuclear collisions are generated by performing binary-nucleon inelastic collisions in a Monte-Carlo procedure. This is visualized in the three panels of Fig. \ref{fig:trento_red_thickness}. 
The density distribution of a nucleon is modeled by a three-dimensional Gaussian function with a width parameter $w$. 
Each participant nucleon's deposited energy is 
sampled from a $\Gamma$-distribution with unit mean and standard deviation parameter $\sigma_k$.
This additional source of fluctuations is introduced to model the large multiplicity fluctuations observed in minimum-bias proton-proton collisions. A more thorough discussion of this model can be found in Ref.~\cite{Moreland:2019szz}.

\section{Pre-hydrodynamic transport with kinetic theory}
\label{ch3:pre_hydro}

Many phenomenological studies of heavy-ion collisions include a brief period of expansion before the onset of hydrodynamics, as visualized by the pre-hydrodynamic QGP stage in Fig. \ref{fig:little_bang}. In this section, we describe a few models which are all based on simplifying limits of kinetic theory. At this time, I have only applied the freestreaming approximation towards global Bayesian inference for heavy-ion collisions, but we expect that more flexible frameworks will be essential in reducing theoretical biases in the future.

The Boltzmann equation, which describes the evolution of the one-particle distribution function $f(x;p)$ of a system of particles, is given in general curvilinear coordinates by
\begin{equation}
    p^{\mu} \partial_{\mu} f + \Gamma^{i}_{\mu\nu}p^{\mu}p^{\nu} \partial_{p^i}f = C[f], 
\end{equation}
where $\Gamma^{i}_{\mu\nu}$  were defined in Eq.~(\ref{eq:metric_connection}).
Without simplifying approximations, this is an integro-differential equation in a high-dimensional phase-space. Given the current state-of-the-art computing power of both CPUs and GPUs, numerically propagating this equation deterministically for realistic microscopic collision kernels is still too costly to be possible in global phenomenological statistical analyses.\footnote{%
Monte Carlo methods for propagating the equations tend to be much faster, but may introduce difficulties related to finite-particle statistical fluctuations and causality ~\cite{Molnar:2019yam}. 
} Consequently, we will investigate models of this equation which simplify the scattering kernel $C[f]$ and impose boost-invariant symmetries to reduce the dimensionality.

\subsection{Freestreaming approximation}

The freestreaming limit of the Boltzmann equation is given by fixing $C[f]=0$, yielding 
\be
\label{eq:fs_boltzmann}
    p^{\mu} \partial_{\mu} f + \Gamma^{i}_{\mu\nu}p^{\mu}p^{\nu} \partial_{p^i}f = 0.
\ee
This equation is often solved using the method of characteristics.
The solution of Eq.~(\ref{eq:fs_boltzmann}) in Cartesian coordinates is given by
\be
    f(t,\boldsymbol{x};\bm{p}) = f\bigl(t_0, \boldsymbol{x}{-}\boldsymbol{v}(t{-}t_0); \bm{p}\bigr).
\label{free-streaming-soln}
\ee
For massless particles we have $p^0 = |\boldsymbol{p}|$, and it is convenient to define a unitless vector which encodes the direction of each particles momenta $\hat{p}^{\mu} \equiv p^{\mu} / p^0 = (1, \boldsymbol{v})$, where $|\boldsymbol{v}|=c$. Since all massless particles move with the speed of light, their velocity vectors are completely determined by a set of angles. It is convenient to define a moment $F(x; \bm{\Omega}_p)$ of the distribution function by
\begin{equation}
\label{eq:F_moment}
    F(x;\bm{\Omega}_p) \equiv \frac{g}{(2\pi)^3} \int p_0^3\, dp_0\, f(x;p) 
\end{equation}
where $g$ is a degeneracy factor. Then the stress tensor at any time $t > t_0$ is given by
\begin{eqnarray}
\label{Tmn_fs}
    T^{\mu\nu}(t,\boldsymbol{x}) &=&  \int d\bm{\Omega}_p\, \hat{p}^{\mu} \hat{p}^{\nu}\, F(t, \boldsymbol{x}; \bm{\Omega}_p) \nonumber \\
    &=& \int d\bm{\Omega}_p\, \hat{p}^{\mu} \hat{p}^{\nu}\, F\bigl(t_0, \boldsymbol{x}{-}\boldsymbol{v}(t{-}t_0\bigr) ; \bm{\Omega}_p),
\end{eqnarray}
where $\bm{\Omega}_p$ is the solid angle in momentum space and the second equality was obtained by inserting the free-streaming solution (\ref{free-streaming-soln}).

%

In Milne coordinates, the solution to Eq.(\ref{eq:fs_boltzmann}) appears more complicated due to the nonzero metric connection. If we assume that the longitudinal pressure vanishes identically $T^{zz} = \tau^2 T^{\eta\eta}=0$, equivalent to the particles having identically zero longitudinal momenta $p^z=0$, as well as assuming 
the initial momentum distribution is isotropic in $\phi_p$ 
and boost-invariance, then the stress tensor at any proper time $\tau = \tau_0 + \Delta \tau$ is given by 
\be
    T^{\mu\nu}(\tau, \bm{r}_T) = \frac{\tau_0}{\tau}\int \frac{d\phi_p}{2\pi} \hat{p}^{\mu} \hat{p}^{\nu} T^{\tau\tau}(\tau_0, \bm{r}_T - \hat{\bm{p}}_T \Delta \tau).
\ee
Let's consider the energy conservation law Eq.~(\ref{eq:conservation_laws_milne}) in the case that the longitudinal pressure and transverse gradients vanish $T^{\eta\eta} = \partial_x T^{\tau x} =  \partial_y T^{\tau y} = 0$: 
\be
    \partial_\tau T^{\tau\tau} = -\frac{1}{\tau}T^{\tau\tau}.
\ee
This equation has the solution 
\be
    \tau T^{\tau\tau} = ({\rm const.})
\ee
In particular, we see that this equation implies that $\lim_{\tau \rightarrow 0^+} T^{\tau\tau} = \infty$. However, $\lim_{\tau \rightarrow 0^+} (\tau T^{\tau\tau}) \equiv \tau_0 T^{\tau\tau}(\tau_0)$ is finite. When we employ the \trento{} model to define the initial conditions, we assume that the field defined by the \trento{} output is $\tau_0 T^{\tau\tau}(\tau_0, x, y)$.
A more detailed description of the boost-invariant method for freestreaming can be found in \cite{Liu:2015nwa, Broniowski:2008qk, fs_code}.

When the hybrid model with freestreaming is compared to data in Ch. \ref{ch5} and Ch. \ref{ch6}, we use a particular parametrization of the freestreaming duration
$\tau_{\text{fs}}$. It is common for model calculations to assume that this hydrodynamic initialization time is the same for all centralities and/or for different collision systems.\footnote{%
    See, e.g., Ref.~\cite{Oliinychenko:2015lva} for examples and exceptions.}
However, there are reasons to expect that systems with higher energy densities ``hydrodynamize'' faster \cite{Basar:2013hea}. In hopes of approximating this effect, the free-streaming time $\taufs$ is parametrized to include a dependence on the initially deposited transverse energy density $\bar\epsilon$ from Eq.~(\ref{eq:TrentoEd}):
\begin{equation}
   \taufs = \tau_R \left(\frac{\langle \bar\epsilon \rangle}{\bar\epsilon_R}\right)^{\alpha}.
   \label{eq:taufs}
\end{equation}
Here $\tau_R$ is a normalization factor for the duration of the free-streaming stage, and the parameter $\alpha$ controls its dependence on the average initial energy density in the transverse plane, defined by
\begin{equation}
\label{eps_ave}
    \langle \bar\epsilon \rangle \equiv \frac{\int d^2x_\perp\, {\bar\epsilon}^2(\mathbf{x}_\perp)}{\int d^2x_\perp\, \bar\epsilon(\mathbf{x}_\perp)}.
\end{equation}
The quantity $\bar\epsilon_R = 4.0$\,GeV/fm$^2$ is fixed as an arbitrary reference scale when performing Bayesian inference; the resulting posteriors in Ch. \ref{ch5} and Ch. \ref{ch6} depend on this choice. 

\subsection{Isotropization time approximation}

The absence of any mechanism to isotropize the stress tensor in the LRF during the freestreaming evolution may quantitatively bias the model parameters estimated via Bayesian inference. The need for better pre-hydrodynamic models is a problem of which the community is aware. 
However, there is a significant challenge in that deterministic microscopic transport methods are often numerically expensive.\footnote{%
Besides the spatial dependencies, the momentum-dependence of the microscopic distribution must also be propagated, increasing the dimensionality of the numerical problem.
} Large scale Bayesian inference requires the modeling of upwards of one million fluctuating events. Consequently, each event needs to run in a reasonable amount of time (less than 20 minutes) on a single CPU core.

There are models of the Boltzmann equation which replace the expensive collision kernel $C[f]$ (often  a high-dimensional integral) with simplified forms. The relaxation time approximation~\cite{ANDERSON1974466} collision kernel specifies
that the microscopic distribution function $f$ relaxes to the local-equilibrium distribution $f_0$ on a timescale given by the relaxation time $\tau_r$:
\be
    C_{\rm RTA}[f] = -\frac{u \cdot p}{\tau_r}(f-f_0),
\ee
where $u$ denotes the timelike eigenvalue of the stress-tensor $u^{\mu} T^{\nu}_{\mu} = \epsilon u^{\nu}$, with $\epsilon$ the energy density. In its simplest form, the relaxation time $\tau_r = \tau_r(x)$ is a function of spacetime only, usually depending parametrically on the local temperature. For a conformal system, the only dimensional parameter is the temperature $T$, and we are required to fix $\tau_r = c_r/T$, where $c_r$ is a dimensionless parameter defining the strength of interactions.

To reduce further the momentum degrees of freedom, the authors of Ref.~\cite{Kurkela:2018ygx, Kurkela:2019kip} designed a model based on the RTA Boltzmann equation for massless particles, but with the momentum magnitude integrated out.
In Cartesian coordinates, the moment $F(x; \bm{\Omega}_p)$ defined in Eq.~(\ref{eq:F_moment}) integrates out the length of the momentum vector, leaving only the momentum-space angles as degrees of freedom. 
The Isotropization Time Approximation is a conformal model which evolves the moment $F(x;\bm{\Omega}_p)$ dynamically in time via
\be
    v^{\mu} \partial_{\mu} F = C[F] = -\frac{u \cdot v}{\tau_{\rm iso}} (F - F_{\rm iso}),
\ee
where $v^{\mu} \equiv p^{\mu}/p^{0}$, $F_{\rm iso}$ is defined to be the moment of a distribution which is isotropic in the LRF and the isotropization time $\tau_{\rm iso}$ is a timescale on which the collisions isotropize the momentum distributions. 

I have written a boost-invariant computational implementation of the isotropization time approximation model which can be found at Ref.~\cite{ita_code}. The implementation is efficient enough to be run event-by-event in a global phenomenological analysis, occupying roughly the same runtime as 2+1D viscous hydrodynamics. We plan to report the numerical implementation and some details about the possible physics that can be explored in a forthcoming manuscript.

\subsection{Matching conditions}

The pre-hydrodynamic model is stopped at a longitudinal proper time $\tau_{\text{fs}}$, and the energy-momentum tensor Eq.~(\ref{Tmn_fs}) is matched to a viscous hydrodynamic tensor decomposition via Landau matching conditions. The energy density $\epsilon$ in the local rest frame (LRF) and the flow velocity $u^{\mu}$ are the eigenvalue and time-like eigenvector of $T^{\mu\nu}$, satisfying
\be
    u_{\mu}T^{\mu}_{\nu} = \epsilon u_{\nu},
\ee
\be
    u^2=1.
\ee
The shear stress tensor is given by the traceless and transverse projection of the stress tensor, satisfying
\be
    \pi^{\mu\nu} = \Delta^{\mu\nu}_{\alpha\beta}T^{\alpha\beta}.
\ee
The transverse-traceless projector is defined by 
\begin{equation}
    \Delta^{\mu\nu}_{\alpha\beta} \equiv \frac{1}{2}(\Delta^{\mu}_{\alpha} \Delta^{\nu}_{\beta} + \Delta^{\nu}_{\alpha} \Delta^{\mu}_{\beta}) - \frac{1}{3}\Delta^{\mu\nu}\Delta_{\alpha\beta}
\label{eq:delta_munu_alphabeta}
\end{equation}
and
\be
    \Delta^{\mu\nu} \equiv g^{\mu\nu} - u^{\mu}u^{\nu}
\ee
projects onto the spatial directions in the local rest frame. The LRF is defined by $u^{\mu}_{\rm LRF} = (1, \bf{0})$. Because the free-streaming dynamics continuously drive the system out of momentum-space isotropy, the size of the initial shear stress tensor grows with the free-streaming time $\tau_{\text{fs}}$. 

The total isotropic pressure, $p+\Pi$, is given by
\begin{equation}
    p(\epsilon)+\Pi=\frac{\epsilon-T^{\mu}_{\phantom{\mu}\mu}}{3}.
\end{equation}
Because we assume massless degrees of freedom (i.e. conformal symmetry) during the pre-hydrodynamic stage, the energy-momentum tensor (\ref{Tmn_fs}) is traceless, $T^{\mu}_{\phantom{\mu}\mu}{\,=\,}0$.
The QGP fluid described by viscous hydrodynamics, on the other hand, is characterized by an equation of state that breaks conformal symmetry primarily by interactions: $p_{_{\text{QCD}}}(\epsilon) < \epsilon/3$. Matching of the conformal pre-hydrodynamic energy-momentum tensor to the non-conformal hydrodynamic one thus entails a non-zero, positive initial fluid bulk viscous pressure at $\taufs$: 
\begin{equation}
\label{Pi_init}
    \Pi=\frac{\epsilon}{3}-p_{_{\text{QCD}}}(\epsilon) \ge 0.
\end{equation}
Note that for an expanding system we would expect a negative bulk viscous pressure in the Navier-Stokes limit. 
Persistence effects from the initially positive bulk viscous pressure depend on the the bulk relaxation time and is studied in \Appendix{app_bulk_relax}. Different pre-hydrodynamic evolution models which break conformal symmetry may lead to different initial conditions for the bulk viscous pressure, even in sign. The effect of this discontinuous matching of the bulk pressure has been studied in other works and found to meaningfully effect final state observables~\cite{NunesdaSilva:2020bfs}, indicating that is a non-negligible source of model uncertainty.

We note that the pre-hydrodynamic model employed in Ref.~\cite{Nijs:2020ors, Nijs:2020roc} introduced a model parameter $v_{\rm FS}$ which breaks conformal symmetry for $v_{\rm FS} < 1$. However, this model is not equivalent to the freestreaming of massive degrees of freedom with a non-trivial transverse momentum distribution because it is assumed that all particles move with identically the same transverse velocity $v_{\rm FS}$. 
This implies that in the local rest frame of each cell the transverse pressure is zero.\footnote{%
This is the stress-tensor describing dust, an ensemble of particles which have zero relative momentum in their local rest frame.
} 
Although this model potentially allows a smoother matching of the pre-hydrodynamic isotropic pressure in the lab frame to the pressure given by Lattice QCD, I am not sure if the transverse dynamics it describes are sensible when $v_{\rm FS} < 1$. 

There is no requirement that particles be massless to employ the \textit{microscopic} freestreaming solution Eq.(\ref{free-streaming-soln}). Similarly, the stress tensor for massive degrees of freedom can be propagated forward in according to a similar method. However, an additional integral over momentum-space is required, because massive particles with a distribution of transverse momenta can have any velocity magnitude between $0 < |\bm{v}|< 1$. We require an assumption for the initial distribution's dependence on the momentum magnitude (or velocity magnitude). 
In this case, in Cartesian coordinates for simplicity,
\begin{equation}
\label{Tmn_fs_massive}
    T^{\mu\nu}(t,\boldsymbol{x}) = \frac{g}{(2\pi)^3} \int \frac{d^3p}{p_0} p^{\mu}p^{\nu} f(t,\boldsymbol{x};\bm{p}) = \frac{g}{(2\pi)^3} \int \frac{d^3p}{p_0} p^{\mu}p^{\nu} f\bigl(t_0, \boldsymbol{x}{-}\boldsymbol{v}(t{-}t_0); \bm{p}\bigr) 
\end{equation}
Again, with the assumptions of isotropy and an assumed initial dependence on $|\bm{v}|$, the initial microscopic distribution can be related to the initial energy density. 
Although I haven't personally implemented this in a computer model simulator, it should be straightforward. This would yield a non-conformal freestreaming model with the correct microscopic dynamics and without unphysical behavior. The cost of the additional integration over momentum space, compared to the massless freestreaming model, will render this model computationally slower, although probably by a manageable factor.\footnote{%
The massless and boost invariant freestreaming model takes $\mathcal{O}(1 {\rm s})$ to run. An additional grid and quadrature in $|\bm{p}|$ or $|\bm{v}|$ may only incur a slow-down of a factor of $\lesssim 50$. 
} 

\section{Viscous hydrodynamic transport}
\label{ch3:hydro}

Hydrodynamics is an effective theory for describing the long-wavelength (low-energy) macroscopic behavior of a medium and is often motivated by an expansion in gradients. 
The stress tensor of a fluid in local-equilibrium is without loss of generality decomposed via 
\be
\label{eq:tmunu_eig}
    T^{\mu\nu}_{(0)} = \epsilon u^{\mu\nu} - p_{\rm eq} \Delta^{\mu\nu}.
\ee
%
%
%
%
%
The equilibrium pressure $p_{\rm eq}$ generally depends on the microscopic physics of the medium and is a function of the energy density\footnote{%
The pressure can also depend on other conserved charge densities. For example, the QCD pressure is also a function of baryon density. Again, our discussion is restricted to the case that the energy and momentum are the only conserved charges.
}
\be
p_{\rm eq} = p_{\rm eq}(\epsilon).
\label{eq:hydro_eos}
\ee
This relationship $p_{\rm eq}(\epsilon)$ will be referred to as the hydrodynamic equation of state (EoS).

The formulation of viscous hydrodynamic theories, which have historically been understood to describe fluids near local-equilibrium, often proceeds by an expansion around local-equilibrium in a series organized by gradients of the fundamental variables $\{ \epsilon, p_{\rm eq}, u^{\mu} \}$. Specifically, we suppose that the full stress-tensor $T^{\mu\nu}$ is given by an expansion
\be
    T^{\mu\nu} = T^{\mu\nu}_{(0)} + T^{\mu\nu}_{(1)} + T^{\mu\nu}_{(2)} + \cdots
\ee
and require that energy-momentum be conserved at all orders:
\be
\label{eq:tmunu_cons_hydro}
    D_{\mu} T^{\mu\nu} = 0.
\ee

As an effective field theory, at each order in the series we are \textit{required} to write down all possible terms which satisfy the necessary symmetries. 
Typically, to satisfy angular-momentum conservation, we require the stress-tensor to be symmetric $T^{\mu\nu} = T^{\nu\mu}$. Furthermore, any gradient of the equilibrium pressure can be related to a gradient of the energy density via the equation of state, $D_{\mu} p_{\rm eq} = \frac{\partial p_{\rm eq}}{\partial \epsilon} D_{\mu} \epsilon$. 

A tensor $T_{\mu\nu}^{(1)}$ constructed from the scalars $\{\epsilon, D_{\mu} u^{\mu} \equiv \theta \}$, vectors $\{ D_{\mu}, u_{\mu} \}$, and symmetric tensors $\{ \Delta_{\mu\nu}, u_{\mu}u_{\nu} \}$ with exactly one gradient operator in every term can be expressed 
\be
\label{eq:tmunu_gradients}
    T_{\mu\nu}^{(1)} = a D_{(\mu}u_{\nu)} + b \Delta_{\mu\nu} \theta + c u_{\mu}u_{\nu} \theta + d u_{(\mu} D_{\nu)} \epsilon,
\ee
where $\{a,b,c,d\}$ are yet-undetermined scalars.
Parentheses around a pair of indices denotes symmetrization, e.g.
\begin{equation}
    a^{(\mu}b^{\nu)} \equiv \frac{1}{2}(a^{\mu}b^{\nu} + a^{\nu}b^{\mu}).
\end{equation}
%
Upon contraction of Eq.~(\ref{eq:tmunu_gradients}) with $u^{\mu}u^{\nu}$, the first term vanishes as a consequence of the fixed length of the flow vector, $u^2=1$, from which it follows
\be
    D_{\mu}(u_{\nu}u^{\nu}) = 2 u_{\nu} D_{\mu} u^{\nu} = 0.
\ee
In order to satisfy Eq.~(\ref{eq:tmunu_eig}) at all orders, we are find the requirement $c=d=0$. 

It is useful to decompose the covariant derivative into terms parallel and orthogonal to the flow velocity:
\be
    D_{\mu} = g_{\mu}^{\nu} D_{\nu} = (\Delta^{\nu}_{\mu} + u^{\nu}u_{\mu}) D_{\nu} = \nabla_{\mu} + u_{\mu}D
\ee
where we have defined the spatial gradient in the LRF as
\be
    \nabla_{\mu} \equiv \Delta^{\nu}_{\mu} D_{\nu}
\ee
and the temporal derivative in the LRF as
\be
    D \equiv u^{\mu}D_{\mu}.
\ee
Usually the gradients are reorganized by defining a symmetric and tracelesss tensor $\sigma^{\mu\nu}$, called the velocity shear tensor, 
\be
    \sigma^{\mu\nu} \equiv \partial^{(\mu}u^{\nu)} - \frac{1}{3}\Delta^{\mu\nu}\theta \equiv  \partial^{\langle \mu} u^{\nu \rangle, }
\ee
such that 
\be
\label{eq:tmunu_NS}
    T^{\mu\nu}_{(1)} = 2\eta \sigma^{\mu\nu} - \zeta \Delta^{\mu\nu} \theta. 
\ee
%

The quantities $\eta$ and $\zeta$, which we call the shear and bulk viscosities, have arisen as unknown scalars multiplying the only possible tensor structures. If we accurately knew the microscopic physical theory of the medium we are describing, these coefficients could be calculated in in a microscopic approach~\cite{Kubo:1957mj, doi:10.1063/1.1740082}. 
Whether we calculate the transport coefficients in fixed microscopic theory, or treat them as unknown parameters in a model-data comparison, the conservation law Eq.~(\ref{eq:tmunu_cons_hydro}) and equation of state Eq.(\ref{eq:hydro_eos}) provide us five equations with which we can dynamically evolve the five fundamental variables $\{\epsilon, p_{\rm eq}, u^{i}\}$. These equations are referred to as the relativistic Navier-Stokes (NS) equations. 

Let's investigate these equations briefly in the local rest frame (LRF)
to better understand the action of the viscosities. 
The conservation of energy requires
\be
    D \epsilon = - (\epsilon + p_{\rm eq} + \Pi)\theta + \pi_{\mu\nu}\sigma^{\mu\nu}
\label{ch3:eq:ns_e}
\ee
and the conservation of momentum
\be
   (\epsilon + p_{\rm eq} + \Pi) D u^{\mu} = \nabla^{\mu}(p_{\rm eq} + \Pi) - \Delta^{\mu\nu}\nabla^{\sigma}\pi_{\nu\sigma} + \pi^{\mu\nu}Du_{\nu}.
\label{ch3:eq:ns_u}   
\ee
Consider a system with a positive expansion rate $\theta > 0$. In this case, the Navier-Stokes bulk pressure $\Pi = -\zeta \theta \leq 0$ because we require $\zeta \geq 0$. Let's also assume that $\epsilon + p_{\rm eq} >  |\Pi|$, and therefore $(\epsilon + p_{\rm eq} + \Pi) > 0$. Thus, the action of the bulk pressure is to reduce the magnitude of the time-rate-of-change of the energy density, which for a positive expansion rate $\theta$ is decreasing in time: The bulk pressure slows down the dilution of energy density. 
The shear stress tensor has a similar effect because $\pi_{\mu\nu}\sigma^{\mu\nu} = 2 \eta \sigma_{\mu\nu} \sigma^{\mu\nu} \geq 0$. Both of these effects are usually called `viscous heating' -- the viscous pressures heat up the system, increasing its energy density relative to an ideal expansion.  

From Eq.~(\ref{ch3:eq:ns_u}) we see that the quantity $(\epsilon + p_{\rm eq} + \Pi)$ acts as the \textit{inertia} of the fluid's acceleration $D u^{\mu}$. The spatial gradient of the total isotropic pressure $(p_{\rm eq}+\Pi)$ acts to increase the acceleration of the fluid in the direction of the gradient; however, the shear stress `deflects' the acceleration through both it's gradient $\Delta^{\mu\nu}\nabla^{\sigma}\pi_{\nu\sigma}$ and its contraction with the fluid acceleration vector itself $\pi^{\mu\nu}Du_{\nu}$.

This discussion of the relativistic Navier-Stokes theory was intended as a heuristic to guide physical understanding of the hydrodynamic theories which we employ in this thesis. However, the Navier-Stokes theory in the reference frame we have chosen (the Landau frame) yields equations with both acausal and unstable small-wavelength modes. This doesn't spoil the Navier-Stokes theory in principle, nor in certain applications for which these small-wavelength modes can be carefully removed; the theory was never intended to describe these modes at all -- the truncation of the gradient series assumed long-wavelengths. But, in practice, heavy-ion collisions lack sufficient symmetries to be solved analytically. Rather, the equations must be solved numerically on a spacetime grid in which the acausal and unstable modes can arise and spoil the solution. 

Until very recently, the commonplace method of curing the relativistic Navier-Stokes equations involved promoting the dissipative stresses to dynamical variables. That is, the shear stress $\pi^{\mu\nu}$,
%
\be
\label{eq:shear_pi_def}
    \pi^{\mu\nu} \equiv \Delta^{\mu\nu}_{\alpha\beta} T^{\alpha\beta},
\ee
and the bulk viscous pressure $\Pi$,
\be
\label{eq:bulk_Pi_def}
    \Pi \equiv -\frac{1}{3}\Delta_{\alpha\beta} T^{\alpha\beta} - p_{\rm eq},
\ee
are promoted to independent degrees of freedom.
The shear-stress tensor $\pi^{\mu\nu}$ 
has five independent degrees of freedom in the absence of symmetries, and the scalar bulk pressure $\Pi$ has a single degree of freedom. 
Therefore, propagation of the fields $\pi^{\mu\nu}$ and $\Pi$ in spacetime require additional equations of motion beyond the conservation laws Eq.~(\ref{eqn:zero_div_T}) and EoS. 
A widely-used class of theories to propagate the dynamical variables $\{\epsilon, u^{\mu}, \pi^{\mu\nu}, \Pi\}$ will be referred to as Muller-Israel-Stewart (MIS) type theories~\cite{Muller:1967zza,Israel:1976tn, Israel:1979wp} in this thesis for the reasons explained below. 
MIS-type theories assume that the shear-stress and bulk pressure are described by a set of relaxation equations: 
\begin{align}
    \tau_{\Pi }\dot{\Pi}+\Pi &= -\zeta \theta + (\text{second order terms})\;,
\label{eq:IS_eqn_PI}
\\
    \tau_{\pi }\dot{\pi}^{\left\langle \mu \nu \right\rangle }+\pi ^{\mu \nu }
    &= 2\eta \sigma ^{\mu \nu } + (\text{second order terms}).
\label{eq:IS_eqn_pi}
\end{align}
According to these theories, the bulk pressure and shear-stress relax to their Navier-Stokes limits $\Pi_{\rm NS} = -\zeta \theta$ and $\pi^{\mu\nu}_{\rm NS} = 2 \eta \sigma^{\mu\nu}$, according to the relaxation times $\tau_{\Pi}$ and $\tau_{\pi}$, respectively. The second order terms differ theory-by-theory, but can be understood as 
terms which drive the bulk and shear stress from their Navier-Stokes limits. 

Analysis of the linearized relaxation-time equations reveal that such theories contain non-hydrodynamic modes~\cite{Romatschke:2017ejr}, which enter via the time-scales $\tau_{\pi/\Pi}$. These modes are interpreted in Ref.~\cite{Romatschke:2017ejr} as the ultraviolet (small wavelength)-completion of the hydrodynamic effective field theory, which are necessary to make the theories causal and stable. In practice we should remember that MIS-type theories do not propagate length and time-scales $L \lesssim \tau_{\pi/\Pi}$ \textit{hydrodynamically}. If the system created in a heavy-ion collision has very large spatial gradients at early times, for example, the non-hydrodynamic modes may actually dominate the proceeding dynamics. 
The reader is warned that MIS-type theories are typically referred to simply as `viscous hydrodynamic theories' in the literature, despite the presence of non-hydrodynamic modes.

We now return to the consideration of physical scales in hydrodynamics introduced in Ch.~\ref{ch3:units_scales}. As we earlier stated, our trust in the hydrodynamic theory is founded on a separation of microscopic and macroscopic scales. The microscopic scales present in our second-order hydrodynamic equations can be considered to be the relaxation times $\tau_{\pi/\Pi}$. Therefore, we expect our hydrodynamic theory to accurately describe spacetime scales $L > \tau_{\pi/\Pi}$. When these scales become competitive $L \sim \tau_{\pi/\Pi}$ we are in the regime of large Knudsen number (often called `rarefied gas'), and the predictions of the hydrodynamic theory become questionable. Moreover, in this regime the dynamics of MIS-type theories may be dominated by the non-hydrodynamic modes.

The smallest length scales present in our initial conditions are fixed by the nucleon width parameter $w$ in \trento{}, and throughout these analysis $w \gtrsim 0.5$ fm.
As proxy for the largest microscopic timescales evolved hydrodynamically, we consider
the shear relaxation time $\tau_\pi = b_{\pi} \frac{\eta}{sT}$, which is large when the specific shear viscosity $\eta/s$ is maximal. The largest specific shear viscosities probed in this work are approximately $\eta/s \sim 0.3$, and at early times we have $T \sim 0.3$ GeV $ \sim 1.5$ fm$^{-1}$. Furthermore we require $b_{\pi} \sim 5$, and at early times we can encounter $\rm{Kn} \sim \tau_{\pi} w  \sim 1$.
In practice, we first allow the transverse energy density to diffuse for approximately $1$ fm according to the freestreaming model, which tends to smear-out smaller scale structures in the fields, at the expense of potentially large inverse Reynolds numbers $\sqrt{\pi^{\mu\nu}\pi_{\mu\nu}}/(\epsilon + p_{\rm eq})$ and $|\Pi|/p_{\rm eq}$ at the initialization of hydrodynamics.

The second-order hydrodynamic theory employed in this thesis uses equations which were motivated in the Grad approximation of the Boltzmann equation~\cite{Denicol:2014loa}. These equations are implemented in MUSIC~\cite{hydro_code}, and are given
%
%
%
\begin{equation}
\label{relax_eqn_PI}
    \tau_{\Pi }\dot{\Pi}+\Pi = -\zeta \theta -\delta _{\Pi \Pi }\Pi \theta
    + \lambda _{\Pi \pi }\pi ^{\mu \nu }\sigma_{\mu \nu }\;,
\end{equation}

\begin{equation}
    \tau_{\pi }\dot{\pi}^{\left\langle \mu \nu \right\rangle }+\pi ^{\mu \nu }
    = 2\eta \sigma ^{\mu \nu }-\delta _{\pi \pi }\pi ^{\mu \nu }\theta
    +\varphi_{7}\pi _{\alpha }^{\left\langle \mu \right. }\pi ^{\left. \nu \right\rangle \alpha }
    -\tau _{\pi \pi }\pi _{\alpha }^{\left\langle \mu \right. }\sigma^{\left. \nu \right\rangle \alpha }+\lambda _{\pi \Pi }\Pi \sigma ^{\mu\nu}.
\label{relax_eqn_pi}
\end{equation}
Here $\dot{\Pi} = u^\lambda \partial_\lambda \Pi$, $\dot{\pi}^{\langle\mu\nu\rangle} = \Delta^{\mu\nu}_{\alpha\beta} u^\lambda \partial_\lambda \pi^{\alpha\beta}$, $\theta = \partial_\lambda u^\lambda$, and $\sigma^{\mu\nu} = \Delta^{\mu\nu}_{\alpha\beta}\partial^\alpha u^\beta$, with $\Delta_{\alpha\beta}^{\mu\nu}$ defined in Eq.~(\ref{eq:delta_munu_alphabeta}).

The equilibrium properties of QCD matter enter the hydrodynamic transport Eq.~(\ref{eq:tmunu_cons_hydro}) through the equilibrium pressure $p_{\rm eq}{\,=\,}p_{\rm eq}(\epsilon)$. The near-equilibrium dynamics of QCD matter are controlled by the first and second-order transport coefficients that enter in Eqs.~(\ref{relax_eqn_PI},\ref{relax_eqn_pi}). The first-order transport coefficients are the shear and bulk viscosities, $\eta$ and $\zeta$, which we have already discussed. Second-order transport coefficients entering into our hydrodynamic equations are $\delta _{\Pi \Pi }$, $\lambda _{\Pi \pi }$, $\delta _{\pi \pi }$, $\varphi _{7}$, $\tau _{\pi \pi }$, and $\lambda_{\pi\Pi }$, as well as the shear and bulk relaxation times $\tau _{\pi }$ and $\tau _{\Pi }$.

For the equilibrium properties the equation of state is matched to (i) a lattice calculation ~\cite{Bazavov:2014pvz} at high temperatures and (ii) a hadron resonance gas at lower temperatures (see Refs.~\cite{Bernhard:2018hnz, eos_code} for details). The hadron content of the resonance gas is chosen to be consistent with that of the hadronic afterburner \SMASH ~\cite{Weil:2016zrk} used in this work.\footnote{
Specifically, it is the hadrons occupying the SMASH `box' list of hadrons, which excludes certain exotic species and light-nuclei.
} While this consistency in the thermal pressure is important, the matching procedure does carry some uncertainties (see. e.g., Ref.~\cite{Auvinen:2020mpc}) which are not explored in this work.

\begin{figure*}[!htb]
\centering
\includegraphics[width=0.7\textwidth]{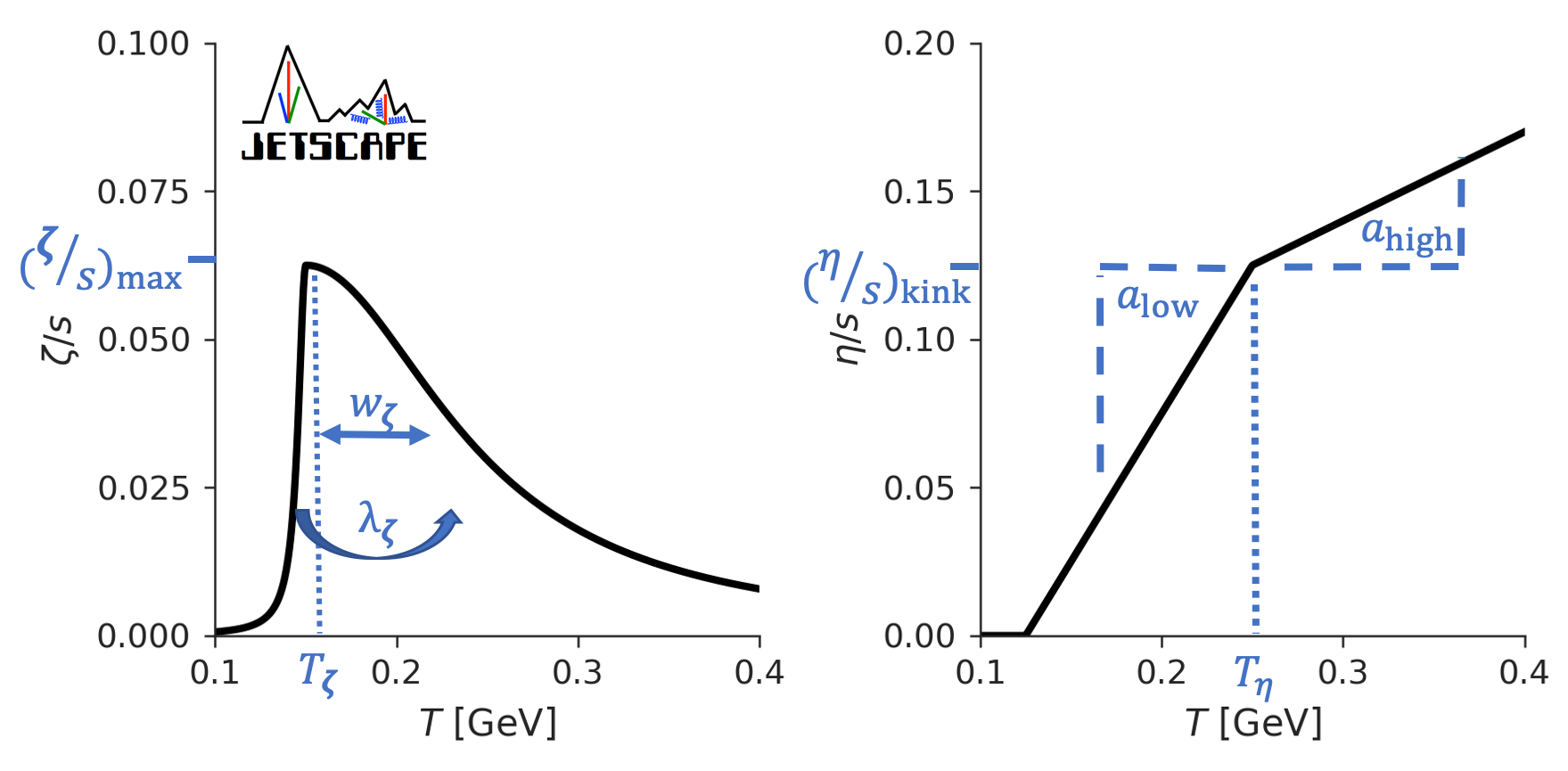}
    \caption{Depictions of the parametrizations of specific bulk (left) and shear (right) viscosity as functions of temperature. The specific bulk viscosity has the form of a skewed Cauchy distribution, while the specific shear viscosity is piecewise-linear with, in general, two different slopes. Both shear and bulk viscosities are required to be positive-definite to satisfy the second law of thermodynamics. The example for $(\eta/s)$ shown here has a positive low-temperature and high-temperature slope ($a_{\rm low},a_{\rm high}$>0).}
\label{fig:eta_zeta_pzations}
\end{figure*}

The shear and bulk viscosities, $\eta$ and $\zeta$, are parametrized as functions of temperature, and measurements are used to estimate the parameters.\footnote{\label{fn10}%
In general, if conserved charges are taken into account (which is not done here), the transport coefficients also depend on chemical potentials.}
They are discussed in more detail below.
The second-order transport coefficients should similarly be parametrized in order to marginalize over their uncertainty. 
In this work we apply this strategy only to the shear relaxation time $\tau _{\pi}$, while all other second-order transport coefficients are required to satisfy parameter-free relations~\cite{Denicol:2014vaa}.
%
The ratios of shear and bulk viscosity to entropy density --- the unitless specific viscosities --- are parametrized, instead of the viscosities themselves. A depiction of the parametrizations for the specific bulk and shear viscosities is shown in Fig. \ref{fig:eta_zeta_pzations}. 

For the specific shear viscosity, $\eta/s$, the parametrization assumes that it may have a single inflection point at or above the deconfinement transition~\cite{Niemi:2015qia}. The position of this inflection point in temperature, $T_{\eta}$, is a parameter, as is the value of $\eta/s$ at this point, $(\eta/s)_{\rm kink}$. A linear dependence of $\eta/s$ on temperature is assumed, with slopes $a_{\rm low}$ below and $a_{\rm high}$ above the inflection point, with both positive and negative slopes allowed. Negative values for $\eta/s$ are not allowed. The formula for this parametrization is
\begin{equation}
\label{positivity}
    \frac{\eta}{s}(T) = \max\left[\left.\frac{\eta}{s}\right\vert_{\rm lin}\!\!\!(T),0\right],
\end{equation}
with
\begin{equation}
\label{eq:eta_s_lin}
    \left.\frac{\eta}{s}\right\vert_{\rm lin}\!\!\!(T) = a_{\rm low}\, (T{-}T_{\eta})\, \Theta(T_{\eta}{-}T)+ (\eta/s)_{\rm kink}
    +\, a_{\rm high}\, (T{-}T_{\eta})\, \Theta(T{-}T_{\eta}).
\end{equation}
We may \textit{expect}, based on the behavior of other non-QGP fluids, a negative slope at temperatures below $T_{\eta}$, i.e. $a_{\rm low}{\,<\,}0$ and a positive slope at temperatures above $T_{\eta}$, i.e. $a_{\rm high}{\,>\,}0$~\cite{Csernai:2006zz}. Nevertheless, in this work the slopes are allowed to take negative and positive values: The aim is to ascertain whether the data themselves have sufficient information to constrain such a temperature dependence.

For the specific bulk viscosity, it is assumed to peak near the deconfinement temperature, and is parametrized by a skewed Cauchy distribution:
\begin{equation}
\label{eq:zeta_s_cauchy}
    \frac{\zeta}{s}(T) = \frac{(\zeta/s )_{\max}\Lambda^2}{\Lambda^2+ \left( T-T_\zeta\right)^2},
\end{equation}
where
\begin{equation}    
    \Lambda = w_{\zeta} \left[1 + \lambda_{\zeta} \sign \left(T{-}T_\zeta\right) \right].
\end{equation}
Here $T_\zeta$ is the temperature and $(\zeta/s)_{\max}$ the value of the peak; $w_\zeta$ and $\lambda_{\zeta}$ control the width and skewness of the Cauchy distribution, respectively. Allowing for a non-vanishing skewness is a generalization compared to Ref.~\cite{Bernhard:2019bmu}, and, in particular, this parameter allows the magnitude of the bulk viscosity to be much larger at higher temperatures.

Previous theoretical studies ~\cite{Kharzeev:2007wb, Karsch:2007jc, NoronhaHostler:2008ju, Rose:2020lfc, Arnold:2006fz} suggest that $\zeta/s$ for QCD may peak near the deconfinement transition, but the functional form of its temperature-dependence is still not well understood. Below the transition ($T\lesssim 150$\,MeV), the bulk viscosity is understood to be non-zero. We emphasize that we do not attempt to describe the dependence of the bulk viscosity below the particlization temperature of our model (discussed in the next section), which is never smaller than 135\,MeV. The fact that our parametrization of $(\zeta/s)(T)$ rapidly approaches zero at low temperature should therefore not be read as a physical feature: This low temperature range is not described by the hydrodynamic model, but microscopically by a hadronic transport model. While we thus cannot make any statements about the bulk viscosity of hadronic matter at these low temperatures it has recently been estimated in the \SMASH{} transport model ~\cite{Rose:2020lfc}.

Previous theoretical work ~\cite{Baier:2007ix, Bhattacharyya:2008jc, Denicol:2014vaa,Florkowski:2015lra, Czajka:2017wdo, Ghiglieri:2018dgf} suggests that, in the absence of conserved charges, the shear relaxation time may be well captured by the following temperature dependence:
\begin{equation}
    T \tau_\pi(T)= b_{\pi}\frac{\eta}{s}(T)
\end{equation}
where $b_\pi$ is a dimensionless constant that we consider unknown. The linearized causality bound ~\cite{Pu:2009fj} requires $b_\pi{\,\ge\,}(4/3)/(1{-}c_s^2){\,\ge\,}2$. Refs.~\cite{Baier:2007ix, Bhattacharyya:2008jc, Denicol:2014vaa, Florkowski:2015lra, Czajka:2017wdo} showed for a variety of weakly and strongly coupled theories other than QCD that this causality bound is respected, with $b_\pi$ varying between ${\sim}2$ and ${\sim}6$; we use these values to motivate the prior range explored for $b_\pi$. 

Previous investigations of the effects of the shear relaxation time and other second-order transport coefficients on soft hadronic observables have found them to be of modest phenomenological importance ~\cite{Song:2008si, Liu-thesis, Bernhard:2015hxa, Schenke:2019pmk}.
However, those studies employed different initial condition and pre-hydrodynamic models than this thesis. 
In particular, sensitivity to the shear-relaxation time is increased in our model because we have used a freestreaming pre-hydrodynamic model, which can drive the shear stress far from its Navier-Stokes limit. 
Varying the shear relaxation time in this work provides additional quantitative insights into the typical magnitude of effects from a second-order coefficient on the estimates for the first-order transport coefficients. Additionally, varying the relaxation times propagates a measure of theoretical uncertainty regarding the non-hydrodynamic physics of the system.

\section{Converting fluids to particles}
\label{ch3:particlization}

\begin{figure*}[!htb]
\centering
\includegraphics[width=0.6\textwidth]{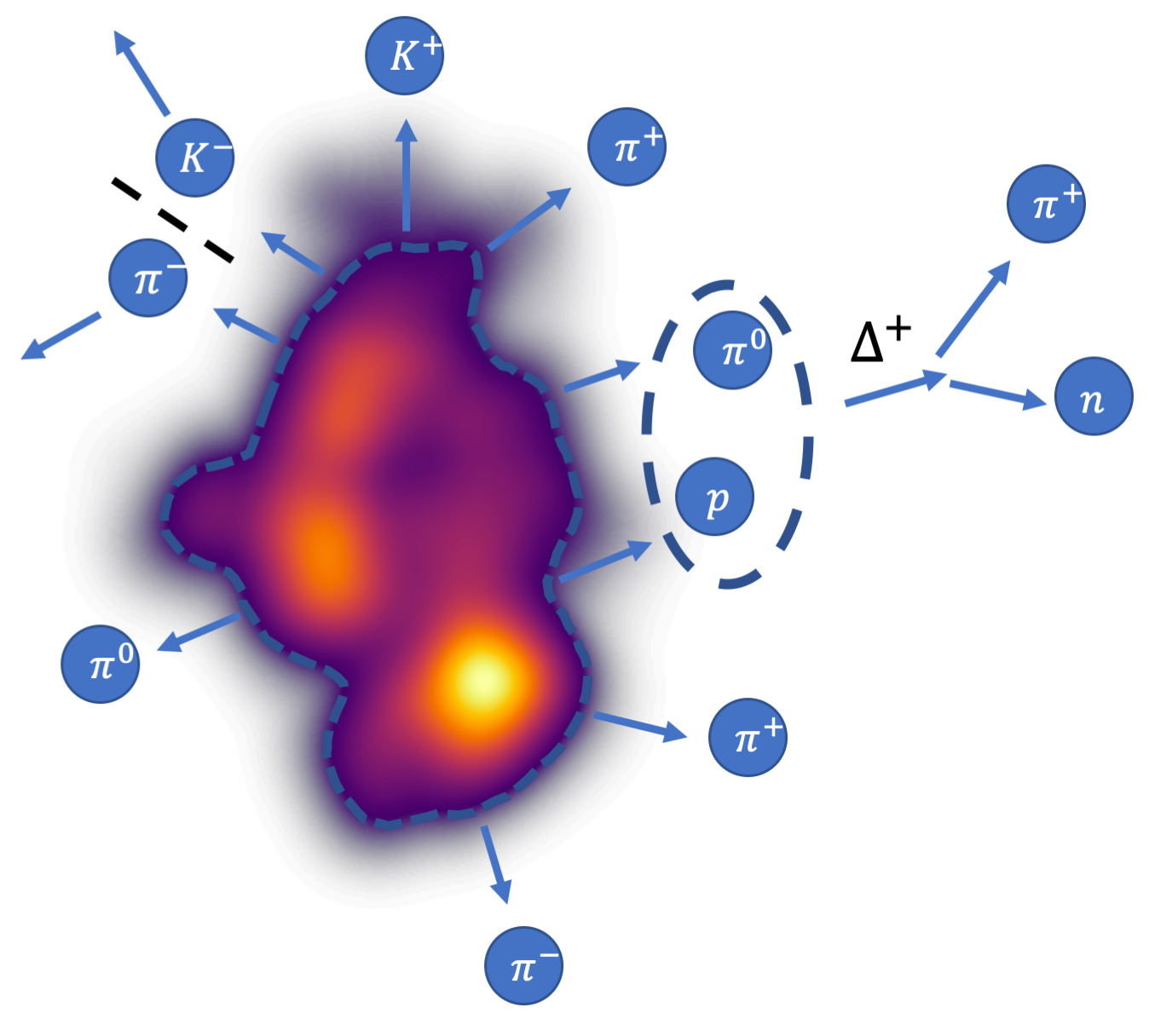}
\caption{An illustration of the conversion of fluid dynamical fields into hadron resonances on a isothermal surface (dashed light blue contour) at a fixed slice in proper time. The emitted hadrons can further scatter, form resonances, and decay. A very small number of hadrons are shown to be emitted for visualization purposes. }
\label{fig:hydro_cf_afterburn}
\end{figure*}

\textit{In principle}, particlization should not be thought of as a physical process, but as a change of description from macroscopic fluid dynamical degrees of freedom to microscopic quasi-particle degrees of freedom. In this work, particlization is implemented on a spacetime surface of constant ``switching'' or ``particlization'' temperature $\Tsw$. If \textit{in practice} this matching is really only a change of language, rather than a sudden change in the dynamical evolution, it requires the simultaneous applicability of both fluid and kinetic descriptions. 
The hydrodynamic description likely breaks down below the reconfinement transition because the mean-free path increases as a consequence of color neutralization, increasing the Knudsen number. On the other hand, the strongly-coupled nature of the color confinement process itself likely makes kinetic theory (the neglect of correlations in Boltzmann's stosszahlansatz assumption) inapplicable during the hadronization phase transition. These conditions probably conspire to yield a narrow window of time during which the system can be reasonably approximated by both fluid dynamics and kinetic theory. The conversion of fluid cells into hadronic degrees of freedom at a particular slice in proper time is illustrated in Fig.~\ref{fig:hydro_cf_afterburn}. 

The Cooper-Frye ~\cite{Cooper:1974mv, Cooper:1974qi} prescription for particlization~\cite{Huovinen:2012is} is used to convert all the energy and momentum of the fluid into hadrons on the switching hypersurface $\Sigma$. The formula for the Lorentz-invariant particle momentum spectrum of particles of species $i$ with degeneracy $g_i$ in terms of their kinetic phase-space distribution $f_i(x;p)$ is given by
\begin{equation} 
\label{CFEqn}
    p^0 \frac{dN_i}{d^3p} = \frac{g_i}{(2\pi)^3} \int_{\Sigma} d^3\sigma_{\mu} p^{\mu} f_i(x;p).
\end{equation}
The integral goes over the switching hypersurface $\Sigma$ with normal vector $\sigma_{\mu}(x)$. The distribution function $f_i(x;p)$ is typically chosen such that it reproduces the hydrodynamic energy-momentum tensor of the fluid on the particlization surface, 
\be
\label{Tmunu_kinetic}
    T^{\mu\nu}(x) = \sum_i g_i \int \frac{d^3p}{(2\pi)^3p^0} p^{\mu}p^{\nu} f_i(x;p). 
\ee
However, viscous hydrodynamics propagates no information about the higher moments of the microscopic distribution, 
leaving infinitely many choices for the microscopic distribution functions $f_i(x;p)$. If the QGP fluid were an ideal fluid in perfect local kinetic and chemical equilibrium, the choice for $f_i(x;p)$ would be unambiguous: It would be of local equilibrium form~\cite{Cooper:1974mv, Cooper:1974qi}, with the local rest frame velocity provided by hydrodynamics and the temperature
fixed by the local-rest-frame energy density. 
However, our model of the QGP fluid is dissipative.
Since hydrodynamics does not provide any microscopic information on how the system evolved to this surface, we are left with a large and irreducible ambiguity as to the choice of local momentum distributions and chemical abundances of the different hadron species~\cite{Dusling:2009df, Molnar:2014fva, Damodaran:2017ior, Damodaran:2020qxx}.
The macroscopic dissipative currents $\pi^{\mu\nu}$ and $\Pi$ reflect deviations of the hadrons'\ microscopic momentum distributions and yields from local thermodynamic equilibrium. To specify these \textit{microscopic} deviations, one may want to require the distribution functions $f_i(x;p)$ to solve a set of coupled Boltzmann equations, but this requires specifying the collision terms and the initial conditions, both of which are expected to be strongly affected by the proximity of the hadronization process (a challenging process to describe microscopically). 

In such a situation of irreducible theoretical ambiguity, we will investigate whether the experimental data have sufficient information to discriminate between different theoretical models. Therefore, we will consider four different models of viscous corrections to the local equilibrium distribution functions when comparing with experimental data:
\begin{enumerate} 
    \item Grad's method ~\cite{Grad} (a.k.a. 14-moments method in the relativistic context ~\cite{Israel:1976tn, Israel:1979wp, Teaney:2003kp, Dusling:2009df, Monnai:2009ad, Dusling:2011fd, Denicol:2012cn});
    \item the first-order Chapman-Enskog (CE) expansion in the time-independent Relaxation Time Approximation ~\cite{chapman1990mathematical, 
    ANDERSON1974466, Jaiswal:2014isa};
    \item the Pratt-Torrieri-Bernhard (PTB) modified equilibrium distribution ~\cite{Pratt:2010jt, Bernhard:2018hnz};
    \item the Pratt-Torrieri-McNelis (PTM) modified equilibrium distribution ~\cite{Pratt:2010jt, McNelis:2019auj, McNelis:2021acu}.
\end{enumerate}

Given the same values of the energy-momentum tensor, these models each assume a different prescription to determine how energy and momentum are distributed among hadronic species and across momentum. 
By performing Bayesian inference using these possible models, we aim to estimate the theoretical biases in the estimation of the transport coefficients resulting from the viscous corrections at particlization. 
We briefly describe the four models individually. For a more in-depth review and comparison of these models we refer the reader to Ref.~\cite{McNelis:2019auj}.

\subsubsection{Linearized viscous corrections: Grad \& Chapman-Enskog}

The Grad and Chapman-Enskog methods have both been used extensively in hybrid models of heavy ion collisions. They give microscopic corrections which are linear in the dissipative stresses $\pi^{\mu\nu}$ and $\Pi$. Both ans\"atze depend on the smallness of these corrections to the thermal equilibrium distribution. In practice this approximation is often pushed to the limit or beyond. In the following we describe the Grad and Chapman-Enskog methods in turn. We then discuss regularization that is applied similarly to both approaches when large viscous corrections are encountered.

\paragraph{Grad (or 14-moments) approximation:}
What we refer to as ``Grad's method'' assumes that the correction to the local equilibrium distribution function can be expanded in powers of hadronic momentum. Including only the terms relevant for a system without conserved charges yields 
\begin{equation}
    \delta f_i =  f_{\text{eq}, i}  \bar{f}_{\text{eq}, i} c_{\mu\nu}p^{\mu}p^{\nu}
\end{equation}
where $\bar{f}_{\text{eq}, i} \equiv 1{-}\Theta f_{\text{eq}, i}$, and $\Theta$ is 1 for fermions and ${-}1$ for bosons. Assuming that the coefficients $c_{\mu\nu}$ are species-independent and requiring the Landau-matching conditions yields the following expression for the viscous correction in terms of the dissipative stresses: 
\begin{eqnarray}
    \delta f^{\text{Grad}}_i = f_{\text{eq}, i} \bar{f}_{\text{eq}, i} \bigl[ \Pi \left( A_{T}m_i^2{+}A_E(u{\cdot}p)^2  \right) 
    + A_{\pi}\pi^{\mu\nu}p_{\langle \mu}p_{\nu \rangle}  \bigr]&&. 
\end{eqnarray}
Here $A_{T}$, $A_{E}$, and $A_{\pi}$ are functions only of spacetime and are combinations of thermodynamic moments of the equilibrium distribution described in Ref.~\cite{McNelis:2019auj}, $m_i$ is the mass of the hadron species $i$, $p_{\langle \mu}p_{\nu \rangle} \equiv \Delta_{\mu\nu}^{\alpha\beta}\, p_\alpha p_\beta$, and $\Delta_{\mu\nu}^{\alpha\beta}$ is defined in Eq.~(\ref{eq:delta_munu_alphabeta}). 

\paragraph{Linearized Chapman-Enskog expansion in the relaxation time approximation (CE RTA):}
The Chapman-Enskog (CE) expansion is a method to solve the Boltzmann equation by expanding in Knudsen number. Although this series can be written down for a more general collision kernel, we introduce the expansion assuming the simpler relaxation-time
approximation (RTA) ~\cite{Bhatnagar:1954zz, ANDERSON1974466},
\begin{equation}
    p^{\mu}\partial_{\mu}f = -\frac{u \cdot p}{\tau_r} (f - f_{\text{eq}}),
\end{equation}
where $f_{\text{eq}}$ is the local equilibrium distribution function, and the relaxation time $\tau_r$ is assumed to be species- and momentum-independent. Expanding the distribution function in the Chapman-Enskog series around local equilibrium, and keeping only the first-order correction, one finds
\begin{equation}
    f = f_{\text{eq}} - \frac{\tau_r}{u \cdot p} p^{\mu}\partial_{\mu}f_{\text{eq}} + \mathcal{O}\left(\partial^2 \right).
\end{equation}
Using the zeroth order conservation laws to rewrite derivatives of the temperature and flow velocity, as well as the Navier-Stokes relations $\Pi = -\zeta \theta $ and $\pi^{\mu\nu} = 2 \eta \sigma^{\mu\nu}$ we finally obtain
\begin{equation}
    \delta f^{\text{CE RTA}}_i = f_{\text{eq}, i} \bar{f}_{\text{eq}, i}
    \left[  \frac{\Pi}{\beta_{\Pi}}\left( \frac{(u{\,\cdot\,}p) \mathcal{F}}{T^2} - \frac{p{\,\cdot\,}\Delta{\,\cdot\,}p}
    {3(u{\,\cdot\,}p)T}\right) \right.\  
    \left. + \frac{ \pi_{\mu\nu}p^{\langle \mu}p^{\nu \rangle}}{2\beta_{\pi}(u \cdot p)T} \right].\quad
\end{equation}
Again we refer to Ref.~\cite{McNelis:2019auj} for the definitions of $\mathcal{F}$, $\beta_\pi$ and $\beta_\Pi$.

\paragraph{Handling large viscous corrections:}
The Grad and Chapman-Enskog momentum distributions discussed above assume $|\delta f| \ll f_{\rm eq}$. The viscous correction $\delta f$ scales linearly with the shear stress $\pi^{\mu\nu}$ and the bulk viscous pressure $\Pi$. It also scales either quadratically or linearly with the hadron four-momentum $p$. There are thus values of $\pi^{\mu\nu}$ and $\Pi$ for which $|\delta f| > f_{\rm eq}$ even for moderate (thermal) momenta. Moreover, even for small values of $\pi^{\mu\nu}$ and  $\Pi$, $|\delta f| > f_{\rm eq}$ at sufficiently large momenta.

In hydrodynamic simulations of heavy-ion collisions it is thus not uncommon to encounter $|\delta f| > f_{\rm eq}$ in certain phase-space regions. Even though these regions are usually small enough to not contribute significantly to experimental observables, from a practical point of view one needs to specify a hadronic momentum distribution even when $|\delta f| \sim f_{\rm eq}$.
This is commonly achieved by regulating the Grad or Chapman-Enskog viscous corrections to prevent $|\delta f| > f_{\rm eq}$. In this work this is achieved locally by setting 
\begin{equation}
    \delta f \rightarrow \sign(\delta f) \min(f_{\rm eq}, |\delta f|)
\end{equation}
in every cell. 
The need for regulation of the linearized viscous corrections has motivated models that attempt to resum the viscous corrections to all orders. We now discuss two such prescriptions. 

\subsubsection{Exponentiated viscous corrections: Pratt-Torrieri-McNelis and Pratt-Torrieri-Bernhard}

The approaches described in this subsection rely on the development of positive definite ``modified equilibrium'' distributions ~\cite{Pratt:2010jt, Bernhard:2018hnz, McNelis:2019auj}. These models are designed to include the effects of the viscous pressures in the argument of an exponential function with similar structure to the local-equilibrium distribution. 

\paragraph{Pratt-Torrieri-McNelis (PTM):}
The Pratt-Torrieri-McNelis (PTM) distribution ~\cite{Pratt:2010jt, McNelis:2019auj, McNelis:2021acu} is defined as follows:
\begin{equation}
    f_{\text{PTM}} = \mathcal{Z} \left[ \exp\left(\frac{ \sqrt{|\mathbf{p}'|^2 + m^2} }{T + \beta_{\Pi}^{-1} \Pi \mathcal{F}}\right) + \Theta \right]^{-1}.
\end{equation}
Here the spatial momentum components have been transformed as $p_i = A_{ij}p'_j$ where
\begin{equation}
    A_{ij} \equiv \left(1+ \frac{\Pi}{3\beta_{\Pi}}\right)\delta_{ij} + \frac{\pi_{ij}}{2\beta_{\pi}}.
\end{equation}
The PTM ansatz has the feature that expanding to first order in the dissipative currents yields the usual linear Chapman-Enskog viscous correction discussed above. The yield of each hadron is corrected from its equilibrium yield by a scaling factor $\mathcal{Z}$, which depends on the bulk viscous pressure as well as the hadron mass as described in Ref.~\cite{McNelis:2019auj}. We note that the yield of each hadron in this method is fixed to exactly reproduce the yields given by the linearized CE RTA method. For this reason, we find that both models yield very similar $p_T$-integrated observables.

\paragraph{Pratt-Torrieri-Bernhard (PTB):}
The Pratt-Torrieri-Bernhard (PTB) distribution ~\cite{Pratt:2010jt, Bernhard:2018hnz} is defined by
\begin{equation}
    f_{\text{PTB}} = \frac{\mathcal{Z}_{\Pi}}{\text{det} \Lambda} \left[\exp\left(\frac{ \sqrt{|\mathbf{p}'|^2 + m^2} }{T}\right) + \Theta \right ]^{-1},
\end{equation}
where $\mathcal{Z}_{\Pi}$ is a scaling factor described in Ref.~\cite{Bernhard:2018hnz, McNelis:2019auj}, which again depends on the bulk viscous pressure, but is species-independent. $\Lambda$ is a momentum-transformation matrix operating on the spatial momentum components as $p_i = \Lambda_{ij}p'_j$ with
\begin{equation}
    \Lambda_{ij} \equiv \left(1+ \lambda_{\Pi}\right)\delta_{ij} + \frac{\pi_{ij}}{2\beta_{\pi}}.
\end{equation}
In particular, $\lambda_{\Pi} \neq \Pi/(3\beta_{\Pi})$; instead, this quantity is adjusted such that the total isotropic pressure and energy density of the system are matched. This method parametrizes the effect of the bulk viscous pressure on the particle yields and momentum spectra, and it does not reduce to the linear Chapman-Enskog correction in the limit of small $\Pi$. Moreover, because the factor $\mathcal{Z}_{\Pi}$ in this method is assumed to be species independent, the ratios of hadronic abundances are not corrected from the equilibrium ratios. 

The PTB distribution was used in several recent Bayesian analyses~\cite{Bernhard:2019bmu, Nijs:2020ors, Nijs:2020roc}. It should be noted that, in contrast to the (unregulated) linearized Grad and Chapman-Enskog distributions, for both PTB and PTM distributions, the matching condition (\ref{Tmunu_kinetic}) is not satisfied exactly when the viscous stresses are large ~\cite{McNelis:2019auj}. The slight matching inconsistencies introduced by the different regulation schemes discussed above were quantitatively studied in ~\cite{McNelis:2019auj} and found to be acceptable in practice. For other approaches to regulate the viscous corrections to the distribution functions during particlization we refer the interested reader to Refs.~\cite{Romatschke:2003ms, Romatschke:2004jh, Martinez:2012tu, Florkowski:2013lya, Florkowski:2014bba, Tinti:2015xwa, Molnar:2016vvu, Tinti:2018nrp, Tinti:2018qfb}.


\subsubsection{Maximum Entropy viscous corrections}

In Ref.~\cite{Everett:2021ulz} we introduced an alternative particlization prescription to those described above. The key idea was to employ the Maximum Entropy principle given by Jaynes~\cite{Jaynes:1957zza} as an unbiased prescription for the microscopic hadronic distributions $f_i(x;p)$ given only the macroscopic moments $T^{\mu\nu}$. The method requires the specification of the entropy density current $s^{\mu}(x)$,

\begin{equation}
\label{eqn:entropy_current}
    s^{\mu}(x) = - \sum_i \frac{g_i}{(2\pi)^3}\int \frac{d^3p}{p_0} p^{\mu} \phi[f_i],
\end{equation}
The function $\phi[f]$ depends on the quantum-statistical nature of the particles and is defined by
\begin{equation}
\label{eqn:phi}
    \phi[f] \equiv f \ln (f) - \frac{1 + \theta f}{\theta} \ln(1 + \theta f),
\end{equation}
where $f = f(x;p)$ is the one-particle distribution function, $p$ the momentum four-vector, and $x$ the position four-vector.
The distributions $f_i$ that maximize the entropy density in the local rest frame $u \cdot s$, subject to the matching conditions of the full stress tensor, were found by the usual variational method with Lagrange multipliers. 
The details and predictions of this method can be found in Ref.~\cite{Everett:2021ulz}. Because this method requires additional work to be computationally fast enough for phenomenology, 
we won't be able to compare this method to the four other methods described above in a meaningful way throughout the rest of this thesis. However, I hope that this method can be useful in future investigations.\footnote{%
I was informed that this prescription may also be useful for sampling the miscroscopic distributions of partonic degrees of freedom at earlier times/higher temperatures of the medium, which is necessary for models describing jet propagation and electrodynamic emissions.}

\section{Hadronic transport}
\label{ch3:hadronic_transport}

In our hybrid model we transition to microscopic hadronic Boltzmann dynamics, simulated with the kinetic evolution code \SMASH{} ~\cite{Weil:2016zrk,smash_code}, by imposing particlization at the switching temperature $\Tsw$ as described above. After particlization of the fluid, the resulting hadrons are allowed to scatter, form resonances, and decay. \SMASH{} solves a tower of coupled Boltzmann equations for a system of hadronic resonances:
\be
    p^{\mu} \partial_{\mu} f_i(x;p) = C[f_i],
\ee
where $f_i$ is the distribution function for hadronic species $i$ and $C[f_i]$ is the collision term describing all scattering, resonance formation, and decays involving particle species $i$.

Past phenomenological studies~\cite{Nonaka:2006yn, Hirano:2007ei, Petersen:2008dd, Song:2010aq, Heinz:2011kt, Song:2013qma, Zhu:2015dfa, Ryu:2017qzn} have found inclusion of a hadronic afterburner improves the model's description of spectra of heavier hadronic states, such as protons. This transport approach allows different species to reach chemical and kinetic freezeout dynamically. This contrasts with other approaches where chemical and kinetic freezeout are enforced at specific temperatures.\footnote{%
    For example, the partial chemical equilibrium approach ~\cite{Hirano:2002ds} enforces chemical freezeout at a given temperature in ideal hydrodynamics, by introducing chemical potentials to conserve all hadronic multiplicities to a chosen chemical freeze-out values. This was a popular procedure before the widespread availability of hybrid codes (see e.g. ~\cite{Song:2010aq, Heinz:2011kt} for comparisons of these two approaches).}
At particlization, the momentum distributions and particle yields already deviate from their equilibrium relations at that temperature due to shear and bulk viscous stresses. After switching to the afterburner, they continue to evolve until yields (chemical freezeout) and momentum distributions (kinetic freezeout) cease changing. Most hadronic yields vary by less than 20\% as a consequence of inelastic collisions in the afterburner phase, and the particlization temperature $\Tsw$ is therefore sometimes associated with a chemical freeze-out temperature~\cite{Song:2010aq}. However, baryon and anti-baryon yields may change more significantly, due to the large annihilation cross section~\cite{Bass:2000ib,Steinheimer:2012rd}. 


We note that none of the parameters in the \SMASH{} afterburner are varied in this work. We did validate, however, that the afterburner used in this work (\SMASH{}) agrees well with the popular \URQMD{} implementation which has been used extensively in the past. This comparison is discussed in Appendix \ref{app:smash}. 

\paragraph*{Treatment of the $\sigma$ meson:}

At particlization, the hydrodynamic energy-momentum tensor is converted into hadrons while assuming the system has the thermodynamic properties of a hadron resonance gas. Though the $\sigma$ meson can be formed as a resonance in the $\pi+\pi$ scattering channel, it has been shown in Ref.~\cite{Broniowski:2015oha} that the contribution to the partition function from $\sigma$ meson exchange is almost perfectly canceled by a repulsive channel in $\pi+\pi$ scattering. Based on this observation, usually the $\sigma$ meson is be omitted from isospin-averaged hadron resonance gas models ~\cite{Broniowski:2015oha}. This is the approach used in this work: The $\sigma$ meson is \emph{not} sampled at particlization, and correspondingly it is also omitted in the construction of the equation of state in the hadronic phase.\footnote{%
    More details about the construction of the equation of state are provided in Appendix \ref{appendix:eos}. The physical effects on observables from excluding the $\sigma$ meson from the hadron gas are studied in Appendix \ref{app:sigma_effect}.}
In the hadronic afterburner, we still allow \SMASH{} to dynamically form and decay $\sigma$ resonances because they are an essential ingredient in fitting the $\pi+\pi$ cross section in \SMASH{}.  We note for reference that the Bayesian analysis in Ref.~\cite{Bernhard:2019bmu} did include the $\sigma$ meson in both the sampling at particlization and the construction of the hadronic equation of state, making this one of the potentially large differences with the current analysis.

\section{Physical model simulator}
\label{ch3:physics_simulator}

This section is included to provide an overview of the physical model simulator that is used for Bayesian model-data comparison in later chapters. 
Given a point in the multidimensional parameter space $\bm{x}$, the model simulator does the following:

\begin{enumerate} 
    \item generates $2,500$ minimum bias \trento{} initial conditions according to the parameters $\bm{x}$; each initial condition defines a different event 
    \item propagates each \trento{} event with freestreaming according to the parameters $\bm{x}$
    \item propagates each event with viscous hydrodynamics according to the parameters $\bm{x}$
    \item particlizes each event's hydrodynamic switching surface with temperature fixed by $\bm{x}$ into a hadron gas according to a specific particlization model (Grad, CE RTA, PTM or PTB)
    \item propagates the hadronic rescatterin and decays in each event with SMASH
    \item defines centrality classes for each event by ordering the $2500$ minimum bias events according to $dN_{\rm ch}/d\eta$
    \item computes observables in each centrality class, with the same centrality bins used by the experimental data 
    
\end{enumerate}
\chapter{Methods in model emulation and posterior inference}
\label{ch4}

In this section we describe the statistical and numerical methods used to perform Bayesian inference for our heavy-ion model in comparison with the experimental data. 
This problem is tackled by a physical model surrogate or ``emulator'', which can be necessary when the physical simulation is computationally intensive. Performing a Bayesian inference requires evaluating the  model's prediction on arbitrary points in the relevant region of the parameter space. 
The model simulator described in Ch.~\ref{ch3:physics_simulator} can take $\gtrsim 1000$ CPU-hours to make centrality-averaged predictions at a single point in parameter space, which is far too slow to allow the direct simulation of millions of points in the parameter space. The model surrogate or `emulator' is designed to tackle this problem.

The emulator can be understood as a computationally fast interpolator of the physical model simulator, which includes an estimate of the interpolation uncertainty. The model simulator is evaluated on a finite sample set of points in the parameter space, and the model simulator's predictions at these points are used to infer the predictions at other points in parameter space. Such an emulator dramatically reduces the numerical cost of estimating the posterior. 
However, our emulators, which employ Gaussian processes, introduce an additional source of predictive uncertainty. In addition to describing Bayesian parameter estimation in general, we also discuss specifically the design of the emulator. The discussion in this section presumes familiarity with Refs.~\cite{Petersen:2010zt, Novak:2013bqa, Sangaline:2015isa, Bernhard:2015hxa, Bernhard:2016tnd, Moreland:2018gsh, Bernhard:2019bmu} where many of these techniques were previously applied to Bayesian parameter estimation in relativistic heavy-ion physics. Since many aspects of these methods have already been explained in those articles, the attention in the proceeding sections is focused on methods and potential pitfalls which may have been previously unexplored or unstated. 

\section{Physical model emulator} 
\label{ch4:emu}

Throughout this study, we define an emulator as a map from a point in the multidimensional parameter space to the mean vector and covariance matrix of the distribution of all the predicted model observables of interest. Because we use Gaussian processes, this map provides a \textit{non-parametric} estimation of the physical model simulator predictions at arbitrary points in the region of the parameter space of interest. The Gaussian process is a non-parametric statistical model because predictions at novel points in parameter space are not made by constructing explicit functional interpolations, but rather by modeling the \textit{correlations} between predictions as a function of their parameters. The sample of points in parameter space where we know the physical model simulator predictions are called the \textit{design points} ($\boldsymbol{x}_i; i=1,\dots, m$) or \textit{training set}. 

The parameter design samples have been chosen using the Latin hypercube sampling technique, which \textit{uniformly} fills the volume of parameter space, while maximizing the distance between adjacent points. For models with sufficient smoothness, the number of design points necessary to achieve a certain level of prediction accuracy is expected to scale linearly with the dimension of the parameter space\footnote{%
This scaling of interpolation uncertainty with design size is explored in Ref.~\cite{Nijs:2020roc} for a different set of observables.}
~\cite{Loeppky}. In this work we have evaluated a Latin hypercube design of $500$ points. The number of design points was selected based on the expected similarities between this analysis and Ref.~\cite{Bernhard:2018hnz}, as well as considerations regarding the finite computing allocation. At each design point, the full model simulator runs $2500$ events for each collision system (see Ch.~\ref{ch3:physics_simulator}). The parameter design points and the physical model simulators outputs define the training set with which we will fit our model emulators, according to the following steps.

\subsection{Dimensionality reduction via Principal Component Analysis}
\label{ch4:pca}

When comparing the model simulator outputs with experimental data, we are faced with the large dimensionality of the outputs. Many of the model observables carry correlated information, and training an independent Gaussian process for each output, while possible, may be a waste of computational effort. 
As a simple example, increasing the normalization of the initial energy density increases the pion multiplicity in \textit{all} centrality bins. Therefore, the predicted pion multiplicity in different centralities is effectively tied to a single response. A small linear subspace of the full model simulator output carries nearly all of the information about the model parameters. Therefore, we apply ordinary principal component analysis as a dimensionality reduction method.

\begin{figure*}[!htb]
\noindent\makebox[\linewidth]{%
  \centering
  \begin{minipage}{0.33\textwidth}
    \includegraphics[width=\textwidth]{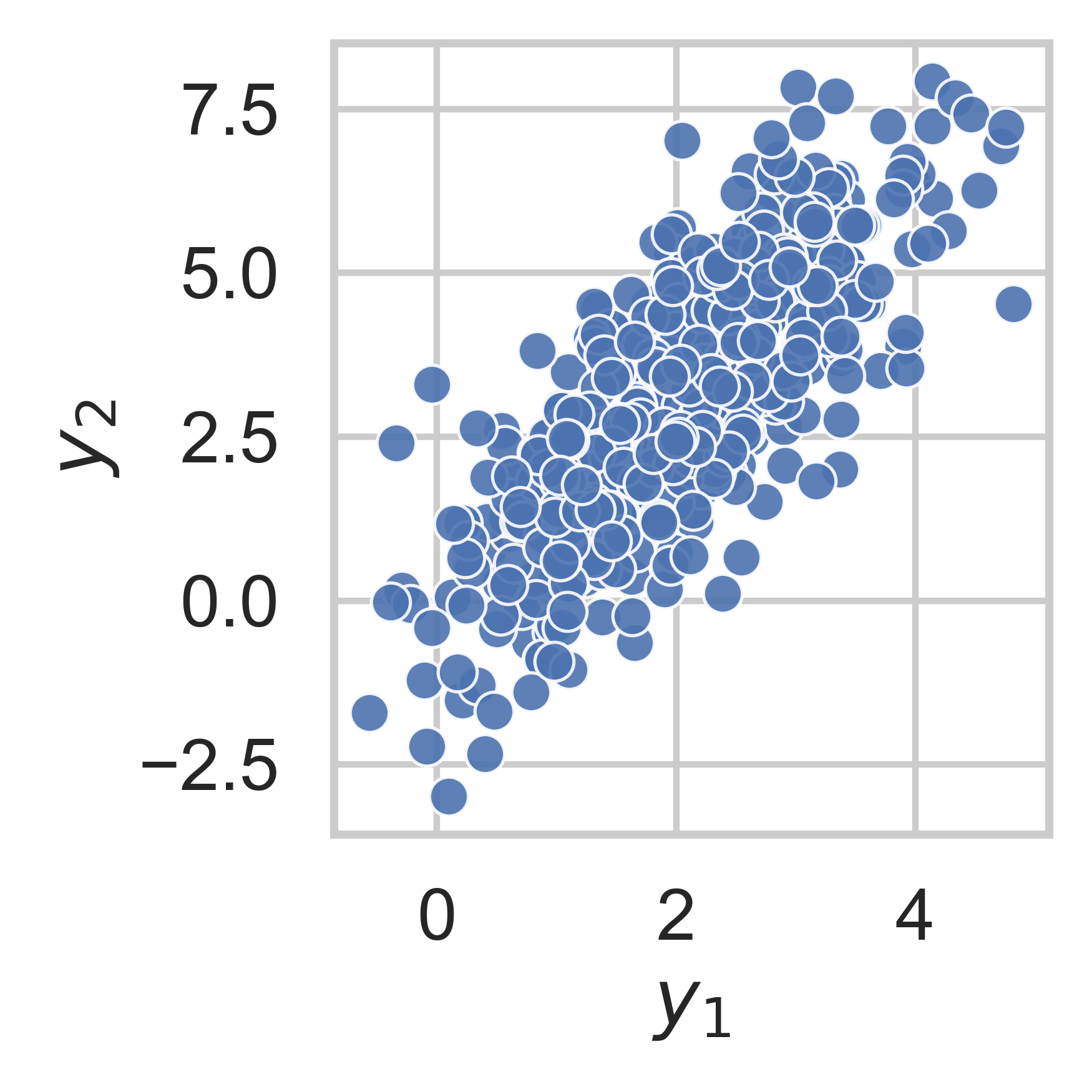}
  \end{minipage}
  \begin{minipage}{0.33\textwidth}
    \includegraphics[width=\textwidth]{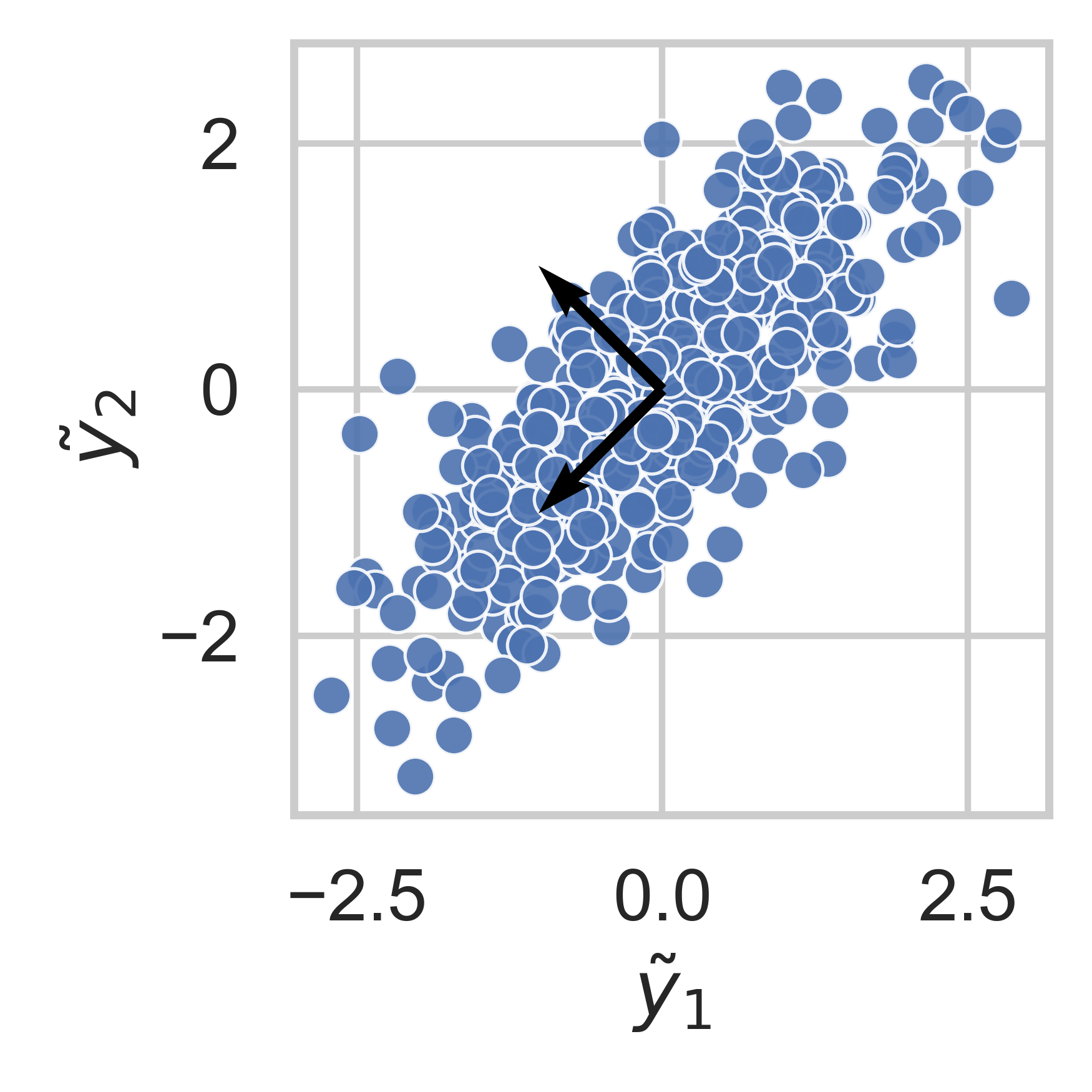}
  \end{minipage}
  \begin{minipage}{0.33\textwidth}
    \includegraphics[width=\textwidth]{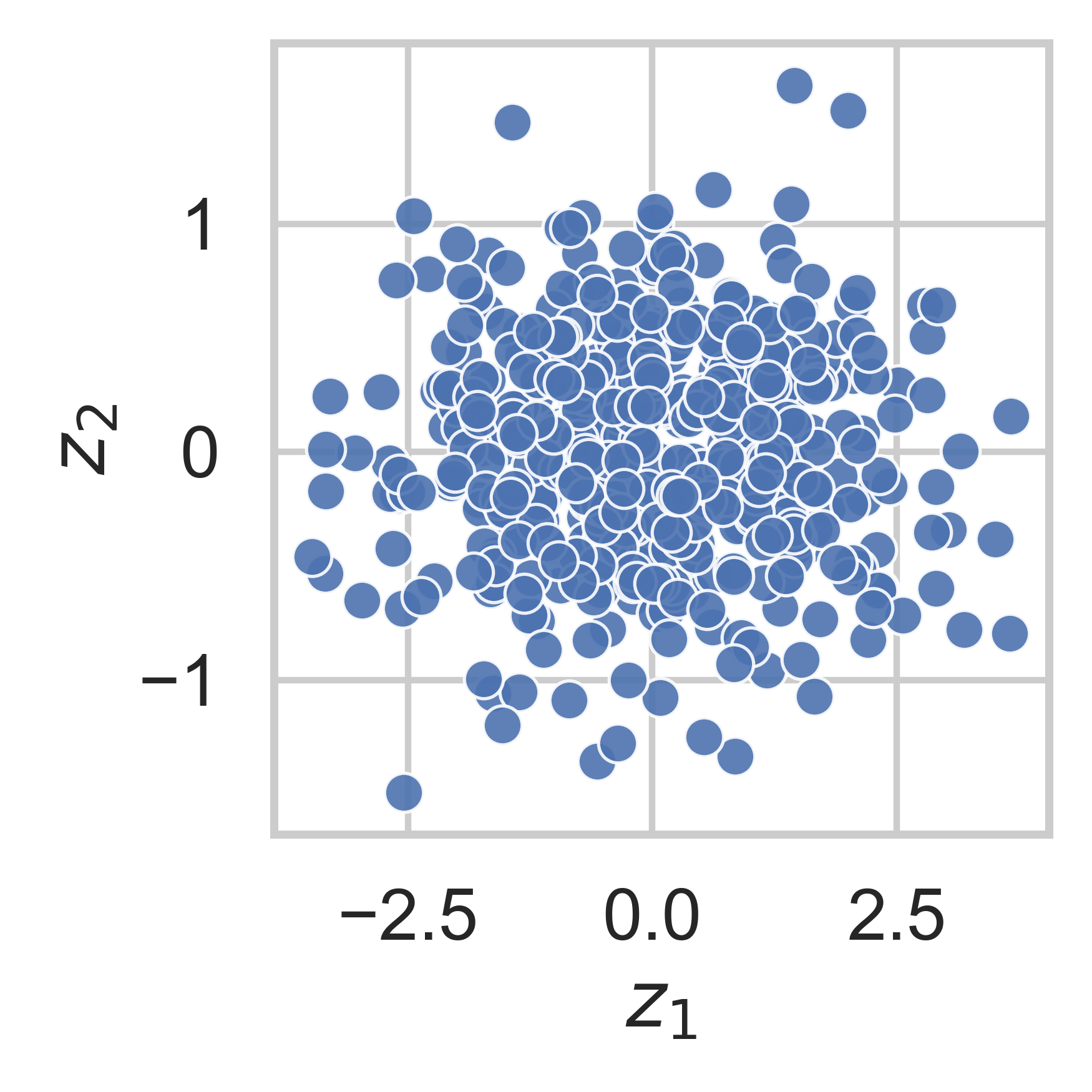}
  \end{minipage}
  }
  \caption{(Left) 500 samples drawn from a correlated bivariate Gaussian distribution of two variables $\{y_1, y_2\}$ with nonzero mean and different variances. (Center) The distribution of $\tilde{y}_1, \tilde{y}_1$, which have been standardized by subtracting their respective mean values and dividing by their variances. The black arrows denote the directions of the principal component vectors. (Right) The distribution of principal component values. }
  \label{fig:pca_visual}
\end{figure*}

Suppose an array of observations $y_i$ ($i=1,\dots,n$) are calculated at each of the $m=500$ design points $j$, and are organized as an $n\times m$ matrix $Y$ with elements $y_{ij}$. First, for each of the observables $y_i$, we compute its mean $\mu_i$  and standard deviation $\sigma_i$ over the sample of $m$ design points. Then, each of the $n$ observables is standardized by subtracting the mean and dividing by the standard deviation, yielding an $n\times m$ matrix $\tilde Y$ with elements $\tilde{y}_{ij} = (y_{ij}{-}\mu_j)/\sigma_j$ for $j=1,\dots, m$. Secondly, we define a new set of ``observables'' $z_{i}$ which are linear combinations of the standardized observables: $z_{i} = O_{ik}\tilde{y}_{k}$. 
In particular, it is desirable to construct the set of $z_i$ such that the linear correlations between different $z$-observables vanish:
\begin{equation}
    \langle\delta z_i \delta z_j\rangle = \frac{1}{m}\sum_{k=1}^m (O\tilde{Y})_{ik} (O\tilde{Y})_{jk} = \frac{1}{m} \bigl(O (\tilde{Y}\tilde{Y}^T) O^T\bigr)_{ij}
    = \lambda_i \delta_{ij} \equiv \textrm{diag}\{\lambda_1, \cdots, \lambda_n\}, 
\end{equation}
where $\delta z_i$ denotes the deviation of the $z_i$ from their mean. Therefore, the coefficients $O_{ij}$ that define $z_i$ are simply the elements of the orthogonal matrix that diagonalizes the covariance matrix of $\tilde{y}_i$. This optimized set of $z_i$ are called the principal components.

The rows of $O$ are organized such that the eigenvalues $\lambda_i$, which are the variances of the $z_i$, have a descending order in magnitude. In this way, each successive principal component explains less variance in the standardized observables. This allows us to reduce the standardized observable space to a much smaller subspace, which captures most of the information about the parameters.
This process of data standardization and PCA is visualized for a set of pseudo-data $y_1$ and $y_2$ in Fig.~\ref{fig:pca_visual}. For the visualization shown, the first principal component $z_1$ explains $\sim 90\%$ of the total variance. 

It is crucial to point out that ordinary principal component analysis can only remove linear correlations among observables. Thus it is important to check that there are no significant non-linear correlations. This is demonstrated in \Appendix{pca_valid} for a subset of observables used in Ch.\ref{ch5} and Ch.\ref{ch6}. If there are significant non-linear correlations among outputs, firstly one can seek a \textit{non-linear transformation} of the outputs which results in purely linear correlations between the transformed observables. Alternatively, there are other methods designed for non-linear dimensionality reduction, such as Kernel PCA, auto-encoders, etc...~\cite{vanderMaaten}. 

In our experience, a very small fraction of the total number of principal components is generally sufficient to capture most of the model observables' dependence on the parameters. This follows from the strong linear correlations present in many pairs of observables. Pairs of observables with stronger linear correlations carry less mutual information about the parameters; knowledge of one observable is nearly sufficient to know the value of the other. Gaussian processes are only trained on this subset of dominant principal components.
The omission of higher principal components also helps to prevent overfitting. Our model simulator centrality-averaged predictions have a stochastic scatter due to the finite number of events. PCA will tend to relegate these stochastic features in the outputs to higher PCs, which are not fit by Gaussian processes but added as a white-noise variance to the predictive uncertainty. 

\subsection{Interpolating principal components via Gaussian process regression}
\label{ch4:gp}
Each dominant principal component is interpolated with a unique Gaussian process. The spirit of a Gaussian process regressor is to infer the outputs of the target (scalar) function $y=M(x)$\footnote{In this context, the output of the target function is one of the dominant principal components.} by a distribution of functions denoted by $\mathcal{GP}$: $f(x) \sim \mathcal{GP}(\textrm{mean}(x), \textrm{cov}(x, x'))$. This distribution is assumed to be a multivariate normal distribution, and is specified by a mean $\mu(x)$ and a covariance $\textrm{cov}(x,x')$. The expectation value of the output at a given $x$ is
\begin{eqnarray}
    \langle  f(x) \rangle = \textrm{mean}(x),
\end{eqnarray}
and the correlation of the output between two independent inputs $x, x'$ is
\begin{eqnarray}
    \langle \delta f(x) \delta f(x') \rangle = \textrm{cov}(x, x'),
\end{eqnarray}
where $\delta f(x) = f(x) - \textrm{mean}(x)$.

To find the distribution of functions that emulates $M(x)$, one starts with a distribution that is completely agnostic to the target function $M(x)$. In this study this distribution, referred to as the unconditioned Gaussian process, is assumed to have mean $\mu(x) = 0$\footnote{%
    It can happen that near the boundaries of parameter space the model prediction for some principal component is nonzero. In this case it may be beneficial to include a non-zero mean function in the Gaussian Process. We do not explore this in this work.}
and a covariance function $k(x, x')$ (the so-called kernel function). A Gaussian process makes a prediction at $m_{\star}$ novel inputs $X_{\star}$ according to the correlations with known values that have been calculated at the $m$ training inputs $X$. Consistency requires that the joint distribution of outputs at both training and novel inputs is also multivariate normal with zero mean,
\begin{gather}
 \begin{bmatrix} 
 \mathbf{f}(X) \\ 
 \mathbf{f}(X_{\star} ) 
 \end{bmatrix}
 \sim
 \mathcal{N} 
 \left( 
 \mathbf{0},
  \begin{bmatrix}
   K( X, X ) & K( X, X_{\star} ) \\
   K( X_{\star}, X ) & K( X_{\star}, X_{\star} ),
   \end{bmatrix}
  \right)
\end{gather}
where $K( X, X_{\star} )$ is the $m \times m_{\star}$ matrix whose elements are composed of the pointwise covariances $k(\mathbf{x}_p, \mathbf{x}_q)$ between pairs of training points $\mathbf{x}_p$ and prediction points $\mathbf{x}_q$.
Then, one conditions the random vector $\mathbf{f}(X)$ on the training outputs $M(X)$ to obtain the probability distribution of $\mathbf{f}(X_*)$ given training data. 
The mean and covariance can be obtained by the properties of the multivariate normal distribution,
\begin{eqnarray}
\label{eq:conditioned-GP}
    \mathbf{f}(X_*) &\sim& \mathcal{GP}\left(\textrm{mean}(X_*), \textrm{cov}(X_*, X_*)\right)
\\
\label{eq:conditioned-mean}
    \textrm{mean}(X_*) &=& K(X_*,X)\left[K(X,X)\right]^{-1}M(X),
\\\nonumber
    \textrm{cov}(X_*, X_*) &=& K(X_*,X_*)\\
\label{eq:conditioned-cov}
    &-& K(X_*,X)\left[K(X,X)\right]^{-1}K(X,X_*).
\end{eqnarray}
Focusing on a single novel input, the predicted mean and standard deviation of the target function is $M(x_*)\approx \textrm{mean}(x_*) \pm \sqrt{\textrm{cov}(x_*, x_*)}$. 

One must choose a model for the kernel function $k(x, x')$; this is where domain knowledge about the behavior of the model as each parameter is varied enters, as well as assumptions regarding the model's smoothness.  
In this work, an independent kernel function $k(\mathbf{x}_p, \mathbf{x}_q)$ is assigned to each dominant principal component, and is given by the sum of a squared-exponential kernel $k_{\rm exp}(\mathbf{x}_p, \mathbf{x}_q)$ and white-noise kernel $k_{\rm noise}(\mathbf{x}_p, \mathbf{x}_q)$,
\begin{equation}
    k(\mathbf{x}_p, \mathbf{x}_q) = k_{\rm exp}(\mathbf{x}_p, \mathbf{x}_q) + k_{\rm noise}(\mathbf{x}_p, \mathbf{x}_q).
\end{equation}
The squared-exponential kernel is given by 
\begin{equation}
    k_{\rm exp}(\mathbf{x}_p, \mathbf{x}_q) = C^2 \exp \left( -\frac{1}{2} \sum_{i=1}^{s} \frac{|x_{p, i} - x_{q, i}|^2}{l_i^2} \right)
\end{equation}
where $C^2$ is the unknown auto-correlation hyperparameter. The index $i$ runs over all $s$ parameters, and each parameter is assigned an uncertain hyperparameter $l_i$. This length-scale $l_i$ controls the smoothness of the response of the principal component output to a change in the $i^{\rm th}$ parameter. 
The white-noise kernel is given by 
\begin{equation}
    k_{\rm noise}(\mathbf{x}_p, \mathbf{x}_q) = \sigma_{\rm noise}^2 \delta_{p, q}
\end{equation}
where $\delta_{p, q}$ is the Kronecker delta, while $\sigma_{\rm noise}$ is an uncertain hyperparameter controlling the amount of statistical spread present in the principal component. The white-noise kernel is present because our model calculations average over a finite number of initial conditions and a finite number of particles. 

All of the hyperparameters $C, l_i$ and $\sigma_{\rm noise}$ are assigned a possible window, and then simultaneously optimized inside this window such that they maximize the likelihood of fit of the Gaussian process to the training calculations. This likelihood includes a complexity penalty, to reduce the potential for overfitting.\footnote{This is implemented already in the scikit-learn GaussianProcessRegressor~\cite{sklearn_gp}, which is based on algorithm 2.1 in Ref.~\cite{10.5555/1162254}.} This procedure is automated, and performing emulator validation is necessary to check that each kernel function has hyperparameters which are not underfit or overfit~\cite{scikit-learn}.

\begin{figure*}[!htb]
\noindent\makebox[\textwidth]{%
  \centering
  \begin{minipage}{0.5\textwidth}
    \includegraphics[width=\textwidth]{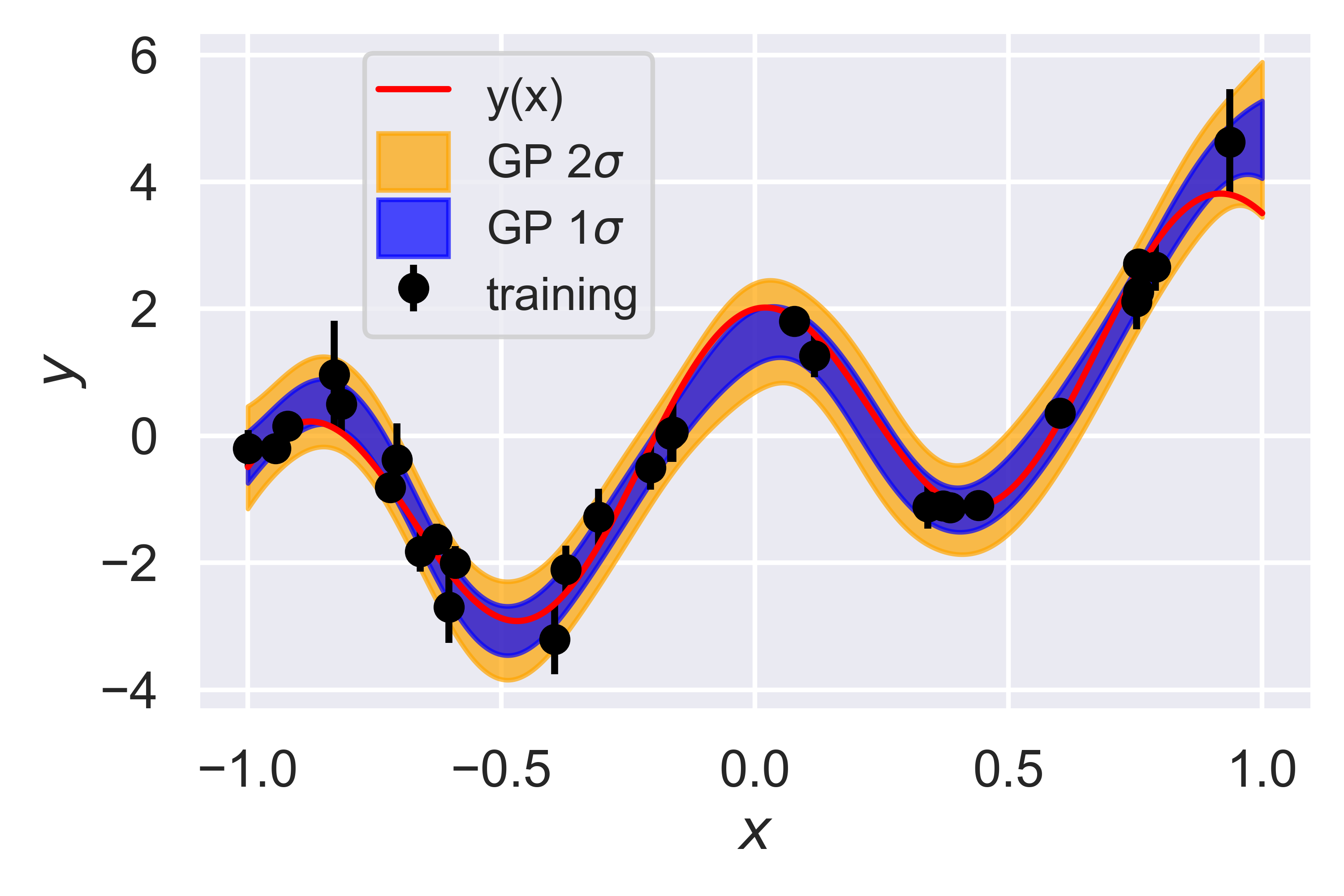}
  \end{minipage}
  \begin{minipage}{0.5\textwidth}
    \includegraphics[width=\textwidth]{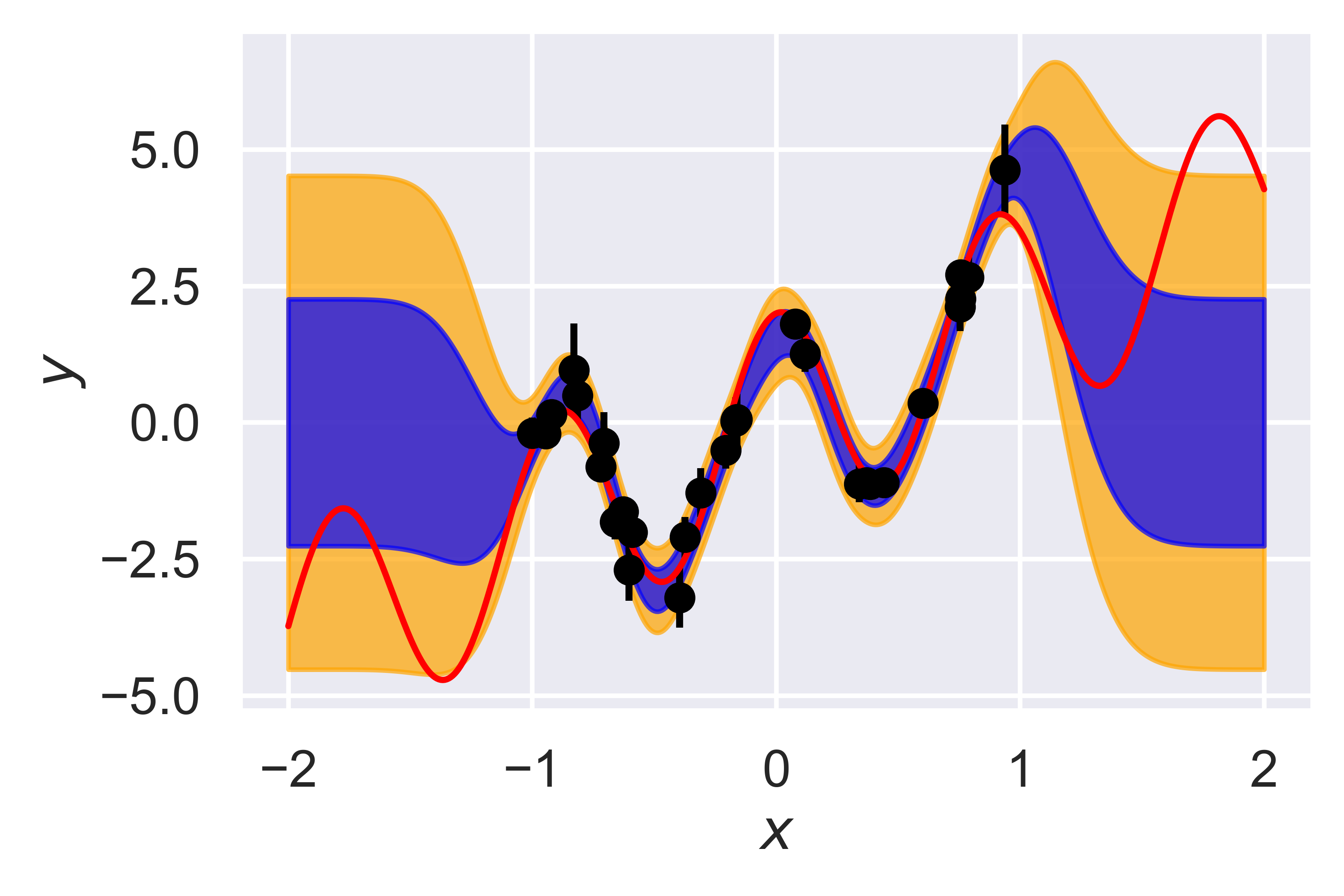}
  \end{minipage}
  }
  \caption{(Left)Training data (black) sampled from a `true' underlying function (red), together with the $1-\sigma$ (blue) and $2-\sigma$ (orange) Guassian process predicted credible intervals. (Right) The same as the plot at left, but on a wider range of input values $x$. }
  \label{fig:gp}
\end{figure*}

Some salient features of a Gaussian process regressor with a squared exponential and white noise covariance functions are shown in Fig.~\ref{fig:gp}. We have taken a trial function $y(x) = mx + A\cos(\omega x)$, and added to this `true' underlying function a homoskedastic error $\epsilon \sim \mathcal{N}(0, \sigma^2)$ to represent model statistical error. The true underlying function, without statistical error, is shown as a red line. We have evaluated the model simulation function (truth $+$ error) at fifteen randomly selected points in the range $[-1, 1]$. These training data are shown as black points in the figure, with error bars equivalent to the sampled statistical error. We have fit a Gaussian process regressor on these training data, and the $1-\sigma$ and $2-\sigma$ credible regions of the predictions are shown as blue and orange bands, respectively. In the right side of the figure, we plot the same functions and data on a wider range of input values $x$. This is to illustrate the behavior of the Gaussian process predictions far from the training data with only a \textit{local} covariance kernel (the squared-exponential kernel). The behavior is often called \textit{mean-reversion}; when the GP is asked to predict many correlation lengths away from any training data, the predictions are reverted to the mean of the Gaussian process prior function, which is zero in our case. In this instance, including either a non-zero mean function or non-local covariance kernel would be essential if we wanted to make robust extrapolations outside the training points.\footnote{See Ref.~\cite{gp_visual_exp, gp_kernel_cookbook, gp_widget} for more illustrative and interactive examples of Gaussian process regression as well as different kernel functions. } 

\subsection{Reconstructing the observables}
\label{ch4:reconst_obs}
The predictions for principal components are then grouped and transformed back into the observables via the inverse PCA transformation. Variances of those non-dominant principal components, for which we did not train Gaussian processes, are included as predictive uncertainty. Because these neglected principal components behave similarly to white noise, we replace them a constant and uncorrelated variance to propagate their contributed uncertainty. 
A more detailed description of the above procedure can be found in Ref.~\cite{Bernhard:2018hnz}. We note that our use of transverse-momentum-integrated observables, principal component analysis, and Gaussian process model emulation 
for heavy-ion collisions is very similar to the methodologies put forward in the seminal study Ref.~\cite{Novak:2013bqa}. 

\section{Treatment of uncertainties} 
\label{ch4:uncertainties}

We divide our uncertainties into three different sources: experimental uncertainties, interpolation and statistical model uncertainties, as well as systematic model discrepancies. 

\subsection{Experimental uncertainties}
\label{ch4:exp_uncertainties}
%
In general, experimental collaborations do not report the error covariance matrix between different observables. 
As such, we only have access to the systematic uncertainties of individual observables, with limited or no information on possible correlations. Assuming no correlations among the errors associated with the $n$ observables results in a diagonal covariance matrix for the experimental systematic covariance:
\begin{equation}
    \Sigma_{\textrm{sys}} = \textrm{diag}(\sigma_{\textrm{sys},1}^2, \cdots, \sigma_{\textrm{sys},n}^2).
\end{equation}

In principle, the systematic uncertainties have nonzero correlations. 
Without knowledge of the experimental covariance matrix we can only make assumptions regarding the form and magnitude of the correlations. We have tested the effect of this approach on the parameter posteriors in \Appendix{app_exp_cov}; however, we did not use this approach in general in the body of this work.
More generally, there is no guarantee that the systematic experimental errors in heavy-ion measurements are multivariate normal in nature; they may be described by different distributions, for example having heavier tails. This remains a significant and outstanding challenge. 

\subsection{Predictive uncertainties}
\label{ch4:pred_uncertainties}
%
The statistical uncertainty which is present in our model calculations results primarily from averaging over a finite number of fluctuating initial conditions, and to a lesser extent sampling a finite number of particles during particlization. These result in a statistical spread in each of the principal components (recall from Ch.~\ref{ch4:gp} that it is the principal components that are interpolated, not the individual observables). 

The total interpolation uncertainty is
\begin{eqnarray}
   \Sigma_{\text{interp}} = \Sigma_{\text{trunc.}} + \Sigma_{\text{GP}}.
\end{eqnarray}
The covariance $\Sigma_{\text{GP}}$ contains the total covariance of all the Gaussian Processes (one for each dominant principal component), including both interpolation and statistical uncertainties. The covariance $\Sigma_{\text{trunc.}}$ contains the total covariance of all the remaining principal components to which Gaussian processes were not fit and which were replaced by noise terms. 

\paragraph{Additional systematic model discrepancy:}
%
Our model of heavy ion collisions is imperfect; there exist additional sources of systematic discrepancy in our model when we use it to describe real physical observations. Quantifying and \textit{interpreting} the associated discrepancies presents a challenging problem~\cite{Brynjarsdottir_2014}. 

In Ref.~\cite{Bernhard:2019bmu} a parametrized systematic model discrepancy was included; this single uncertain parameter was included as a proxy for all systematic model discrepancies. The parameter was added in quadrature to the covariance matrix of the Gaussian process for each principal component, in the form of a diagonal matrix parametrized by $\sigma_m$. That is, to every principal component of the final state observables was added the same systematic uncertainty in percentage. This results in a complicated distribution of the uncertainty across observables, depending on the linear transformation from principal components to observables.
This type of discrepancy function, which is added only to the final state observables, also introduces challenges with interpretability. Given the posterior of this parameter, it isn't clear how to assign the discrepancy to any of the particular submodel components; for example, we do not know whether the problem lies with the initial conditions, the prehydrodynamic expansion, hydrodynamics, etc...

Because our model is multistage, and physical discrepancies/inadequacies can enter in any given substage of the dynamical evolution, an interpretable method would include model discrepancy functions as physical effects in each submodel.
As an example, suppose that we consider the conformal/non-conformal mismatch between our freestreaming and hydrodynamics models. This mismatch results in a positive initial bulk pressure, which we speculate causes a discrepancy between our model and a realistic expansion of QCD matter. To propagate this potential discrepancy, we could include a parametrized discrepancy function during pre-hydro/hydro matching conditions:
\be
    \Pi_0(x) \rightarrow \delta \cdot \Pi_0(x)
\ee 
where $\delta$ is a continuous parameter bounded by $[-1, 1]$, with a prior $\mathcal{P}(\delta|I)$ guided by physical considerations. Then, consider calibrating all model parameters, including $\delta$, against the observed data. The other model parameters will have posterior distributions marginalized over all $\delta$ consistent with the data and our prior, which naturally introduces a parametric source of uncertainty reflecting the deficiency of the model. Furthermore, the posterior of $\delta$ is likely to inform us regarding the sensitivity of our observables to this effect. Finally, we can check whether the inclusion of the discrepancy can improve the model's ability to describe the observed data, by plotting the model-data discrepancies for both models (the models with and without this discrepancy function). In this case, contrary to the method in Ref.~\cite{Bernhard:2019bmu}, we know exactly how to interpret the physical meaning of the discrepancy \textit{by design}. However, this method can not be introduced \textit{after} the simulator calculations have been performed, because the discrepancy function is acting as an extension of our simulator model which must be performed concurrently.

Motivating the sources of model discrepancy requires us to consider the specific outputs/observables which we desire to calibrate against.
In our case, all of the hadrons produced by our model arise from the hydrodynamically generated switching surface, but
non-hydrodynamic physical processes become increasingly important for describing the hadronic spectra at intermediate and larger values of transverse momentum $p_T \gtrsim 2$ GeV. 
To calibrate a \textit{hydrodynamic model} against such observables, because they should be described by non-hydrodynamic processes, can bias the results of our model calibration unless an attendant measure of theoretical predictive uncertainty is included. Throughout this manuscript we have only calibrated to the soft hadronic observables, integrated over $p_T$, which helps to minimize the impact of this particular source of model deficiency during calibration.

\section{Validation of model surrogate predictions} 
\label{ch4:emu_validation}

The entire emulation procedure, including the principal components reduction, their interpolation via Gaussian processes, and the recombination of all trained and untrained principal components into observables, can be validated using a set of validation points. 
Each model emulator is trained on a Latin hypercube design of five-hundred points uniformly filling the parameter space. An additional Latin hypercube of one-hundred (different) points was also generated, and the model simulator run, to generate a model validation set. We note that the model which was run on the validation points had fewer fluctuating events per design point; only $1,000$ fluctuating initial conditions were run at each validation point, compared with $2,500$ events for each training point. Therefore, the statistical uncertainties in the model simulator calculations are larger in the validation set then in the training set. 
This compromise was necessary given the finite computing resources, and in general a training set with the same magnitude of statistical scatter and statistical uncertainty would be preferred.

\begin{figure*}[!htb]
\noindent\makebox[\textwidth]{%
  \centering
  \hspace{-0.1\textwidth}
  \begin{minipage}{0.6\textwidth}
    \includegraphics[width=\textwidth]{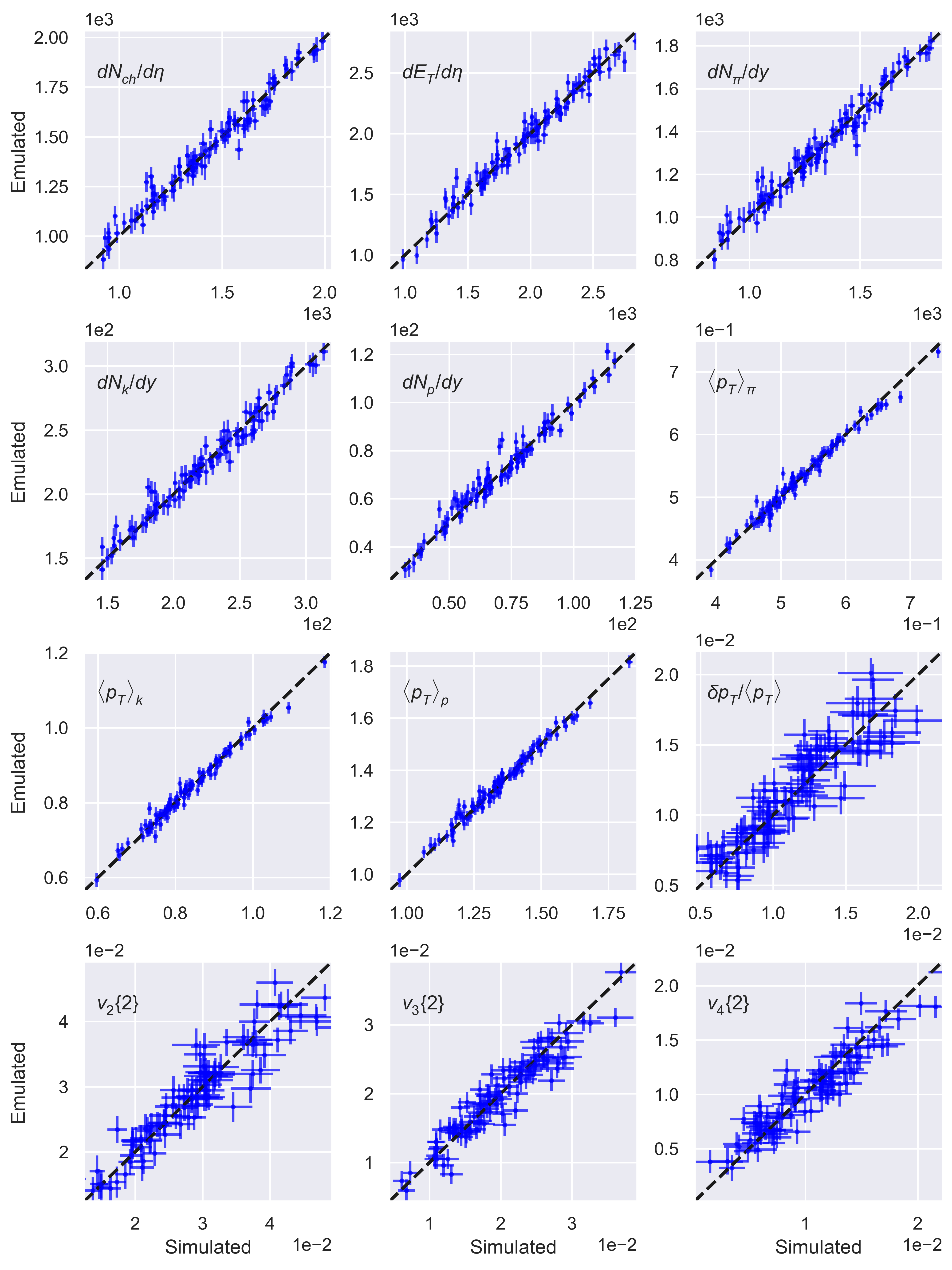}

  \end{minipage}
  \begin{minipage}{0.6\textwidth}
    \includegraphics[width=\textwidth]{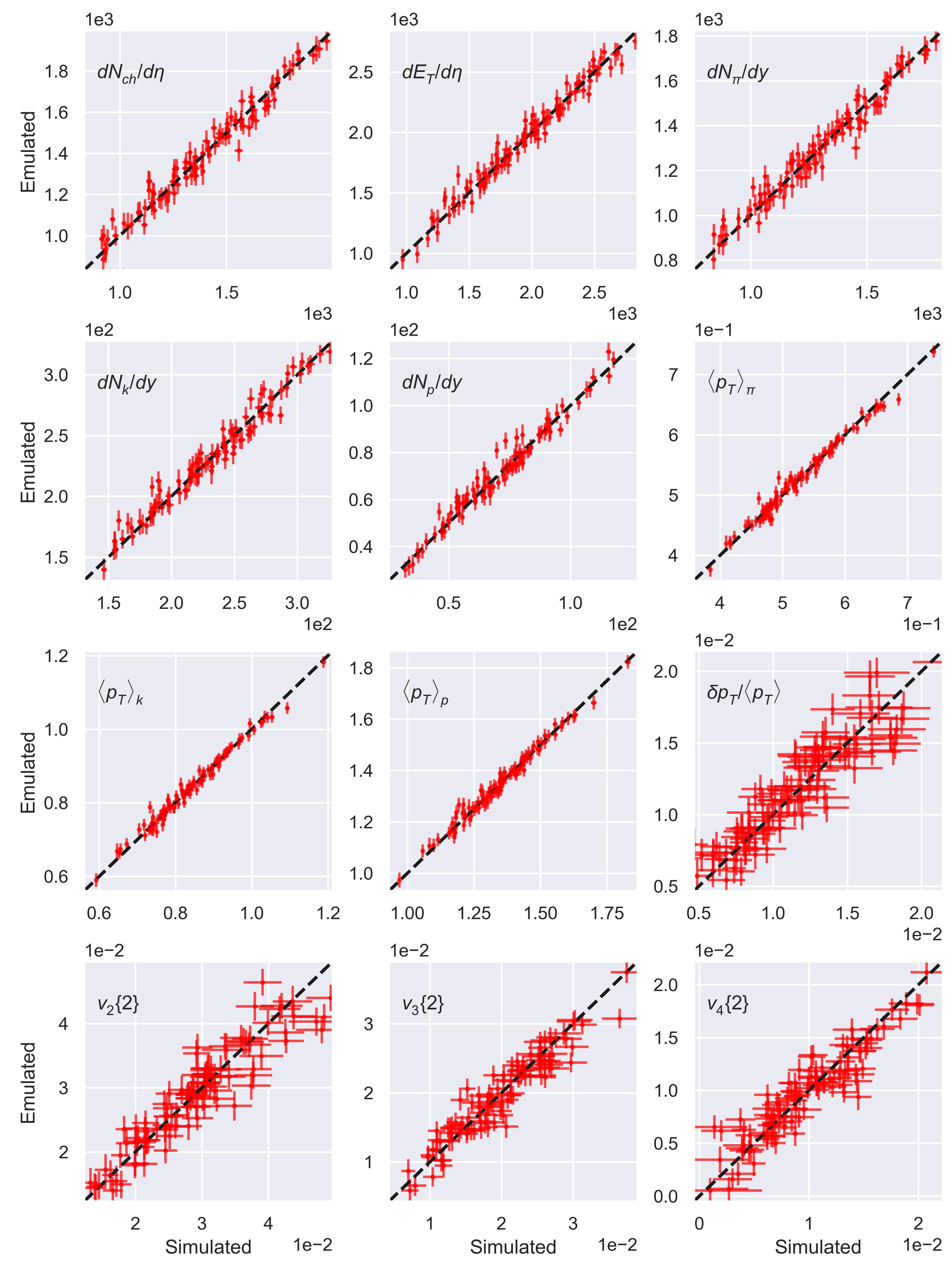}
  \end{minipage}
  }
  \caption{
  Diagnostics of the predictive accuracy and uncertainty of the Grad (left, blue) and Chapman-Enskog RTA (right, red) emulators for Pb-Pb $\sqrts{}=2.76$ TeV collision model. The Pratt-Torrieri-Bernhard model emulation is not shown, but was found to have similar performance.  }
  \label{ch4:emu_valid_scatter}
\end{figure*}

In Fig.~\ref{ch4:emu_valid_scatter} are shown scatter plots of the emulator predictions for the Grad and Chapman-Enskog RTA models at each of the one-hundred validation points, against the `true' simulator predictions, in the most central centrality bins. We again note that the `true' simulated calculations have statistical uncertainties. Moreover, we see that the mean-$p_T$ fluctuation and harmonic flow $v_n\{2\}$ observables are plagued by large statistical simulator uncertainties. This large statistical scatter, which was also present in the training calculations although to a slightly lesser degree, is manifested in large emulation uncertainties, as it should be. 

In general, the performance of the each emulator is good given the large statistical scatter in training and validation points. There does not appear to be any significant bias, and large predictive uncertainties are mostly a consequence of large simulator statistical uncertainties. We note that Ref.~\cite{Bernhard:2018hnz} also used 500 design points, but at each design point was averaged over a larger number (nearly $20,000$) of fluctuating initial conditions. The model simulator used in that work had been carefully optimized; more importantly, the afterburner which was employed, \texttt{UrQMD}, is significantly faster than \texttt{SMASH} ver. 1.7, used in this work. Moreover, we had to run four different models, for each of the four viscous correction models -- effectively scaling the runtime of our simulations by a factor of four.\footnote{The evolution of the initial conditions through freestreaming and hydro were reused among the four different viscous correction models, and only the Cooper-Frye sampling and afterburner \texttt{SMASH} needed to be run separately for each model. However, the runtime of \trento{}, \texttt{freestream-milne} and \texttt{MUSIC} were small compared to the runtime required for \texttt{SMASH} to run multiple oversampled events, even considering only a single viscous correction model. Therefore, the majority of the simulator runtime was occupied by \texttt{SMASH}. }

\section{Sampling the posterior} 
\label{ch4:mcmc}

For the models employed in this work, our posterior is often an 18-dimensional probability distribution.\footnote{There are usually 16 shared parameters and one additional parameter per collision system (the \trento{} normalization).} Estimation of the posterior is accomplished via Markov Chain Monte Carlo algorithms~\cite{Hogg:2017akh}. These algorithms are usually able to estimate the shape of the posterior without knowledge of its normalization. 
Efficient and accurate Markov Chain Monte Carlo algorithms are now readily available, thanks to their widespread use in other fields (e.g. in cosmology). This includes nested sampling, Hamiltonian methods, and parallel tempering~\cite{brooks2011handbook}, among others. In this work, we used an implementation of parallel tempering~\cite{Vousden_2015}; the algorithm showed good convergence in sampling our posterior, and at the same time made possible the estimation of the Bayesian evidence, discussed in Ch.~\ref{ch4:num_bayes_evidence}. While this algorithm does provide an estimate for the evidence, we will use this information only for performing model comparison; for parameter estimation it is always the unnormalized posteriors which are shown.

\begin{figure*}[!htb]
\noindent\makebox[\textwidth]{%
  \centering
  \hspace{-0.1\textwidth}
  \begin{minipage}{0.65\textwidth}
    \includegraphics[width=\textwidth]{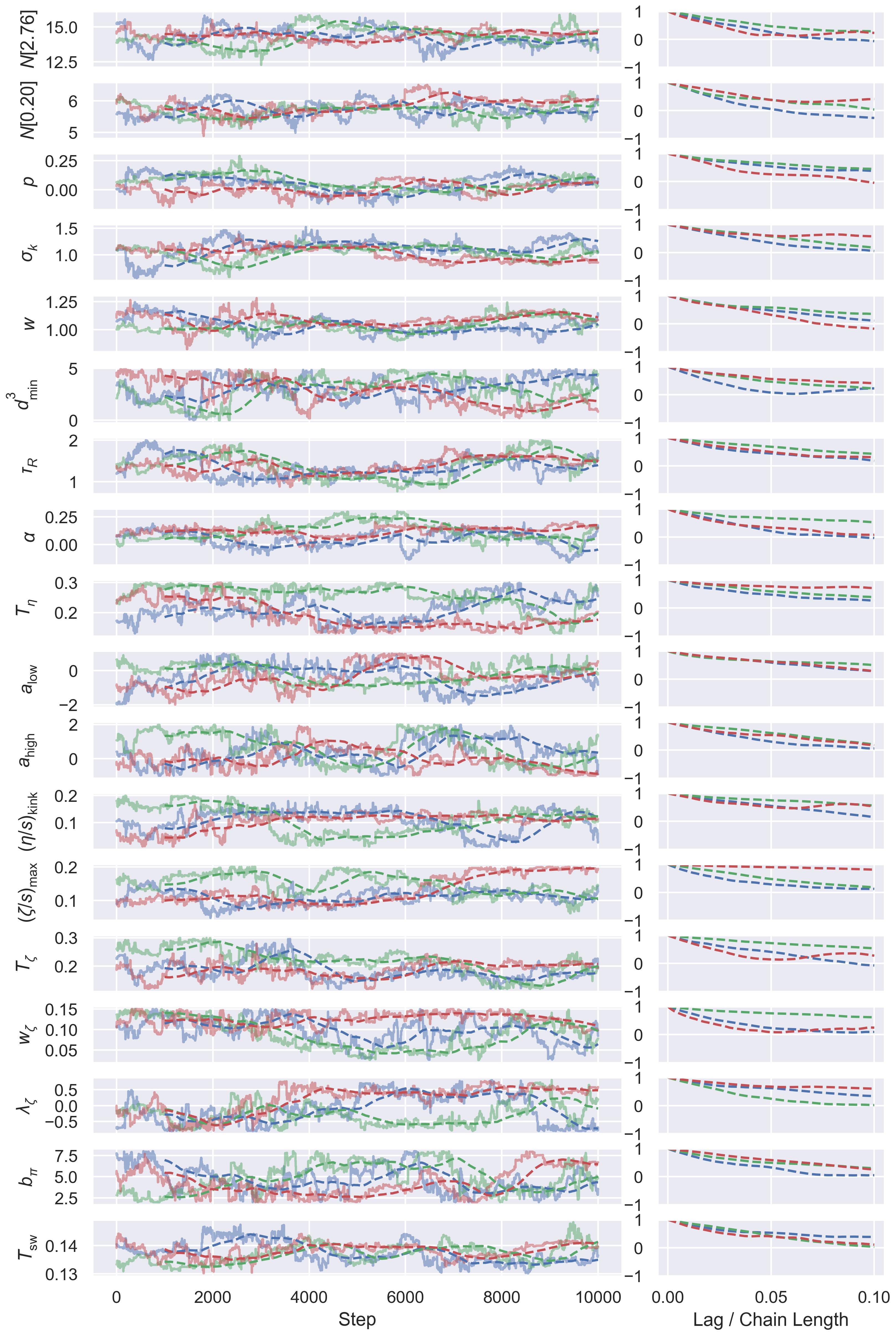}

  \end{minipage}
  \begin{minipage}{0.65\textwidth}
    \includegraphics[width=\textwidth]{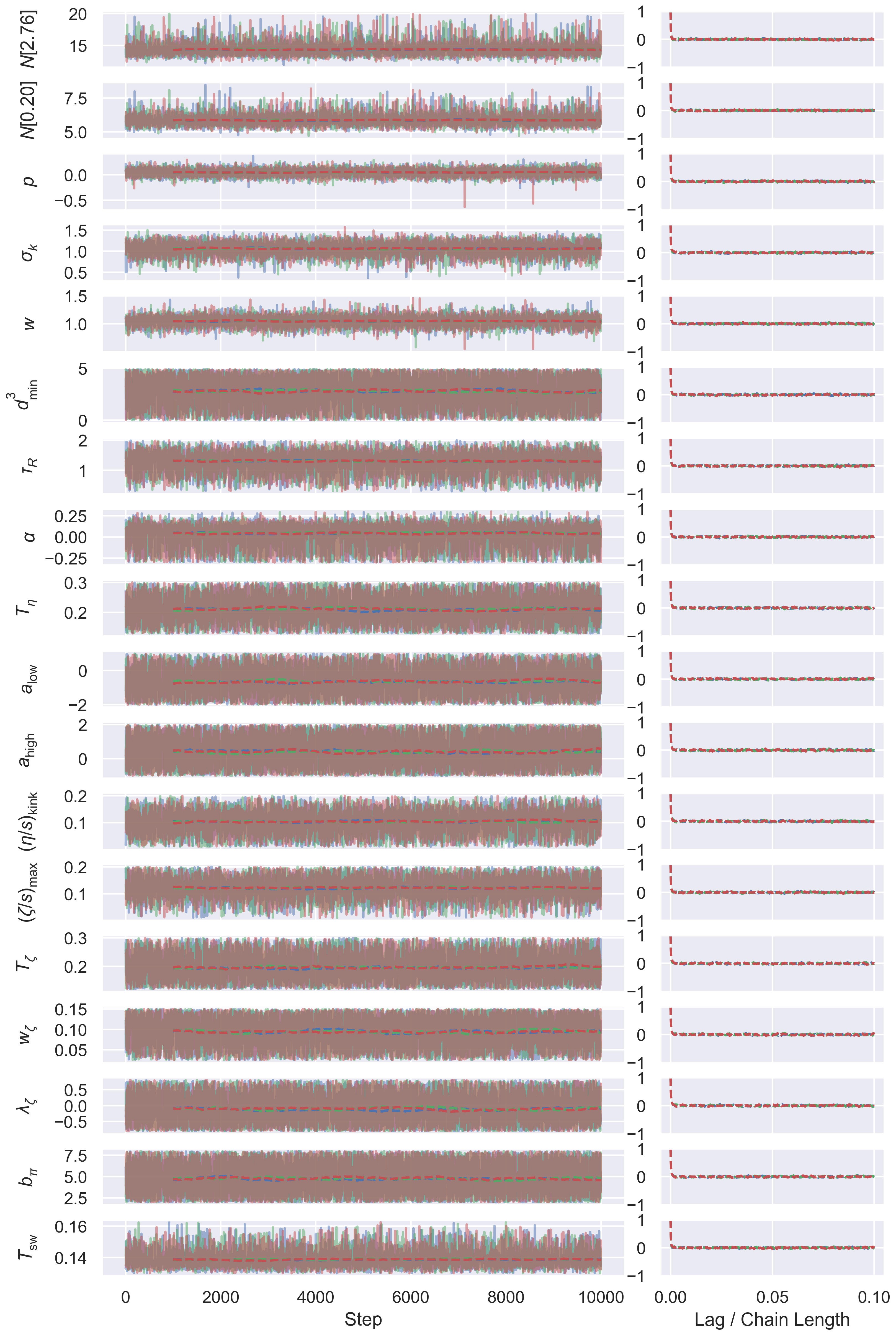}
  \end{minipage}
  }
  \caption{
  Diagnostics of the Markov chain monte carlo sampling with either the \texttt{emcee} ensemble sampler (left panel) or \texttt{ptemcee} parallel-tempered sampler (right panel). In each panel, the rows on the left are the projections onto each parameter the traces of three walkers (red, blue and green solid lines) and their thousand-step rolling averages (red, blue, and green dotted lines). The rows on the right are each walker's autocorrelation in units of the total chain length. }
  \label{ch4:mcmc_trace}
\end{figure*}

In Fig.~\ref{ch4:mcmc_trace} are shown the trace and autocorrelation plots of random walkers from the \texttt{emcee} ensemble sampler and \texttt{ptemcee} parallel-tempered sampler. 
The trace records each walkers trajectory through the parameter space. The autocorrelation $\rho(h)$ of a stationary series $\{s_t\}_{t=1}^{N}$ can be defined
\begin{equation}
    \rho(h) \equiv \frac{\text{Cov}(s_t, s_{t-h})}{\text{Var}(s_t)}
\end{equation}
where $h$ denotes the lag. 
This test was performed using the Grad viscous correction model emulator, and each sampler was first given a burn-in period (unshown) of two thousand steps. If the sampler's performance is robust, the chain represents a sample of the target distribution, which is in this case Grad model's posterior calibrated against both the LHC and RHIC data. However, we see that the \texttt{emcee} ensemble sampler has a very long autocorrelation length, reducing the effective number of samples of the target posterior~\cite{Hogg:2017akh}. Furthermore, we can see large variations in each walker's rolling averages over long time scales (a large number of steps), again suggesting that the chain has not thermalized. On the other hand, the parallel-tempered sampler displays very short autocorrelation length, and a very stable rolling average.


Although the primary result of parameter estimation is the posterior distribution, it is also useful to calculate the point in parameter space which maximizes the posterior. This is referred to as the Maximum a Posteriori (MAP) set of parameters. Because throughout this work we use priors which are uniform distributions, the MAP parameters are those which maximize the likelihood function; that is, the parameters which optimize the fit to the experimental data. 
Please see Ch.~\ref{ch2:ex_bayes_param_est} for a discussion of the interpretation of the MAP parameters.

\section{Estimating the Bayes evidence}
\label{ch4:num_bayes_evidence}

The integral necessary to compute the Bayes evidence
is very high-dimensional and does not lend itself to elementary methods. Fortunately, there exist methods for estimating the evidence in the existing Markov Chain Monte Carlo implementation~\cite{ptemcee_code} used throughout this work. A `parallel-tempered' Markov Chain Monte Carlo routine defines a ladder of inverse `temperatures' $\beta_i$, and then evolves an ensemble of walkers by sampling from a set of distributions defined by
\begin{equation}
   \{ \left[\mathcal{P}( \mathbf{y}_{\exp} | \mathbf{x}_A, A ) \right]^{\beta_i}  \mathcal{P}(\mathbf{x}_A) \}_{i}.
\end{equation}
We see that in the limit $\beta \rightarrow 0$, we recover our prior $\mathcal{P}(\mathbf{x}_A)$. At regular intervals walkers inside of each tempered distribution have the opportunity to swap positions with walkers at adjacent temperatures. Walkers at very high temperatures are not strongly affected by peaks in the likelihood function, while walkers at $\beta = 1$ are sampling from the target posterior. This gives this algorithm the advantage that it can efficiently sample multimodal distributions, which can be more difficult for other algorithms, including the ordinary Metropolis-Hastings, to sample accurately. 

Besides these advantages, the ladder of tempered distributions also gives an estimation of the Bayes evidence by the following trick. Defining the Bayesian evidence as a function of inverse temperature:
\begin{equation}
    Z(\beta) = \int d\mathbf{x}_A \left[\mathcal{P}( \mathbf{y}_{\exp} | \mathbf{x}_A, A ) \right]^{\beta}  \mathcal{P}(\mathbf{x}_A)
\end{equation}
we note that it satisfies a differential equation
\begin{eqnarray}
\nonumber
    && \frac{d \ln Z}{d\beta}  \\\nonumber
    &=&  \frac{1}{Z(\beta)} \int d\mathbf{x}_A \mathcal{P}(\mathbf{x}_A) \ln[\mathcal{P}( \mathbf{y}_{\exp} | \mathbf{x}_A, A )] [\mathcal{P}( \mathbf{y}_{\exp} | \mathbf{x}_A, A ) ]^{\beta} \\ 
     &\equiv&  \langle \ln\left[ \mathcal{P}\left( \mathbf{y}_{\exp} | \mathbf{x}_A, A \right)\right] \rangle_{\beta}.
\label{eq:thermo_int_evidence}
\end{eqnarray}
Therefore, $\ln Z(\beta = 1)$ can be estimated by integrating by quadrature the average at each temperature. The uncertainty in this estimate $\delta \ln Z$ is primarily from using a finite number of points in the quadrature (finite grid in `temperature').

\section{Empirical coverage tests}
\label{ch4:closure}

Tests of empirical coverage, sometimes called `closure tests', are required to ensure that, in a situation with known model parameters, the numerical Bayesian inference workflow correctly reproduces them from a set of pseudo-data. These data are outputs of the model simulator at known values of the parameters, for the observables which one intends to use for the model calibration. Ideally, if the inverse map from observables $\bm{y}$ to parameters $\bm{x}$ is single-valued, the posterior of a closure test should approach a delta-function around the true value of the model parameters, $\mathcal{P}(\boldsymbol{x}|\mathbf{y}_{\hat{\boldsymbol{x}}}) \rightarrow \delta(\boldsymbol{x}{-}\hat{\boldsymbol{x}})$. In practice, the posterior is always smeared by the uncertainties present in the Bayesian parameter estimation methods, and can be multi-modal if the inverse map is exactly or approximately degenerate. 
We note that if there are problems in any component of the workflow, including the physical model simulator itself, the principal components reduction and training of Gaussian processes, or as estimation of the posterior via MCMC, they can be manifested in a closure test. Therefore it represents a `sanity check' of the entire workflow. 

A first source of uncertainties is in the pseudo-data model calculations: since the initial conditions of heavy-ion collisions fluctuate stochastically and running the model is expensive, statistical uncertainties in the pseudo-data are often large, and these will propagate non-trivially and contribute to the width for the parameter posterior. Additional uncertainties are contributed by the emulator: (i) statistical uncertainties from the calculations used to train the emulator; (ii) interpolation uncertainty from the limited number of parameter samples used to train the emulator; and (iii) the limited number of principal components that are interpolated via Gaussian processes. Finally, partial degeneracies in the model can make the inverse map, from observables to parameters, multiple-valued. 
Even if a sufficiently large set of observables can break any exact degeneracies, approximate degeneracies can persist until all the uncertainties decrease below a certain threshold. 

Closure tests provide a way to identify these potential issues and, for a chosen set of observables, quantify the effect of these types of uncertainties on the parameter estimation before any comparison with measurements is performed. Closure tests can also help clarify the level of constraint on the model parameters that can be expected given the emulator predictive uncertainties. These two aspects of closure tests are not independent; however they are sufficiently different objectives that they benefit being discussed separately.

\subsection{Validating Bayesian inference with closure tests}
\label{ch4:valid_closure}

The following demonstrates a sample set of closure tests. They employ the same emulator that is used for calibration with experimental data.

\begin{figure*}[!htb]
\noindent\makebox[\textwidth]{%
  \centering
  \begin{minipage}{0.47\textwidth}
    \includegraphics[trim=0 0 0 20, clip, width=\textwidth]{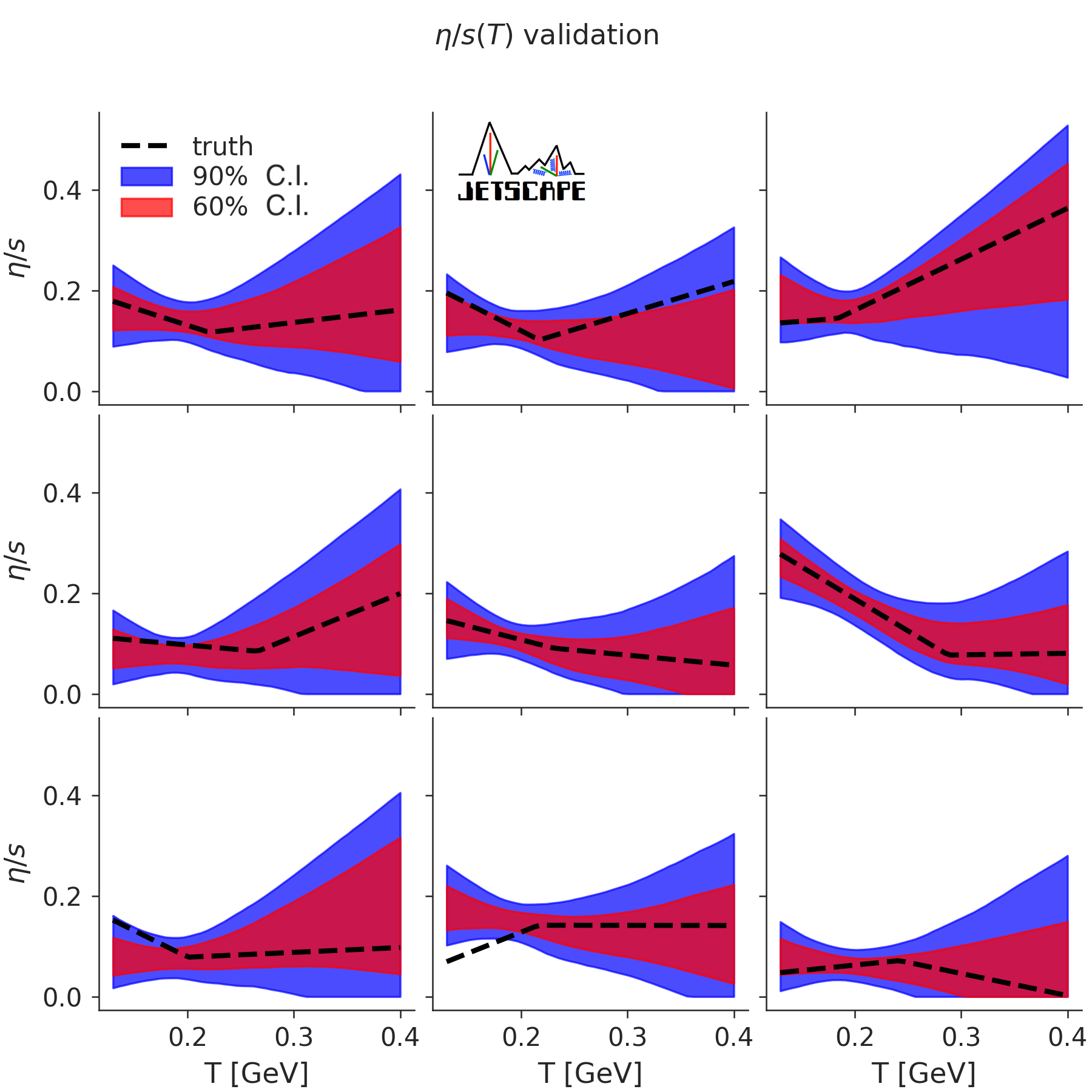}
  \end{minipage}
  \begin{minipage}{0.47\textwidth}
    \includegraphics[trim=0 0 0 20, clip, width=\textwidth]{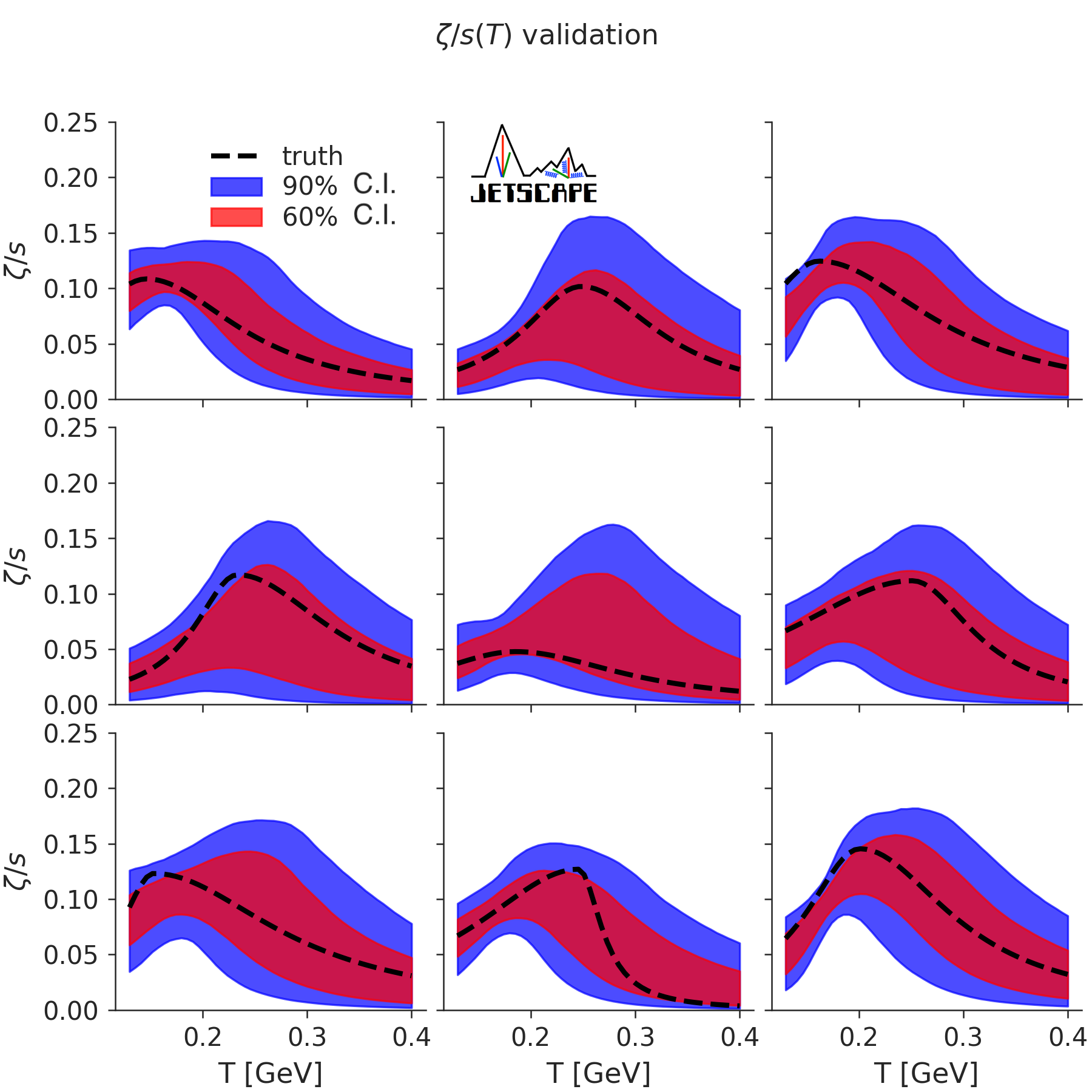}
  \end{minipage}
  }
  \caption{
  Closure tests of the specific shear (left) and bulk (right) viscosities using 9 validation points in the parameter space. Performed with the emulator for Pb-Pb $\sqrts{} = 2.76$ TeV collisions. Shown in blue and red are the $90$\% and $60$\% credible intervals, respectively. The true temperature dependent specific shear and bulk viscosities are shown as black dashed lines. }
  \label{closure_plots}
\end{figure*}

We proceed as follows:
\begin{enumerate}
    \item We generate a set of design points ($\boldsymbol{x}_{i}; i=1,\dots, m_v$) for training the model emulator, and a separate set of design points for validation ($\hat{\boldsymbol{x}}_{i}; i=1,\dots, \hat{m}_v$).
    \item We perform full model calculations at both the training and validation design points and compute final state observables.
    \item We perform principal component analysis on the \emph{training} calculations, and fit a Gaussian process to each retained principal component.
    \item For each point $i$ in the \emph{validation} set, we use the trained emulator to perform parameter estimation using the calculated model observables at validation point $\hat{\boldsymbol{x}}_{i}$ as the ``data''. 
    \item We compare the posterior $\mathcal{P}(\boldsymbol{x}|\mathbf{y}_{\hat{\boldsymbol{x}}_{i}})$ to the known true values $\hat{\boldsymbol{x}}_{i}$.
\end{enumerate}

Our emulator uses 500 design points. At each design point we use the full model simulator to compute predicted values for all observables that will also be used in the calibration with real data (see Ch.~\ref{ch5:bayes_param_est_overview}). As discussed previously, our model includes statistical fluctuations, which arise from averaging over a finite number of initial conditions (2500 hydrodynamic events per design point), as well as Cooper-Frye sampling each particlization hypersurface a finite number of times (at least $10^5$ particles sampled per hydrodynamic event). We use $10$ principal components, which explain approximately $98\%$ of the model variance for Pb-Pb data at $\sqrts{}=2.76$\ TeV. These uncertainties, combined with the emulator uncertainty discussed above, lead to a finite spread of our posterior $\mathcal{P}(\boldsymbol{x}|\mathbf{y}_{\hat{\boldsymbol{x}}_{i}})$. What can be verified is how often the truth lies within given regions of inferred posterior credibility.

Figure~\ref{closure_plots} shows the result of our closure tests for $9$ sets of validation points. We focus on the specific shear and bulk viscosities of the QGP, $\eta/s$ and $\zeta/s$. Because the parametrization of these physical quantities involves non-linearly correlated parameters,
we focus in Fig.~\ref{closure_plots} on the resultant posterior for $\eta/s$ and $\zeta/s$ as functions of temperature $T$, compared to the underlying truth values (shown as dashed black lines). Red and blue bands show the 60\% and 90\% credible intervals of the estimation; at different temperatures these credible intervals are nominally calculated independently; note, that does not mean that there are not correlations across temperature, there are such correlations induced by the parametrization. The results demonstrate that the functional shapes of the ``true'' viscosity-to-entropy ratios are well enclosed by the inferred 60\% and 90\% credible regions. An ensemble roughly fifty of such plots was inspected visually, from which it was concluded that the workflow provides an unbiased estimator of model parameters. No quantitative metrics were calculated with respect to the incidences of truth lying in certain regions of credibility; such investigations were however explored in Ref.~\cite{Cao:2021keo}.

\subsection{Experimental design using closure tests}
\label{ch4:guiding_closure}

Figure~\ref{closure_plots} provides convincing evidence that the emulators and the computational Bayesian inference are performing well. Importantly, it also provides insights regarding the eventual results of Bayesian parameter estimation performed with real data.

Recall that our posterior inferences are conditional on a variety of factors, including (i) the set of observables, (ii) the values and uncertainties of these observables, (iii) the model and its priors and (iv) the uncertainty of the emulator. 
In an ideal (and hypothetical) scenario, the emulation predictive uncertainty would be much smaller than the possible experimental uncertainties on the observables.
In such a scenario, there would be a minimal or negligible amount of information loss in the use of the emulator, and we maximize the utility of the experimental data.  
This is usually the goal one would strive for; however, in practice is difficult to achieve. 

For the case shown in Fig.~\ref{closure_plots}, emulation uncertainties are not negligible. However, given the model, observables and emulator, 
a comparison of the closure test in Fig.~\ref{closure_plots} with the prior from Fig.~\ref{viscous_prior} demonstrates that the current methodology 
and observables 
have the best constraining power for $\eta/s$ and $\zeta/s$ at low temperatures. This is expected, since these temperatures are closer to the switching temperature between hydrodynamics and the hadronic transport model, and much of the space-time volume explored by the expanding medium is characterized by such moderate temperatures ~\cite{Shen:2013vja}. 
Furthermore, the finite relaxation times $\tau_\pi$ and $\tau_\Pi$ reduce the sensitivity of the model to changes in the viscosities. For example, in the limit $\tau_\pi \rightarrow \infty$, the equations describing the relaxation of the shear stress tensor would essentially become \textit{independent} of $\eta/s$. This problem can compounded by the assumed relations between the first order viscosities and their second order relaxation times; for example, the shear relaxation time is related to the shear-viscosity via
\begin{equation}
    \tau_\pi = b_{\pi} \frac{\eta}{s} \frac{1}{T}.
\end{equation}
We might expect observable signatures of a large specific shear viscosity to manifest in the observables, but a large specific shear viscosity results in a large relaxation time. If the duration of the collision described by hydro $\tau_h$ is sufficently short $\tau_h \lesssim \tau_{\pi}$ then we can expect the observables to lose sensitivity to $\eta/s$.\footnote{Moreover, $\tau_h \lesssim \tau_{\pi}$ signals a breakdown of our hydrodynamic theory, and we should probably not be using hydrodynamics at all!} These same effects can manifest for $\tau_{\Pi}$, the bulk relaxation equation, and $\zeta/s$.

Let's take a moment to consider a scenario with a model which is computationally fast, and therefore requires no emulation. Moreover, we suppose that this model has a parameter $x$ for which none of its outputs are sensitive; all of the observables $\bm{y}$ in this model are essentially independent of $x$. Then, consider trying to infer the value of $x$ given a set of observed data. Of course, this exercise is doomed from the start. The likelihood function will be completely flat in $x$, and the posterior should return the prior distribution $\mathcal{P}(x | \bm{y}) \propto \mathcal{P}(x)$. Said differently, the selected observables contain no information about $x$, because the model outputs have no sensitivity to $x$. This same situation will approximately manifest in more realistic models; models may have certain parameters for which none of the observables have significant sensitivity. In these cases, finding an uninformed or relatively flat likelihood should be expected. If indeed this is a property of the underlying physical model, for which we build an emulator, this property should propagate to performing inference with the model emulator.


Additional observables or collision energies may help improve these constraints on the viscosities of QCD. For example, emission of electromagnetic radiation puts somewhat stronger weight on the earlier and shorter-lived hot fireball regions than hadrons~\cite{Shen:2013vja}. On the other hand, electromagnetic observables are plagued by larger statistical and systematic uncertainties. Closure tests can be used exactly for the purpose of assessing the value of adding such additional measurements even before such data are available: they allow for quantifying the contribution of different observables towards constraining the properties of the quark-gluon plasma. In the future this could be an important tool to guide the priorities of experimental campaigns. Observables contribute differently to constraining different model parameters: by quantifying the effect of adding a new observable, or reducing the uncertainty on an existing one, one can provide meaningful feedback which measurements should be prioritized. These methods are closely related to those employed in ``Bayesian Experimental Design''~\cite{chaloner1995}. 

One caveat to keep in mind in is that closure tests evidently rely on the correctness of the underlying physics model. When we compare to experimentally observed data, we cannot assume that our model provides a perfect description of the observables given the `best' choice of parameters~\cite{Brynjarsdottir_2014}. The systematic model discrepancy, whether quantified or not, must not be forgotten in principle. Hence, the result of a closure test should not be taken as the final word: the importance of a given observable in constraining model parameters may need to be revisited when physics tested by this observable is modified in the model. 
In spite of these unavoidable limitations, closure tests can provide important guidance to experimental collaborations to help determine which observables can best constrain physical parameters.

\chapter{Heavy-Ion Model Parameter Estimation and Exploration}
\label{ch5}

\section{Prior specification}
\label{ch5:priors}

\begin{table*}[!hbt]
\noindent\makebox[\textwidth]
{%
\footnotesize
\begin{tabular}{||p{0.2\linewidth}|p{0.13\linewidth}|p{0.15\linewidth}||p{0.2\linewidth}|p{0.13\linewidth}|p{0.15\linewidth}||}
\hline
\hline
Norm. Pb-Pb 2.76 TeV & $N$[2.76 TeV] & {[}10, 20{]} & temperature of $(\eta/s)$ kink & $T_{\eta}$ & {[}0.13, 0.3{]} GeV \\
\hline
Norm. Au-Au 200 GeV & $N$[0.2 TeV] & {[}3, 10{]} & $(\eta/s)$ at kink & $(\eta/s)_{\rm kink}$ & {[}0.01, 0.2{]} \\
\hline
generalized mean & $p$ & {[}--0.7, 0.7{]} & low temp. slope of $(\eta/s)$ & $a_{\text{low}}$ & {[}--2, 1{]} GeV$^{-1}$ \\
\hline
nucleon width & $w$ & {[}0.5, 1.5{]} fm & high temp. slope of $(\eta/s)$ & $a_{\text{high}}$ & {[}--1, 2{]} GeV$^{-1}$ \\
\hline
min. dist. btw. nucleons & $d_{\text{min}}^3$ & {[}0, 1.7$^3${]} fm$^3$ & shear relaxation time factor & $b_{\pi}$ & {[}2, 8{]} \\
\hline
multiplicity fluctuation & $\sigma_k$ & {[}0.3, 2.0{]} & maximum of $(\zeta/s)$ & $(\zeta/s)_{\text{max}}$ & {[}0.01, 0.25{]} \\
\hline
free-streaming time scale & $\tau_R$ & {[}0.3, 2.0{]} fm/$c$ & temperature of $(\zeta/s)$ peak & $T_{\zeta}$ & {[}0.12, 0.3{]} GeV \\
\hline
free-streaming energy dep. & $\alpha$ & {[}--0.3, 0.3{]} &  width of $(\zeta/s)$ peak & $w_{\zeta}$ & {[}0.025, 0.15{]} GeV \\
\hline
particlization temperature & $\Tsw$ & {[}0.135, 0.165{]} GeV & asymmetry of $(\zeta/s)$ peak & $\lambda_{\zeta}$ & {[}--0.8, 0.8{]}\\
\hline
\hline
\end{tabular} 
}
\caption{A table of all prior hyperparameters. All prior distributions are assumed to be uniform and nonzero within the range listed above, and zero outside. The table excludes the step functions that enforce positivity of the shear viscosity.}
\label{prior_table}
\end{table*}

When choosing priors for our heavy-ion model, we will consider the constraints imposed by a combination of theoretical physics and `common sense', and take an empirical Bayesian approach for the model parameters which are not straightforward to constrain a priori. 

\paragraph{Initial conditions:} Physical constraints motivate the prior for the width parameter $w$ in \trento{}: the electric charge radius of the proton is about $0.9$\,fm. Therefore, we do not allow the width parameter $w$ in \trento{} to be much smaller or larger than this value. The prior for $d_{\rm min}^3$ is also motivated by low-energy models of nuclei, in which we expect the inter-nucleon distance to be less than about $1.5$ fm.
The remaining parameters in \trento{}, the energy normalization $N$, multiplicity fluctuations $\sigma_k$, and generalized thickness parameter $p$ have prior ranges which have been shown to provide wide coverage of the experimental data in previous analyses~\cite{Bernhard:2019bmu}.  

\paragraph{Pre-hydrodynamics:} The range for the freestreaming time scale $\tau_{\rm FS}$ was specified considering theoretical scenarios of hydrodynamization in heavy-ion collisions, which often find viscous hydrodynamics to be applicable around $1-2$ fm after the collision~\cite{Bernhard:2018hnz}. The energy dependence parameter $\alpha$ had a prior range empirically fixed such that simulations of peripheral events could not have unreasonably long freestreaming times.

\paragraph{Hydrodynamic Transport coefficients:} To satisfy Boltzmann's H-Theorem both the specific shear and bulk viscosities must be non-negative. 
The finite-order hydrodynamic approximations upon which our model is derived breaks down when the shear and bulk viscosities are too large, because they drive the inverse-Reynolds numbers to be large. For self-consistency of our hydrodynamic approximation, we thus use the prior to exclude large values of $\eta/s$ and $\zeta/s$. 
The shear relaxation time also has a strong lower limit imposed by causality, and weaker constraints motivated by simple microscopic theories. The minimum value for $b_\pi=T \tau_\pi/(\eta/s)$ is set by requiring the linearized equations be causal, yielding $b_\pi \gtrsim 2$. Theoretical calculations of $b_\pi$ within different microscopic theories ranging from weakly to strongly coupled provide a window in which we expect the relaxation time, however this prior could likely be relaxed in the future provided the equations are stable for very large values of the relaxation time. On the other hand, our hydrodynamic theory also breaks down when the Knudsen number is large, which in this context we can approximate by $\text{Kn} \approx \tau_\pi L^{-1}$, where $L$ denotes the smallest macroscopic length scale in our medium. So very large shear-relaxation times also push our second-order hydrodynamic theories into regions where the predictions become dubious.

\paragraph{Particlization switching temperature:}
The switching temperature $T_{\rm sw}$ between hydrodynamics and hadronic transport is assigned a reasonable window of temperatures bracketing the pseudo-critical hadronization temperature $T_c \sim 150$ MeV. This is motivated by the expectation that the hadronization process itself may cause a sudden change in the Knudsen number, rendering hydrodynamic transport inapplicable. 

\begin{figure*}[!htb]
\centering
\includegraphics[trim=0 0 0 15, clip, width=0.7\textwidth]{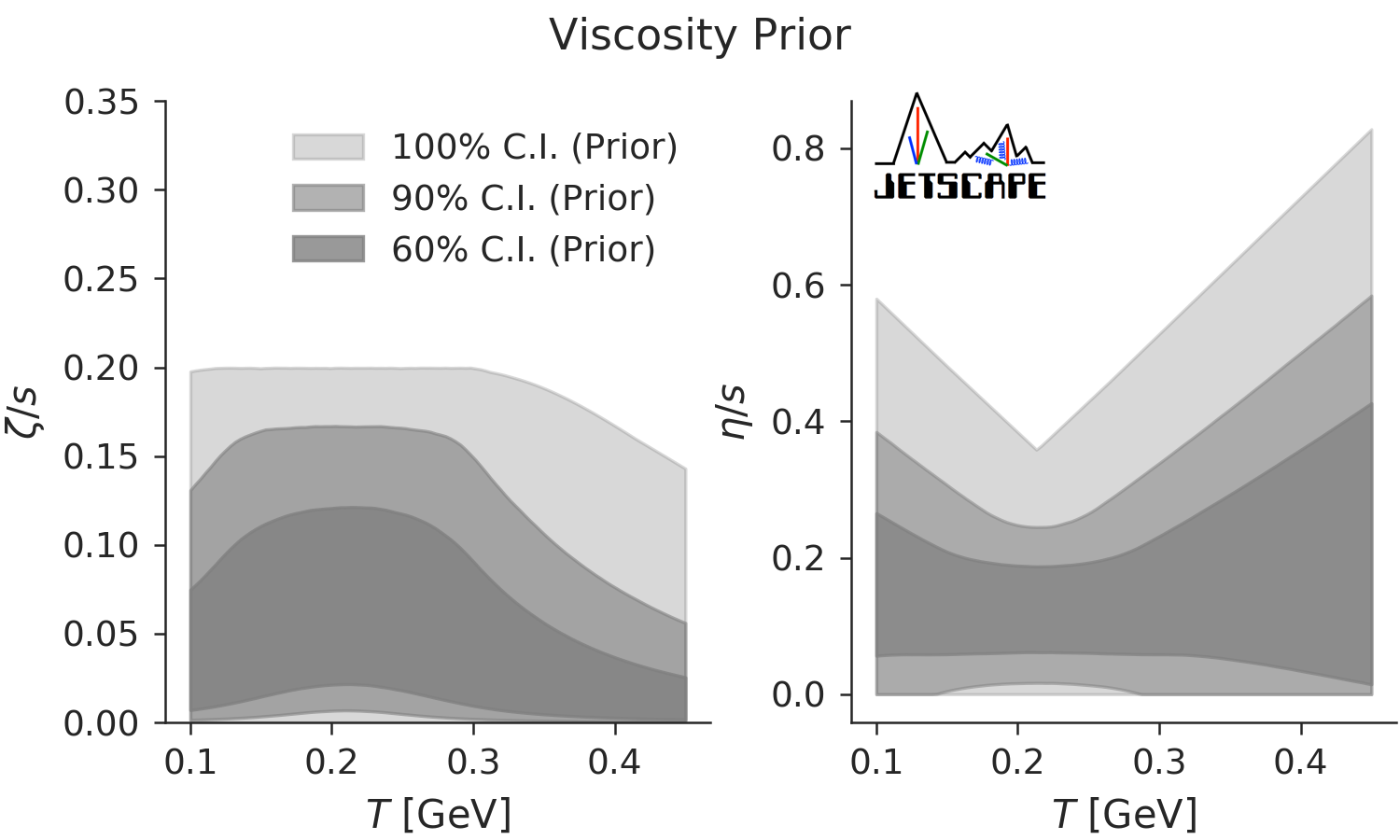}
\caption{Credible intervals of the prior probability density for the specific bulk (left) and shear (right) viscosities that we use when performing Bayesian parameter estimation. The 60\%, 90\% and 100\% credible intervals (C.I.) are shown.}
\label{viscous_prior}
\end{figure*}

In the present analysis, for simplicity all of the parameters (denoted by the vector $\boldsymbol{x}$) are assigned a uniform prior probability density $\mathcal{P}(\boldsymbol{x})$ on a finite range. These ranges are listed in Table~\ref{prior_table}; as discussed above, they have been chosen according to various considerations. The priors for different parameters are assumed to be independent, so that the joint prior is simply given by their product,
\begin{eqnarray}
   \mathcal{P}(\boldsymbol{x}) \propto \prod_{i} \Theta(x_i-x_{i,\min})\Theta(x_{i,\max}-x_i),
\end{eqnarray}
where $i$ runs over all the model parameters in $\boldsymbol{x}$. Note that uniform priors are not uninformative priors. Moreover, the choice of priors in principle affects the results of the Bayesian parameter estimation, especially in situations where the data do not have sufficient information to correct prior prejudice. For instance, in this work, we require $(\eta/s) (T)$ and $(\zeta/s) (T)$ to be given by specific parametrizations, with each of the parameters sampled from a uniform prior. The resulting prior for $(\eta/s)(T)$ is, however, not uniform as a function of temperature; thus, our choice of parametrization informs our prior. A plot showing credible intervals \emph{for the prior} for the shear and bulk viscosities is shown in Fig.~\ref{viscous_prior}.
%
%
We see that this prior encapsulates our belief that the bulk viscosity should have a peak somewhere near the deconfinement transition temperature, and that the specific shear viscosity may have a minimum in that region. 

Nevertheless, we used a broad prior for $\eta/s$, allowing it to take either a maximum or a minimum in the deconfinement region. By doing so we tried to limit the theoretical bias of our prior for $(\eta/s)(T)$. When selecting the priors for the remaining model parameters we followed similar considerations, with the goal of ensuring that our posterior parameter constraints will be guided as much as possible by the heavy-ion data and not by prior prejudice. 

It is important, however, to understand that in practice theoretical bias can never be fully avoided; in certain cases they can be helpful. If highly constraining data are lacking, exploring the reaction of the posterior distribution to different prior theoretical assumptions can yield useful insights into the variability and reliability of model predictions. The Bayesian theory of probability accepts the reality of theoretical bias; remember, the \textit{only} probabilities which we are able to quantify systematically are conditional probabilities. This is not a weakness of the methodology, but the strength; we are very rarely in a position where we have absolutely no prior information, and ignoring the information at our disposal general leads to poorer results. Moreover, the methodology forces us to make \textit{explicit} the biases and information we bring to the analysis, rather than perhaps leaving them implicit or neglected.  
Sensitivity to our prior assumptions is further explored in Ch.\ref{ch5:prior_sensitivity} and Ch.\ref{ch6:bulk_posterior_prior_sens}. 

\section{Bayesian parameter estimation with a statistical emulator}
\label{ch5:bayes_param_est_w_emu}

\subsection{Overview of Bayesian parameter estimation}
\label{ch5:bayes_param_est_overview}

Bayesian parameter estimation is a systematic approach to infer the probability distribution of model parameters ($\boldsymbol{x}$) by comparing theoretical calculations ($\mathbf{y}_{\boldsymbol{x}}$) to experimental data ($\mathbf{y}_{\exp}$). The starting point is the prior distribution $\mathcal{P}(\boldsymbol{x})$ that encodes the current state of knowledge regarding the model parameters $\boldsymbol{x}$ before making comparison with data. 
The posterior distribution of model parameters $\mathcal{P}(\boldsymbol{x}|\mathbf{y}_{\exp})$ which updates our prior based on the observed data, is given by Bayes' theorem,
\begin{equation}
    \mathcal{P}(\boldsymbol{x}|\mathbf{y}_{\exp}) =  \frac{ \mathcal{P}(\mathbf{y}_{\exp}|\boldsymbol{x}) \mathcal{P}(\boldsymbol{x}) }{ \mathcal{P}(\mathbf{y}_{\exp}) },
    \label{eq:likelihood}
\end{equation}
where $\mathcal{P}(\mathbf{y}_{\exp}|\boldsymbol{x})$ is the ``likelihood'' that the model agrees with experimental measurement, given the parameters $\boldsymbol{x}$, and the normalization $\mathcal{P}(\mathbf{y}_{\exp})$ is called the ``Bayesian evidence''. The exact form of the likelihood is often unknown, as it depends on the probability distribution of the experimental and theoretical uncertainties. In this work, we follow the common assumption that the likelihood can be approximated to be a multivariate normal distribution. This choice is justified when uncertainties are normally distributed. The reader should note that there may be many sources of systematic experimental errors in heavy-ion measurements which are not multivariate normal in nature. Quantifying the entire distributions of systematic errors and incorporating better informed likelihood functions requires a large effort, and is left as an outstanding problem. 
With the assumption of multivariate normal likelihood function, the logarithm of $\mathcal{P}(\mathbf{y}_{\exp}|\boldsymbol{x})$ contains the quadratic form of the difference between the measurement and the prediction $\Delta \mathbf{y}_{\boldsymbol{x}} =\mathbf{y}_{\boldsymbol{x}} - \mathbf{y}_{\exp}$,
\begin{equation}
    \label{eq:log_likelihood}
    \ln \left[\mathcal{P}(\mathbf{y}_{\exp}|\boldsymbol{x})\right] = -\frac{1}{2}\ln  \left[(2\pi)^{n} \det\Sigma\right]-\frac{1}{2}\Delta \mathbf{y}_{\boldsymbol{x}}^T\Sigma^{-1}\Delta \mathbf{y}_{\boldsymbol{x}}.
\nonumber
\end{equation}
Here, $n$ is the number of observation points (i.e. the length of the vector $\mathbf{y}_{\exp}$), and $\Sigma$ is a covariance matrix that encodes both experimental and model uncertainties, as well as correlations among uncertainties. These correlations are generally not readily available experimentally. As such the treatment of uncertainties can become a relatively complex question. 


In principle, in order to calculate the posterior, one is faced with the task of calculating the evidence $\mathcal{P}(\mathbf{y}_{\exp})$. For many problems of interest the required high dimensional integration can be numerically challenging or even intractable. Fortunately, when performing Bayesian parameter estimation, knowledge of the relative probability of different points in parameter space is sufficiently interesting in itself. That is, as the evidence $\mathcal{P}(\mathbf{y}_{\exp})$ does not depend on the parameters $\boldsymbol{x}$, it is sufficient to consider the proportionality
$$\mathcal{P}(\boldsymbol{x}|\mathbf{y}_{\exp}) \propto   \mathcal{P}(\mathbf{y}_{\exp}|\boldsymbol{x}) \mathcal{P}(\boldsymbol{x}).$$ 
Methods for estimating the posterior which take advantage of this include Markov Chain Monte Carlo. Therefore, when we discuss or plot the posterior of parameter estimates throughout this section, we implicitly mean the unnormalized posterior. Hence, we are interested in the relative probability density of each parameter set, and {\it not} the absolute probability. 

Because the plotted posterior for the model parameters in general does not contain information about this normalization, it is imperative to check the level of agreement between the posterior prediction of \textit{observables} to assess quantitatively how well the model can describe the experimental data. It is meaningless to ponder on the posterior parameter estimates of a model which poorly explains the observed data.
Thus, we will also explore how well the model observables sampled from the posterior describe the experimental data. An estimation of the evidence $\mathcal{P}(\mathbf{y}_{\exp})$ becomes necessary if we want to compare models in a Bayesian framework and this will be discussed in \ref{ch5:model_selection}. 

\paragraph*{Simultaneous constraints from multiple collision systems:}
%
When combining constraints from different experiments, Au-Au collisions at RHIC and Pb-Pb collisions at the LHC for example, the joint likelihood function is assumed to be the product of the individual likelihoods for each system:
\begin{equation}
    \mathcal{P}(\mathbf{y}^{\rm Pb}_{\exp}, \mathbf{y}^{\rm Au}_{\exp}  |\boldsymbol{x}) = \mathcal{P}(\mathbf{y}^{\rm Pb}_{\exp} |\boldsymbol{x}) \mathcal{P}(\mathbf{y}^{\rm Au}_{\exp} |\boldsymbol{x}).
\end{equation}
The parameter values that maximize the joint likelihood strike a compromise between maximizing the individual likelihoods.

Importantly, one must make an assumption regarding which parameters are shared for the different collision systems.
Comparisons with measurements can always help determine if model assumptions need to be relaxed. If RHIC and LHC measurements could be described independently by the model but not simultaneously, it would be an indication that the $\sqrts{}$ dependence of certain parameters needs to be revisited, i.e., that enforcing the same value of certain parameters at RHIC and the LHC puts the model under too much tension. 
Such tension should be visible in the parameter posteriors arising from calibrating the model to each set of observables \textit{separately}, as well as posterior predictive distributions of the simultaneously calibrated model. 
We will compare more complex models which relax some of these assumptions by estimating Bayes factors in \ref{ch5:model_selection}. 
Inclusion of data at two very different collision energies raises the question where and how we make allowance for $\sqrts{}$ dependence of the model parameters.

\paragraph{Initial stage model:} Because \trento{} is a parametric initial condition model, not a dynamical one, many of its parameters should, in principle, be beam-energy dependent.\footnote{%
    For example, in the color glass condensate effective theory for QCD at very high energies, the only relevant scale is the saturation scale $Q_s$, which controls correlations in the transverse direction and which runs with the energy of the collision system ~\cite{Gelis:2014qga}. This suggests that the nucleon width in \trento{} should perhaps have a similar $\sqrts{}$ dependence.}
Generically, we assume that at high collision energies the parameters that we try to estimate with experiment data evolve sufficiently slowly with $\sqrts{}$ that their change from RHIC to LHC can be ignored. As an exception we retain the $\sqrts{}$ dependence of the normalization $N$ of the energy density in \trento{}, because it is directly responsible in our model for the large increase of mid-rapidity particle and energy production from RHIC to LHC. Rather than parametrizing its $\sqrts{}$ dependence, we simply use two independent normalizations at $\sqrts{}=200$ and 2760\ GeV, labeled by $N$[0.2\,TeV] and $N$[2.76\,TeV], respectively. We also point out that in \ref{ch5:model_selection} we use Bayesian Model Selection to explore whether experimental data would prefer a dependence of the nucleon width $w$ in \trento{} on $\sqrts{}$. The free-streaming time (\ref{eq:taufs}) is allowed to depend on $\sqrts{}$ parametrically, through the deposited energy density.

\paragraph{Transport coefficients:} The specific shear and bulk viscosities, as well as the second-order transport coefficients in our hydrodynamic approach, are medium properties that (for systems without conserved charges) depend only on the temperature of the plasma. Their parametrizations as functions of temperature, $(\eta/s)(T)$ and $(\zeta/s)(T)$, are therefore assumed independent of $\sqrts{}$.

\paragraph{Particlization:} We use the same particlization temperature $\Tsw$ at RHIC and at the LHC. Although we justify particlization with the assumption of a nonzero window of mutual applicability of both viscous hydrodynamics and the Boltzmann transport of a hadron resonance gas, the validity of this assumption depends on the \textit{dynamical} properties of the expanding system. 
As we've pointed out in Ch.~\ref{ch3:hydro}, propagation of viscous hydrodynamics assumes that Knudsen number is small $\text{Kn} \equiv \tau_{\rm micro} / \tau_{\rm macro} < 1$.
However, in the hadron resonance gas phase of the collision, the expansion rate is large and the microscopic interaction times sufficiently small that $\text{Kn} \equiv \tau_{\rm micro} / \tau_{\rm macro} > 1$. Moreover, in the hadron resonance gas there is not a single relevant microscopic timescale for all species, rather different species have different reaction rates, e.g. processes that change the number of protons fall out of equilibrium very quickly. In any case, an ideal fluid dynamical property such as the temperature in the fluid rest frame may not be the most appropriate particlization criterion. Rather, a dynamical property such as the Knudsen number may be preferred based on theoretical considerations. As a simple proxy we use the temperature, and allow this temperature to vary in a wide range surrounding the pseudocritical temperature.

\section{Bayesian parameter estimation using RHIC and  \\LHC measurements}
\label{ch5:param_est_RHIC_LHC}

In this Section we perform Bayesian parameter estimation against RHIC Au-Au $\sqrts{}=0.2$ TeV and LHC Pb-Pb $\sqrts{}=2.76$ TeV measurements. We focus on constraints for the shear and bulk viscosities provided by transverse-momentum-integrated data. We perform these first analyses for a specific model of viscous corrections at particlization, the Grad model. The effect of using different viscous corrections as well as other systematic uncertainties of the model are quantified in the next section.

\subsection{Calibrating $\eta/s$ and $\zeta/s$ to Pb-Pb measurements at\\ $\sqrts{}=2.76$\ TeV}
\label{ch5:eta_zeta_Pb}

We first study the parameter estimates including only the data from Pb-Pb collisions at $\sqrts{} = 2.76$\ TeV. We use the following measurements from the ALICE collaboration:
\begin{itemize}
    \item the charged particle multiplicity $dN_{\text{ch}}/d\eta$~\cite{Aamodt:2010cz} for bins in $0{-}70$\% centrality;
    \item the transverse energy $dE_T/d\eta$ ~\cite{Adam:2016thv} for bins in $0{-}70$\% centrality;
    \item the multiplicity $dN/dy$ and mean transverse momenta $\langle p_T \rangle $ of pions, kaons and protons ~\cite{Abelev:2013vea} for bins in $0{-}70$\% centrality;
    \item the two-particle cumulant harmonic flows $v_n\{2\}$ for $n=2,3,4$, for bins in $0{-}70$\% centrality for $n{\,=\,}2$, and for bins in $0{-}50$\% centrality for $n=3$ and $4$ ~\cite{ALICE:2011ab};
    \item the fluctuation in the mean transverse momentum $\delta p_T / p_T$ ~\cite{Abelev:2014ckr} for bins $0{-}70$\% centrality.
\end{itemize}

Before being reduced by principal component analysis this data set represents 123 ``independent observables'', given that measurements at different centralities are treated as separate observables. We found that 10 principal components (linear combinations of observables) are sufficient to capture most of the sensitivity of these observables to the full set of parameters: they capture more than $98$\% of the variance. This number of dominant principal components represents only 8\% of the total number of observables. Thus there is a significant amount of redundant information in the observables with respect to our model parameters. We tested the effect of reducing the number of principal components: we determined that our results are robust with respect to the number of principal components used. The results of this test are presented in \Appendix{pca_valid}. 

We remind that all observables used in the calibration analyses are $p_T$-integrated. Observables which are differential in transverse momentum undeniably carry additional microscopic information about the medium ~\cite{Nijs:2020ors, Nijs:2020roc}. But there is reasonable evidence that low-$p_T$ ($p_T\lesssim 1.5$\,GeV) information is included in $p_T$-integrated observables~\cite{Novak:2013bqa}. The higher-$p_T$ range ($p_T\gtrsim 1.5$\,GeV) tends to have larger modeling uncertainties, if only from viscous corrections at particlization which can be very significant at higher transverse momenta. At sufficiently high $p_T$, hadron production is beyond the realm of hydrodynamics altogether; this threshold is not known precisely, but even a breakdown at $p_T\gtrsim 2-3$ GeV would not be wholly surprising. 
Because of these limitations, there is a risk that posterior inferences using hydrodynamically modeled observables in the higher-$p_T$ range  ($p_T\gtrsim 1.5$\,GeV) lead to less robust constraints on the parameters. While both avenues are worth exploring, in the present analysis we opt for the more conservative approach of using $p_T$-integrated observables that introduce less model bias, while also studying in detail model uncertainties.

\begin{figure}[!htb]
\centering
\includegraphics[trim=0 0 0 15, clip, width=0.7\textwidth]{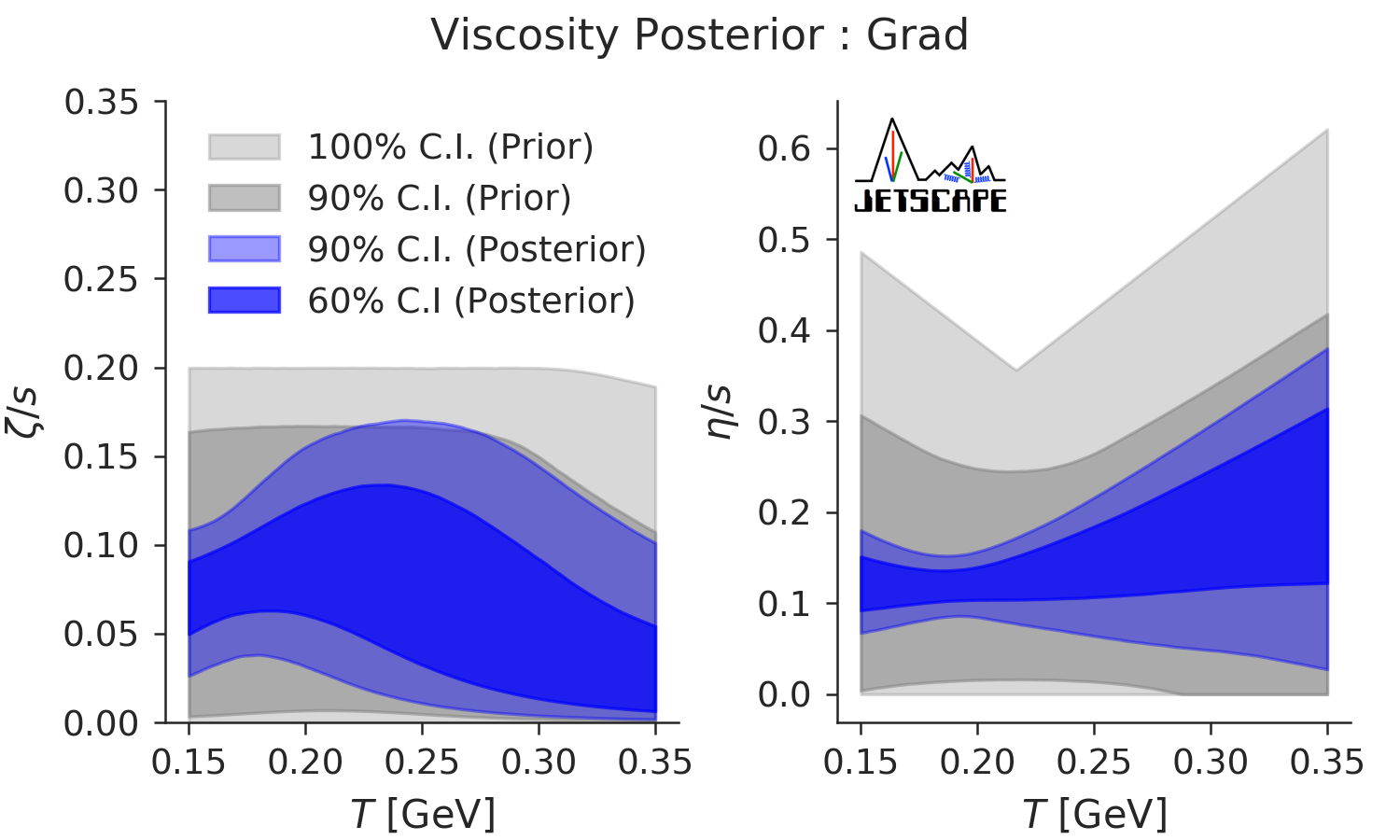}
\caption{The posterior for specific bulk (left) and shear (right) viscosities resulting from a Grad viscous correction model parameter estimation using ALICE data for Pb-Pb collisions at $\sqrts{} = 2.76$ TeV}
\label{grad_lhc_posterior}
\end{figure}

The posteriors for the shear and bulk viscosities are shown in Fig.~\ref{grad_lhc_posterior}. Recall that this result is for a {\it single} viscous correction model, the Grad viscous correction.
We first note a general feature which will remain when we examine other viscous corrections and include more systems: the constraint on the shear and bulk viscosities is best near the switching temperature $\Tsw$. This was already observed in the closure tests performed in Ch.~\ref{ch4:closure}. The viscous corrections in the particlization procedure depend on the magnitude of shear stress $\pi^{\mu\nu}$ and bulk pressure $\Pi$ on the switching surface, making the model predictions sensitive to the viscosities near these temperatures. As we have discussed in the closure test, the uncertainties in $\zeta/s$ and $\eta/s$ are larger in the high temperature region. We see that for the bulk viscosity in particular, our 90\% posterior credible interval is only slightly smaller than our prior above 250 MeV. 

\subsection{Calibrating $\eta/s$ and $\zeta/s$\\
to Au-Au measurements at $\sqrts{}=0.2$~TeV}
\label{ch5:eta_zeta_Au}

We also examine the constraints on the viscosities provided by the existing data for Au-Au collisions at $\sqrts{} = 200$\,GeV. Heavy-ion collisions at RHIC provide complimentary information, having smaller temperatures and a shorter lifetime than collisions at the LHC. We use the following experimental measurements from the STAR Collaboration:
\begin{itemize}
    \item the yields $dN/dy$ and mean transverse momenta $\langle p_T \rangle $ of pions and kaons for bins in 0--50\% centrality ~\cite{Abelev:2008ab}; 
    \item the two-particle cumulant harmonic flows $v_n\{2\}$ for $n=2,3$ for bins in 0--50\% centrality ~\cite{Adams:2004bi, Adamczyk:2013waa}.
\end{itemize}
We remark that because of the tension between STAR and PHENIX measured proton yields at mid-rapidity in Au-Au collisions at $\sqrts{} = 200$\,GeV ~\cite{Abelev:2008ab,Adler:2003cb}, we have \textit{excluded} the proton yield and mean transverse momentum measured at RHIC from the calibration.\footnote{%
Moreover, both measurements ~\cite{Abelev:2008ab,Adler:2003cb} show a notable excess of proton production over anti-proton production, suggesting the importance of including a non-zero baryon chemical potential ($\mu_B$) in our calculation. The current study assumes $\mu_B=0$ in both initial condition and dynamical evolution, and improvements should be considered in future studies.}
The data above includes 29 observables, again counting centrality bins as separate observables. After performing principal component analysis, we kept 6 principal components (equivalent to 21\% of the total number of observables), which explain more than 98\% of the variance of the observables across the parameter space.

\begin{figure}[!htb] 
\centering
\includegraphics[trim=0 0 0 15, clip, width=0.7\textwidth]{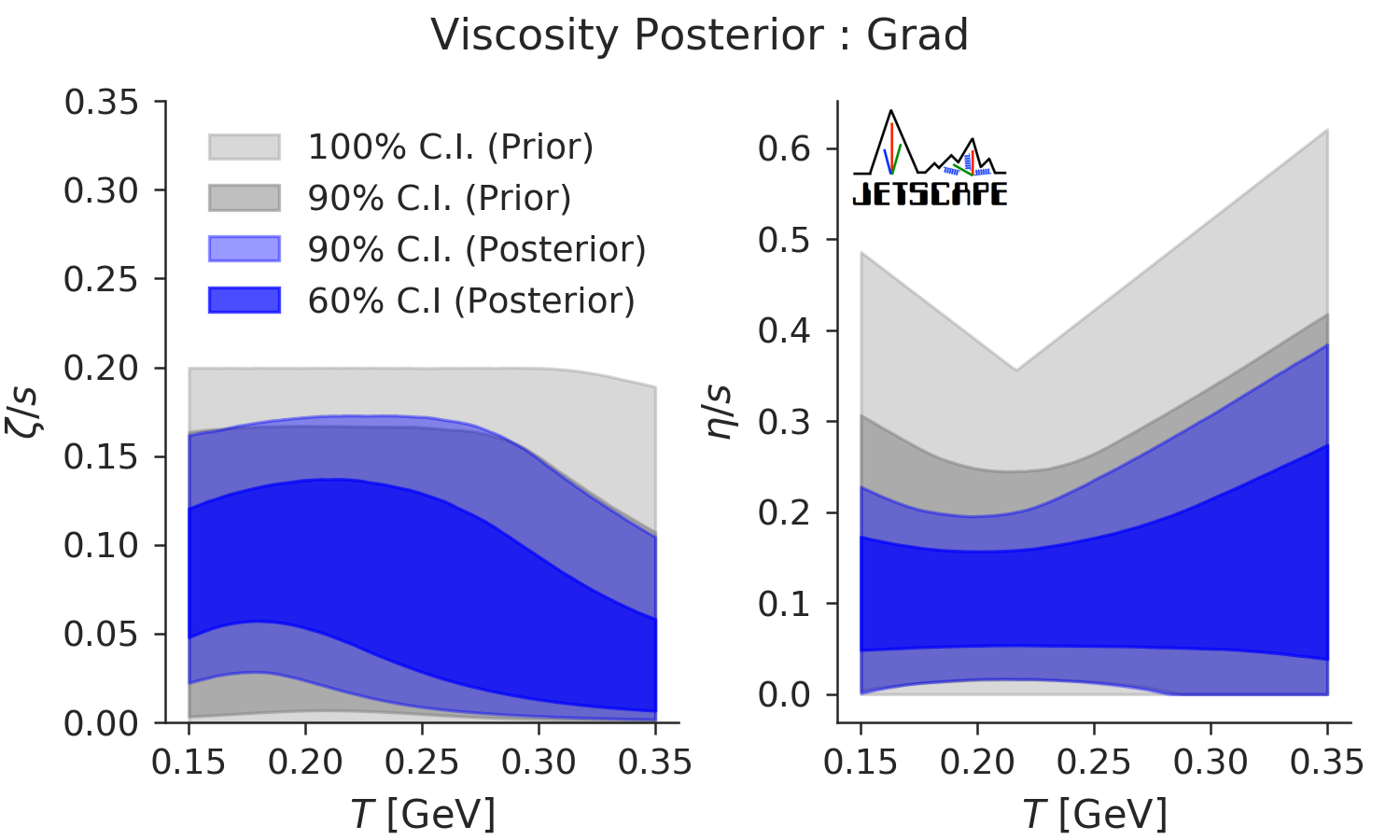}
\caption{The posterior for specific bulk (left) and shear (right) viscosities resulting from a Grad viscous correction model parameter estimation using STAR data for Au-Au collisions at $\sqrts{} = 200$\,GeV.}
\label{grad_rhic_posterior}
\end{figure}

The estimated viscosities using only these measurements from RHIC, again for the Grad viscous correction, are shown in Fig.~\ref{grad_rhic_posterior}. 
%
%
The posteriors for specific bulk and shear viscosity when calibrating against only RHIC data have in general different features than those given by the LHC data. For instance, we see that a large specific bulk viscosity is allowed near the switching temperature. Also, the 90\% credible interval for the specific shear viscosity extends to lower values for these data than the LHC data; only using these RHIC observables, a specific shear viscosity which is nearly zero ($\eta/s < 0.03$) is consistent with the data. In general, the uncertainties on the viscosities are larger using only these RHIC data, likely because there are far fewer measurements included than at the LHC. 

It is important to note that not only the specific bulk and shear viscosity parameters have different posteriors, but in general the entire parameter posterior will be different when we use RHIC observables rather than LHC observables. The two are compared for a different subset of model parameters in \Appendix{app:post_LHC_RHIC_separate}. 

\subsection{Viscosity estimation and model accuracy for combined RHIC \& LHC data}
\label{ch5:eta_zeta_combined}

\begin{figure}[!htb] 
\centering
\includegraphics[trim=0 0 0 17, clip, width=0.7\textwidth]{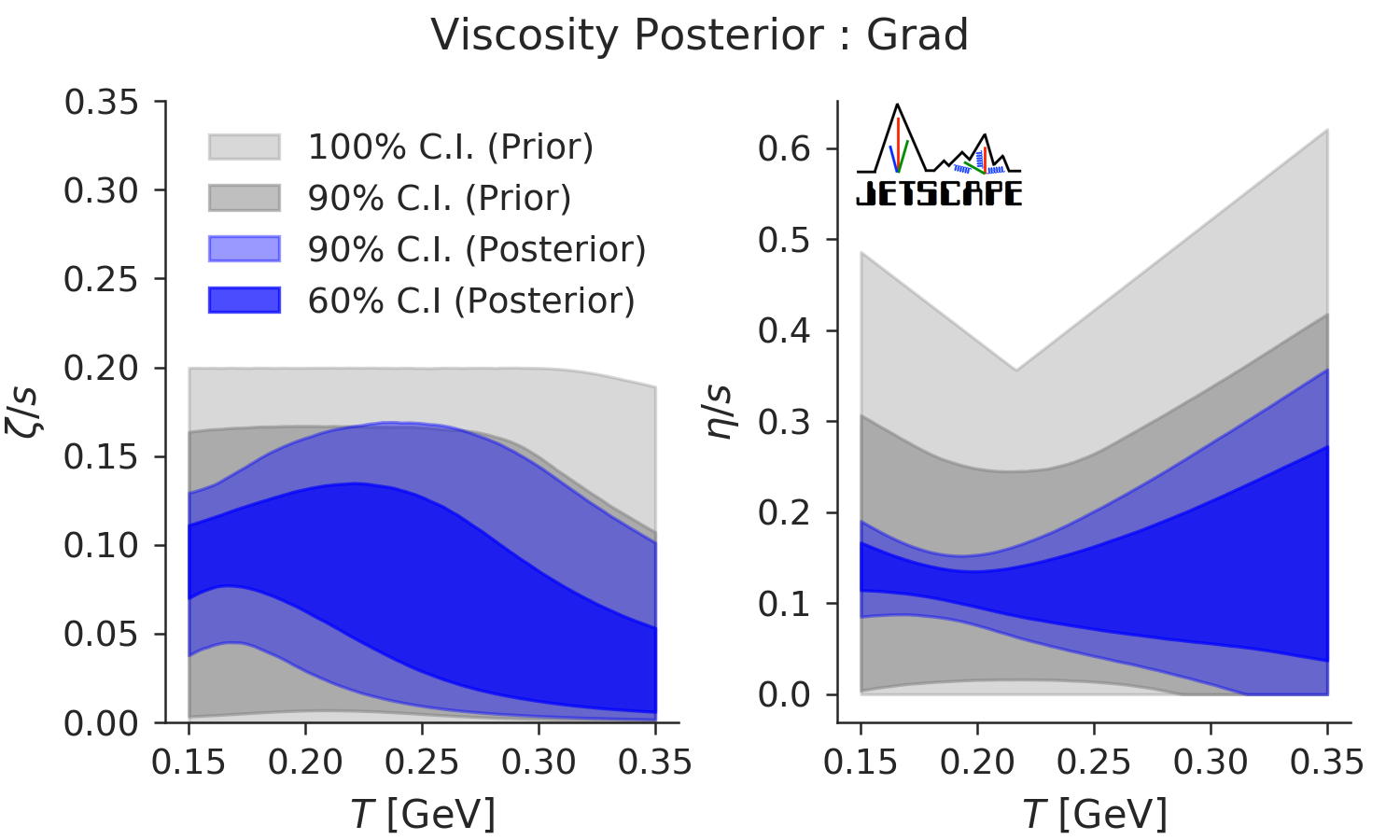}
\caption{The posterior for specific bulk (left) and shear (right) viscosities resulting from a Grad model parameter estimation using combined data for Au-Au collisions at $\sqrts{} = 200$ GeV and Pb-Pb collisions at $\sqrts{} = 2.76$ TeV. }
\label{grad_lhc_rhic_posterior}
\end{figure}

Reviewing Figs.~\ref{grad_lhc_posterior} and \ref{grad_rhic_posterior} we find that the observables measured in Pb-Pb collisions at $\sqrts{} = 2.76$ TeV give stronger constraints on the slope of the specific shear viscosity at large temperature. It is expected that higher $\sqrts{}$ collisions are more sensitive to the transport coefficient at high temperature. This conclusion was verified quantitatively in previous Bayesian parameter estimation ~\cite{Pratt:2015zsa,Sangaline:2015isa}. For the present analysis, we do caution that we currently use a different number of observables at RHIC and the LHC; consequently, we are not in a position to compare systematically the constraining power of the two collision energies at the moment. We do expect RHIC and LHC data to be complementary, and we proceed to a combined Bayesian parameter estimation for Pb-Pb at $\sqrts{} = 2.76$\,TeV and Au-Au at $\sqrts{} = 200$\,GeV collisions. For this combined analysis, the viscosity posterior for the Grad viscous correction is shown in Fig.~\ref{grad_lhc_rhic_posterior}. 

As discussed in Ch.~\ref{ch5:bayes_param_est_overview}, all parameters are held the same for the two systems except for their overall normalizations of the initial conditions --- $N$[2.76\,TeV] and $N$[0.2\,TeV]. Recall that model parameters being held constant does not imply that the effective physical quantities are the same at the two collision systems. For example, the transport coefficients are temperature dependent, and the free-streaming time depends on $\sqrts{}$ and centrality through the initial average energy density of the event.

The information gained by fitting both systems slightly reduces the width of the credible intervals for the specific shear and bulk viscosities at temperatures above 250 MeV; the 90\% confidence band in the posterior for specific shear and bulk viscosity is slightly smaller than the credible intervals given by calibrating against either one of these two systems alone. This illustrates the added constraining power accessed by combining the two data sets.

%
\begin{figure}[!htb] 
\centering
\includegraphics[trim=0 0 0 25, clip, width=0.6\linewidth]{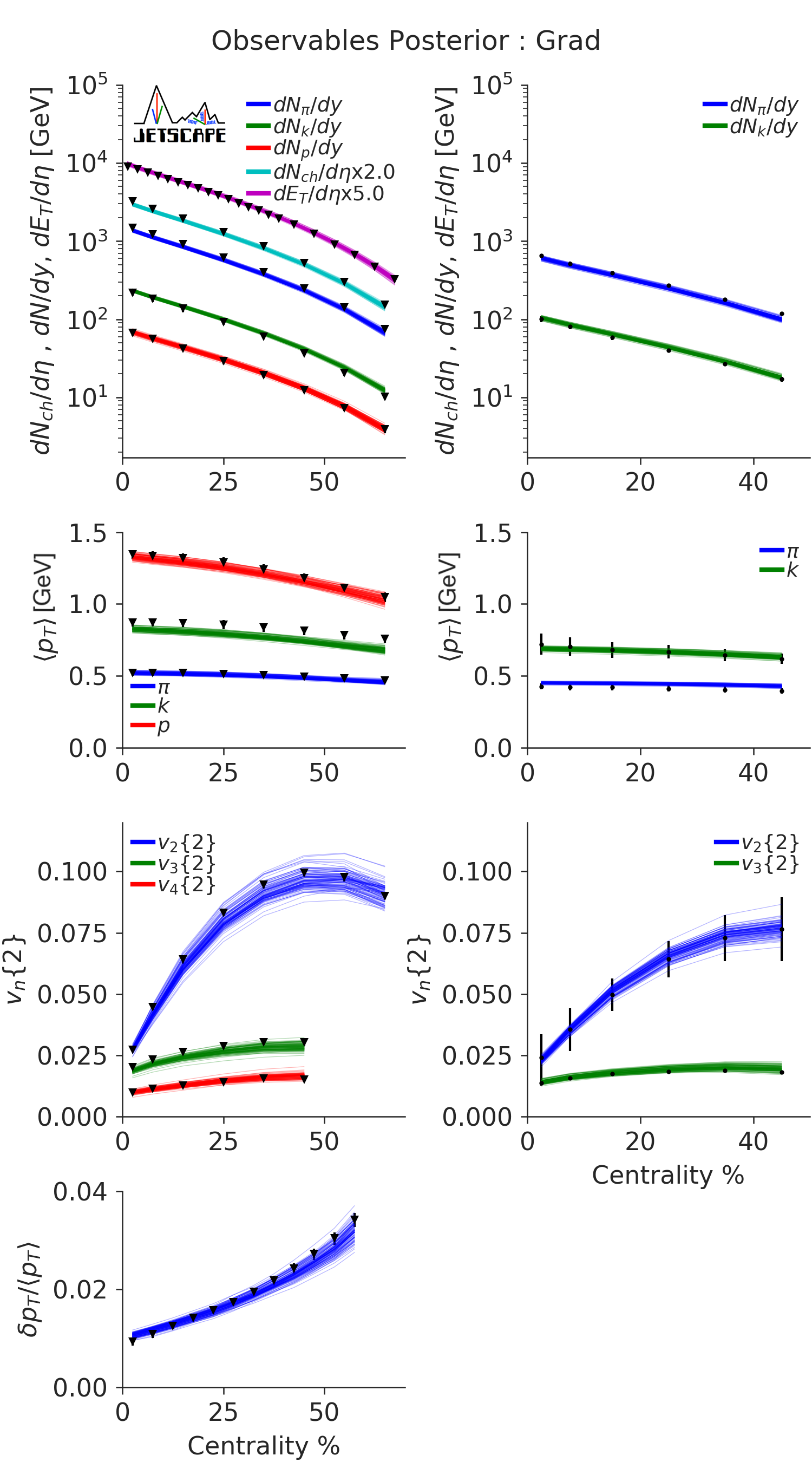}
\caption{The observables predicted by the Grad viscous correction emulator, drawn from the posterior resulting from the combined fit of ALICE data (left) for Pb-Pb collisions at $\sqrts{} = 2.76$ TeV and STAR data (right) for Au-Au collisions at $\sqrts{} = 200$ GeV. The simultaneous calibration yields model observables which agree within ${\sim}20\%$ of experimental measurements.}
\label{grad_lhc_rhic_expt_fit}
\end{figure}
%

The posterior predictive distribution for the model simultaneously calibrated to both collision systems is shown in Fig.~\ref{grad_lhc_rhic_expt_fit}, where we have plotted the emulator predictions of the observables at one hundred samples drawn from the posterior. 
Note that
our hybrid model can describe simultaneously all of the observables we considered for the two systems to within 20\% of the experimental results. As discussed earlier, this is important: our confidence in the significance of this section's parameter estimates rests on a good posterior description of the experimental data.  

\subsection{Maximum a posteriori (MAP)}
\label{ch5:MAP}

\begin{table}[!htb]
\centering
\begin{tabular}{|l|l|l|l|}
\hline
Parameter & Grad & CE & PTB \\ \hline
$N$[2.76\,TeV] & 14.2 & 15.6 & 13.2\\ \hline
$N$[0.2\,TeV] &  5.73 & 6.24 & 5.31\\ \hline
$p$ &  0.063 & 0.063 & 0.139\\ \hline
$\sigma_k$ & 1.05 & 1.00 & 0.98\\ \hline
$w$\,[fm] &  1.12 & 1.19 & 0.81\\ \hline
$d_{\text{min}}^3$\,[fm$^3$] &  2.97 & 2.60 & 3.11\\ \hline
$\tau_R$\,[fm/$c$] &  1.46 & 1.04 & 1.46\\ \hline
$\alpha$ &  0.031 & 0.024 & 0.017\\ \hline
$T_\eta$\,[GeV] & 0.223 & 0.268 & 0.194\\ \hline
$a_\text{low}$\,[GeV$^{-1}$] & --0.776 & --0.729 & --0.467\\ \hline
$a_\text{high}$\,[GeV$^{-1}$] &  0.37 & 0.38 & 1.62\\ \hline
$(\eta/s)_{\text{kink}}$ & 0.096 & 0.042 & 0.105 \\ \hline
$(\zeta/s)_{\text{max}}$ & 0.133 & 0.127 & 0.165\\ \hline
$T_\zeta$\,[GeV] &  0.12 & 0.12 & 0.194\\ \hline
$w_{\zeta}$\,[GeV] &  0.072 & 0.025 & 0.026\\ \hline
$\lambda_{\zeta}$ & --0.122 & 0.095 & --0.072\\ \hline
$b_{\pi}$ &  4.65 & 5.62 & 5.54\\ \hline
$\Tsw{}$\,[GeV] & 0.136 & 0.146 & 0.147\\ \hline
\end{tabular}
\caption{Table of MAP parameters of the Grad, Chapman-Enskog (CE) and Pratt-Torrieri-Bernhard (PTB) viscous correction models, from combined RHIC and LHC data. }
\label{table_MAP_grad}
\end{table}

\begin{figure}[!htb] 
\centering
\includegraphics[trim=0 0 0 25, clip, width=0.6\linewidth]{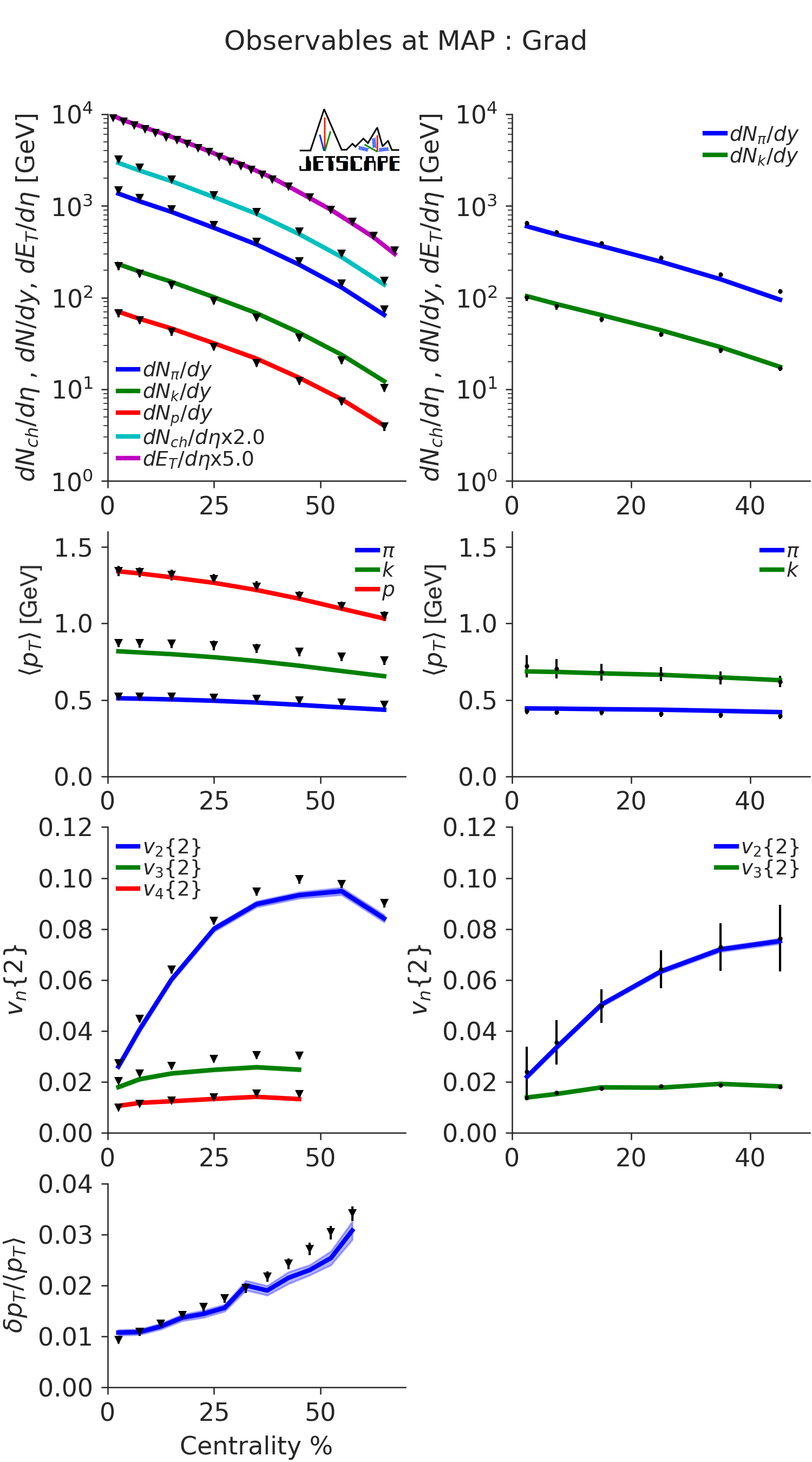}
\caption{The observables resulting from averaging over five thousand fluctuating events for each system, run with the MAP parameters of the combined calibration of ALICE data for Pb-Pb collisions at $\sqrts{} = 2.76$ TeV and STAR data for Au-Au collisions at $\sqrts{} = 200$ GeV. Results are shown for the Grad viscous correction. Shaded bands around model predictions reflect the variance arising from initial state fluctuations and finite particle statistical fluctuations. Pb-Pb $\sqrts{}=2.76$ TeV events are shown at left and Au-Au $\sqrts{}=0.2$ TeV events at right.}
\label{grad_lhc_rhic_MAP_observables}
\end{figure}

We have calculated the Maximum A Posteriori (MAP) predictions of the Grad viscous correction model using the model simulator. Using these parameters, we simulated 5,000 fluctuating events and performed centrality averaging. The comparison between the hybrid model prediction at the MAP parameters and the experimental data are shown in Fig.~\ref{grad_lhc_rhic_MAP_observables}, and MAP parameters for the Grad, Chapman-Enskog and Pratt-Torrieri-Bernhard models are listed in Table~\ref{table_MAP_grad}.\footnote{%
For reasons explained in Ch.~\ref{ch5:compare_df}, the Pratt-Torrieri-McNelis model (PTM) is omitted from Table~\ref{table_MAP_grad}.}

Because our prior for each of these parameters was uniform on a finite range, the parameters which maximize the posterior also maximize the likelihood function; this means that they also optimize the fit to the experimental data (i.e. minimize $\chi^2$). 

\clearpage

\section{Exploration of posteriors, model-dependencies and uncertainties}
\label{ch5:param_est_model_unc}

In this section, we continue our exploration of the estimated parameter posterior for the combined LHC Pb-Pb $\sqrts = 2.76$ TeV and RHIC Au-Au $\sqrts = 0.2$ TeV data. We identify and discuss some of the largest sources of theoretical bias and uncertainty in the physical model, and the influence these uncertainties have on constraining the viscosities of QGP. 
The first source of bias that we investigate in Ch.~\ref{ch5:particlization} originates from mapping the hydrodynamic fields to hadronic momentum distributions, the ``viscous corrections'' at particlization. This can be considered to be related to an uncertainty in the space of particlization models.
Recall that the results from the previous section were for a specific choice of viscous corrections, the Grad model.
A second source of uncertainty is the determination of the particlization hypersurface, which in this work is defined at a fixed switching temperature $\Tsw{}$. This is a parametric source of uncertainty for each model.
We discuss the dependence of our results on this switching temperature in \ref{ch5:init_cond_Tsw}. We discuss at the same time the transition between the early stage of the model and hydrodynamics, which we find exhibits clear correlation with the switching temperature.
Finally, we discuss the effect of second-order transport coefficients, an additional parametric source of uncertainty, as quantified with the shear relaxation time in \ref{ch5:taupi}.

\subsection{Mapping hydrodynamic fields to hadronic momentum distributions}
\label{ch5:particlization}

\begin{figure}[!htb] 
\centering
\includegraphics[trim=0 0 0 17, clip, width=0.8\textwidth]{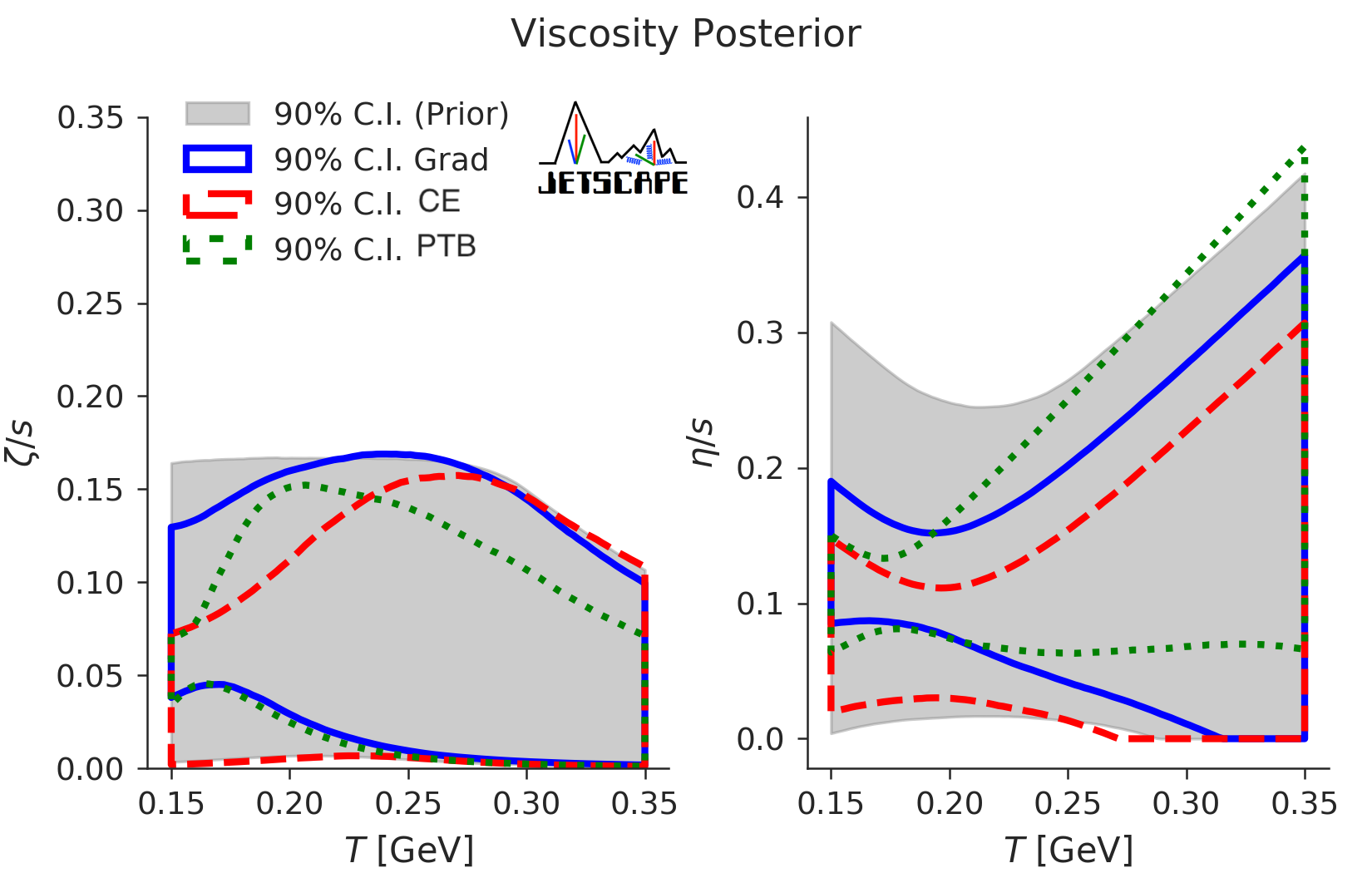}
\caption{The 90 \% credibility intervals for the prior (gray shaded area) and for the posteriors (colored outlines) of the specific bulk (left) and shear (right) viscosities, for three viscous correction models: Grad (blue), Chapman-Enskog (CE, red) and Pratt-Torrieri-Bernhard (PTB, green). The Pratt-Torrieri-McNelis (PTM) posterior is not shown, but is nearly identical with the Chapman-Enskog result. 
}
\label{compare_visc_posteriors}
\end{figure}

As discussed in Ch.~\ref{ch3:particlization}, there are still significant uncertainties in matching the energy-momentum tensor from hydrodynamics to hadronic momentum distributions. Inevitably, using simple and approximate models may bias our estimates of the shear and bulk viscosity of QCD, and all features of the model; the joint posterior of every model parameter is conditional on the specification of viscous correction model at particlization. Recall that in this work, we chose to study four different models of viscous corrections 
: 
(i) Grad (``14-moments''); (ii) Chapman-Enskog in Relaxation Time Approximation; (iii) an exponentiated version of the Chapman-Enskog model referred to as ``Pratt-Torrieri-McNelis''; and (iv) an additional exponentiated model of viscous corrections referred to as ``Pratt-Torrieri-Bernhard''. In our tests, we found that the posteriors for the exponentiated Chapman-Enskog ansatz called ``Pratt-Torrieri-McNelis'' were always very similar to the results for the linearized Chapman-Enskog ansatz. To simplify exposition and reduce clutter in each plot, we therefore decided not to show the posteriors for the Pratt-Torrieri-McNelis model, neither in this section nor anywhere else in this work. We begin with the marginalized posteriors for the QGP viscosities, shown in \fig{compare_visc_posteriors}.
%
%
The figure illustrates clear differences in the experimentally preferred shear and bulk viscosities for the different viscous correction models. Remember that it is essential to read the posteriors in \fig{compare_visc_posteriors} with respect to the prior, whose 90\% credibility region is indicated by the gray shaded area. 
If and where the posterior covers the same area as the prior means that experimental information, which enters the likelihood, is weak. On the other hand, where the posterior systematically excludes certain regions of the prior provides evidence that parameter values in these excluded regions are disfavored by data.

For the bulk viscosity (left), we can see in \fig{compare_visc_posteriors} that each of the different viscous correction models excludes only relatively small regions of the prior. For all particlization models the constraints on $\zeta/s$ are tighter at lower temperatures than at higher ones. However, the $\zeta/s$ regions favored by each model at low temperature differ from each other: the Grad viscous correction model favors a larger $\zeta/s$ where the Chapman-Enskog model favors lower values, with the Pratt-Torrieri-Bernhard model lying in between. We note in particular that the Pratt-Torrieri-Bernhard posterior is very narrow at low temperature. We understand this to be a consequence of mean transverse momenta, pion yield and harmonic flows being very sensitive to the bulk pressure on the switching surface for the Pratt-Torrieri-Bernhard viscous correction model. We quantify and revisit this difference in sensitivity of the viscous correction models in Ch.\ref{ch5:model_sensitivity}.
Overall, only large values of $\zeta/s$ at low temperature are excluded by all three viscous correction models. As such, our constraints on the bulk viscosity are limited, especially after accounting for the model uncertainty introduced by the viscous corrections.

For the shear viscosity shown in the right panel of \fig{compare_visc_posteriors} we encounter a similar situation: limited constraints on $\eta/s$ at higher temperatures, and exclusion of large values of $\eta/s$ at low temperature by all viscous correction models. Overall, shear viscosity is best constrained at temperatures around 200\,MeV.

From the results of this section, we see that viscous corrections represent a considerable source of model uncertainty in constraining the QGP shear and bulk viscosities. It is important to remember that all viscous correction models studied in this work are based on relatively simple assumptions. The capability of any of these models to describe correctly the momenta and chemistry of a realistic out-of-equilibrium system of hadrons is still under investigation (see Ref.~\cite{Damodaran:2017ior, Damodaran:2020qxx} and references therein for a recent overview). For instance, all of these particlization models assume that the hydrodynamic shear stress is shared ``democratically'' among the hadronic species. This approximation greatly simplifies the models, but microscopic transport theory suggests that it may not be suitable for heavier hadrons such as protons ~\cite{Molnar:2014fva}. Additional theoretical efforts (see e.g. Refs.~\cite{Dusling:2009df, Molnar:2014fva, Wolff:2016vcm, Damodaran:2017ior, Damodaran:2020qxx}) may be able to shed more light on this question and provide additional insights that exclude certain particlization models or build more realistic ones. Until this happens the particlization model uncertainty must be considered as ``irreducible'' and can be propagated for by Bayesian Model Averaging as reported in Ch.\ref{ch6:bma_df}, or model-mixing more broadly. 

\subsection{Transition to and from hydrodynamics: initial state and switching temperature}
\label{ch5:init_cond_Tsw}

The previous section focused on the uncertainty originating from transitioning from a hydrodynamic to kinetic description of the system. This transition occurs on a hypersurface defined by a temperature $\Tsw{}$. Recall that this switching temperature is also a model parameter, allowed to vary between $135$ and $165$~MeV.
The other transition point to hydrodynamics is the time at which hydrodynamics is initialized with the energy-momentum tensor from the preceding free-streaming evolution (Ch.~\ref{ch3:pre_hydro}). This hypersurface is defined at a constant proper time $\taufs{}$, the value of which depends on two parameters as defined in \eq{eq:taufs}. The hydrodynamic initial conditions on this hypersurface further depend on the initial condition parameters of the \trento{} ansatz. In this section, we discuss the posterior of $\Tsw{}$, $\taufs{}$ and the \trento{} parameters, how they are correlated, and how they are affected by the viscous correction models discussed in the previous section.

\begin{figure*}[!htb] 
\centering
\includegraphics[width=\textwidth]{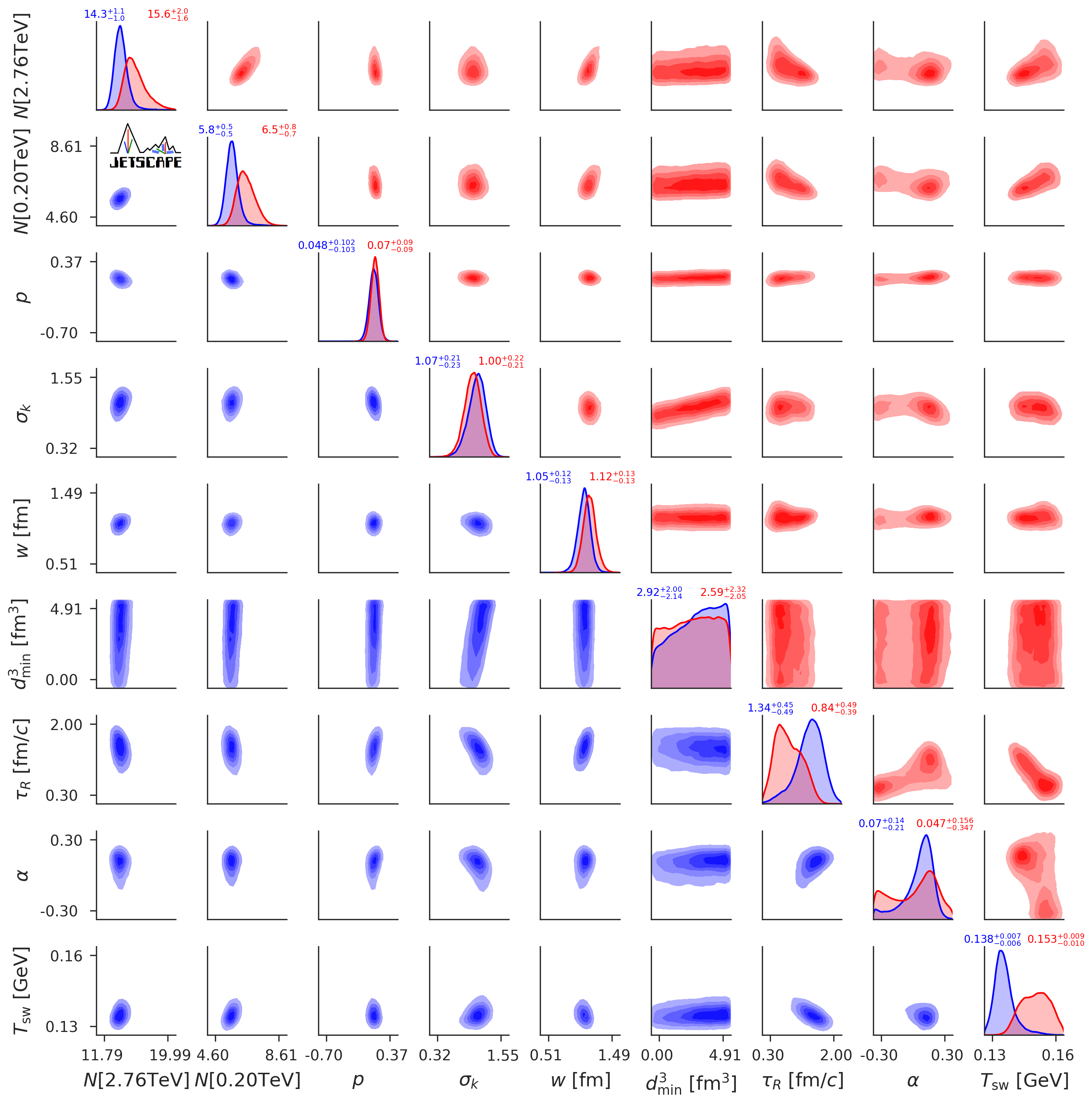}
\vspace*{-3mm}
\caption{
The posterior for Grad (blue) and Chapman-Enskog (red) viscous corrections for select parameters related to the initial state, pre-hydrodynamic evolution and switching temperature. The histograms on the diagonal are the marginal distributions for each parameter, with appended numbers denoting the median and the left and right limits of the $90$\% credible interval. Off-diagonal histograms display the joint-posterior of each pair of parameters, marginalized over all others.}
\label{initial_state_posterior}
\end{figure*}

\fig{initial_state_posterior} provides a dimensionally reduced representation of the joint posterior probability distribution for all model parameters except those related to shear and bulk stress, for two viscous correction models, Grad (blue) and Chapman-Enskog (red). Densities on the diagonals are the marginalized one-dimensional posteriors for each parameter. The off-diagonal densities are the joint posterior densities for each pair of parameters, marginalized over all others. 

One observes that, within the chosen prior range, the normalizations of the initial energy density for the two systems $N[2.76$\,TeV$]$ and $N[0.2$\,TeV$]$ are well constrained by the observables, for both viscous correction models, but with slightly shifted peak values. Note that, since the final multiplicities are fixed by experiments, lower normalization factors for the initial energy (and hence entropy) density reflect larger viscous heating effects during the subsequent dynamical evolution. The amount of viscous heating is also affected by the particlization temperature $\Tsw$, with lower values of $\Tsw$ corresponding to longer lifetimes of the hydrodynamic stage. 

We further find that the estimation of the generalized mean parameter $p$ is nearly the same for the two viscous correction models, close to $p=0$. The estimate of $p$ is also close to zero ($p\approx 0.1$) with the Pratt-Torrieri-Bernhard viscous model (see Ch.\ref{ch6:bma_df}). These posteriors for $p$ are consistent with previous studies which also used the Pratt-Torrieri-Bernhard viscous correction model but differed in other model details~\cite{Bernhard:2018hnz}. The result $p\approx 0$ seems to be robust across all existing Bayesian inference analyses of high-energy heavy-ion collision data ~\cite{Bernhard:2016tnd, Bernhard:2018hnz, Moreland:2018gsh, Bernhard:2019bmu}.\footnote{Perhaps even more robust considering that the first study ~\cite{Bernhard:2016tnd} used \trento{} to define the initial \textit{entropy density}, \textit{not} the \textit{energy density} as did subsequent studies.} 

We note that \trento{} with $p=0$ shares important aspects of fluctuating collision geometry with phenomenologically successful initial condition models based on saturation physics. For example, $p=0$ predicts that the energy deposition is proportional to $\sqrt{T_A T_B}$ as discussed in Ch.~\ref{ch3:initial_conditions}. This 
feature is also found for the \textit{entropy density} predicted by the pQCD+saturation based EKRT initial condition model~\cite{Niemi:2015qia}.\footnote{%
    Though the EKRT model used a different parametrization for the relation between entropy density and $T_A T_B$, its functional form agrees very well with the $\sqrt{T_A T_B}$ relation for typical nuclear thickness functions obtained for lead nuclei.}
In models with approximate longitudinal boost-invariance, the $\sqrt{T_A T_B}$ dependence can be motivated by arguments based on conservation of energy and momentum during the initial energy deposition process~\cite{Shen:2020jwv}. Earlier studies ~\cite{Moreland:2014oya, Bernhard:2016tnd} further noted that for $p\approx 0$ \trento{} can reproduce the centrality dependent 2-particle cumulant eccentricity $\epsilon_2$ and triangularity $\epsilon_3$ of the IP-Glasma initial condition model ~\cite{Schenke:2012wb}. However, one should keep in mind that the two models have very different participant scaling of local energy deposition. According to Eqs.~(\ref{eq:TrentoEd},\ref{eq:TR}), \trento{} for $p=0$ sets the initial local energy density proportional to $\sqrt{T_A T_B}$, but the IP-Glasma model predicts a $T_A T_B$ scaling immediately after the collision ~\cite{Lappi:2006hq}. The two models also have different levels of granularity and fluctuation in the energy deposition ~\cite{Schenke:2012wb}. Moreover, studies ~\cite{Gale:2012rq, McDonald:2016vlt} that used the IP-Glasma model to initialize the hydrodynamics defined centrality differently from the present and earlier studies using the \trento{} model  ~\cite{Bernhard:2016tnd,Moreland:2018gsh}. All these differences are convoluted in the comparison of centrality dependent $\epsilon_2$ and $\epsilon_3$ between the two models. Therefore, \trento{} ($p=0$) should not be considered a substitute for these theories based on saturation physics but rather taken as an efficient parametrization of general geometric features shared by these initial state models that is evidently preferred by the experimental data.

The nucleon width $w$, which controls the transverse length scale of energy fluctuations in the initial state, is also well constrained by the data and found to be about $1.1$\,fm, nearly independent of the Grad or Chapman-Enskog RTA viscous correction models. A smaller value for this nucleon width ($w \approx 0.9$~fm) was found, however, with the Pratt-Torrieri-Bernhard viscous correction (see Ch.\ref{ch6:bma_df}).

In general, our conclusions for the \trento{} parameters is that they do not appear to be highly sensitive to the choice of viscous correction model at particlization. However, the particlization uncertainty should not be ignored as it can be larger than the width of the posteriors for each of these parameters. We will use model-averaging to marginalize over this source of uncertainty in Ch.\ref{ch6:bma_df}.

\begin{figure}[!htb] 
\centering
\includegraphics[trim=0 0 0 17, clip, width=0.5\textwidth]{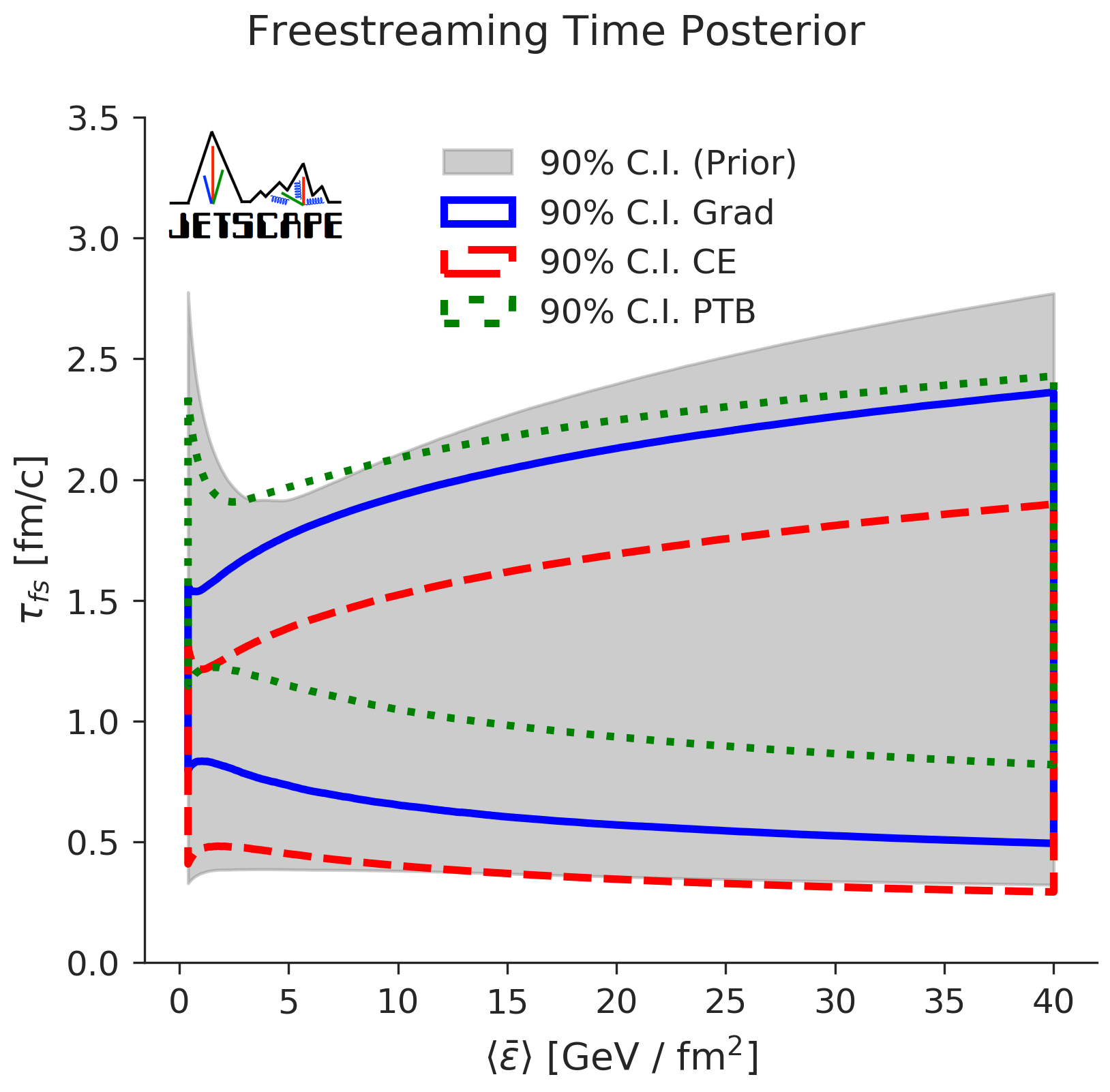}
\caption{
The 90\% posterior credible intervals for the free-streaming time, as a function of the initial average transverse energy density defined in Eq.~(\ref{eps_ave}), resulting from parameter estimation using combined data for Au-Au collisions at $\sqrts{} = 200$\,GeV and Pb-Pb collisions at $\sqrts{} = 2.76$\,TeV. The Grad model is shown with solid blue, Chapman-Enskog with dashed red, and Pratt-Torrieri-Bernhard with dotted green lines.}
\label{tau_fs_posterior}
\end{figure}

The posteriors for the free-streaming time scale $\tau_R$ and the associated energy dependence parameter $\alpha$ are not easily interpreted; they are correlated by our parametrization (\eq{eq:taufs}) of the effective free-streaming time $\taufs$. We can point out, however, that the posteriors for the Grad and Chapman-Enskog models are quite different for these two parameters; in the case of the Chapman-Enskog model, the posterior for $\alpha$ is bimodal. It is not clear whether the peak near $\alpha \sim -0.3$ is a local maximum or if there exists a global maximum in the posterior for values of $\alpha < -0.3$. We cannot currently differentiate between these two scenarios.

We also plot in \fig{tau_fs_posterior} the posterior of the free-streaming time as a function of the physical scale $e_0$, which is the magnitude of the average initial energy density in the transverse plane.
We see that the 90\% credible interval for the energy dependence of the free-streaming time is not well constrained, and that it is consistent with having no energy dependence. What is constrained is the overall magnitude of the free-streaming time. The Chapman-Enskog model has a posterior which prefers smaller free-streaming times, while the Pratt-Torrieri-Bernhard model prefers the largest free-streaming time of all particlization models studied. 

It is expected that collisions with higher energy density will hydrodynamize more rapidly ~\cite{Basar:2013hea}. In our model this would correspond to $\alpha<0$. The peak at $\alpha>0$ in the posterior for $\alpha$ is at variance with this expectation. One should remember, however, that our pre-hydrodynamic model does not actually lead to hydrodynamization, or even isotropization, at $\taufs$. As such, it is conceptually problematic to associate the free-streaming time $\taufs$ with a hydrodynamization time. As discussed in connection with \eq{Pi_init}, matching an energy-momentum tensor from a conformally invariant pre-hydrodynamic evolution model without thermalization to dissipative hydrodynamics with a non-conformal EoS leads to a (possibly large) {\it positive} initial value for the bulk viscous pressure, whose subsequent decay can have counter-intuitive effects on the hydrodynamic flow and its dependence on $\taufs$. Recent studies demonstrate that this problem persists when the free-streaming module is replaced by a thermalizing but conformal effective kinetic theory, and that the magnitude of the mismatch depends on centrality ~\cite{NunesdaSilva:2020bfs}. Although we have not been able to fully dissect the mechanisms leading to positive preferred values for $\alpha$ in our analysis, we strongly suspect that these issues play a role. 

Turning to the later stages of the collision, we now look at the posterior for the switching temperature in \fig{initial_state_posterior}. Its marginalized posterior turns out to be quite different for the two particlization models. We find that for the selected experimental observables the effects of increasing the magnitude of the bulk viscous pressure or increasing $\Tsw{}$ are qualitatively similar. We verified that if we hold all other parameters fixed while increasing the switching temperature from 135 MeV to 165 MeV, the mean transverse momenta of pions and protons is reduced and the number of protons is increased. On the other hand, holding $\Tsw{}$ fixed and increasing $(\zeta/s)_{\text{max}}$ has the same effect. Because the Grad model prefers a large specific bulk viscosity near switching, it also prefers a lower switching temperature. 

Although some of the parameters which define the \trento{} model are well constrained, their interpretation is not always straightforward. To what extent $p\approx0$ in the \trento{} model provides support for saturation physics or is mostly a consequence of energy-momentum conservation combined with approximate boost invariance at high collision energies deserves further study. Similarly, we found our posterior for the energy density dependence of the free-streaming time difficult to understand. Further theoretical understanding of these issues is likely to find its way into improved models. For example, there is considerable room for improving the description of the pre-hydrodynamic evolution stage. It is encouraging that, using the likely values for the \trento{} model parameters, we are able to describe our experimental observables with good accuracy. Still, there is obvious value in seeking models which can not only fit the experimental data, but at the same time offer a coherently and consistently interpretable physical picture.

\subsection{Second-order transport coefficients: shear relaxation time}
\label{ch5:taupi}

As discussed in Ch.\ref{ch3:hydro}, second-order transport coefficients are treated differently in this work than first-order ones: the shear and bulk viscosities are parametrized, while the second-order transport coefficients are related to the first-order ones through relations derived in kinetic theory. The one exception is the shear relaxation time $\tau_\pi$, whose normalization is allowed to vary in a theoretically motivated range. This allows for a quantification of the contributed uncertainty of a second-order transport coefficient on phenomenological estimations on $\eta/s$ and $\zeta/s$.

\begin{figure}[!htb] 
\centering
\includegraphics[trim=0 0 0 17, clip, width=0.7\textwidth]{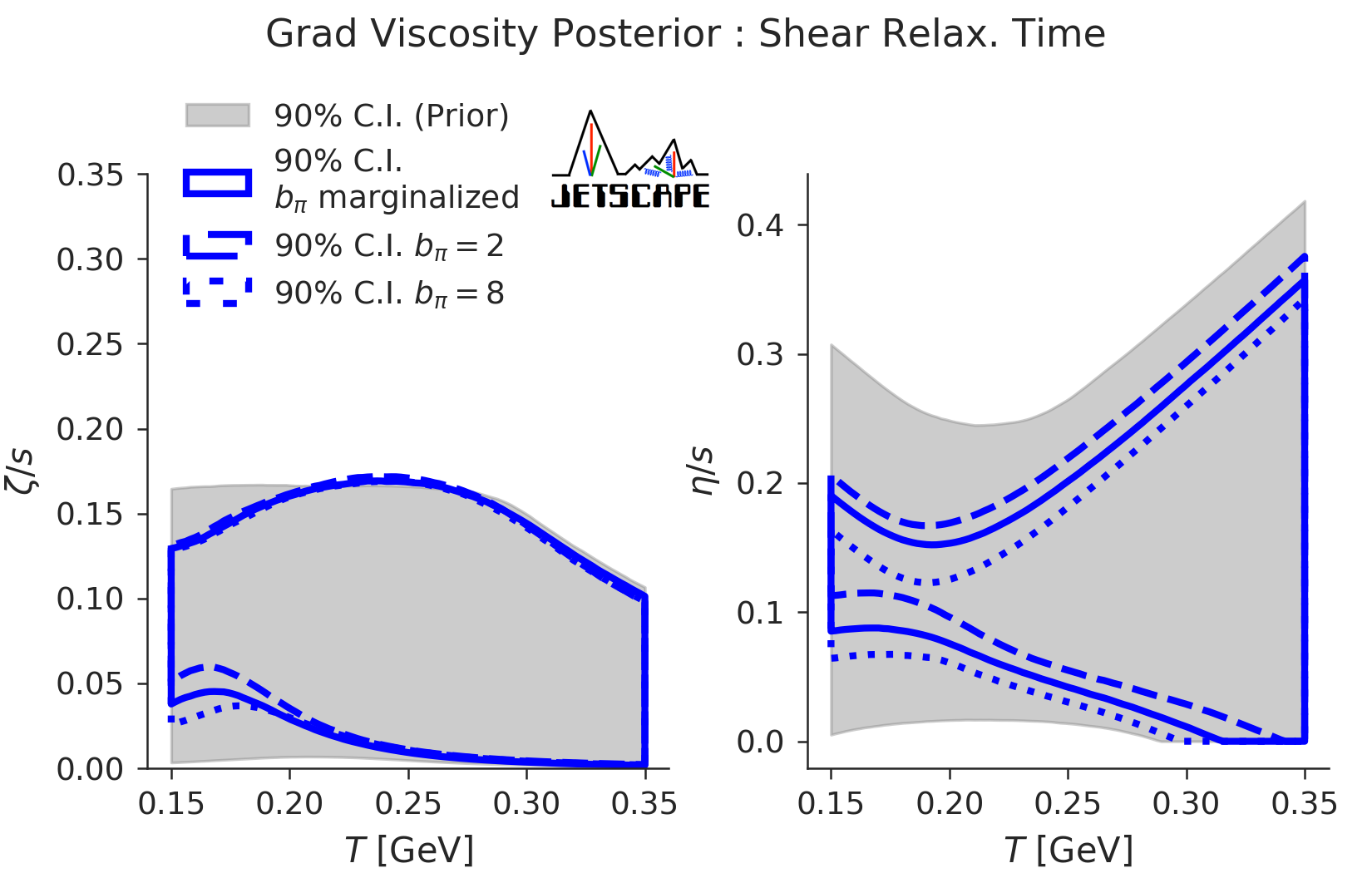}
\caption{The posterior of specific bulk(left) and shear(right) viscosities, depending on whether one marginalizes over the shear relaxation time factor $b_{\pi}$ (solid blue) or fixes it (dashed or dotted blue). The shear relaxation time and magnitude of $\eta/s$ are seen to be inversely related when fitting the LHC and RHIC data. }
\label{visc_post_shear_relax}
\end{figure}

Figure~\ref{visc_post_shear_relax} examines the extent to which the shear relaxation time normalization factor $b_{\pi}$ affects the posterior of first-order transport coefficients. 
%
%
When trying to fit the experimental data, we see that in general smaller values of the shear relaxation time are correlated with larger shear viscosities, and vice versa. This is because increasing either $b_{\pi}$ or $\eta/s$ tends to reduce the harmonic flows, for instance. 

Some sensitivity to the shear relaxation time factor $b_{\pi}$ may be caused by the use of free-streaming as a pre-hydrodynamic model, which can generate large initial values of $\pi^{\mu\nu}$. The subsequent relaxation of $\pi^{\mu\nu}$ to its Navier-Stokes limit $2\eta\sigma^{\mu\nu}$ is governed by $\tau_{\pi}$.
Previous viscous hydrodynamics studies, using different initial conditions (not including free-streaming), found a smaller sensitivity to $b_\pi$~\cite{Luzum:2008cw, Song:2009gc}. Note however that these studies also did not consider the higher harmonic flows ($v_3$, $v_4$) which we find to have a stronger sensitivity to $b_{\pi}$ than $v_2$. This will be discussed in Ch.~\ref{ch5:model_sensitivity}.

Since there is significant uncertainty induced by the shear-relaxation time on $\eta/s$ (and to a lesser extent on $\zeta/s$), future efforts should consider varying other second-order transport coefficients. Among those, the most important is likely the bulk relaxation time.
Performing a systematic analysis of all second-order transport coefficients would be a significant future undertaking, in part because their parametric dependence must be specified, and a prior needs to be fixed before performing parameter estimation.\footnote{%
    Existing microscopic calculations such as Ref.~\cite{Denicol:2014vaa} can help constrain the parametric dependence and the priors.  For example, Ref.~\cite{Denicol:2014vaa} finds $\delta_{\pi \pi} = \frac{4}{3} \tau_{\pi} + \mathcal{O}((m/T)^2)$. One could assume $\delta_{\pi \pi} \propto \tau_{\pi}$ with a parameter being a proportionality factor of order 1.}
Non-conformal hydrodynamic has a large number of second-order transport coefficients ~\cite{Denicol:2012cn}; relatively little is known for many of them. There is value in simultaneously studying these transport coefficients theoretically and phenomenologically. Even if these coefficients cannot be constrained from measurements, their influence on other model parameters should be studied.
We note that a recent study~\cite{Nijs:2020ors, Nijs:2020roc} included several second order transport coefficients.

\section{Full posterior of model parameters}
\label{app:full_post}

\begin{figure}[!htb] 
\hspace{-0.5cm}
\makebox[\textwidth][c]{\includegraphics[width=1.2\textwidth]{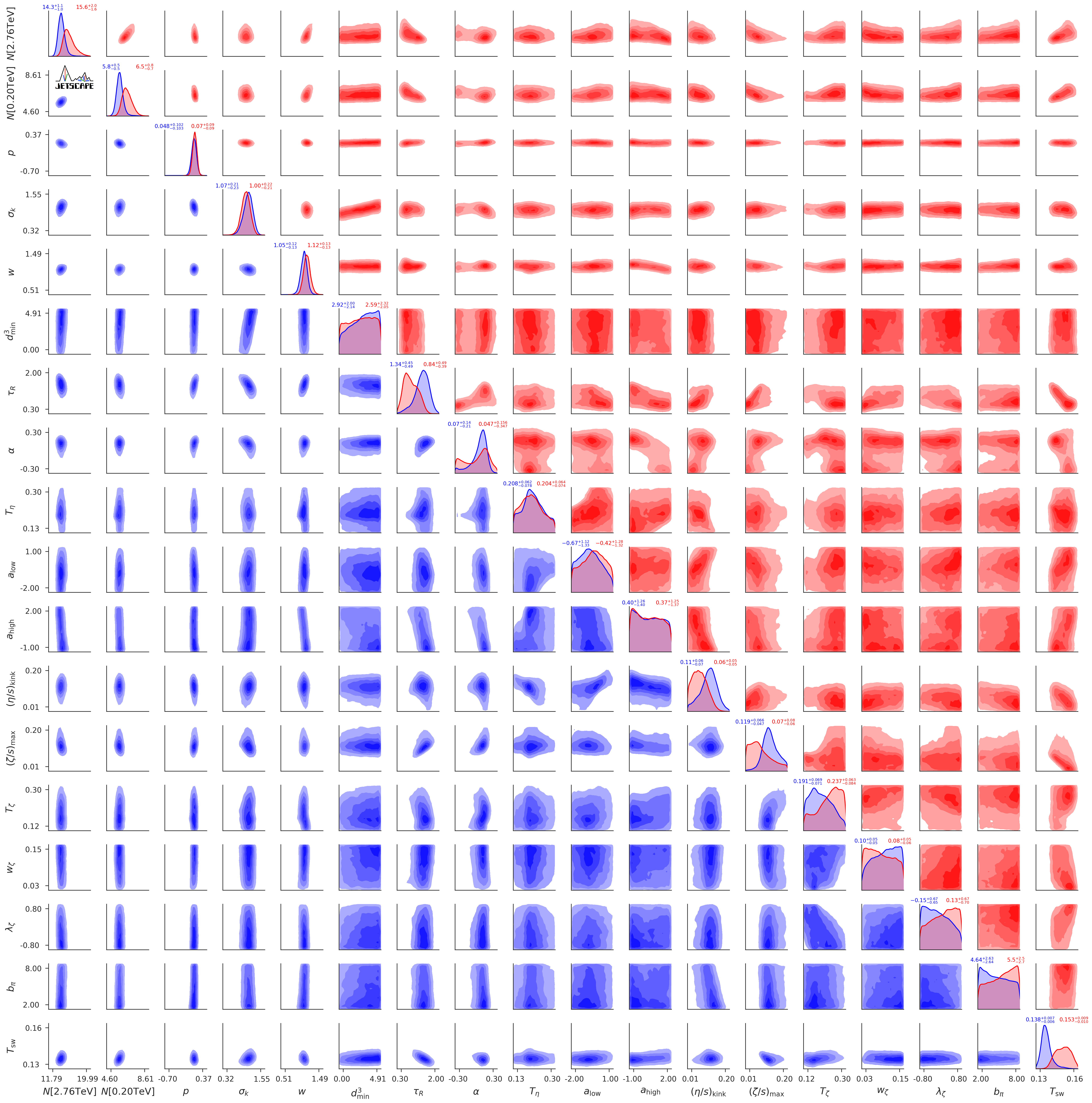}}%
\caption{
The posterior for Grad (blue) and Chapman-Enskog (red) viscous correction models for all model parameters, combining both RHIC and LHC experimental results. Units for dimensionful quantities are those given in Table~\ref{prior_table}.
}
\label{fig:full_posterior}
\end{figure}

For completeness, we show in Fig.~\ref{fig:full_posterior} the posterior of all model parameters single and joint-parameter marginal distributions for the Grad (blue) and Chapman-Enskog (red) viscous correction models, calibrated to both RHIC and LHC experimental results. In general, there is an enormous amount of information in this corner plot of all parameters for the two viscous correction models. We will discuss a few insights which we find important to point out. 

The parameters which are best constrained by the likelihood within their priors tend to be the \trento{} initial conditions. On the other hand, the parameters defining the shear and bulk viscosities are ill-constrained by the likelihood. The only parameters with relatively well-defined peaks are $(\eta/s)_{\rm kink}$, which adjusts the magnitude of the shear viscosity at all temperatures, and $(\zeta/s)_{\rm max}$ which controls the normalization of specific bulk viscosity at the peak. This is consistent with theoretical studies ~\cite{Paquet:2019npk}, which in simpler geometries, and with simpler models, demonstrate that the hadronic observables are more sensitive to the \textit{integrated} effect of the viscosities, rather then their values at specific temperatures. 

The anti-correlation between $(\zeta/s)_{\rm max}$ and $T_{\rm sw}$ is clearly visible in the joint-posterior, as discussed in Ch.\ref{ch5:compare_df}. 
A pronounced positive correlation between $\tau_R$ and $(\zeta/s)_{\rm max}$ can be understood by considering the mean transverse momenta of particles. The freestreaming quickly increases the radial flow, and fitting the experimental data requires a negative bulk pressure. 

There may be certain correlations present which are not related to any underlying physical process, but are rather manifestations of the parametrizations we have chosen. 
Future studies should aim to develop \textit{non-parametric} ansatze of both the specific shear and bulk viscosities, at least to avoid such correlations which are induced by a relatively ad hoc parametrization.

\section{Posteriors for independent calibrations to LHC Pb-Pb and RHIC Au-Au}
\label{app:post_LHC_RHIC_separate}

\begin{figure}[!htb]
\centering
\includegraphics[width=0.7\textwidth]{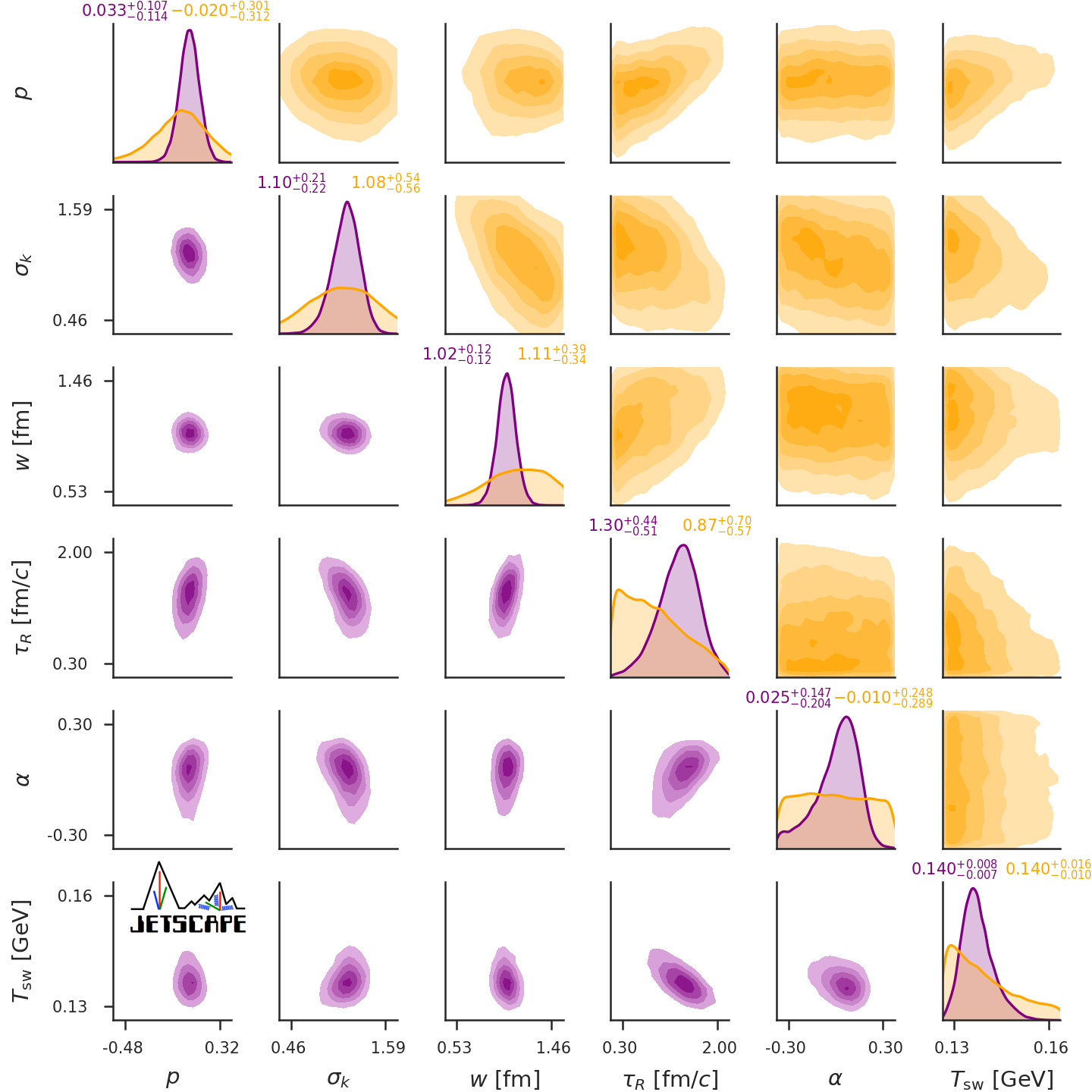}
\caption{The posterior of initial conditions and switching temperature for the Grad viscous correction model using only LHC Pb-Pb $\sqrts{}=2.76$ TeV data (purple) or only RHIC Au-Au $\sqrts{}=0.2$ TeV data (orange). }
\label{fig:post_LHC_RHIC_separate}
\end{figure}

In Fig.~\ref{fig:post_LHC_RHIC_separate}
we show the parameter estimates for select \trento{} initial condition parameters, as well as the switching temperature. Each posterior was estimated using only observables from a single system; LHC Pb-Pb $\sqrts{}=2.76$ TeV observables (purple) or RHIC Au-Au $\sqrts{}=0.2$ TeV observables (orange).
In this study, we have included more observables for Pb-Pb $\sqrts{} =2.76$ TeV collisions at LHC, so the likelihood functions are more tightly constrained, while those for Au-Au $\sqrts{} =0.2$ TeV collisions at RHIC are broader. In addition, the switching temperature for RHIC posterior is poorly constrained because we have omitted the proton yield measured at RHIC, which would be among the more sensitive observables. Overall, we see that there is good agreement in the estimates of these parameters whether one uses Pb-Pb $\sqrts{} = 2.76$ TeV observables or Au-Au $\sqrts{} = 0.2$ TeV observables, and no evidence of statistically significant tension.

\section{Parameter estimation: including Xe-Xe at $\sqrts{}=5.44$ TeV data}
\label{ch5:xe_data}

In this section, we explore the additional constraining power of the centrality dependent charged particle yield $dN_{\rm ch}/d\eta$~\cite{Acharya:2018hhy} and elliptic $v_2\{2\}$ and triangular $v_3\{2\}$ flows~\cite{Acharya:2018ihu} measured in a brief two-day run of Xe-Xe collisions at $\sqrts{} = 5.44$ TeV at the LHC. 
The analysis proceeded following all of the methods described in Ch.\ref{ch4} and Ch.\ref{ch5} for modeling both Pb-Pb $\sqrts{} = 2.76$ TeV and Au-Au $\sqrts{} = 0.2$ TeV collisions.   
An emulator for the Xe observables using the Grad viscous correction model was constructed using a parameter design of one thousand points, and running approximately 2500 events per design point. After validating the Grad model emulator for Xe-Xe $\sqrts{}=5.44$ TeV, we proceeded to perform Bayesian parameter estimation.

Our modeling assumptions allowed only the energy-density normalization parameter $N$ in \trento{} to differ from the corresponding normalizations of Pb-Pb $\sqrts{} = 2.76$ TeV and Au-Au $\sqrts{} = 0.2$ TeV collisions. 
However, we required all other model parameters to be shared between all three systems when calibrating the model simultaneously against all three sources of data. Importantly, in \trento{} we defined the Xenon ($A=129$) nucleus by a deformed Woods-Saxon ansatz, with radius $R=5.36$ fm, thickness $a=0.59$ and deformation parameters ${\beta_2 = 0.162, \beta_4=-0.003}$~\cite{Moller:2015fba}. These Woods-Saxon parameters defining the Xenon nucleus in \trento{} were fixed, not varied or estimated.

\begin{figure*}[!htb]
\centering
\includegraphics[width=\textwidth]{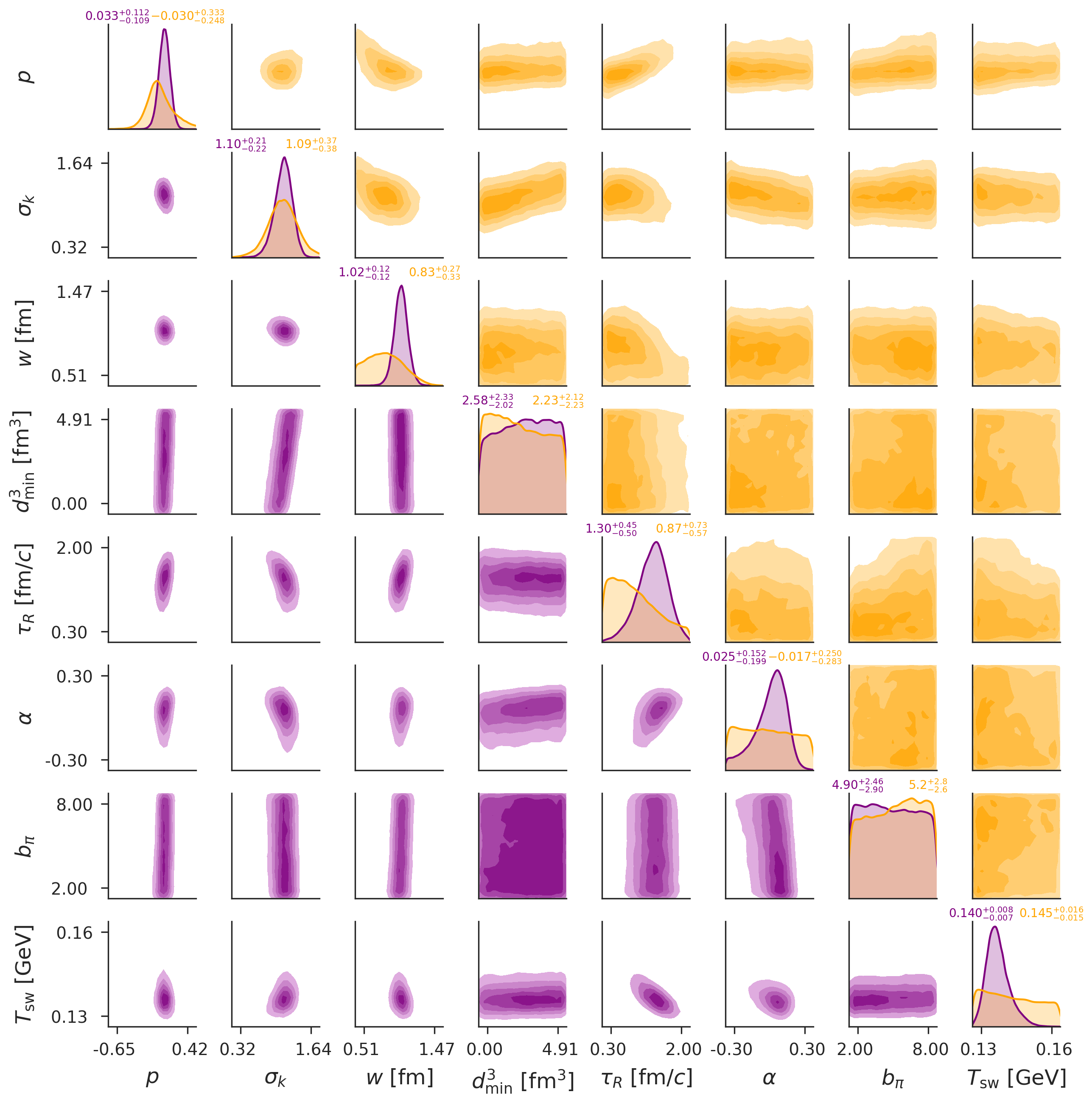}
\caption{Posteriors for the initial conditions and switching temperature, depending on whether one includes data from Pb-Pb $\sqrts{}=2.76$ TeV collisions(purple) or Xe-Xe $\sqrts{}=5.44$ TeV collisions(orange) in the likelihood. }
\label{fig:ic_tsw_pb_xe}
\end{figure*}

Because these data are limited in comparison to e.g. the wealth of observables at Pb-Pb $\sqrts{}=2.76$ TeV, there is a significant volume of parameter space in which the model can describe the Xe-Xe data without tension. Therefore, standing alone these Xe-Xe data are not very constraining on the specific shear and bulk viscosities. 
In Fig. \ref{fig:ic_tsw_pb_xe} is shown the corner-plot of the posterior for select model-parameters excluding the specific shear and bulk-viscosities. In this case, the model is either calibrated against only the Pb-Pb $\sqrts{}=2.76$ TeV data (purple) or only the Xe-Xe $\sqrts{}=5.44$ TeV data (orange), to compare the constraining power of each set of data separately. In general, we find the Pb-Pb $\sqrts{}=2.76$ TeV to offer more sharply-peaked likelihood functions for nearly all of the model parameters. We believe this follows simply from the larger number of experimental data, including identified particle yields and mean $p_T$, transverse energy, mean $p_T$ fluctuations and quadrangular flow $v_4\{2\}$ which have not been measured in Xe collisions. 

In particular, we note that the absence of the proton-yield leaves the switching temperature for Xe-Xe collisions unconstrained. The \trento{} generalized mean parameter $p$ estimated by either data set are consistent within uncertainties, and roughly $p \sim 0$. Similarly, the energy-density fluctuation parameter $\sigma_k$ is very consistently estimated across the two systems. Only a small amount of tension can be seen in the estimation of the nucleon width $w$ and the freestreaming time $\tau_R$, although the estimates are consistent within the $90\%$ credible intervals. 

\begin{figure}[!htb]
\centering
\includegraphics[width=0.7\textwidth]{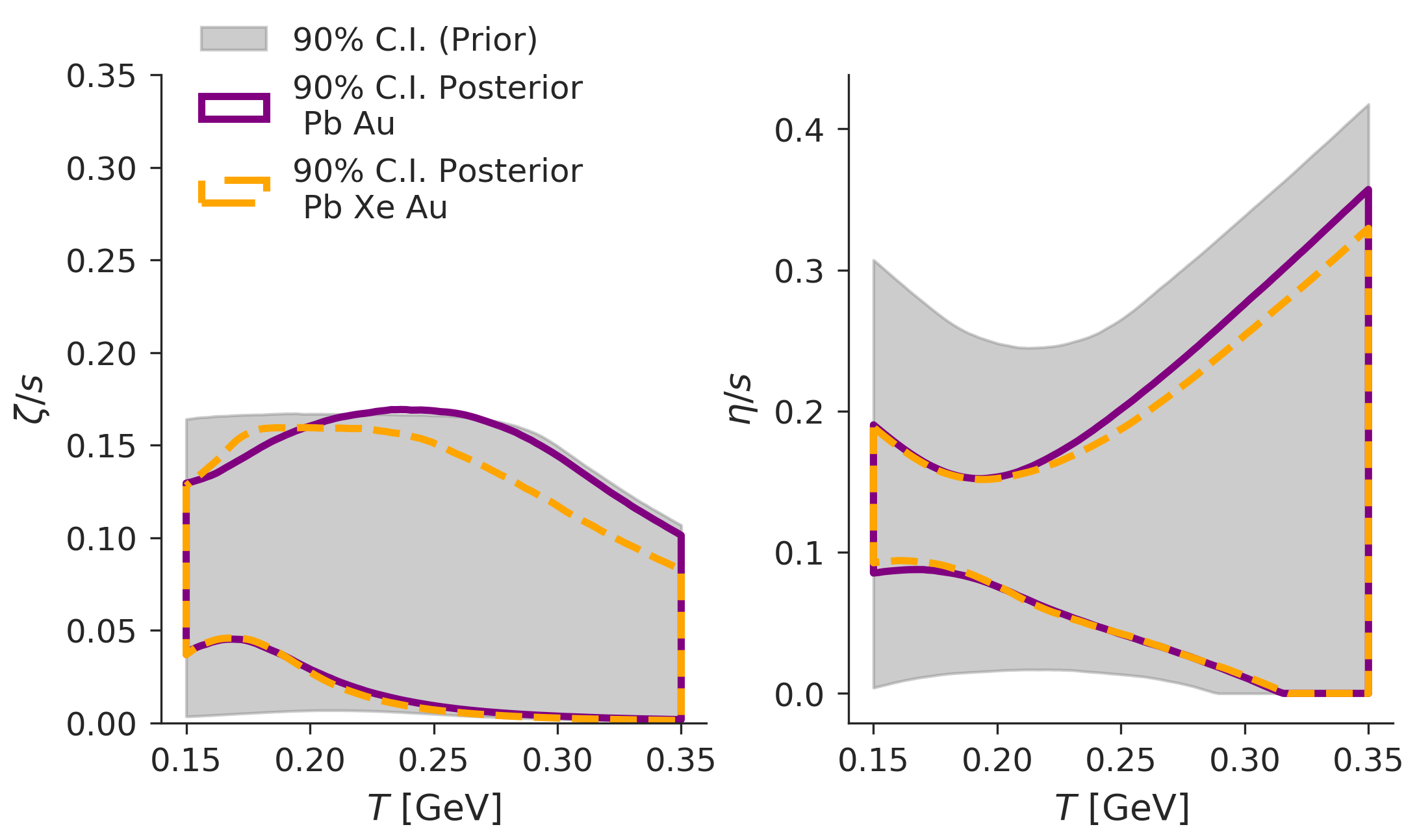}
\caption{Posteriors for the specific bulk(left) and shear(right ) viscosities, for the Grad viscous correction model, including data from both Pb-Pb $\sqrts{}=2.76$ TeV collisions and Au-Au $\sqrts{}=0.2$ TeV collisions, whether one jointly calibrates against Xe-Xe $\sqrts{}=5.44$ TeV observables. }
\label{fig:shear_bulk_pb_au_xe}
\end{figure}

We also consider fitting our model simultaneously against all three sources of data, to consider whether tension or additional information among the three systems can reduce the inferred specific shear and bulk viscosities. The specific shear and bulk viscosities calibrated against all three collision systems observables are shown in Fig. \ref{fig:shear_bulk_pb_au_xe}. We find that the inclusion of Xe can nominally reduce the uncertainty of the bulk viscosity at high-temperatures $T \gtrsim 0.25$ GeV, with minimal information gained regarding the specific shear-viscosity.

\begin{figure*}[!htb]
\centering
\includegraphics[width=\textwidth]{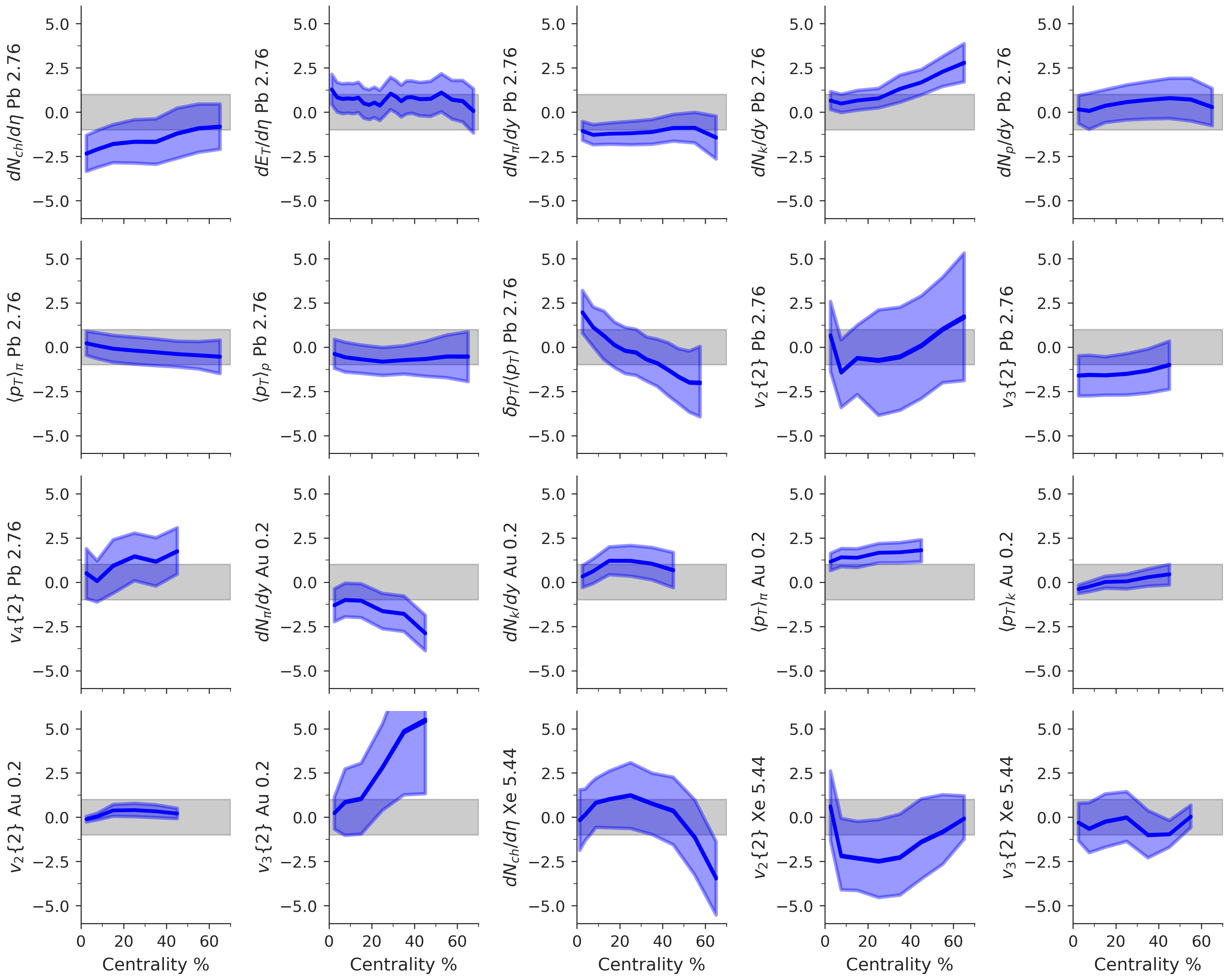}
\caption{Posterior predictive distributions of the model-data discrepancy for nearly all of the calibration observables for the Grad viscous correction model. The light shaded blue region encloses 90\% of the posterior distribution; the dark blue line is the median.   }
\label{fig:observables_posteriors}
\end{figure*}

As noted by Box ~\cite{box2011bayesian}, while the investigator performs Bayesian parameter estimation with their model, they do so under the belief that their model is `good enough'. This is not just a philosophical concern -- the posterior probabilities obtained for all model parameters are \textit{conditional} on the model employed (and its priors). Therefore, once useful insights have been made from the posteriors, the investigator must switch gears from a position of relative belief in the model(s) to a position which seeks to criticize the model(s). This requires a different set of techniques.

Perhaps the simplest way to investigate our models' inadequacies is by plotting the residuals (discrepancies) between each model and the observed data. Upon visual inspection, we may be able to see patterns in the residuals that can inform us as to where and why the model may fail.
In Fig. \ref{fig:observables_posteriors} we display the posterior-predictive distribution for our model of all three collision systems, plotted as the $90\%$ credible limits of the model-data discrepancy. The discrepancy is normalized by the experimental uncertainty $\sigma_{\rm exp}$, and therefore the $y$-axis of each plot is $(y_m - y_{\rm exp})/\sigma_{\rm exp}$. The plot includes nearly all of the observables used in model calibration. In general, we find very good performance of our hybrid model in describing the observed data for these collision systems. 

We notice that for all centrality bins, this model (Grad) over-predicts the yield of protons $dN_p/dy$ in Pb collisions, and under-predicts the yield of pions $dN_\pi/dy$ in both Pb collisions and Au collisions. 
This informs us that there is a quantitatively small, but systematic failure of our model to describe the chemistry of the heavy-ion collision. Actually, we will show in Ch.~\ref{ch5:model_selection} that other viscous correction models can do even worse in describing the chemistry. The bulk viscous corrections are one source of theoretical modeling that is able to change the abundances of particles from those of an equilibrium hadron resonance gas. However, there are other physical effects which could also be responsible. The dynamics in the hadronic afterburner stage are also responsible for changes in the (anti)proton abundance in particular, and there may also be some inaccuracy in the rate of proton-antiproton annihilation. 

A systematic overprediction of the yields of kaons in both Pb and Au collisions may point to the absence of physical processes for controlling strangeness in our collision model.
Additional systematic tension can be seen in the description of the charged particle yield $dN_{\rm ch}/d\eta$ for Xe. Although the discrepancy is not too large given our uncertainties, the shape of the discrepancy as a function of centrality suggests that something may be missing in our model description of the centrality dependence of charged particle production, and the problem is exacerbated for smaller systems. 

\section{Sensitivity to prior knowledge and assumptions}
\label{ch5:prior_sensitivity}

Each model parameter in this study is assumed to have a uniform prior probability density over a finite range. This range represents crucial prior knowledge or assumptions. Our wider, less subjective prior (Table \ref{restricted_prior_table}, middle column) almost completely\footnote{%
    We do not include a curvature parameter for the shear viscosity at high temperatures as was done in Ref.~\cite{Bernhard:2019bmu} since there it was found, within the given prior limits, to be rather poorly constrained.} 
encloses as a subspace the narrower, more subjective prior range postulated in Ref.~\cite{Bernhard:2019bmu} (Table \ref{restricted_prior_table}, right column).

\begin{table}[!htb]
\centering
\begin{tabular}{|l|l|l|}
\hline
parameter & full prior range & restricted prior \\   & & range or value \\ \hline
$\alpha$ & $[-0.3, 0.3]$ & $0.0$       \\ \hline
$T_{\eta}$\ [GeV] & $[0.13, 0.3]$ &$0.154$        \\ \hline
$a_{\mathrm{low}}$ [GeV$^{-1}$] & $[-2, 1]$ & $0.0$  \\ \hline
$\lambda_\zeta$ & $[-0.8, 0.8]$ & $0$       \\ \hline
$b_{\pi}$ & $[2, 8]$ & $5$      \\ \hline
$p$ & $[-0.7, 0.7]$ & $[-0.5, 0.5]$     \\ \hline
$w$ [fm] & $[0.5, 1.5]$ & $[0.5, 1.0]$     \\ \hline
$\tau_R$ [fm/$c$] & $[0.3, 2]$ & $[0.3, 1.5]$      \\ \hline
$(\zeta/s)_{\mathrm{max}}$ & $[0, 0.25]$ & $[0.01, 0.1]$       \\ \hline
$T_{\zeta}$ [GeV] & $[0.12, 0.3]$ & $[0.15, 0.2]$       \\ \hline
$w_{\zeta}$ [GeV] & $[0.025, 0.15]$ & $[0.025, 0.1]$ \\ \hline
\end{tabular}
\caption{
Table of full (left) and restricted (right) parameter ranges. The restricted prior is similar to the prior employed in Ref.~\cite{Bernhard:2019bmu}.}
\label{restricted_prior_table}
\end{table}

\begin{figure}[!htb]
  \centering
   \includegraphics[trim=0 0 0 17, clip, width=0.7\textwidth]{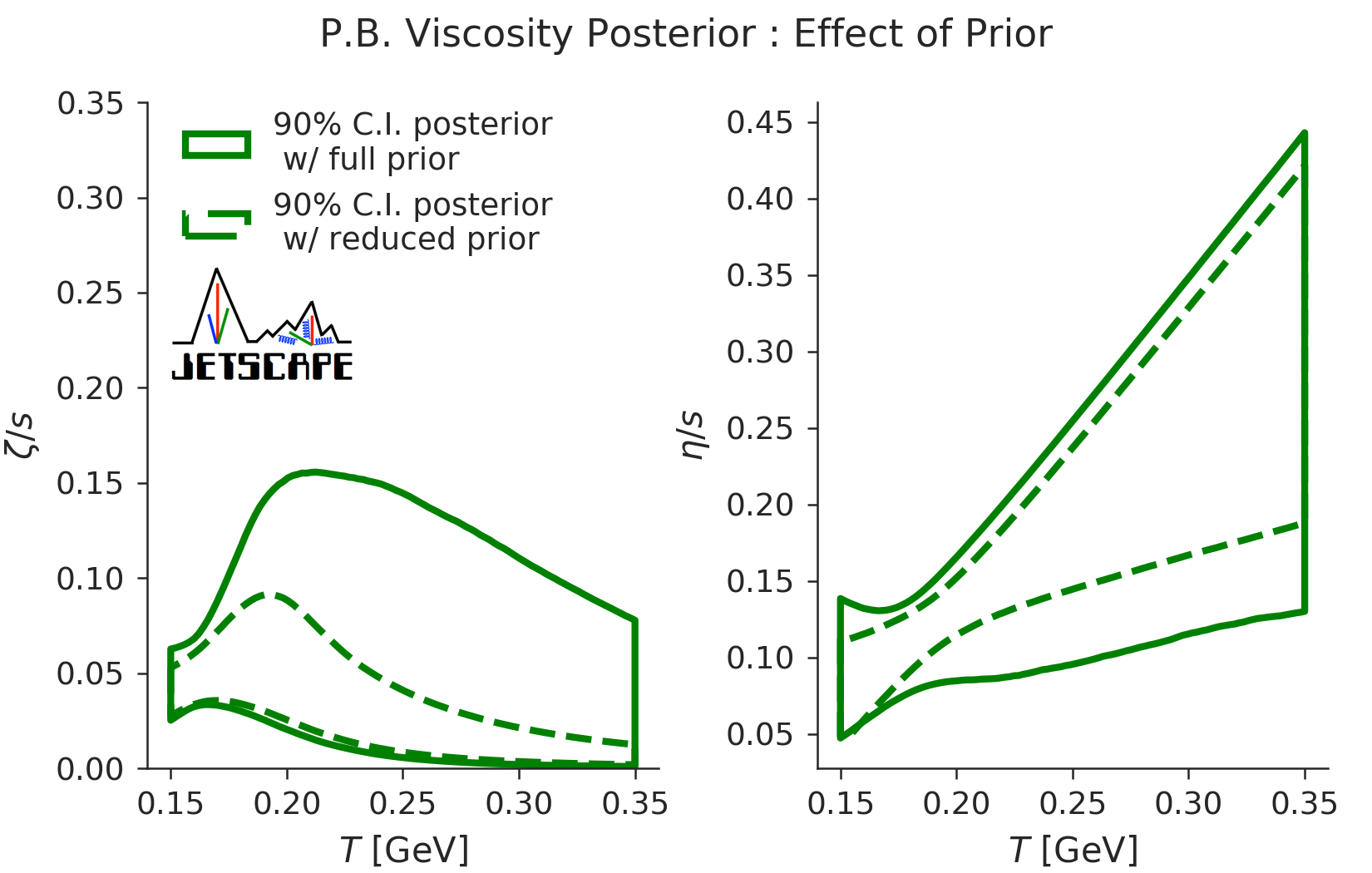}
    \caption{
    The 90\% posterior credible intervals of the specific bulk (left) and shear (right) viscosities for the Pratt-Torrieri-Bernhard viscous correction model, including only observables from LHC Pb-Pb collisions at $\sqrts=2.76$\,TeV, depending on whether one uses a more informed or less informed prior.}
    \label{fig:visc_post_red_prior}
\end{figure}

By comparing the posteriors for these different priors we assess the sensitivity of our inference to prior elicitation. This is illustrated in Fig.~\ref{fig:visc_post_red_prior} for one of the particlization models studied in this work (the Pratt-Torrieri-Bernhard model ~\cite{Pratt:2010jt, Bernhard:2018hnz}). We compare the posteriors for the specific shear and bulk viscosities using either the more or less subjective priors described above. These posteriors were obtained via Bayesian parameter estimation using only the ALICE measurements of Pb-Pb collisions at $\sqrts = 2.76$ TeV. 

%

Clearly the more subjective prior drastically reduces the width of the credible intervals of the posterior for both the shear and bulk viscosities. Table \ref{restricted_prior_table} shows that, in addition to narrower prior ranges for the shear and bulk viscosities, the more restrictive prior assumed additional information about the initial conditions, and the shear relaxation time (which is a second-order transport coefficient). 

Since the posterior of any Bayesian inference is proportional to the product of the prior and likelihood function, a tightening of the prior also causes the posterior to tighten. Insofar the results shown in Fig.~\ref{fig:visc_post_red_prior} are in principle expected. However, the observed large sensitivity of the posterior (in particular for the bulk viscosity) to the prior suggests that the constraining power of the experimental data is still limited, and that for a fully uninformed prior the 90\% confidence intervals for the specific viscosities would be even wider than what is indicated by the solid lines in Fig.~\ref{fig:visc_post_red_prior}. If the data were sufficiently informative and the likelihood sufficiently sharp, small changes in the prior would not significantly change the posterior. 

Future improvements of the precision of our knowledge of the QGP viscosities require progress along at least one of the following two directions: (i) theoretical work leading to more objective priors, if not for the parameters of primary interest (i.e. the viscosities) then at least for the ``nuisance parameters"; (ii) inclusion of additional  measurements into the Bayesian analysis that have the potential to provide tighter likelihood functions for both the parameters of primary interest and the ``nuisance parameters". We note that inclusion of additional observables must be handled with care and consideration for the theoretical model's limitations/discrepancies. 
In the absence of theoretical progress towards tighter first-principles constraints on the viscosities, less subjective priors for these parameters of primary interest should be employed, to minimize sensitivity of the posterior to prior specification.  

\section{Model Sensitivity}
\label{ch5:model_sensitivity}


\begin{figure*}
\makebox[\linewidth][c]{%
\centering
\hspace{-1cm}
\includegraphics[trim=0 0 0 18, clip, width=0.6\textwidth]{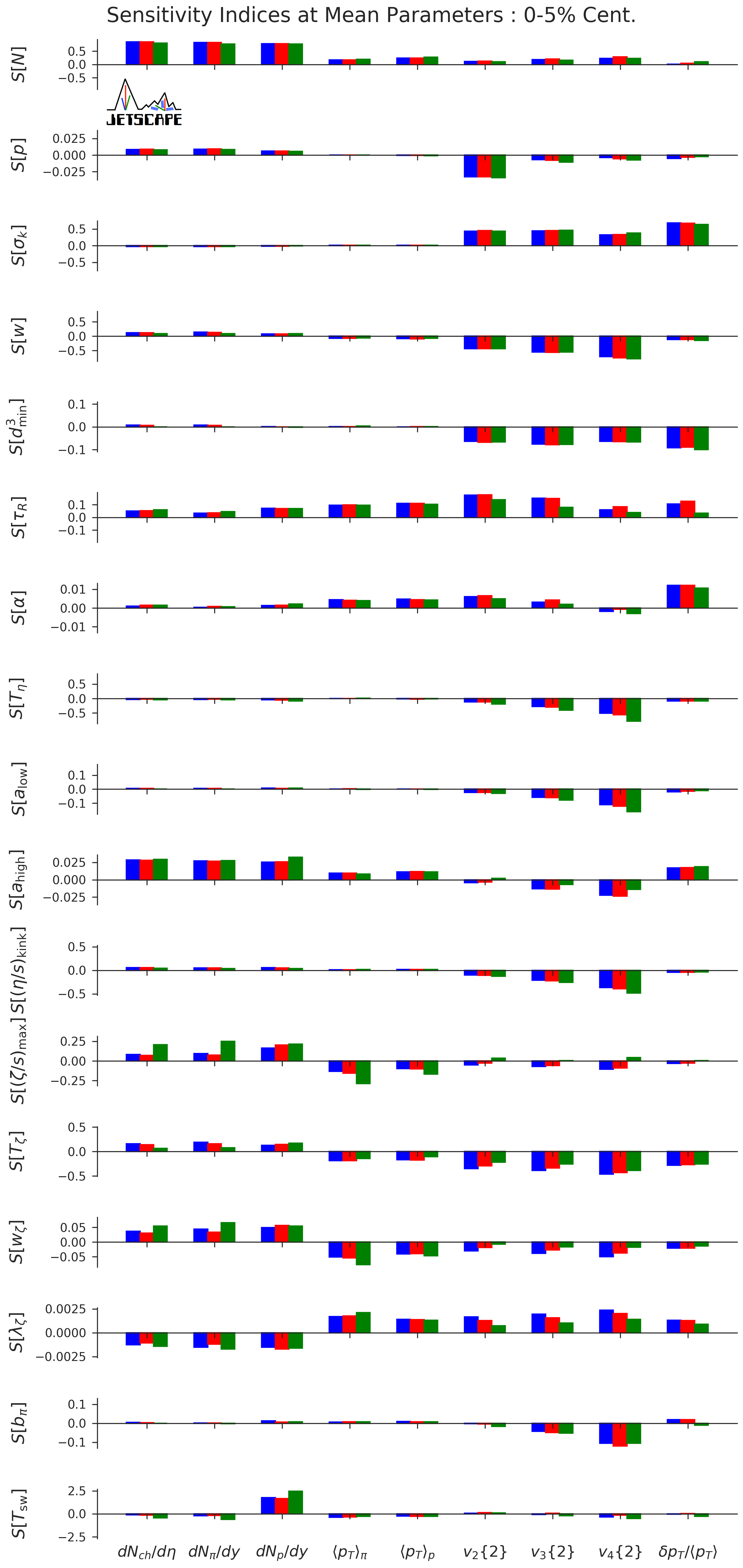}
\includegraphics[trim=0 0 0 18, clip, width=0.6\textwidth]{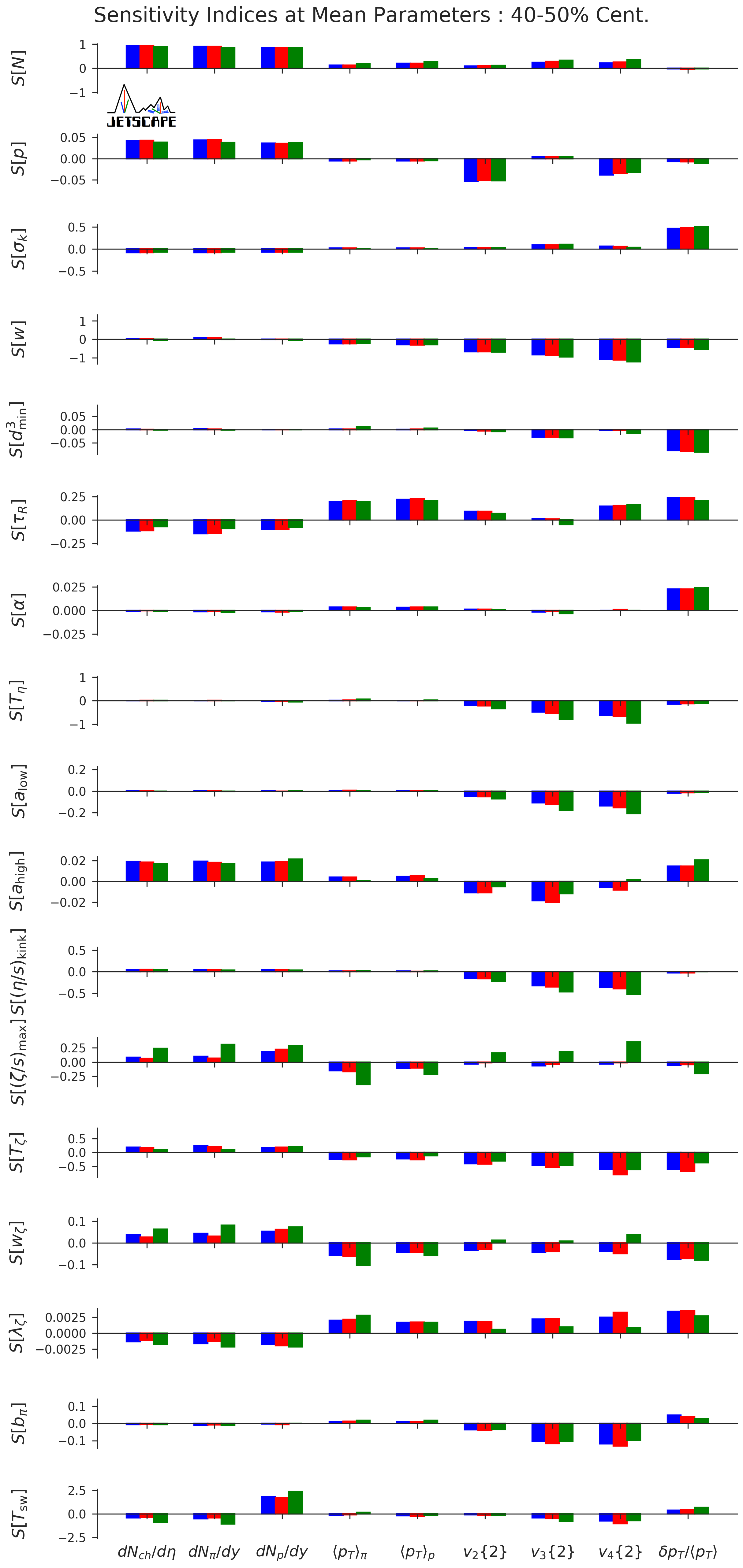}
}
\caption{Local sensitivity indices for LHC observables measured in the $0-5$\% (left) and $40-50$\% (right) centrality bin (except for the mean $p_T$ event fluctuation $\delta p_T / p_T$ for which the $40-45$\% bin is plotted on the right), as a function of all model parameters. Plotted in blue is the Grad viscous correction model, in red the Chapman-Enskog model, and in green the Pratt-Torrieri-Bernhard model. The bars show the sensitivity to a 10\% change in each parameter ($\delta = 0.1$). }
\label{obs_sensitivity}
\end{figure*}

To understand the posterior inferences regarding models and their parameters, it is useful to quantify which observables carry information about which model parameters. We explore two different metrics designed to quantify this information; the first is a local measure while the second a global measure. 

\subsection{The model gradient sensitivity index}

The first (local) measure of model sensitivity is defined as follows:
\begin{itemize}
    \item select a point in parameter space at which the model sensitivity is explored
    \item vary a single parameter at a time and quantify how much each observable responds
\end{itemize}
Note that our model is non-linear, and consequently this is a \emph{local} measure of observables sensitivity, at a given point in the multidimensional parameter space.

Following Ref.~\cite{Hamby} we define a local sensitivity index as follows: define two points in parameter space by $\mathbf{x} = (x_1, x_2, ...,x_j, ..., x_p)$ and $\mathbf{x'} = (x_1, x_2, ..., (1 + \delta)x_j , ..., x_p)$ where $\delta$ is a fixed percent difference. We use our emulator to predict all of the observables at these two points in parameter space. Suppose for some particular observable $O$, the emulator predicts $\hat{O} = \hat{O}(\mathbf{x})$. Then, defining the percent difference in the observable by 
\begin{equation}
    \Delta \equiv \frac{ \hat{O}(\mathbf{x'}) - \hat{O}(\mathbf{x}) }{\hat{O}(\mathbf{x})},
\end{equation}
our ``sensitivity index'' $S[x_j]$ for observable $O$ under a change in parameter $x_j$ is given by 
\begin{equation}
    S[x_j] \equiv \Delta / \delta.
\end{equation}
We chose $\mathbf{x}$ to be defined as the average of the three different Maximum A Posteriori (MAP) parameters (see \ref{ch5:MAP}) of each three viscous correction models, listed in \Table{table_MAP_grad}.

These local sensitivity indices $S[x_j]$ for pairs of observables and parameters are shown in \fig{obs_sensitivity} for select Pb-Pb observables at $\sqrts=2.76$ TeV and a step size $\delta = 0.1$. We verified that we obtain quantitatively similar results with a larger parameter step size $\delta = 0.4$, indicating that the parameter dependence of the model is reasonably close to linear in the region of parameter space studied. Note that the propagation of emulator uncertainties in the sensitivity analysis is left for a future work.

Although a local measure of the response of the model observables to changes in parameters is a strong approximation, it can nonetheless help guide our understanding regarding which observables carry information about each of the parameters. First note that the scale of the sensitivity indices is different for each parameter. Changing the shear relaxation time normalization $b_\pi$ has a very small effect on all observables investigated in this work, with a $10$\% change in $b_\pi$ leading to less than $1$\% change in observables.

On the other hand, very strong dependence on the model parameters can be seen for certain other observables. The proton yield shows strong sensitivity to the switching temperature. Increasing the switching temperature by $10\%$ increases the proton yield by about $20\%$. Consequently, most of the constraining power (or information) about the switching temperature is carried by the proton yield among the observables used herein. 

As noted throughout this study, many of the observables show stronger sensitivity to the maximum of the specific bulk viscosity when the Pratt-Torrieri-Benhard distribution was employed, compared to any other viscous corrections used here. Looking again at \fig{compare_visc_posteriors}, the strong sensitivity of the Pratt-Torierri-Bernhard viscous correction causes the $90$\% posterior credible interval for the specific bulk viscosity to be most tightly constrained among the viscous correction models explored here. The narrower posterior of $\zeta/s$ for this viscous correction model is a direct consequence of these larger sensitivities.

The parameter $w$ in \trento{} is largely responsible for controlling the eccentricities of the initial state. We find that the elliptic, triangular and quadrangular flows $v_{2}\{2\}$, $v_{3}\{2\}$, and $v_{4}\{2\}$ show strongest sensitivity among the observables plotted in \fig{obs_sensitivity}. This may be expected from hydrodynamic response, in which $v_{2}\{2\} \propto \epsilon_2$, $v_{3}\{2\} \propto \epsilon_3$ and $v_{4}\{2\} \propto v_{2}\{2\}^2$. In addition, the initial geometry is more sensitive to the nucleon width $w$ for peripheral collisions: we see that the harmonic flows for 40--50\% centrality bins show close to twice the sensitivity to the width parameter than
for 0--5\% centrality. 

The triangular flow $v_3\{2\}$ and quadrangular flow  $v_4\{2\}$ show strongest  sensitivity to $b_{\pi}$ in our model. This sensitivity remains small however: a $10$\% change in the shear relaxation time leads to a $1$\% change in $v_4\{2\}$ for example. This explains the challenge of constraining the shear relaxation time.

\subsection{Analysis of variance: the Sobol sensitivity index}

The dimensionless model gradient explored above, being a \textit{local} measure of sensitivity in the parameter space, does not quantify the amount of information contained in the \textit{global} variance of each model output corresponding to changes in the inputs. The analysis of variance (ANOVA) are a set of metrics which decompose the total variance of each model output in terms corresponding to variances of model parameters, pairs of model parameters, etc...  In particular, we use the Sobol indices ~\cite{SOBOL2001271, SALTELLI2002280} to characterize the variance in model observables corresponding to the variances in the model parameters. 
Suppose that our model predicts a particular observable output $y$ as a function of all model parameters $\bm{x}$ according to $y=f(\bm{x})$. 
Our priors for all model parameters then induce prior predictive distributions $\mathcal{P}(y)$ of each output, which are marginalized over the parameter-space according to the prior,
\be
    \mathcal{P}(y) = \int d\bm{x} \mathcal{P}(y|\bm{x}) \mathcal{P}(\bm{x}).
\ee
The idea is to decompose the variance of this distribution, $\text{Var}(y)$ into variances associated with each parameter. 
Consider fixing a single model parameter $x_i$ to take the specific value $x_i = x^*$, and computing the variance of the resulting distribution of outputs $y$,
\be
    \mathcal{P}(y|x_i=x^*) = \int dx_1 \cdots dx_{i-1} dx_{i+1} \cdots dx_n \mathcal{P}(y|\bm{x}) \mathcal{P}(\bm{x}).
\ee
We will denote the variance of this distribution by $\text{Var}(y)|_{x^*} \equiv \text{Var}(\mathcal{P}(y|x_i=x^*))$, corresponding to the variance in the model outputs given by varying all parameters \emph{except} $x_i$, evaluated over the subspace $x_i = x^*$. Because the value of $x_i$ is uncertain a priori, we need to marginalize over the possible values of $x_i$. Therefore we compute 
\be
    \text{Var}(y)|_{x_i} \equiv  \int dx^* \text{Var}(y)|_{x^*} \mathcal{P}(x^*).
\ee
Finally, we can define a first-order Sobol index for a particular observable $y$ and parameter $x_j$ according to
\begin{equation}
    S[x_j] \equiv \frac{\text{Var}(y) - \text{Var}(y)|_{x_j}}{\text{Var}(y)},
\end{equation}
which is the variance in the observable which results from variance of parameter $x_j$.

\begin{figure*}[!htb]
  \hspace{-1cm}
  \makebox[\textwidth][c]{\includegraphics[width=1.2\textwidth]{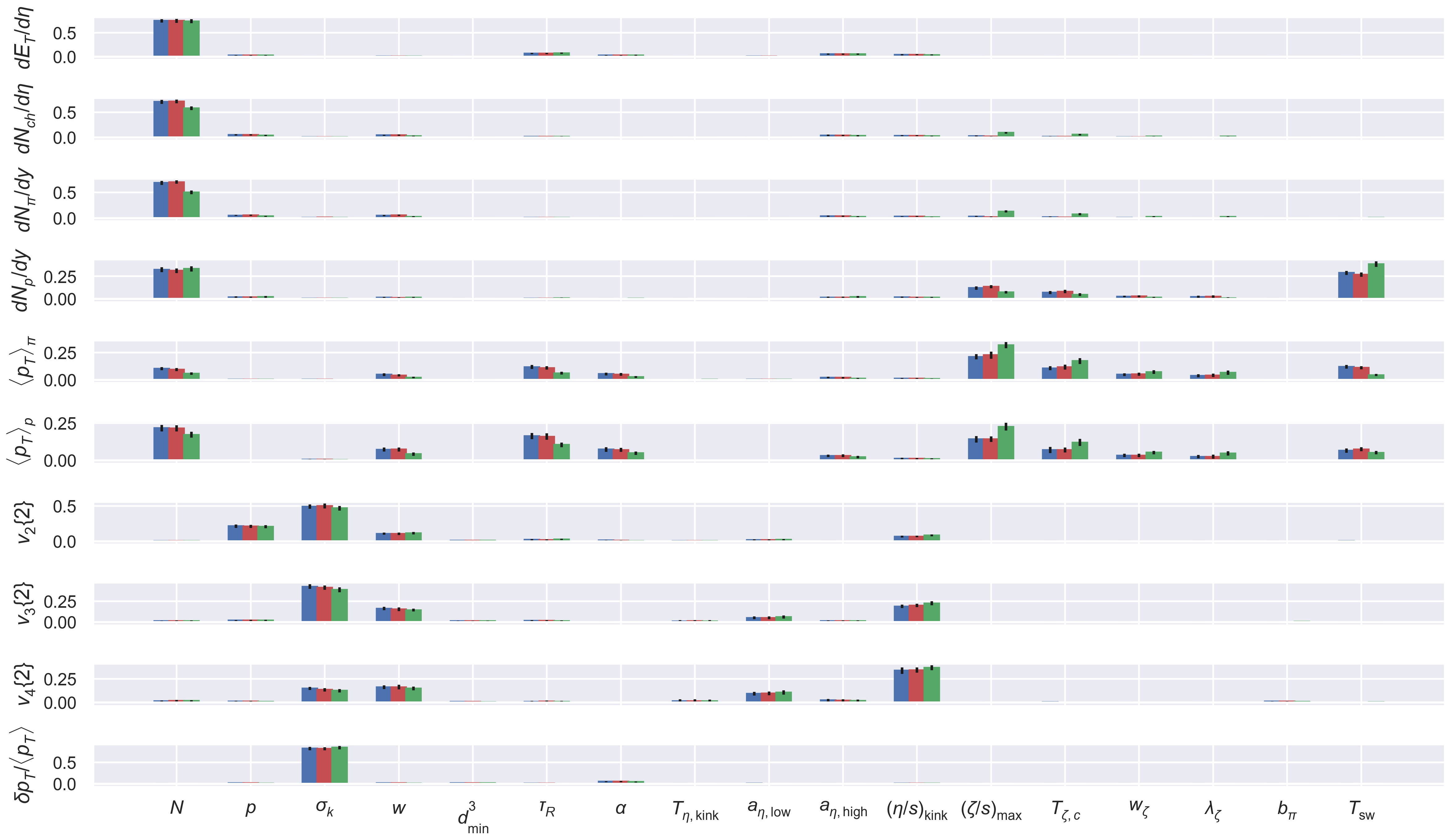}}%
  \caption{The Sobol sensitivity indices among pairs of observables and parameters in the central $0-5\%$ centrality bin. The Grad, CE and PTB models are shown as blue, red and green bars, respectively. Black lines denote statistical uncertainty arising from a finite number of prior samples.}
  \label{fig:sobol_ind_central}
\end{figure*}

\begin{figure*}[!htb]
\hspace{-1cm}
  \makebox[\textwidth][c]{\includegraphics[width=1.2\textwidth]{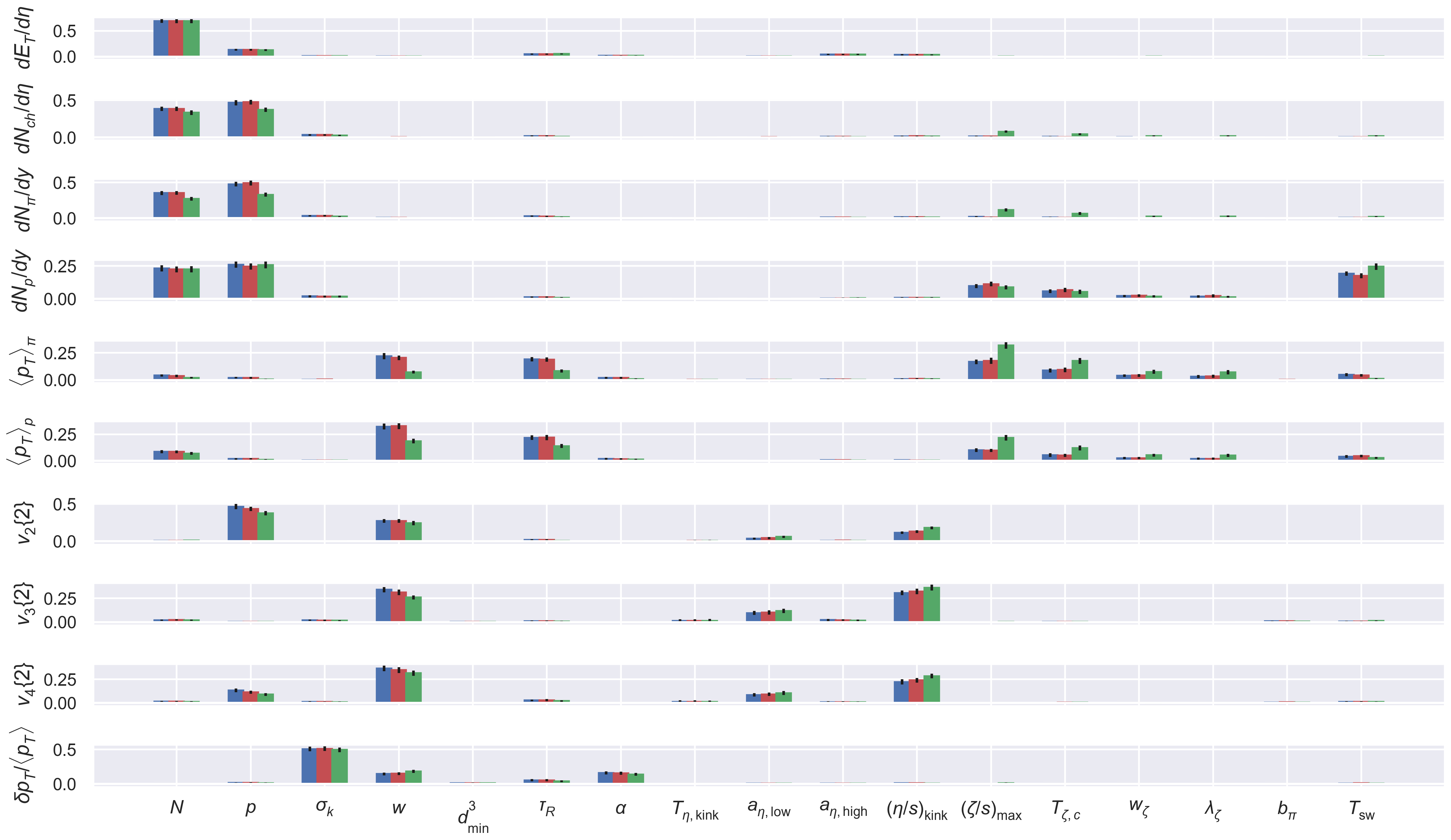}}%
  \caption{The Sobol sensitivity indices among pairs of observables and parameters in the more peripheral $40-50\%$ centrality bin. The Grad, CE and PTB models are shown as blue, red and green bars, respectively. Black lines denote statistical uncertainty arising from a finite number of prior samples.}
  \label{fig:sobol_ind_peripheral}
\end{figure*}

The Sobol indices among various pairs of observables and model parameters were readily estimated using the python library SALib~\cite{SALib}, and are shown in Fig.~\ref{fig:sobol_ind_central} and Fig. ~\ref{fig:sobol_ind_peripheral}. 
We have chosen to write these formulae explicitly with the priors, to remind that these Sobol indices as formulated depend on the specified priors for all model parameters. A model may be very sensitive to small changes in some parameter, but if the prior for this parameter is very narrow, it will only induce a small spread in the predictive distributions. 

This idea can be made clear using a simple example. Suppose we have a single model output $y$ which depends on a single parameter $x$ through a function $y=f(x)$. Moreover suppose that the model has no sources of uncertainty, so that the likelihood of obtaining the outcome $y$ given x is deterministic, i.e. has a probability density
\be
    \mathcal{P}(y|x) = \delta(y - f(x)).
\ee
Furthermore, suppose for simplicity that $x_0 = 0$, and $f(x_0) = 0$, which can always be accomplished by a constant shift our our variables. Then, the Taylor expansion of $f(x)$ around $x_0 = 0$ is given by 
\be
    f(x) = f(x_0) + \frac{df}{dx}|_{x_0} (x-x_0) + \mathcal{O}((x-x_0)^2) \equiv cx + \mathcal{O}(x^2),
\ee
where in the last equality we have used that $x_0=0$ and $f(x_0) = 0$, and defined $c \equiv \frac{df}{dx}|_{x_0}$. 
Then the distribution $\mathcal{P}(y)$ of outputs is given by 
\be
    \mathcal{P}(y) = \int dx \mathcal{P}(y|x) \mathcal{P}(x) \approx \int dx \delta(y - cx) \mathcal{P}(x) = \frac{1}{|c|}\mathcal{P}(x)|_{x=y/c}.
\ee

As an illustrative example, we can consider a Gaussian prior for parameter $x$,
\be
    \mathcal{P}(x) = \frac{1}{\sqrt{2\pi}\sigma} \exp[-\frac{x^2}{2\sigma^2}].
\ee
This induces a prior predictive distribution for outputs $y$
\be
    \mathcal{P}(y) = \frac{1}{\sqrt{2\pi} (\sigma|c|) } \exp[-\frac{y^2}{2(\sigma|c|)^2}],
\ee
recognizable as a Gaussian with a width given by the \emph{product} of the prior width in the parameter $x$ and the gradient sensitivity $|c|$. This product is indeed the relevant quantity to consider when interpreting the first order Sobol indices. Certain model observables may have a large \textit{local} sensitivity $|c|$ to a certain parameter, but if the parameter prior has a narrow width $\sigma$, the Sobol sensitivity can be small. The converse is also true.   

From the figures, we see that the switching temperature $T_{\rm sw}$ accounts for a large fraction of the variance of the proton yield $dN_p/dy$. However, we also see that the \trento{} normalization $N$ accounts for nearly the same fraction. If $\theta$ denotes a parameter and $\delta_{\theta} \equiv (\theta_{\rm max} - \theta_{\rm min}) / \theta_{\rm min}$ denotes a normalized measure of its prior width, for the \trento{} normalization we find $\delta_{N} \sim 2$, while for the switching temperature $\delta_{T_{\rm sw}} \sim 0.2$.
This explains why the two Sobol indices are competitive. 

Our previous insights that the PTB model is more sensitive to bulk viscous pressure are reinforced; the Sobol indices between the peak bulk viscosity $(\zeta/s)_{\rm max}$ and the mean transverse momenta of pions and protons $\langle p_T \rangle_{\pi,p}$
are larger for the PTB model than the Grad or CE models.
Additionally, we see the that \trento{} reduced thickness parameter $p$ explains a large fraction of the variance of the elliptic flow in the $40-50\%$ centrality bin. Again, the parameter $p$ is varied over a relatively wide prior, which induces a wide spectrum of initial conditions. 

The Sobol indices also yield expectations for the \textit{constraining power} of our observables. Returning to our simple example above, we can consider the outcome of calibrating the model parameters against a set of observed data $y_{\rm exp}$. If the experimental data, in the form of the likelihood, constrain the plausible values of $y$ to a distribution which is narrower than the prior predictive distribution $\mathcal{P}(y)$, then we have gained information about the model parameters. On the contrary, we expect the experimental observables to yield no information regarding the model parameters which account for predictive variances in the observables smaller than the variances (uncertainties) in the experimental data. That is, if parameter $x$ accounts for a variance $\text{Var}(y)$ in the model output $y$, and $\text{Var}(y) \lesssim \sigma^2_{y, \rm exp}$, then we gain no information by experimental calibration.\footnote{This discussion neglects the correlations among model parameters induced by conditioning on the observed data; the potential for information gain in the joint posterior could be quantified by the second-order Sobol indices, for example.}
This situation can be observed for the \trento{} parameter $d_{\rm min}^3$, for example, which is observed to have very small Sobol indices for all observables. Indeed, after calibrating against both LHC and RHIC observables, we find in Fig.~\ref{initial_state_posterior} a marginal likelihood for $d_{\rm min}^3$ which is essentially flat -- we haven't learned anything from the data. These conclusions are also true of the shear relaxation time factor $b_{\pi}$, for example.

\section{Model selection and criticism}
\label{ch5:model_selection}

In this section we explore methods for comparing and criticizing differing models to describe the collision. 
We will first illustrate the application of the Bayes factor toward three of the viscous correction models for particlization that were used throughout this work. We then use it to compare the model thus far employed with simpler models which are `nested' inside. Finally, the Bayes factor is applied towards answering whether a consistent model, with the same set of parameters describing the system created in RHIC Au-Au $\sqrts{}=0.2$ TeV and LHC Pb-Pb $\sqrts{}=2.76$ TeV collisions, or more complicated models where some parameters are allowed to differ, is better justified in light of the experimental data.

\subsection{Comparing viscous correction models}
\label{ch5:compare_df}

As a first illustration of Bayesian model selection, we quantify if our experimental data give evidence to prefer one viscous correction model over another. We have estimated the logarithm of the Bayes evidence $\ln Z$ as well as the integration uncertainty $\delta \ln Z$ for three of the four models using the parallel-tempering described in Ch.~\ref{ch4:num_bayes_evidence}. Their mutual Bayes factors are given in Table~\ref{bayes_fac_estimates}. 

\begin{table}[!htb]
\centering
\begin{tabular}{|l|l|l|}
\hline
Model $A$ & Model $B$ & $\ln B_{A/B}$   \\ \hline
Grad & CE & $8.2 \pm 2.3$  \\ \hline
Grad & PTB & $1.4 \pm 2.5$  \\ \hline
PTB & CE & $6.8 \pm 2.4$ \\ \hline
\end{tabular}
\caption{ A table of the logarithm of the Bayes factor $\ln B_{A/B}$ for each pair of viscous correction models, and its integration uncertainty, for the Grad, Chapman-Enskog (CE) and Pratt-Torrieri-Bernhard (PTB) viscous correction models. }
\label{bayes_fac_estimates}
\end{table}

From the Table we see that the Grad and Pratt-Torrieri-Bernhard models have Bayesian evidences that are compatible within the numerical uncertainty. The odds that the Grad model is better than the Pratt-Torrieri-Bernhard model are about 3:1, given our $0.6 \sigma$ observation.\footnote{%
    The probability is given by the left-tailed $p$-value.}
Therefore, the $p_{T}$-integrated calibration observables cannot distinguish which of these two models is more likely. However, we have moderate evidence to conclude that both of these models work better to describe the hadronic observables studied in this work than the Chapman-Enskog model, with the Grad versus Chapman-Enskog comparison being a $3.6 \sigma$ observation (odds about 5000:1), and Pratt-Torrieri-Bernhard comparison a $2.8 \sigma$ observation (odds about 400:1). 

For the Chapman-Enskog model, we note from \fig{initial_state_posterior} that the marginal posterior of the free-streaming energy dependence $\alpha$ has a local maximum for $\alpha \lesssim -0.3$. It is possible that widening our prior to include smaller values of $\alpha$ would also increase the Bayes evidence for the Chapman-Enskog model. Unfortunately, this would require a new set of model calculations at new design points, which is beyond the scope of the present work. 
We also considered the frequentist odds, defined by the maximum likelihood ratio. If $L_A$ is the maximum value of the likelihood function for model $A$, and $L_B$ the same for model $B$, the maximum likelihood ratio is simply defined by $L_A / L_B$. These maximum-likelihood odds were found to be close to 300:1 for the ratio of Grad to Chapman-Enskog models.

\begin{figure}[!htb]
  \centering
    \includegraphics[width=0.7\textwidth]{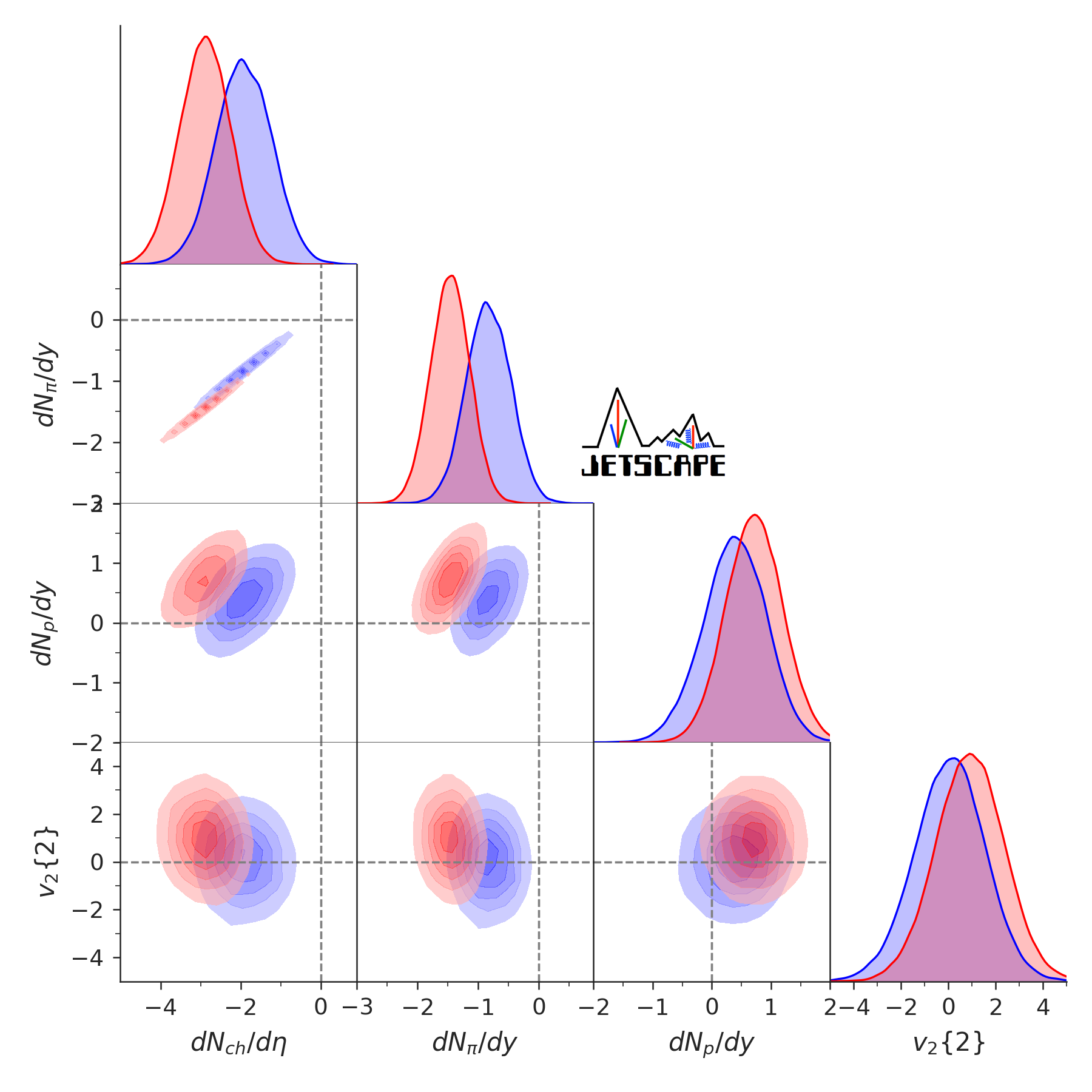}
    \caption{Diagonal and off-diagonal panels show one- and two-dimensional densities of the 
    $n$-dimensional posterior predictive distributions for selected Pb-Pb $\sqrts{}=2.76$ TeV observables at fixed 
    collision centrality of 0--5\%. Plotted are the discrepancies between prediction and 
    measurements in units of the experimental standard deviation; axes are labeled with 
    shorthand notation $y \equiv (y_{\rm model}{-}y_{\rm exp}) / \sigma_{\rm exp}$ where 
    $y$ stands for the observable whose model discrepancy is shown. The Grad model is 
    shown in blue and Chapman-Enskog in red.
    }
    \label{fig:posterior_predictive_discrepancy}
\end{figure}

The Chapman-Enskog model is not able to simultaneously fit the proton multiplicity together with the other observables, such as the pion multiplicity. This puts the model under tension, and reduces the average of the likelihood of the Chapman-Enskog model. This is illustrated by Fig.~\ref{fig:posterior_predictive_discrepancy}, which displays the single and joint posterior predictive distributions of select observables for the most central bin 0--5\% for Pb-Pb collisions at $\sqrts{} = 2.76$\,TeV. For each of the Grad model (blue) and Chapman-Enskog model (red), parameter samples are drawn from the posteriors calibrated to all observables of both LHC Pb-Pb $\sqrts{} = 2.76$\,TeV and RHIC Au-Au $\sqrts{} = 0.2$\,TeV. Then, the model predictive distribution is calculated using the emulator for all observables, and plotted is the model-experiment discrepancy, i.e., the difference between model prediction and experimental mean normalized by experimental standard deviation. That the chemical abundances disfavor the Chapman-Enskog model was further strengthened by recalculating the posteriors and Bayes factor for the Grad and Chapman-Enskog models excluding the LHC Pb-Pb $\sqrts{} = 2.76$\,TeV proton multiplicity from the calibration data. In this case, the odds were greatly reduced to only about 5:1 in favor of Grad.\footnote{The Chapman-Enskog RTA model also has more trouble describing the yield of pions (it underpredicts) simultaneously with other observables, even if the proton is excluded.} For both the linearized Grad and Chapman-Enskog models, it is the bulk viscous correction which changes the chemical abundances from their equilibrium values (the shear viscous correction does not correct the equilibrium yields). Therefore, in light of the chemical abundances being a strong discriminator, it is specifically the bulk viscous correction given by the Chapman-Enskog model which is disfavored by the particle yields. 

In conclusion, the hadronic observables studied in this work favor the Grad and Pratt-Torrieri-Bernhard models of viscous corrections over the Chapman-Enskog model. This is due in large part because the Chapman-Enskog model is worse at simultaneously fitting the chemical abundances. In light of the methodological uncertainties, we do not believe this finding should be taken as a blanket statement on the validity of the Chapman-Enskog viscous correction model in studying heavy ion collisions. Future studies will be necessary to clarify if viscous corrections can be systematically constrained from measurements.

Knowing the relative odds between the different particlization models, a model-averaged posterior which propagates model-space uncertainties to inferred parameters can be estimated using Bayesian Model Averaging. The model averaged posterior is a weighted average of the individual model posteriors, with each model weighed by its evidence (see Ch.~\ref{ch6:bma_df}). 

\subsection{Comparing hydrodynamic models}
\label{ch5:compare_hydro_models}
As another application of Bayesian model selection, we quantify whether simpler models, which are nested within the model described in \ref{ch3:hydro}, are favored or disfavored by the data. We make comparisons against models with simplified assumptions for the shear viscosity. As a reminder, the more complex model will be penalized by the additional parameters (the `Occam penalty') that are constrained by the data, and will only yield a larger evidence than the simpler model if the extra constrained parameters significantly improve the model's description of the data. An additional model parameter that is not well-constrained by the data within the range of the prior will have an insignificant Occam penalty.  

\subsubsection{Temperature independent specific shear viscosity}
\label{ch5:compare_hydro_models:const_shear}

We consider whether our model with a temperature-dependent specific shear viscosity is preferred by the data to a simpler model with a temperature-independent specific shear viscosity. In both cases we use the Grad particlization model. We denote by model $A$ the model with temperature dependent specific shear viscosity, and denote by $B$ the model in which the low-temperature and high-temperature slopes $a_{\rm low}$ and $a_{\rm high}$ are fixed to zero. The temperature of the kink $T_{\eta}$ is irrelevant in this scenario, and is also fixed to an arbitrary value. We find the logarithm of the Bayes factor to be consistent with zero within its uncertainty, $\ln B_{A/B} = -0.2 \pm 2.4$. Hence, given all sources of methodological uncertainties, the selected data provide no evidence in favor of the common theoretical preference for a temperature-dependent specific shear viscosity of QCD matter. As noted above, the Occam penalty for including the additional parameters, which here are the slopes of the specific shear viscosity and position of its inflection, is minimal; this is because these parameters are not well constrained within the range of the prior. In any case, this inconclusive result suggests inclusion of more discriminating observables in future studies. 

\subsubsection{Zero specific shear viscosity}
\label{ch5:compare_hydro_models:zero_shear}

We also study if the calibration data provide strong evidence that the specific shear viscosity is non-zero. This can be quantified in the same way as above, setting the parameters for the specific shear viscosity such that $(\eta/s)(T) \approx 0$. We again use the Grad viscous correction model for this comparison and allow the specific bulk viscosity, as well as all other parameters, within their full prior ranges. We find the logarithm of the Bayes factor $\ln B_{A/B} = 11.7 \pm 2.6$ where model $A$ is the default model with nonzero and temperature-dependent specific shear viscosity, while model $B$ has $\eta/s \approx 0$. We conclude that the data provide strong evidence that the specific shear viscosity is nonzero.

\subsection{Quantifying tension between LHC Pb-Pb and RHIC Au-Au}
\label{ch5:tension_RHIC_LHC}

The Bayes factor is also useful for quantifying if models are under significant tension when trying to simultaneously fit the observables in both collision systems ~\cite{Marshall_2006}. Throughout this work we have assumed that all model parameters are shared between the Pb-Pb $\sqrts{} = 2.76$\,TeV and Au-Au $\sqrts{} = 0.2$\,TeV systems except for their initial energy density normalizations. We can however relax these assumptions, and allow other parameters to differ for the two different systems.

\subsubsection{No common parameters between Pb-Pb and Au-Au collision systems}
\label{ch5:tension_RHIC_LHC:no_common}

Suppose that we allow all of the parameters to be different for the two systems defined by Au-Au $\sqrts{} = 0.2$\,TeV collisions at RHIC and Pb-Pb $\sqrts{} = 2.76$\,TeV collisions at the LHC, including the initial conditions, viscosities, and switching temperature. In this case, the model has a total of 34 parameters. We will compute the Bayes factor $\ln B_{A/B}$, where model $A$ is the default model, while the more complex model $B$ assumes independent sets of model parameters for describing the data collected at different collision energies.

As usual, we take the ratio of our prior beliefs about these two models to be unity, $\mathcal{P}(A) / \mathcal{P}(B) = 1$, such that the Bayes factor reduces to the ratio of marginal evidences:
\begin{equation}
  B_{A/B} = \frac{ \mathcal{P}( \mathbf{y}_{\rm LHC}, \mathbf{y}_{\rm RHIC} | A )  }{  \mathcal{P}( \mathbf{y}_{\rm LHC}, \mathbf{y}_{\rm RHIC} | B )}.
\end{equation}
As a consequence of the assumed statistical independence of measurements performed for different collision systems with different detectors, we can estimate the model evidence in the denominator as follows:
\begin{equation}
    \mathcal{P}( \mathbf{y}_{\rm LHC}, \mathbf{y}_{\rm RHIC} | B ) = \mathcal{P}( \mathbf{y}_{\rm LHC} | B ) \mathcal{P}( \mathbf{y}_{\rm RHIC} | B ).
\end{equation}
Integrating over the model parameters for the LHC model yields
\begin{equation}
    \mathcal{P}( \mathbf{y}_{\rm LHC} | B ) 
     = \int d\mathbf{x}_{\rm LHC} \mathcal{P}( \mathbf{y}_{\rm LHC} | B, \mathbf{x}_{\rm LHC}) \mathcal{P}( \mathbf{x}_{\rm LHC} | B )
\end{equation}
which we have estimated using \eq{eq:thermo_int_evidence}. A similar result holds for the model describing the RHIC data. 

Using these relations we find $\ln B_{A/B} = 24.1 \pm 2.6$, and conclude that these data from the LHC and RHIC yield very strong evidence that a model in which all parameters except the initial energy density normalizations are the same is strongly preferred over a model in which all parameters are allowed to be different. The Occam penalty for nearly doubling the number of model parameters far outweighs the small gain of accuracy in the description of observed data. We take this as strong evidence that a hybrid viscous hydrodynamic model with a single set of parameters provides a coherent physics picture for the experimental data measured at these two collision systems, which differ by over an order of magnitude in center-of-mass energy.

Admittedly, allowing all of the parameters to be different leads to a very extreme comparison, adding far more model complexity than perhaps reasonable, thus entailing an outsized Occam penalty. A more systematic study may try to identify tensions between a few specific observations and their predictions from the calibrated model, 
and introduce a controlled amount of model complexity to relieve tension in that sector. Such an exhaustive analysis is left for a future work, and instead an interesting example which is motivated by physical expectations is studied in the next section.    
It is also worthwhile to note that these conclusions depend heavily on the likelihood function of the experimental data. When the assumed experimental systematic covariance matrix is changed, or the likelihood function is changed to a different (non-normal) distribution, these conclusions may also change. Such an exhaustive theoretical study may not be worthwhile until an exhaustive analysis of the systematic experimental uncertainties has been performed.

\subsubsection{Allowing different transverse length scales in the initial conditions}
\label{ch5:tension_RHIC_LHC:trento_w}
\begin{figure}[!htb]
  \centering
    \includegraphics[width=0.8\textwidth]{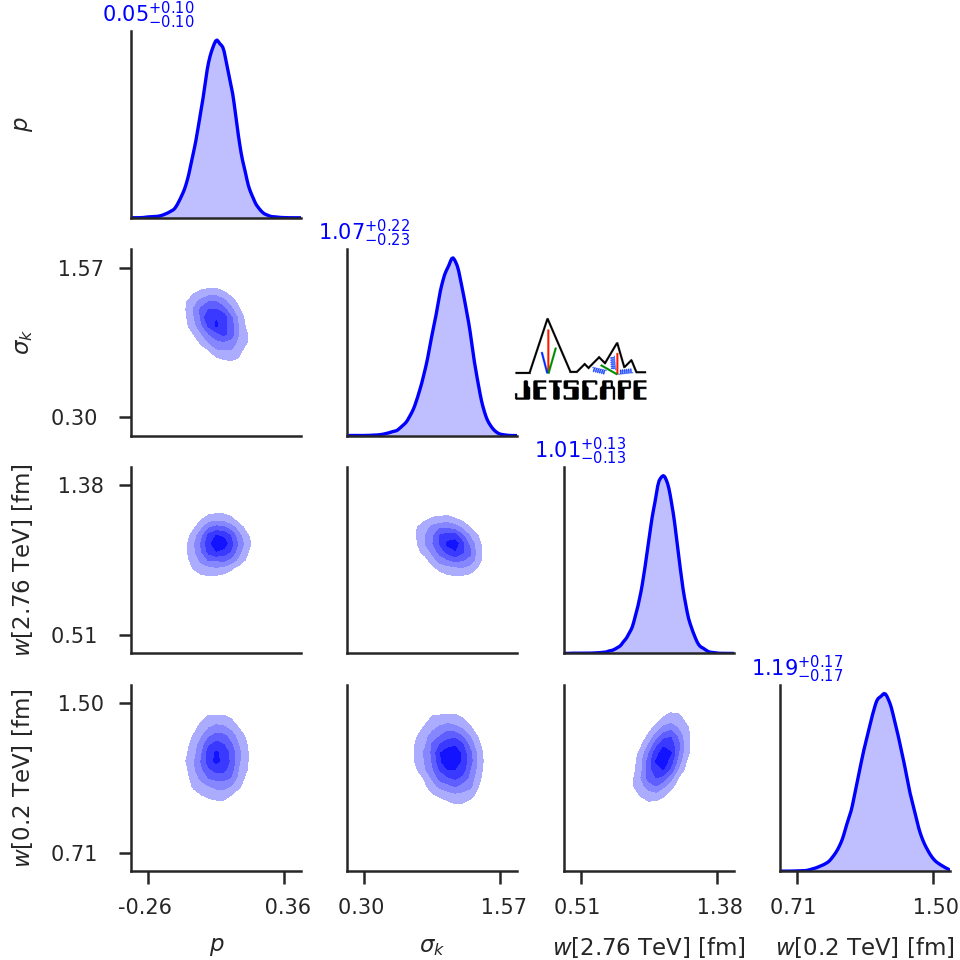}
    \caption{Partial representation of the posterior  for a model that uses the Grad particlization model and allows for different nucleon width parameters $w$ at RHIC and LHC energies. The estimated nucleon widths at the two collision energies inferred from the Bayesian analysis are found to agree within the 90\% confidence limits.}
    \label{fig:posterior_diff_nuc_width}
\end{figure}

%
We mentioned earlier that some theoretical models of the energy deposition in a heavy ion collision feature transverse length scales that depend on the collision energy. In our \trento{} model, it is the ``nucleon width'' $w$ which controls the transverse length scale for fluctuations in the initial conditions, and we have so far assumed that its value is independent of the collision energy. To test this assumption, we calculate the posterior for a model that introduces one additional parameter, allowing the nucleon width $w$ to differ between Pb-Pb $\sqrts{} = 2.76$\,TeV and Au-Au $\sqrts{} = 0.2$\,TeV collisions. The posterior for select initial condition parameters for this model is shown in \fig{fig:posterior_diff_nuc_width}. 

We see that the most probable value for the nucleon width $w[0.2$\,TeV$]$ in Au-Au collisions at RHIC is about $20$\% larger than the width $w[2.76$\,TeV$]$ in Pb-Pb collisions at the LHC, though both agree within uncertainties as shown in Fig.~\ref{fig:posterior_diff_nuc_width}. For reference, the Color Glass Condensate model predicts roughly a factor of two difference between the color flux tube diameters at top RHIC and LHC energies~\cite{Gelis:2014qga}. The measured total inelastic nucleon-nucleon cross section also increases by about a factor two from RHIC to LHC, indicating a possible growth of $w$ by a factor $\sim \sqrt{2}$. If $A$ denotes the default model and $B$ the model where the nucleon widths at the two collision energies are allowed to differ, we find $\ln B_{A/B} = 0.7 \pm 2.5$. Within the uncertainty of the estimate, we can thus not distinguish which model is preferred. The amount of tension that is caused by ignoring energy dependence of the nucleon width is not significant, and any small gains due to better description of observed data is erased by Occam's penalty: the nucleon widths are well constrained within their priors. 

\section{Predicting $p_T$-differential observables}
\label{ch5:pT_predictions}

\begin{figure*}[!htb]
  \centering
    \includegraphics[width=\linewidth]{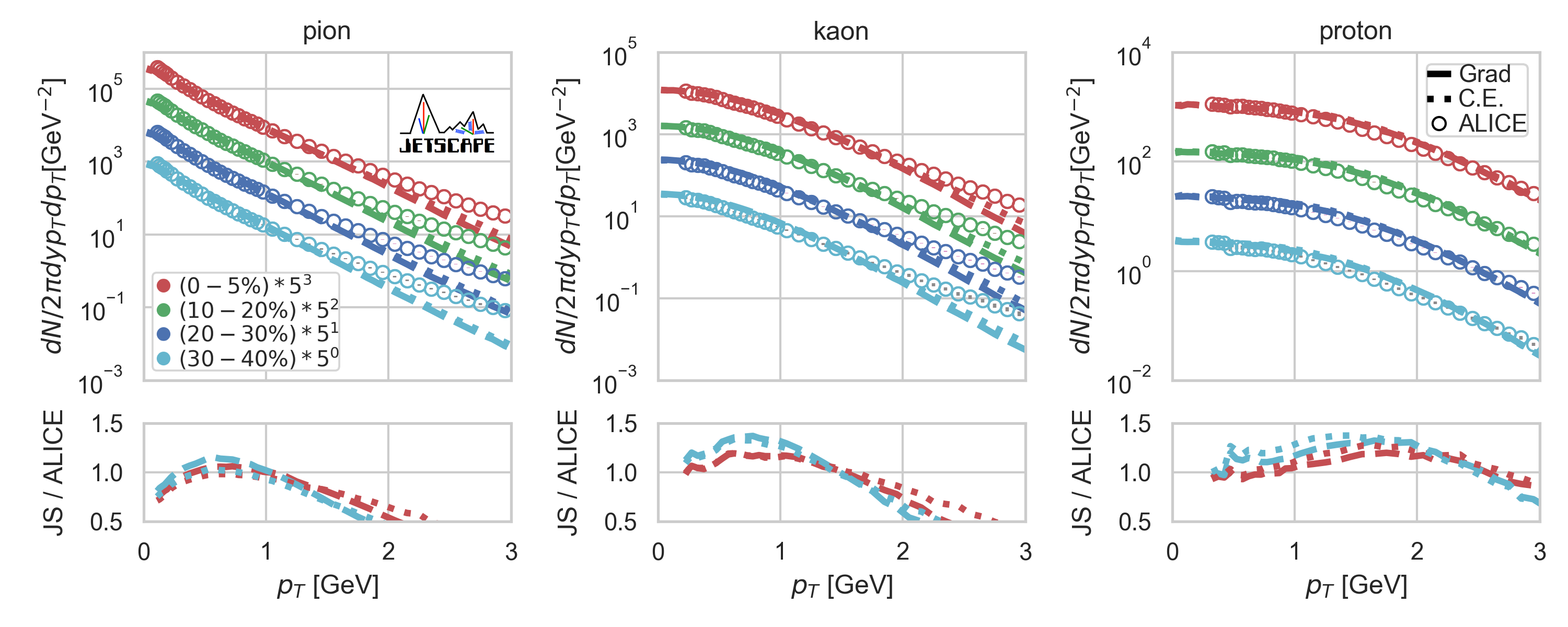}
    \caption{The transverse momentum spectra for pions (left), kaons (center) and protons (right) averaged over five thousand fluctuating events predicted by the Grad (dashed lines) and Chapman-Enskog (dotted lines) models, each run at their respective MAP parameters. Shown are the predictions for the $0-5\%$(red), $10-20\%$(green), $20-30\%$(blue), and $30-40\%$(cyan) centralities, each having been scaled by a power of five for visualization. Also shown are the measurements from ALICE (open circles). The bottom panel shows the ratio of the model prediction `JS' divided by the ALICE data. }
    \label{fig:pT_spectra}
\end{figure*}

\begin{figure*}[!htb]
\centering
    \includegraphics[width=0.7\linewidth]{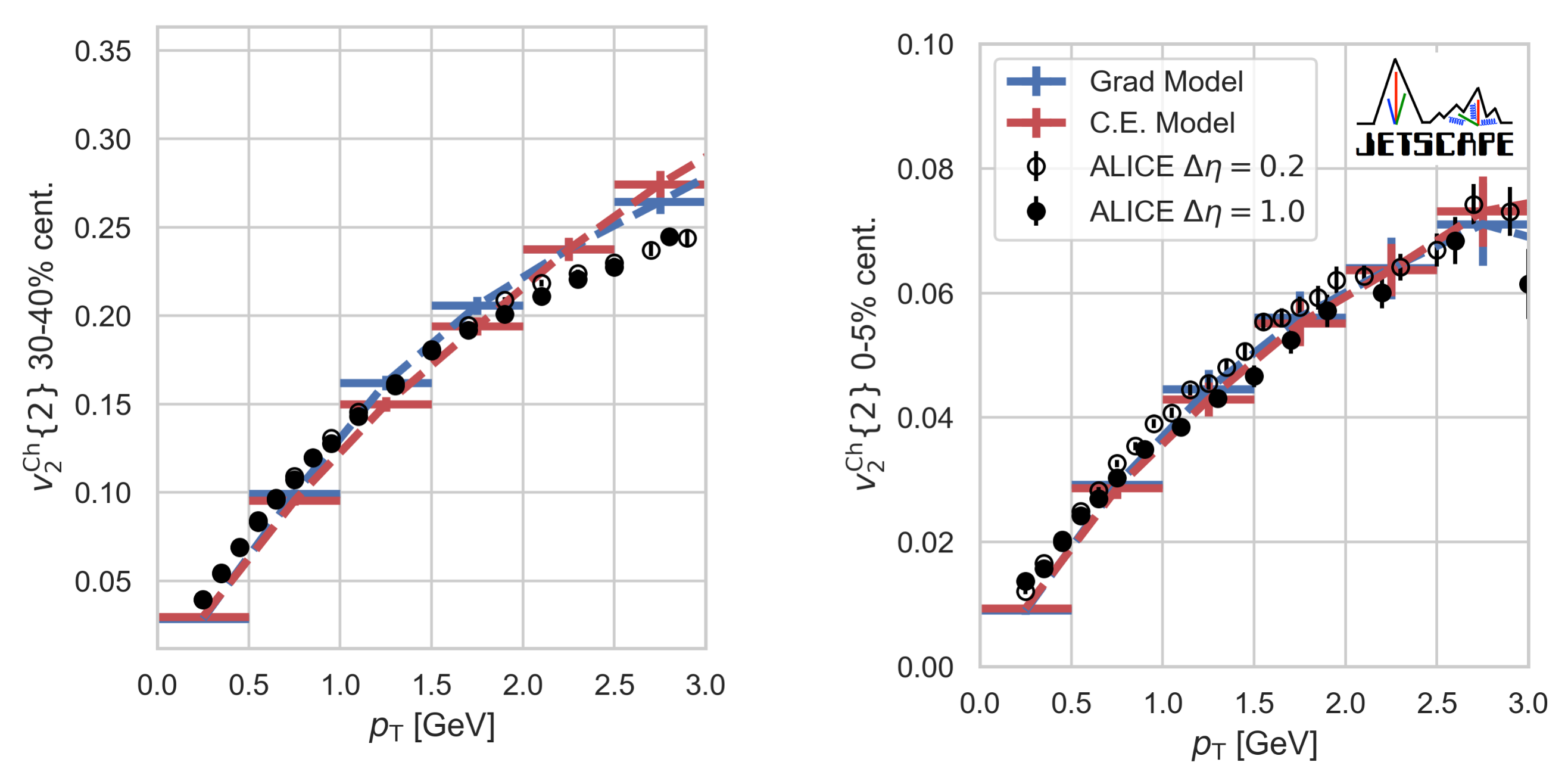}
    \caption{The $p_T$-differential two-particle cumulant elliptic flow of charged particles $v^{\rm Ch}_n\{2\}$ averaged over five thousand fluctuating events predicted by the Grad (blue) and Chapman-Enskog (red) models, run at their respective MAP parameters. Also shown are measurements from ALICE using a pseudorapidity gap $\Delta \eta = 0.2$ (open circles) or $\Delta \eta = 1.0$ (filled circles).}
    \label{fig:pT_v2}
\end{figure*}

A model is more useful if it is capable of accurately describing observables that were not used for its calibration. This fits the physicist's frame of mind in which belief in a model's veracity is increased when the model makes an accurate {\it pre}diction of some observable. Similarly, models that make inaccurate predictions are held in lower esteem. 

We thus check whether our calibrated models for heavy-ion collisions make accurate predictions. We consider as a prediction any observable calculated from the model using the Maximum A Posteriori (MAP) parameters (see \Table{table_MAP_grad}) that has not been used for the model calibration, neither through the prior nor via the likelihood. As our model is intended to describe the physics of particles with soft momenta $p_T \lesssim 2$\,GeV, accurately predicted soft observables should increase our belief in the model, while soft observables that are inaccurately predicted will decrease it. As an example, in this section we use our model to predict the shapes of the $p_T$-differential identified hadron spectra and charged hadron elliptic flow for Pb-Pb $\sqrts{} = 2.76$\,TeV collisions measured by ALICE at the LHC, shown in Figs.~\ref{fig:pT_spectra} and \ref{fig:pT_v2} for the Grad and Chapman-Enskog particlization models.\footnote{%
    We remind the reader that the posterior of our model parameters was estimated using only $p_T$-integrated observables, e.g. the multiplicities and mean transverse momenta for pions, kaons and protons, the $p_T$-integrated harmonic flows, etc.} 

Because the multiplicities and mean transverse momenta are dominated by particles with typical (flow-boosted) thermal momenta, the model tends to fit the slope of the pion differential spectra better at soft momenta $p_T \lesssim 1.5$\,GeV. The stronger boost from radial flow experienced by heavier hadrons ~\cite{Schnedermann:1993ws, Heinz:2004qz} extends this agreement with the model to higher $p_T\lesssim 2.5$\,GeV for protons. This finding is consistent with that of Ref.~\cite{Novak:2013bqa} which showed that the shape of the pion and proton spectra could be characterized well by the mean transverse momenta and yields. It remains to be checked if this conclusion holds also for the Pratt-Torrieri-Bernhard model, which tends to have very non-trivial $p_T$-dependence at both small and large $p_T$ when compared with the Grad or CE models~\cite{McNelis:2021acu}. 

For the differential elliptic flow, the agreement between model prediction and experiment is generally good for both the Grad and Chapman-Enskog models; neither model performs qualitatively better than the other. To what extent each of these models' predictions also agree with additional experimental results that were not used for model calibration will be further explored in future studies. We note that the Chapman-Enskog viscous correction model is not able to fit the experimental multiplicities of pions and protons as well as the Grad model, but in the $p_T$-differential elliptic flow the normalizations of the spectra approximately cancel and only their shapes as a function of $p_T$ matter. Again, it remains to be checked if the PTB model also accurately predicts the differential elliptic flow.

\subsection{Summary}

We briefly review some of the most important insights from this chapter. Firstly, the viscous hybrid hydrodynamic models employed, combining \trento{} initial conditions, freestreaming, viscous hydrodynamics, differing particlization models and hadronic rescattering, are capable of \textit{simultaneously} describing essentially of the $p_T$-integrated hadronic observables measured in Pb-Pb $\sqrts{}=2.76$ TeV, Au-Au $\sqrts{}=0.2$ TeV and Xe-Xe Pb-Pb $\sqrts{}=5.44$ TeV collisions to a level of agreement of roughly $20\%$. The notable exception to this statement is the yield of protons measured in Au-Au $\sqrts{}=0.2$ TeV collisions at RHIC, but the discrepancy between different experimental collaborations measurements (STAR and PHENIX) must be resolved before this should be considered as a model deficiency. 
However, there remains systematic sources of model discrepancy when comparing to the observed data. A notable source of discrepancy is the predicted chemistry of the final state, including the yields of pions, kaons and protons.

Secondly, our estimation of the transport coefficients $\eta/s$ and $\zeta/s$ demonstrate model-dependence regarding the particlization model employed. This should be considered as a leading source of uncertainty in their estimation in the future. 

The estimation of the transport coefficients also shows a strong sensitivity to the elicitation of the prior, which is exacerbated at higher temperatures. The $p_T$-integrated observables carry very minimal information regarding the specific shear and bulk viscosities for temperatures above $250$ MeV. The inclusion of more constraining observables should be handled with caution; foremost, additional observables should not be contaminated by large theoretical model deficiencies (e.g. $p_T$-dependent observables at $p_T \gtrsim 2$ GeV). 
\chapter{Model Averaging, Mixing and Prior Sensitivity}
\label{ch6}

\section{Bayesian model averaging over viscous correction uncertainties}
\label{ch6:bma_df}
%

\begin{figure*}[!htb]
\centering
\includegraphics[width=0.8\linewidth]{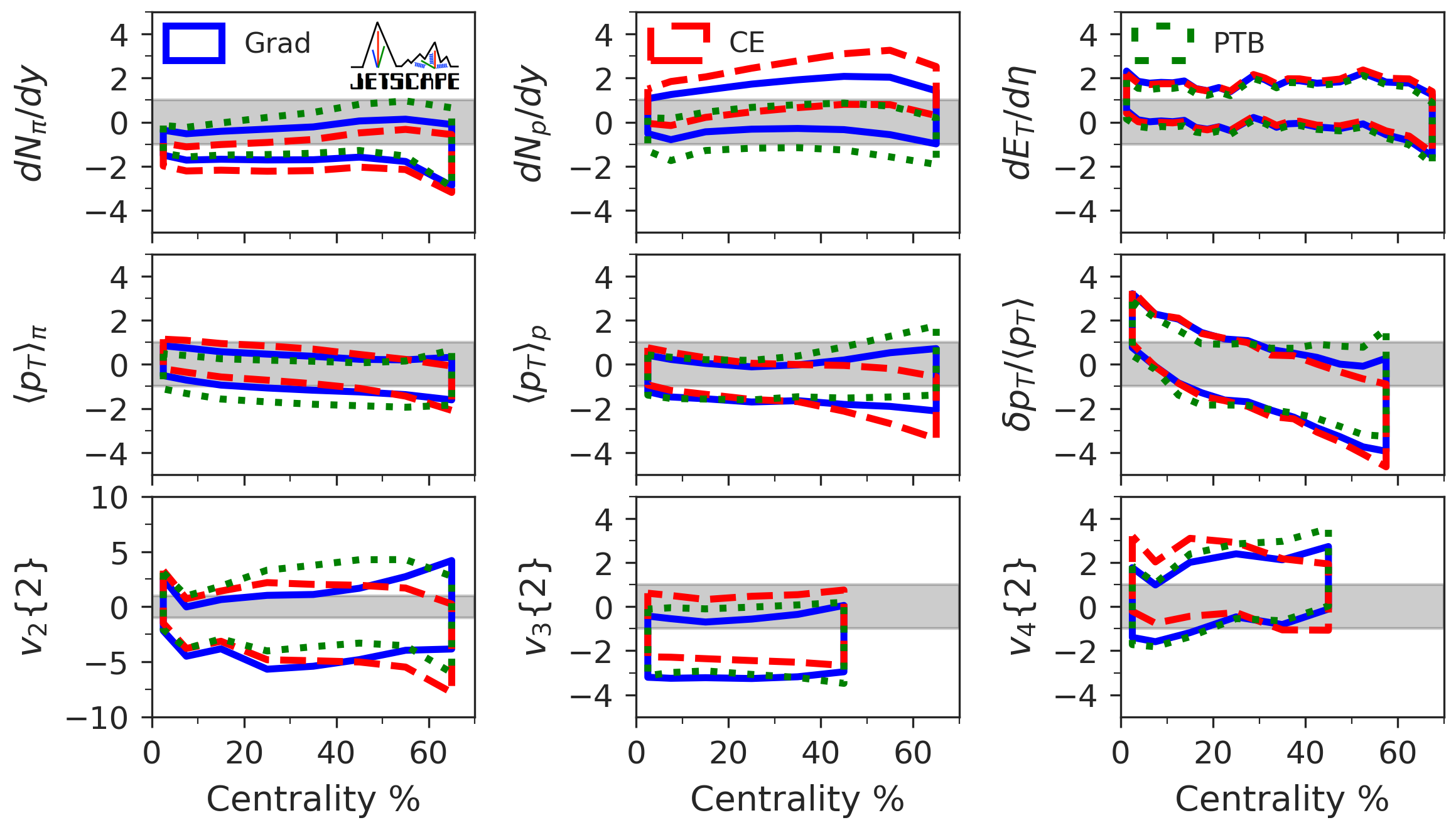}
\caption{The $90$\% credible intervals of the posterior predictive distribution of observables for Pb-Pb collisions at the LHC as functions of centrality, for the Grad (blue), Chapman-Enskog (red) and Pratt-Torrieri-Bernhard (green) particlization models. Plotted is the model discrepancy in units of the experimental standard deviation $\sigma_{\rm exp}$; the vertical axes are labeled with shorthand notation $y \equiv (y_{\rm model}{-}y_{\rm exp}) / \sigma_{\rm exp}$ where $y$ stands for the observable whose model discrepancy is shown. The gray bands represent a discrepancy of one $\sigma_{\rm exp}$ above and below zero. 
\label{F2}
}
\end{figure*}

The level of agreement of each particlization model with a representative subset of measurements is shown in Fig.~\ref{F2}. The bands represent the 90\% credible intervals of the posterior predictive distributions of observables discrepancies with data. All three particlization models show reasonable agreement with the data, giving credence to their respective posterior estimates of the shear and bulk viscosity (and other model parameters) that were inferred from the model-to-data comparison. A closer look at Fig.~\ref{F2} reveals tension within the Chapman-Enskog RTA particlization model, which struggles to describe the pion and proton multiplicities simultaneously. This tension in predicted chemistry is a significant contribution to its small Bayes factor compared to either the Grad or P.T.B. models, which can describe the chemistry more accurately. In \ref{ch5:model_selection} we show that ignoring the proton $dN/dy$ reduces the odds against the Chapman-Enskog particlization model from 5000:1 to 5:1. The key feature behind its failure is the form of its bulk viscous correction to the particle momentum distributions. This highlights the importance of understanding how energy and momentum are distributed across both momentum \emph{and} species at particlization. We remind that our choice of multivariate normal likelihood function, Eq.~(\ref{eq:likelihood}), assumes that probability decreases rapidly away from the mean (it has small tails); this can be unforgiving to tension with the data, resulting in the large ratios of Bayes evidence  encountered in this work. 

In this section, we will build on the analyses presented in the previous chapter. In particular, the three viscous correction models which were calibrated in Ch. \ref{ch5} were shown to yield three differing estimates of the model parameters, and in particular differing estimates of the specific shear and bulk viscosities. 
There is insufficient theoretical evidence at the moment to establish which particlization model is a better description of the interacting hadron gas near the pseudo-critical temperature. In the absence of such theoretical insights, we use experimental measurements to judge the relative performance of each particlization model. The ratio of Bayes evidences was found to be approximately $5000:2000:1$ for the Grad, Pratt-Torrieri-Bernhard and Chapman-Enskog particlization models respectively, clearly disfavoring the Chapman-Enskog model.
We have calculated the Bayesian model-averaged posteriors, averaging over the Grad, Chapman-Enskog RTA, and Pratt-Torrieri-Bernhard viscous correction models using the methods described in section \ref{ch2:ex_bayes_model_avg}.

\subsection{Model-averaged transport coefficients}

\begin{figure}[!htb]
\centering
\includegraphics[width=0.8\linewidth]{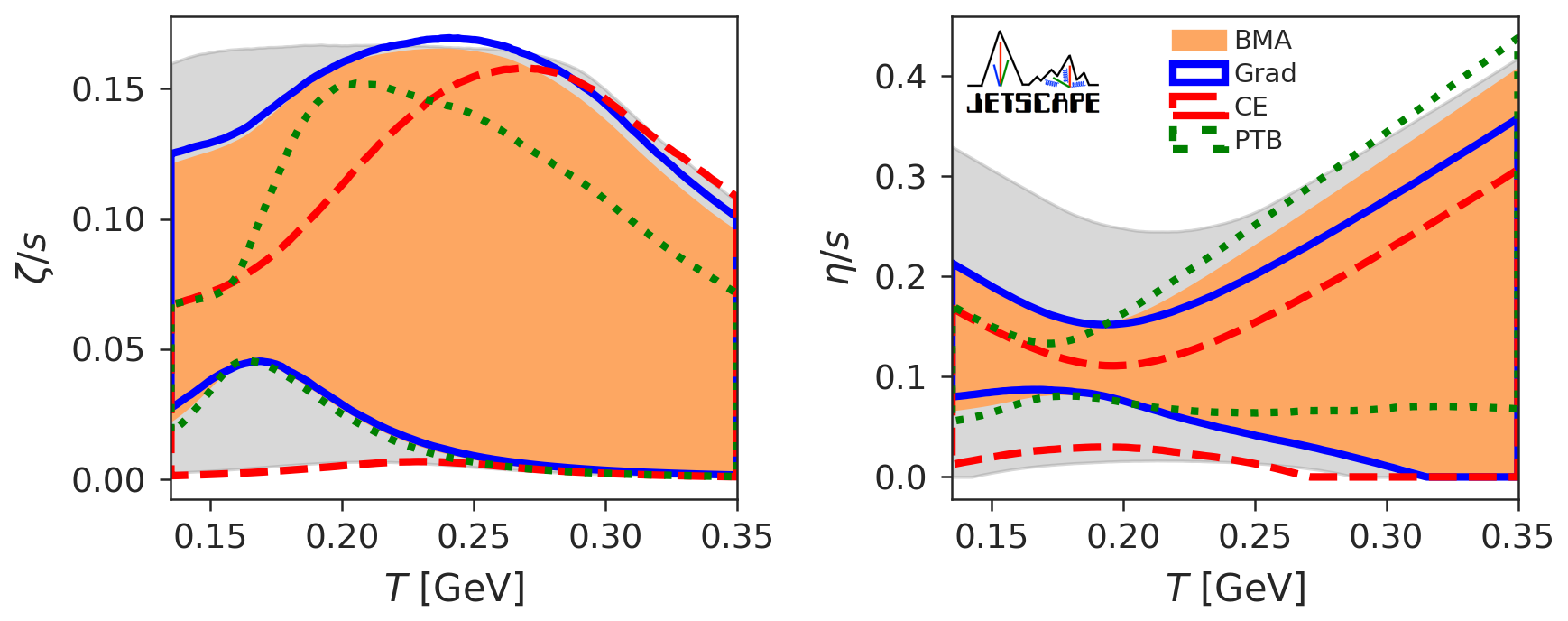}
\caption{The $90$\% credible intervals for the prior (gray), the posteriors of the Grad (blue), Chapman-Enskog (red) and Pratt-Torrieri-Bernhard (green) models, and their Bayesian model average (orange) for the specific bulk (left) and shear (right) viscosities of QGP.
}
\label{F1}
\end{figure}

The temperature dependence of the QGP viscosities favored by the RHIC Au-Au $\sqrts{} = 0.2$ TeV and LHC Pb-Pb $\sqrts{} = 2.76$ TeV data for each of the three viscous correction models are again shown in figure \ref{F1}. The $90$\% credibible intervals are outlined by colored lines. The high-credibility ranges for the different particlization models show similar qualitative features; however they differ significantly in detail, especially in the low-temperature region between $150$ and $250$ MeV. Importantly, at high temperature, the posteriors are close to the $90$\% credibible range of the prior (gray shaded region), which suggests that measurements used in this work do not constrain the viscosities significantly for temperatures~$\gtrsim 250$\,MeV.
The Bayesian model-averaged viscosity estimates are shown as the orange band in Fig.~\ref{F1}. Being strongly disfavored by the Bayesian evidence, the impact of the Chapman-Enskog particlization model on the Bayesian model average is negligible.

\begin{figure*}[!htb]
\centering
\includegraphics[width=0.8\textwidth]{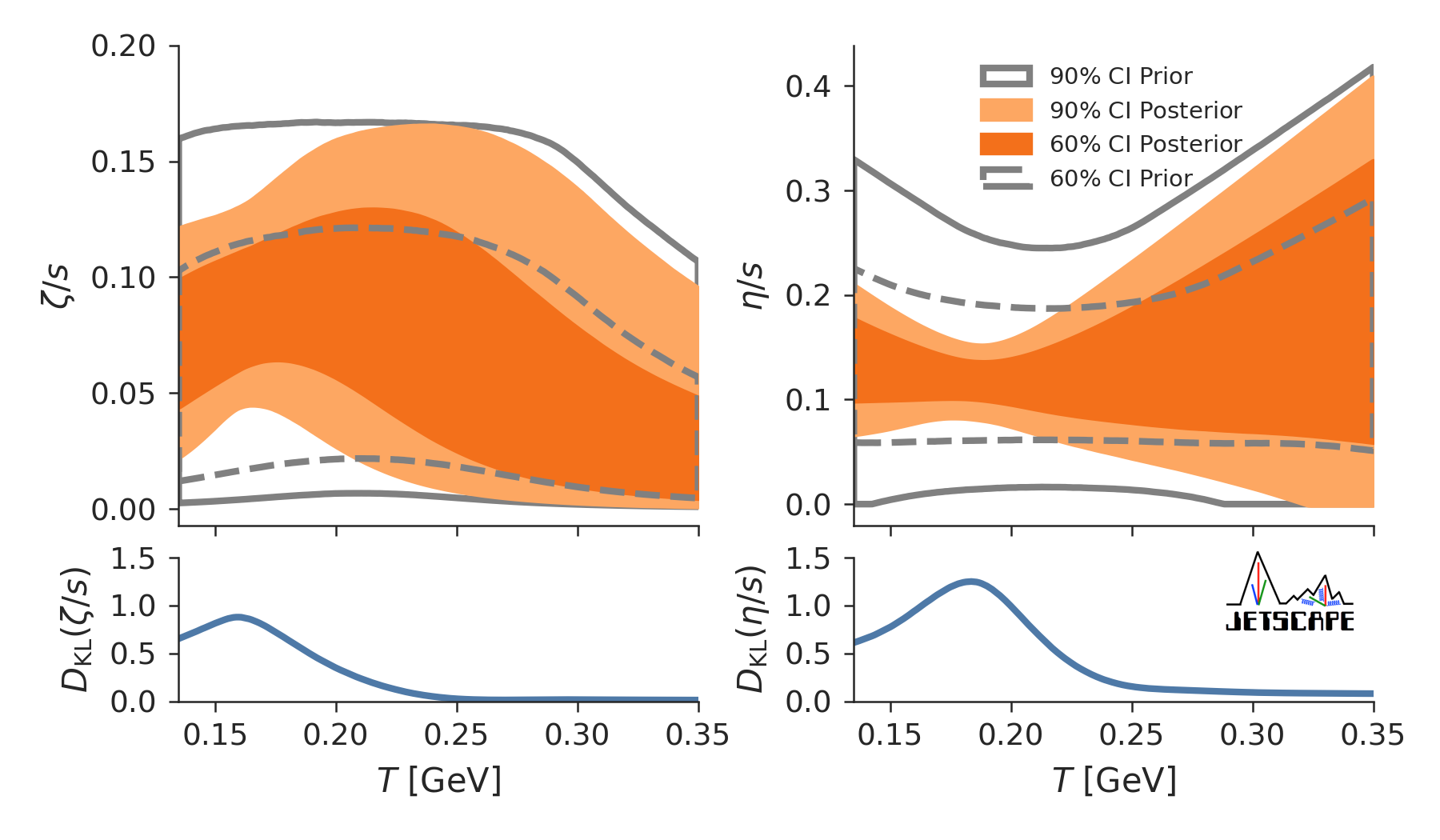}
\caption{(Top) $90$\% and $60$\% credible intervals for the priors (gray) and Bayesian model averaged posteriors of the specific bulk (left) and shear (right) viscosities. (Bottom) The information gain (Kullback-Leibler divergence $D_{KL}$) between the temperature-dependent prior and posterior for specific bulk-viscosity (left) and specific shear-viscosity (right).}
\label{F3}
\end{figure*}

To emphasize the information provided by the experimental data, we plot the Kullback-Leibler divergence ($D_{KL}$) between the temperature-dependent prior and posterior of both the specific shear and bulk viscosity. We show the result in Fig.\,\ref{F3} alongside the 90\% and 60\% prior and Bayesian model averaged posterior credible intervals. While the experimental data are seen to provide significant information for $150\lesssim T\lesssim 250$\,MeV their constraining power rapidly degrades at higher temperatures. In the deconfinement region, the most likely values for $\eta/s$ are of order $0.1$; $\zeta/s$ also favors values around $0.05{-}0.1$ in that region, although constraints are weaker than for $\eta/s$. 
One can impose stronger priors on the viscosities: for example, negative slopes for the shear viscosity at high temperature were excluded based on theoretical expectations in Ref.~\cite{Bernhard:2019bmu}. We elect not to do so, precisely because if a model prior \textit{requires} a positive-definite slope, no other conclusion about the slope could ever be reached, regardless of the amount of information in the experimental data. Again, this is a situation in which we would rather let the data speak for themselves. As it happens, the data do \textit{not} strongly suggest that the slope is positive at high temperatures. 

\subsection{Model-averaged initial condition parameters}

\begin{figure*}[!htb]
\centering
\includegraphics[width=0.6\textwidth]{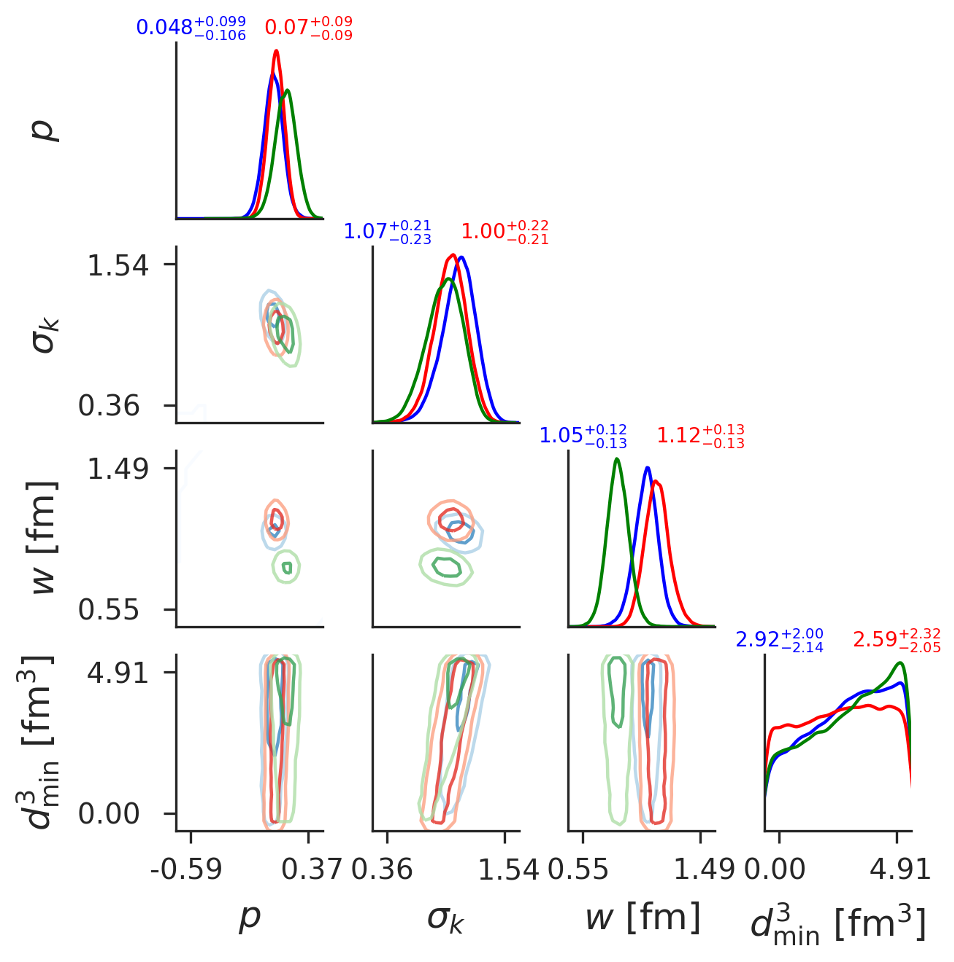}
\caption{Corner-plot posterior densities of select \trento{} initial-condition parameters estimated by simultaneous calibration to both Pb-Pb $\sqrts{}=2.76$ TeV and Au-Au $\sqrts{}=0.2$ TeV collision data, for the Grad (blue), CE RTA (red) and PTB (green) models.}
\label{fig:trento_params_by_df}
\end{figure*}

Similarly to the transport coefficients, the three viscous correction models also yield different estimates of the other model parameters, including the initial conditions defined by \trento{}. 
Shown in Fig.~\ref{fig:trento_params_by_df} are corner-plots of the posterior densities of select initial condition parameters for the three models.
We observe that the estimation of \trento{} reduced-thickness function parameter $p$ is quite robust, with all three models showing good agreement within uncertainties. The same conclusion is also true of the estimation of the multiplicity fluctuation parameter $\sigma_k$. However, there is an observed tension in the estimation of the nucleon width $w$, with the PTB model favoring smaller values than either the Grad of CE models. The PTB model is very sensitive to the bulk viscous pressure, yielding very strong reductions in the mean transverse momenta of pions, kaons and protons. It likely requires a smaller nucleon width $w$ to generate a larger transverse flow to compensate for this effect. 

\begin{figure*}[!htb]
\centering
\includegraphics[width=0.6\textwidth]{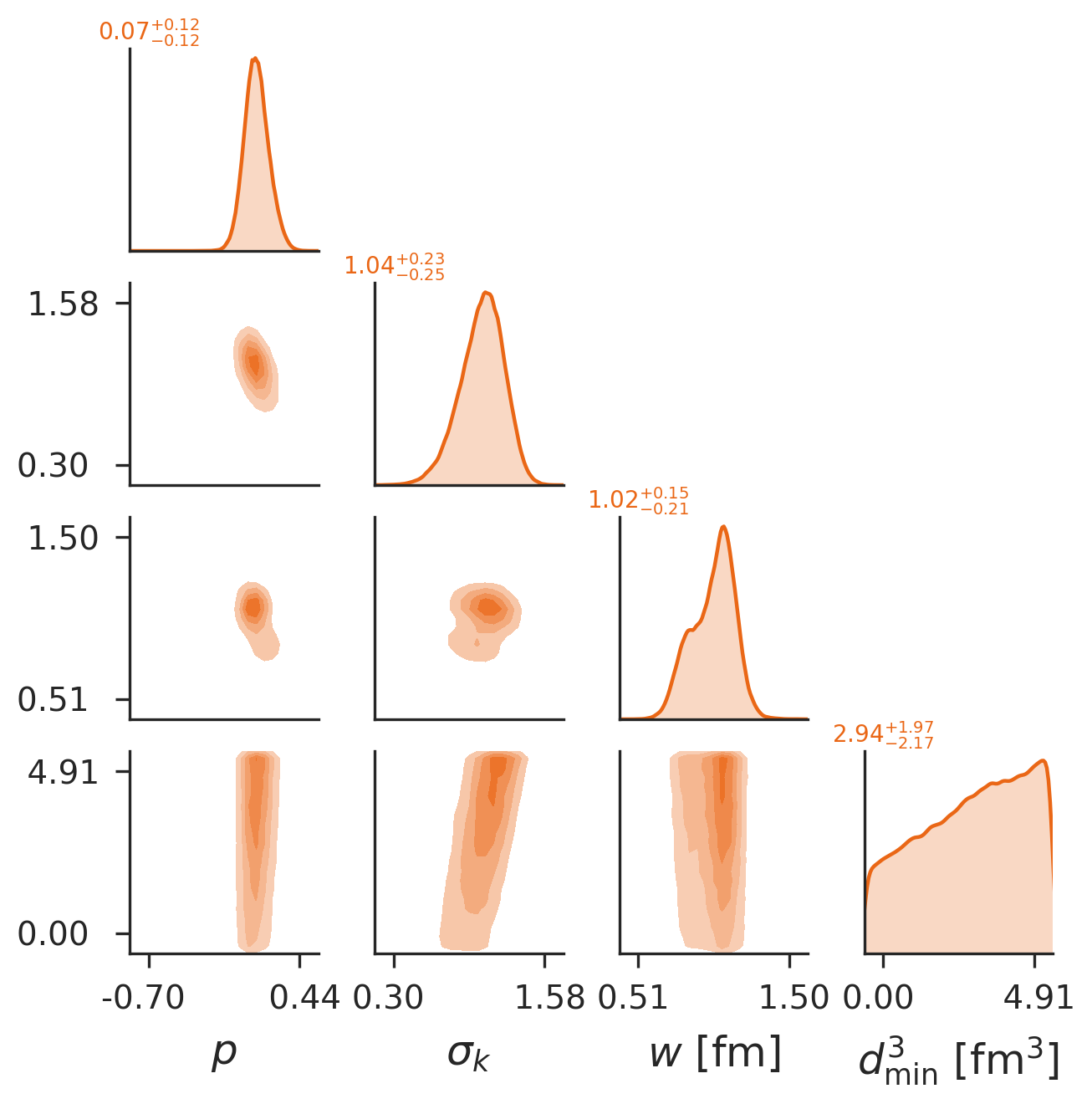}
\caption{Corner-plot posteriors of select \trento{} initial-condition parameters estimated by simultaneous calibration to both Pb-Pb $\sqrts{}=2.76$ TeV and Au-Au $\sqrts{}=0.2$ TeV collision data, having Bayesian model averaged over the Grad, CE RTA and PTB models.}
\label{fig:bma_trento_params}
\end{figure*}

Shown in Fig.~\ref{fig:bma_trento_params} are corner-plots of the posterior densities of the Bayesian model averaged posterior of \trento{} initial conditions. We observe that the estimation of the \trento{} reduced thickness parameter $p$ is indeed very robust. However, the posterior of the nucleon width $w$ is now noticeably broadened with a left-shoulder contributed by the PTB model. 

\section{Quantifying sensitivity to the specification of priors}
\label{ch6:posterior_prior_sens}
%

We remind and caution the reader that the conclusions obtained via Bayesian inference, as a matter of principle, always depend on the prior. In practice, many practitioners develop priors which can minimize the impact of potential subjectivity on the analysis, e.g. the MaxEnt principle~\cite{Jaynes1968}, and hierarchical Bayesian methods~\cite{gelman2013bayesian}. In this section, we explore methods to quantify the sensitivity of our posterior estimates to the specification of our priors.
Usually, priors for the model parameters can be defined according to a set of hyperparameters. As an example, suppose that a priori we are fairly confident that some quantity $r$ should have a value close to $r_0$, but we have a finite uncertainty about how close. A reasonable prior would be given by a Gaussian 
\be
    \mathcal{P}(r) =    \frac{1}{\sqrt{2 \pi \sigma^2}} \exp (-(r-r_0)^2/2\sigma^2)
\ee
with a width $\sigma$ chosen to reflect our uncertainty.\footnote{Actually, Gaussians have very small probability density in the tails; depending on the situation, a distribution with heavier tails may more accurately reflect the current state of knowledge.} In this case, the parameters $(r_0, \sigma)$ are two hyperparameters which define our prior, which we can easily vary.
Similarly, a uniform prior on the parameter $r$ would be defined by two hyperparameters $(r_{\rm min}, r_{\rm max})$ which define the region inside which the prior density is nonzero. 
When a prior is thus formulated in terms of a given functional form with a set of hyperparameters, then we can (and should) check whether any conclusions change as we vary these hyperparameters. We will refer to this as quantifying the `sensitivity' of our posterior inferences to our priors. We will demonstrate the application of this idea to two different questions, both prompted by the inferences made in Ch.~\ref{ch5}. 

In performing Bayesian parameter estimation for the heavy-ion model parameters (in particular the specific shear and bulk viscosities) we found that at low temperatures our data were informative while at high temperatures they `returned our prior'. Specifically, we will focus on the bulk viscosity estimation in subsection \ref{ch5:particlization} and check the extent to which our posterior estimates of the bulk viscosity are robust or sensitive to its prior. 
Secondly, we will quantify the degree to which our relative belief in the particlization models is robust to our specification of the priors for the bulk viscosity. 

\clearpage

\section{Quantifying prior sensitivity: bulk viscosity posterior}
\label{ch6:bulk_posterior_prior_sens}
%

\begin{figure}[!htb]
\centering
\includegraphics[width=\linewidth]{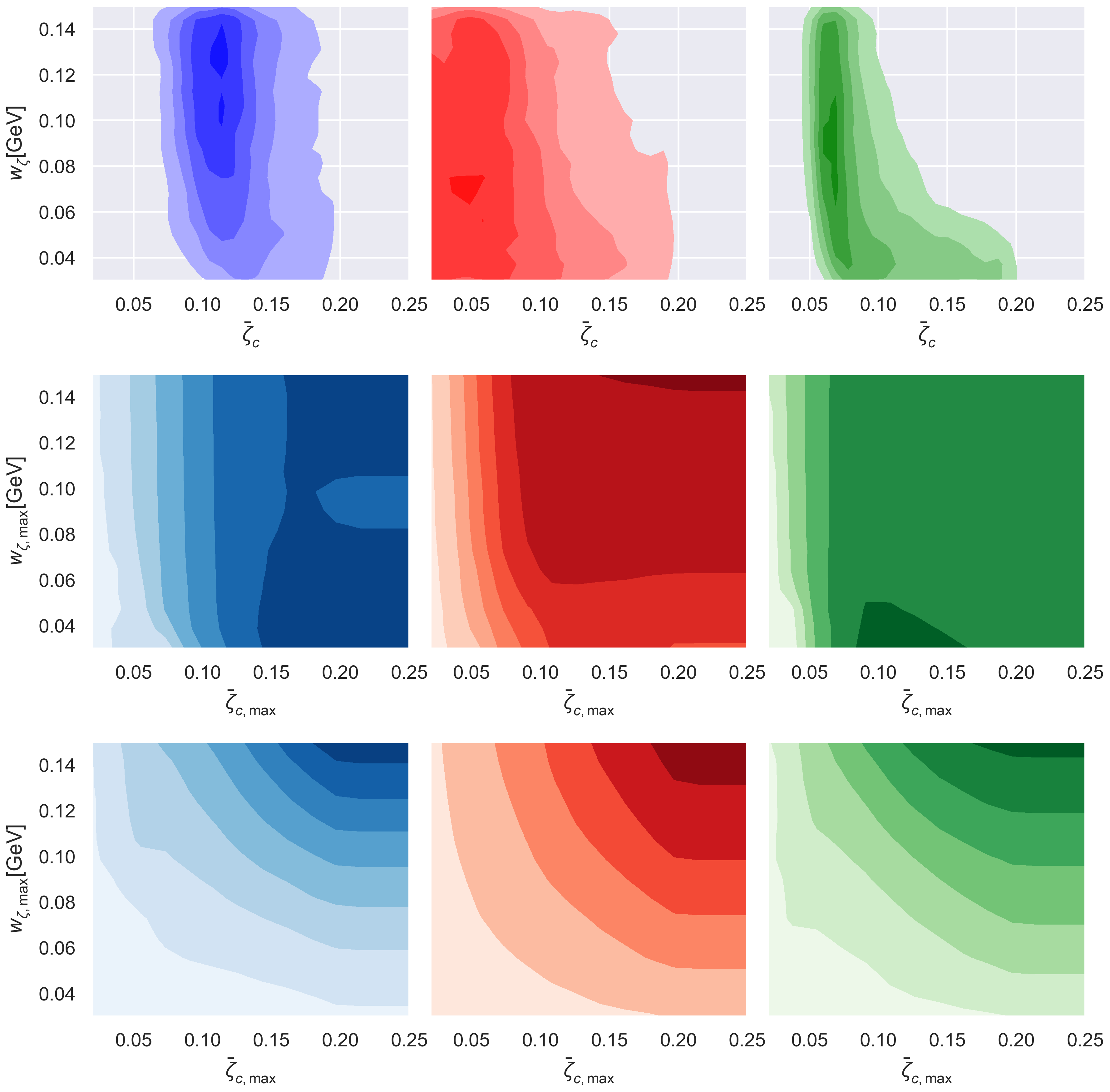}
 \caption{Shown in the top row are the joint posteriors for the peak value $\bar{\zeta}_c$ and width $w_{\zeta}[$GeV$]$ of the specific bulk viscosity for the prior assumed in Ch.~\ref{ch5:priors}. We see that there exist weak non-linear correlations for the Grad model (blue) and stronger non-linear correlations for the P.T.B. model (green). The second and third rows plot the level contours for the upper limit of the $90$\% credible region for $\zeta/s(T=0.15$GeV$)$ and $\zeta/s(T=0.35$GeV$)$, respectively, as functions of the prior hyper-parameters.}
\label{fig:zeta_prior_sensitivity}
\end{figure}

We remind the reader that the temperature-dependence of the specific bulk viscosity is assumed to follow a skewed-Cauchy form with four parameters: 
\begin{equation}
    \frac{\zeta}{s}(T) = \frac{(\zeta/s )_{\max}\Lambda^2}{\Lambda^2+ \left( T-T_{\zeta,c}\right)^2},
\end{equation}
\begin{equation}
    \Lambda = w_{\zeta} \left[1 + \lambda \sign \left(T-T_{\zeta,c}\right) \right].
\end{equation}
We begin by considering a space of hyperparameters which define the prior for the bulk viscosity. We restrict our attention to the two-dimensional space of the maximum allowed value of the bulk viscosity at its peak and the maximum allowed width of the bulk viscosity $w_{\zeta, \rm max}$. For notational convenience, we will here denote the value of the specific bulk viscosity at its peak, previously referred to by $(\zeta/s)_{\rm max}$ in sections \ref{ch5}, by $\bar{\zeta}_c$.
Our prior for the value $\bar{\zeta}_c$ of the bulk viscosity at its peak is given by 
\begin{equation}
  \mathcal{P}(\bar{\zeta}_c|I) =
  \begin{cases}
    1/ [\bar{\zeta}_{c, \rm max} - \bar{\zeta}_{c, \rm min} ] & \text{ if } \bar{\zeta}_{c, \rm min} < \bar{\zeta}_c < \bar{\zeta}_{c, \rm max} \\
    0 & \text{else}.  \\
  \end{cases}
\end{equation}
Similarly, our prior for the width of the bulk viscosity $w_{\zeta}$ is given by 
\begin{equation}
  \mathcal{P}(w_{\zeta}|I) =
  \begin{cases}
    1/ [w_{\zeta, \rm max} - w_{\zeta, \rm min} ] & \text{ if } w_{\zeta, \rm min} < w_{\zeta} < w_{\zeta, \rm max} \\
    0 & \text{else}.  \\
  \end{cases}
\end{equation}

In Fig.~\ref{fig:zeta_prior_sensitivity} we quantify the sensitivity of the posterior of the specific bulk viscosity to two of the prior hyperparameters. We have defined $\bar{\zeta}_{c, \rm max}$ and $w_{\zeta, \rm max}$ as the upper limits below which the prior is non-zero. Therefore, by varying these hyperparameters, we vary the region inside which our prior is nonzero. 
We see that the $T=0.15$ GeV level contours of the upper limit of the $90$\% credible region of the bulk viscosity  have `elbows'; widening the prior further would not significantly widen the $90$\% credible region of the posterior at this temperature. The data, through the likelihood, are informative at this temperature. On the contrary, at $T=0.35$ GeV we find no elbows, and the 90\% posterior credible region monotonically widens as we widen the prior. This confirms that the included data are not strongly informative at these temperatures, leaving a strong sensitivity to prior elicitation. 

\section{Quantifying prior sensitivity: particlization model selection}
\label{ch6:bf_prior_sens}
%

Sensitivity to priors also enters when comparing models. The relevant quantity to determine a model's consistency with observed data, the Bayesian evidence, depends on both the shape and support (consistency with the likelihood) of the specified prior for each model. The Bayesian evidence for model $\mathcal{M}$ is given by
\begin{equation}
    \mathcal{P}(\mathcal{M} |D) = \int_{\mathcal{V}} d \bm{\theta} \mathcal{P}(D| \bm{\theta}, \mathcal{M} )  \mathcal{P}( \bm{\theta} | \mathcal{M} )
\end{equation}
where $\mathcal{P}(D| \bm{\theta}, \mathcal{M} )$ is the likelihood of observing the data $D$ given model $\mathcal{M}$ and its parameters $\bm{\theta}$,  $\mathcal{P}( \bm{\theta} | \mathcal{M} )$ is the prior belief for model $\mathcal{M}$'s parameters $\bm{\theta}$ and $\mathcal{V}$ is the region over which the prior is non-zero. 
Because the Bayes factor $B_{A/B}$ between two models, $\mathcal{M}_A$ and $\mathcal{M}_B$ is defined as the ratio of the evidences:
\begin{equation}
    B_{A/B} = \frac{\mathcal{P}(\mathcal{M}_A |D)}{\mathcal{P}(\mathcal{M}_B |D)}
\end{equation}
it is, therefore, dependent on the priors selected for each model.

The Grad model was shown to be have a large Bayes factor when compared with the Chapman-Enskog model primarily because it showed better agreement with the observed multiplicities of pions and protons.\footnote{When the proton yield was excluded from the experimental data, the Bayes factor was greatly reduced.} 
The abundances of various hadronic species sampled during particlization
are corrected from their equilibrium abundances (given by the Hadron Gas equation of state in equilibrium) by the bulk pressure corrections. Therefore, we are led to investigate whether our preference for the Grad viscous correction model is robust against changes to the priors specified in Ch.~\ref{ch5:priors}.

As in the previous section, we quantify this question by varying the two hyperparameters that control the priors for the width of the bulk viscosity and the value at its peak. 
For any given set of hyperparameters, an exhaustive exploration would require us
to estimate the posterior and Bayes evidence of all models considered on a fine mesh in hyperparameter space. This computation could be considerably expensive. Instead, we will employ Gaussian processes to construct smooth non-parametric interpolations of the behavior of the Bayes factors as functions of the hyperparameters. This was explored in Ref.~\cite{franck2020assessing} and called a \textit{Bayes factor surface}. 

\subsection{Interpolating the Bayes factor surface}

\begin{figure*}[!htb]
\centering
\includegraphics[width=7cm]{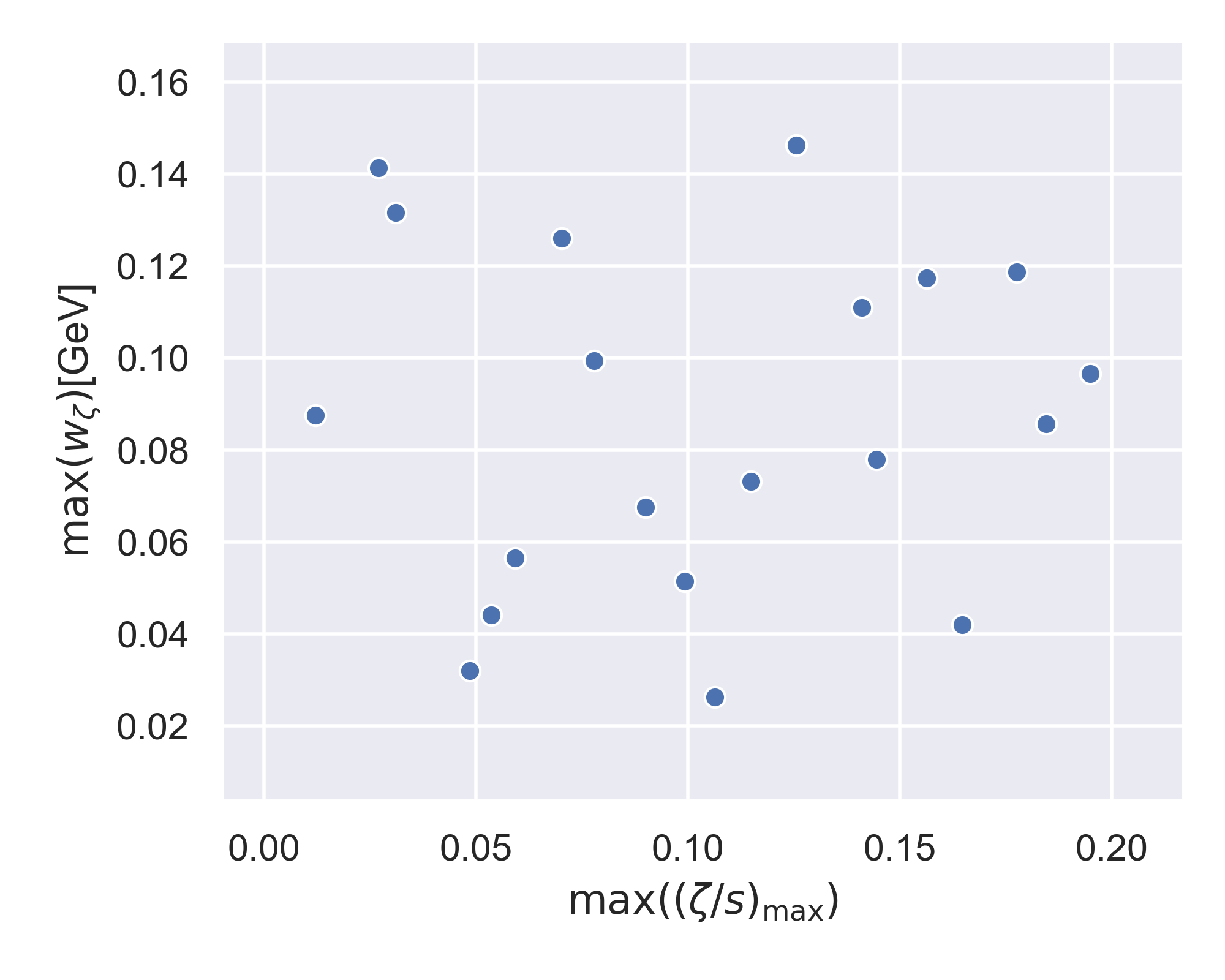}
 \caption{The maximin Latin hypercube samples of the prior hyperparameters.}
\label{fig:lhs}
\end{figure*}

Following Ref.~\cite{franck2020assessing}, we choose a sparse but space-filling maximin Latin hypercube sample of the $(m=2)$-dimensional hyperparameter space, $n = 20$ points. The design is shown in Fig.~\ref{fig:lhs}.
The points define a $n \times m$ design matrix $X$, with each row corresponding to a design point, and each column corresponding to a different hyperparameter.  
For each point $\bm{x}_i$ in this hyperparameter space (remember, each point defines a \textit{different prior}), we compute the posteriors and Bayesian evidences of all three particlization models, and denote the Bayes factor between two models given the hyperparameters $\bm{x}_i$ at the $i$th design point as $B_i$. 
A Gaussian process defines a smooth interpolation of the map $B=B(\bm{x})$, conditioned on the design matrix $X$ and training target values $B_i$.

The kernel function is chosen to be a sum of three separate kernel functions. 
A squared-exponential kernel $k_{\rm RBF}$, often called the Radial Basis Function (RBF) kernel, given by
\begin{equation}
    k_{\rm RBF}(\bm{x}, \bm{x}') = c^2 \exp \left[ \sum_{i=1}^{m} \frac{|x_i - x_i'|^2}{2 l_i^2} \right]
\end{equation}
where $c^2$ is the variance, and $l_i$ length scales for different hyperparameters, controls correlations between outputs which are local in the input parameters. 
A linear kernel $k_{\rm lin}$, given by
\begin{equation}
    k_{\rm lin}(\bm{x}, \bm{x}') = d^2 (\bm{x} - \bm{c}) \cdot (\bm{x}' - \bm{c})
\end{equation}
where $\bm{c}$ defines an offset, 
adds correlations which may arise from global monotonic trends in the outputs as functions of the inputs. 

Finally, a white noise kernel $k_{\rm noise}$, given by
\begin{equation}
    k_{\rm noise}(\bm{x}_i, \bm{x}_j) = \sigma_{\rm noise}^2 \delta_{i, j}, 
\end{equation}
is included to fit the uncorrelated statistical scatter in the evaluation of the Bayes factor. This statistical uncertainty primarily results from the finite sample size of the Markov chain. 

An additional source of uncertainty enters our estimation of the Bayes factor from the finite integration quadrature in the parallel-tempered estimate of the log-evidence. 
For each model $\mathcal{M}_i$, the algorithm yields an expected value of the log-evidence $\ln Z_i$ together with an estimated integration quadrature error $\delta \ln Z_i$. 
In practice, we define the \textit{significance} between two models $\mathcal{M}_i$ and $\mathcal{M}_j$ by 

\begin{equation}
    \sigma_{ij} \equiv \frac{\ln Z_i - \ln Z_j}{ \sqrt{ \delta \ln Z_i^2 + \delta \ln Z_j^2 } }.
\end{equation}
Then the odds between two models are estimated by the ratio of left- and right-tailed $p$-values given the significance level. Thus, strictly speaking, our final result for the odds is not the Bayes factor. However, it is the Bayesian evidences, integrated over their full parameter space, rather than the frequentist point-estimates, which inform these odds. It is only the numerical integration error which is handled via a Frequentist statistic, to finally define the odds.   
A fully-Bayesian alternative would require us to marginalize over the distribution of plausible integration quadrature errors $\xi$.
This would be much more involved, and it is not immediately clear what the assumed distribution of quadrature errors should be without a detailed and exhaustive study of the numerical algorithm.\footnote{For example, there is probably no reason to believe that the error is normally distributed; it is a systematic error which depends on the spacing of the temperature quadrature, the quadrature method, etc...} 
We leave such investigations to future efforts. 

\begin{figure*}[!htb]
\centering
\includegraphics[width=0.9\linewidth]{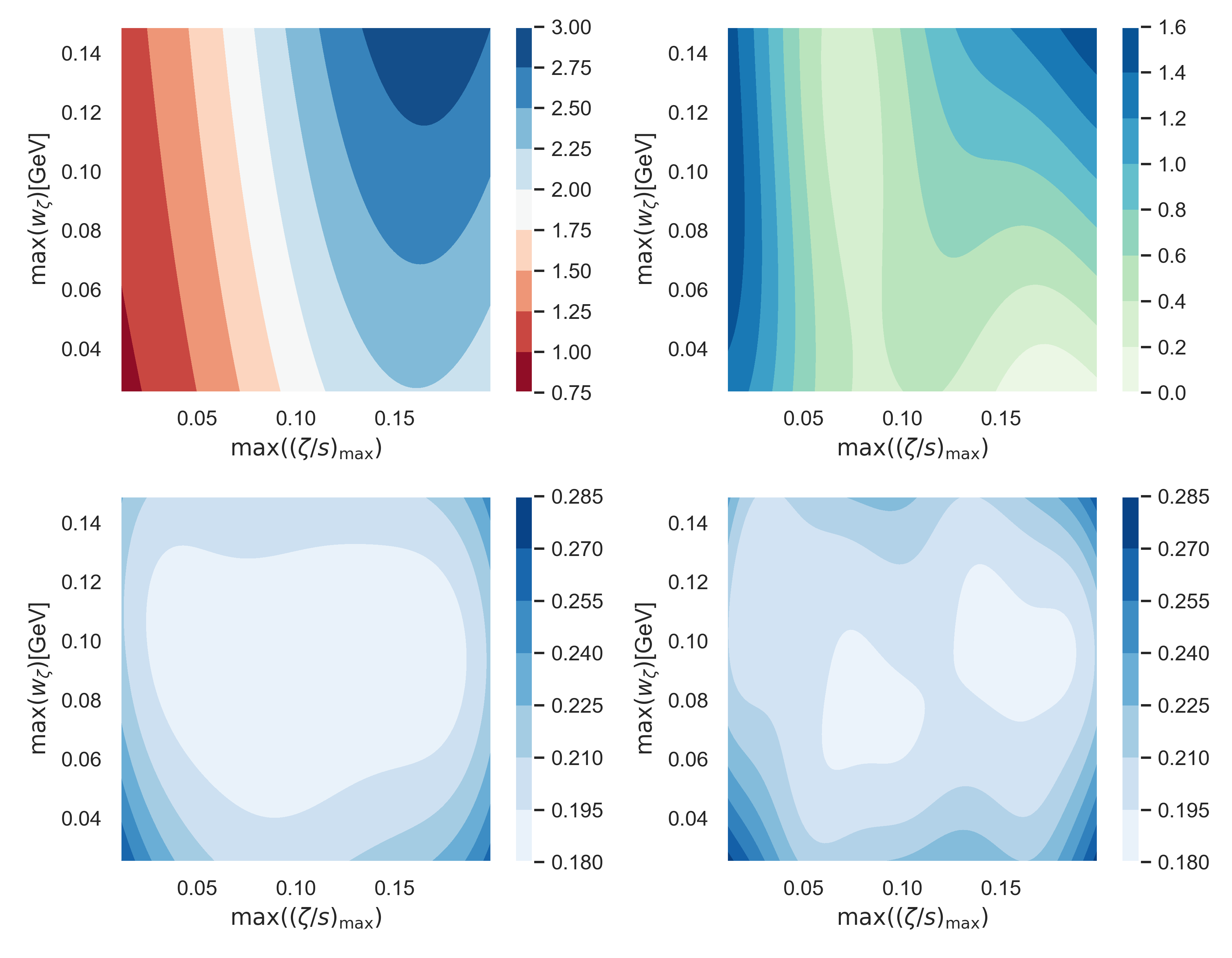}
 \caption{The top row shows the predicted level contours of the log-odds for the Grad vs. Chapman-Enskog (left) and Grad vs. P.T.B (right) models as functions of the bulk viscosity prior hyperparameters. The bottom row displays the predictive uncertainty on the log-odds. Both are estimated by Gaussian processes conditioned on the training set.}
\label{fig:log_odds}
\end{figure*}

Finally, once the Gaussian process has been conditioned, we can employ it to predict the log-odds for pairs of models across the entire range of hyperparameters, together with a measure of the uncertainty on the log-odds.
These are shown in Fig.~\ref{fig:log_odds}. 
We see that the conclusion that the data favor the Grad model over the Chapman-Enskog model are robust against changes in the prior for the specific bulk viscosity. More specifically, the data prefer the Grad model even for priors which restrict the bulk viscosity to be small, which is where the Chapman-Enskog model has its highest posterior density.

When comparing the Grad and P.T.B. models, we see that the log-odds can be quite sensitive to changes in the prior, with a somewhat narrow and nontrivial region of hyperparameter space inside which the odds are approximately unity. This is a consequence of the strong sensitivity of many observables, particularly the hadronic yields and mean transverse momenta in the P.T.B. model to changes in the bulk viscosity. 
Inside of a narrow region in parameter space this model can describe the data well, and it has a large likelihood. But it is penalized by the Occam factor: the likelihood drops sharply outside of this small region of parameter space, reducing its Bayesian evidence. 
In this case, we see how a model which requires parameter fine-tuning to describe the data would also require the researcher to formulate a finely-tuned prior in the context of model comparison. Models which require such fine-tuning tend to make physicists uncomfortable.

\section{Bayesian model mixing viscous correction models}
\label{ch6:model_mixing}
%

Bayesian model mixing~\cite{10.1088/1361-6471/abf1df, gelman2013bayesian} offers an alternative methodology to Bayesian model averaging (see Ch.~\ref{ch2:ex_bayes_model_avg}). Unlike model averaging, model mixing describes a class of modeling assumptions, rather than a unique methodology.
A more thorough introduction to several different additive mixture models can be found in Ref.~\cite{Coleman:2019}. 
A finite additive mixture likelihood is equivalent to the assumption that the observed data are generated from finitely many distinct sub-populations. 
In our case, an additive mixture model generates a model which is not necessarily equivalent to any specific physical model; this allows our generative model, the mixture likelihood, to explore regions of parameter space which are not favored, and generate predictive distributions of observables which are not accessible, by any model alone. Therefore, an additive mixture provides a means of incorporating our uncertainty in our models. 
Following Ref.~\cite{oneill2016bayesian}, we assume that the joint likelihood of all observables $\mathbf{y}_{\rm exp}$ is given by an additive mixture of the individual model likelihoods\footnote{Note that in Eq.~\ref{eq:mix_likelihood} we have a linear combination of the individual model \textit{likelihoods}, while for BMA Eq.~\ref{eq:bma_param_posterior} we had a linear combination of the individual model \textit{posteriors}.} $\mathcal{P}_i(\mathbf{y}_{\rm exp}|\boldsymbol{x})$, given by
\begin{equation}
\label{eq:mix_likelihood}
    \mathcal{P}(\mathbf{y}_{\rm exp}|\boldsymbol{x}, \boldsymbol{\beta}) = \sum_i \beta_i \mathcal{P}_i(\mathbf{y}_{\rm exp}|\boldsymbol{x})    
\end{equation}
where $i$ runs over all models, and the model mixing weights $\beta_i$ lie on the simplex
\begin{equation}
    \sum_i \beta_i = 1.   
\end{equation}

We note that in Ref.~\cite{Coleman:2019} mixture models were explored which allow the mixing parameter to depend on model \textit{inputs}. By inputs, we do not mean the calibration parameters, but an independent set of variables or controls on which the model prediction can depend.\footnote{As an example, a linear model $y(x) = \theta_0 + \theta_1 x$ has as calibration parameters $\theta_0$ and $\theta_1$, and also depends on an input $x$. } However, in the heavy ion model explored in this thesis, every observable is integrated over $p_T$, and is predicted as a function of \textit{only} calibration parameters (there are no corresponding inputs on which we can make the mixing parameters depend). If in the future similar models which instead predict $p_T$-differential observables are explored with model mixing, then the transverse momentum $p_T$ could be considered a kinematic input to the mixing parameters $\beta_i = \beta_i(p_T)$, but this possibility is not explored in this thesis.

In the case that we mix only two models $i \in \{1, 2\}$, the joint mixture likelihood is given by
\begin{equation}
    \mathcal{P}(\mathbf{y}_{\rm exp}|\boldsymbol{x}, \beta) = \beta \mathcal{P}_1(\mathbf{y}_{\rm exp}|\boldsymbol{x}) + (1-\beta) \mathcal{P}_2(\mathbf{y}_{\rm exp}|\boldsymbol{x}).
\end{equation}
The joint posterior of the model parameters $\bm{x}$ and mixing parameter $\beta$ is (as usual) given by Bayes' theorem, 
\begin{equation}
   \mathcal{P}(\boldsymbol{x}, \beta|\mathbf{y}_{\rm exp}) =  \frac{ \mathcal{P}(\mathbf{y}_{\rm exp}|\boldsymbol{x}, \beta) \mathcal{P}(\boldsymbol{x}, \beta) }{ \mathcal{P}(\mathbf{y}_{\rm exp})  }.
\end{equation}
We see a potential advantage of the mixture likelihood model is that by conditioning on the observed data we may learn a \textit{local} (in parameter space) measure of each model's performance. This is in contrast to the Bayes factor, which is by definition marginalized over all model parameters $\bm{x}$.
To proceed, we assume a joint prior such that the mixing parameter is independent of the model parameters, 
\begin{equation}
   \mathcal{P}(\boldsymbol{x}, \beta) = \mathcal{P}(\boldsymbol{x}) \mathcal{P}(\beta),
\end{equation}
and gives no preference to either model,
\begin{equation}
 \mathcal{P}(\beta) = \text{U}(0, 1). 
\end{equation}
\begin{figure*}[htb]
\centering
\includegraphics[width=0.6\textwidth]{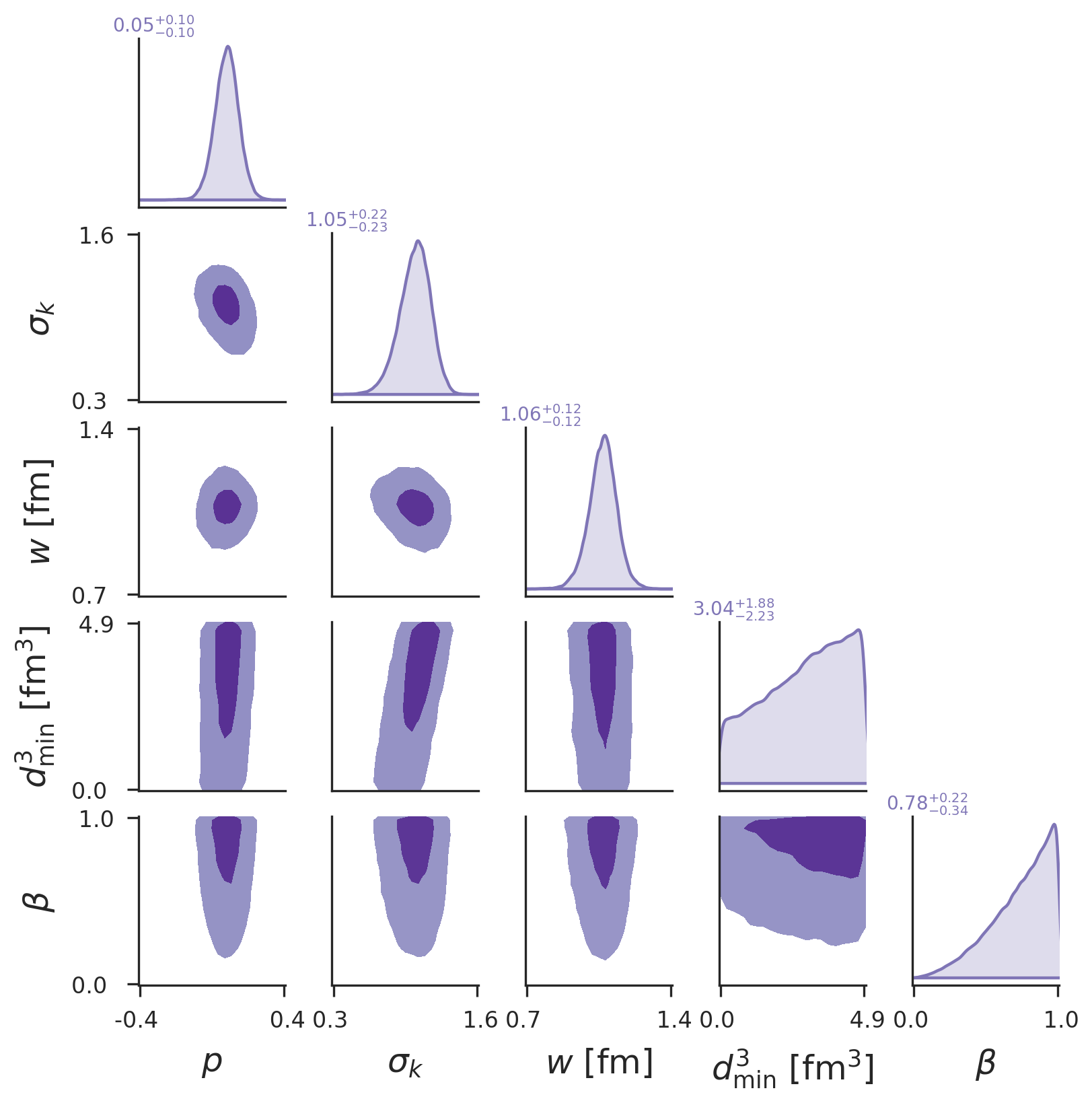}
\caption{The joint posterior of \trento{} initial condition parameters and mixing parameter $\beta$ from the additive mixture model composed of the Grad and Chapman-Enskog viscous correction models. The model has been calibrated to both Pb-Pb $\sqrts{}=2.76$ and Au-Au $\sqrts{}=0.2$ data. $\beta=1$ corresponds to a mixture composed of only the Grad model, while $\beta=0$ only the CE RTA model. }
\label{fig:model_mix_0_1_posterior_ic}
\end{figure*}

\begin{figure*}
\centering
\includegraphics[width=0.6\textwidth]{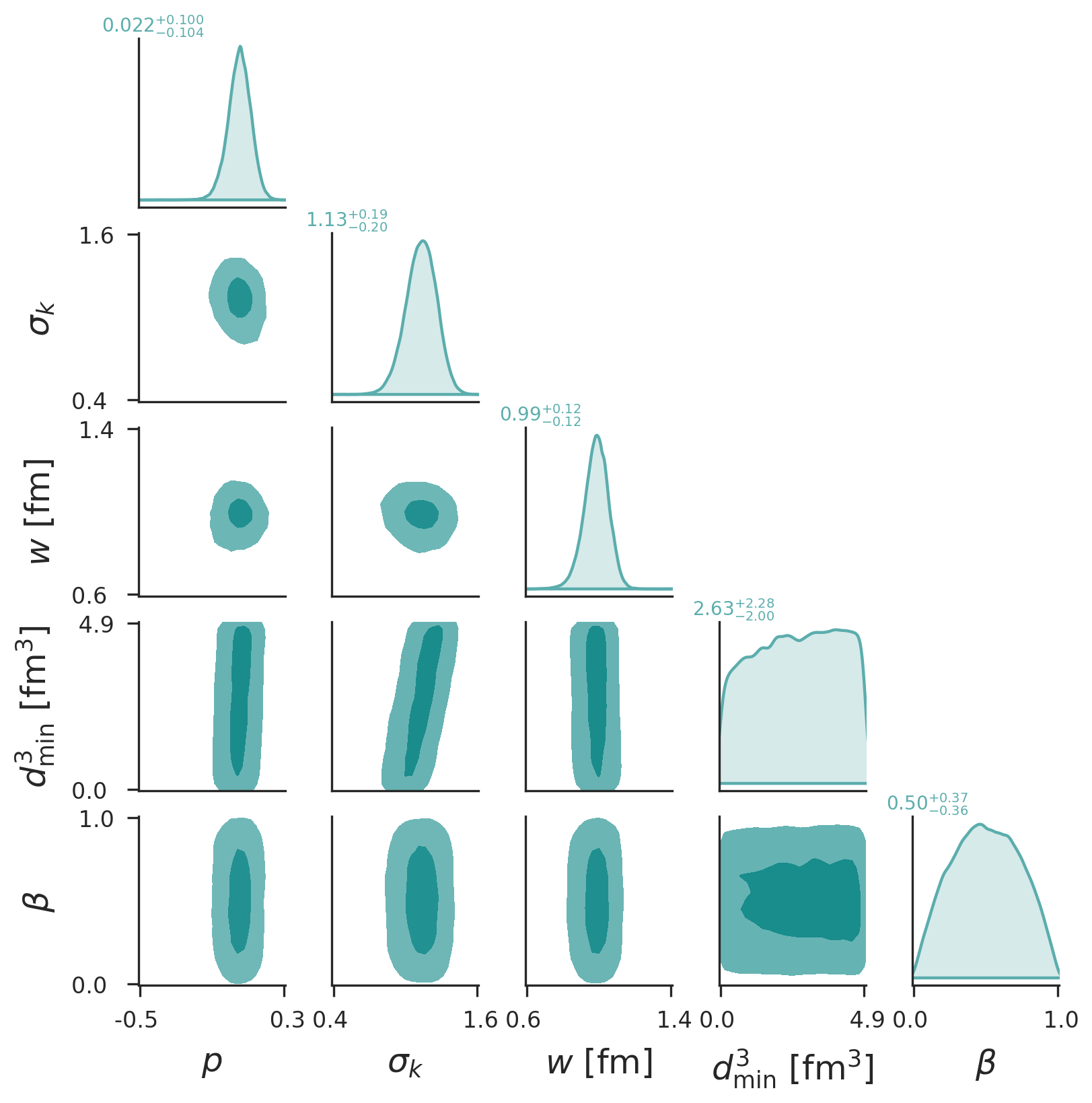}
\caption{The joint posterior of \trento{} initial condition parameters and mixing parameter $\beta$ from the additive mixture model composed of the Grad and Pratt-Torrieri-Bernhard viscous correction models. The model has been calibrated to both Pb-Pb $\sqrts{}=2.76$ and Au-Au $\sqrts{}=0.2$ data. $\beta=1$ corresponds to a mixture composed of only the Grad model, while $\beta=0$ only the Pratt-Torrieri-Bernhard model. }
\label{fig:model_mix_0_3_posterior_ic}
\end{figure*}

The joint posterior can be sampled using the same Markov chain Monte Carlo methods\footnote{A useful and practical introduction to sampling the mixture model is given in Ref.~\cite{model_mixing_ex}.} presented in Ch.~\ref{ch4:mcmc}. In Figs.~\ref{fig:model_mix_0_1_posterior_ic} and ~\ref{fig:model_mix_0_3_posterior_ic} are shown the posteriors of \trento{} initial condition parameters joint with the mixing parameter $\beta$. In Fig.~\ref{fig:model_mix_0_1_posterior_ic} we see that the posterior for $\beta$ still indicates a strong preference towards the Grad model in comparison with the Chapman-Enskog model; however, the left-tail is quite long, indicating that the Chapman-Enskog model has a nontrivial contribution to the posterior. This is in contradistinction to the Bayesian model average posterior; the Bayesian evidence for the Chapman-Enskog model was sufficiently small that its contribution to the BMA posterior was negligible.

In Fig.~\ref{fig:model_mix_0_3_posterior_ic} we see that the posterior for $\beta$ indicates no preference for either the Grad or Pratt-Torrieri-Bernhard models; the calibrated posterior prefers roughly equal mixtures of the two models. This is qualitatively consistent with the Bayes factors found in Ch.~\ref{ch5:model_sensitivity} model average posterior; however the quantitative difference between roughly $3:1$ odds given by the Bayes factor, and roughly $1:1$ odds given by the additive mixture model is certainly noticeable in the parameter posteriors.

\begin{figure*}
\centering
\includegraphics[width=0.9\textwidth]{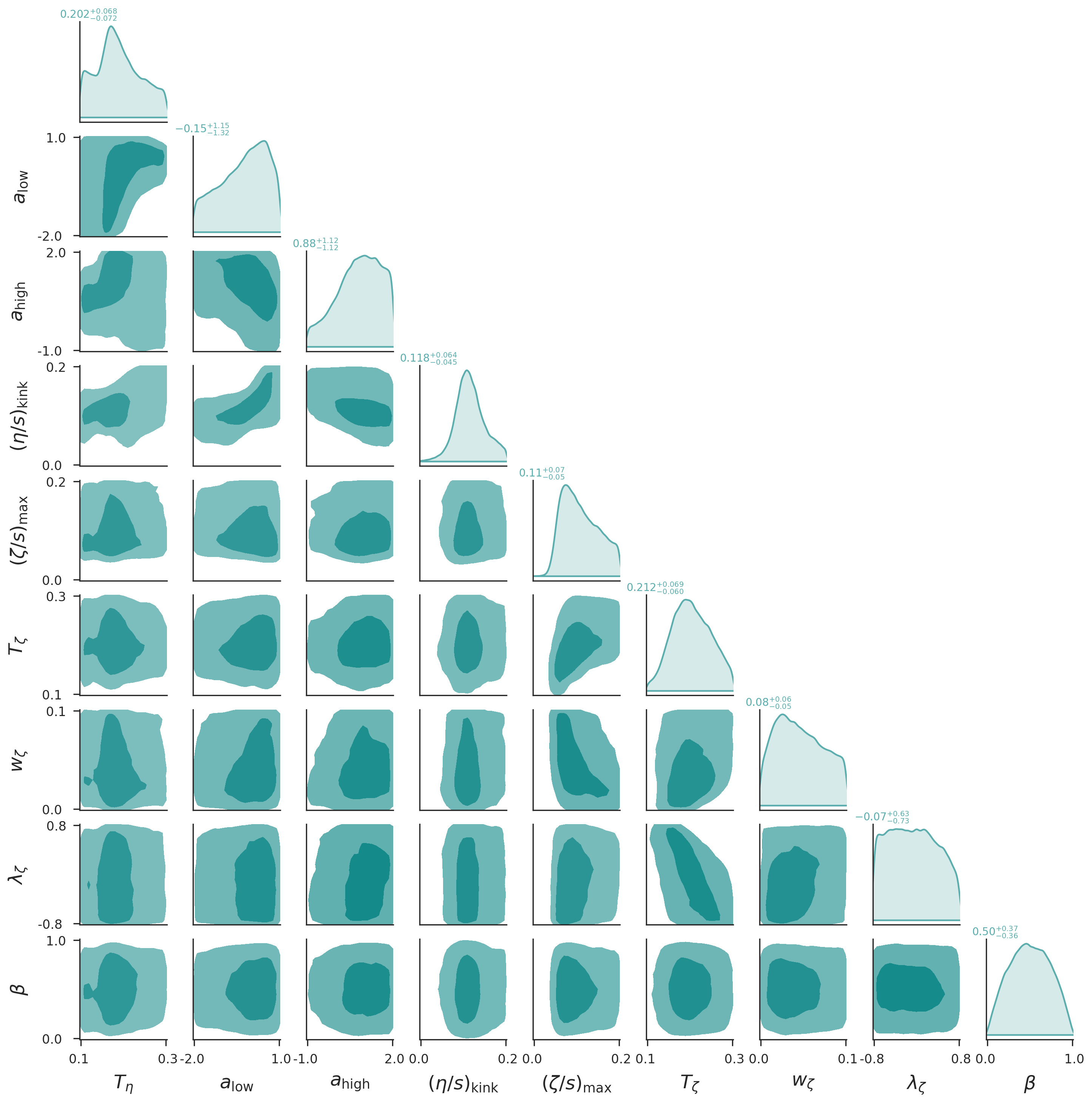}
\caption{The joint posterior of shear and bulk viscosity parameters and mixing parameter $\beta$ from the additive mixture model composed of the Grad and Pratt-Torrieri-Bernhard viscous correction models. The model has been calibrated to both Pb-Pb $\sqrts{}=2.76$ and Au-Au $\sqrts{}=0.2$ data. $\beta=1$ corresponds to a mixture composed of only the Grad model, while $\beta=0$ only the Pratt-Torrieri-Bernhard model. }
\label{fig:model_mix_0_3_posterior_visc}
\end{figure*}

In Fig.~\ref{fig:model_mix_0_3_posterior_visc} is shown the joint posterior of specific shear and bulk viscosities with the mixing parameter $\beta$, for the mixture of Grad and Pratt-Torrieri-Bernhard models. We do not find any particularly illuminating correlations between the viscosities and the mixing parameter, yielding no insights regarding the local performance of each model in describing the data. Differences between the Bayesian model averaged and additive mixture model posteriors would also manifest in posterior predictive distributions, which would be an important consideration for robust experimental design~\cite{10.1088/1361-6471/abf1df}.

\begin{figure*}
\centering
\includegraphics[width=0.8\textwidth]{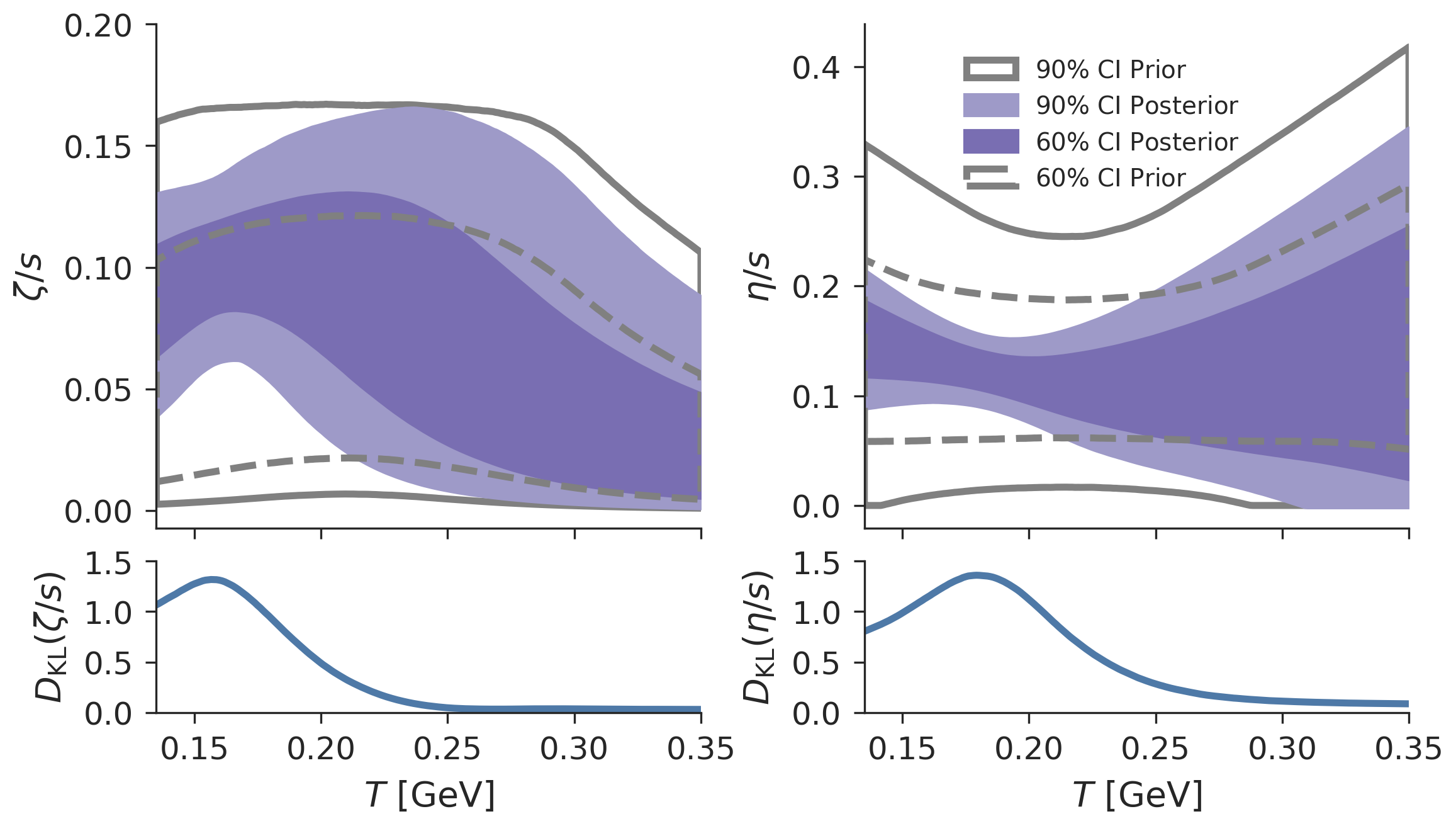}
\caption{The posteriors of specific shear (left) and bulk (right) viscosities for the mixture of Grad and Chapman-Enskog models, marginalized over all other model parameters and the mixing parameter $\beta$. The mixture model has been calibrated to both Pb-Pb $\sqrts{}=2.76$ and Au-Au $\sqrts{}=0.2$ data.}
\label{fig:model_mix_0_1_shear_bulk}
\end{figure*}
\begin{figure*}
\centering
\includegraphics[width=0.8\textwidth]{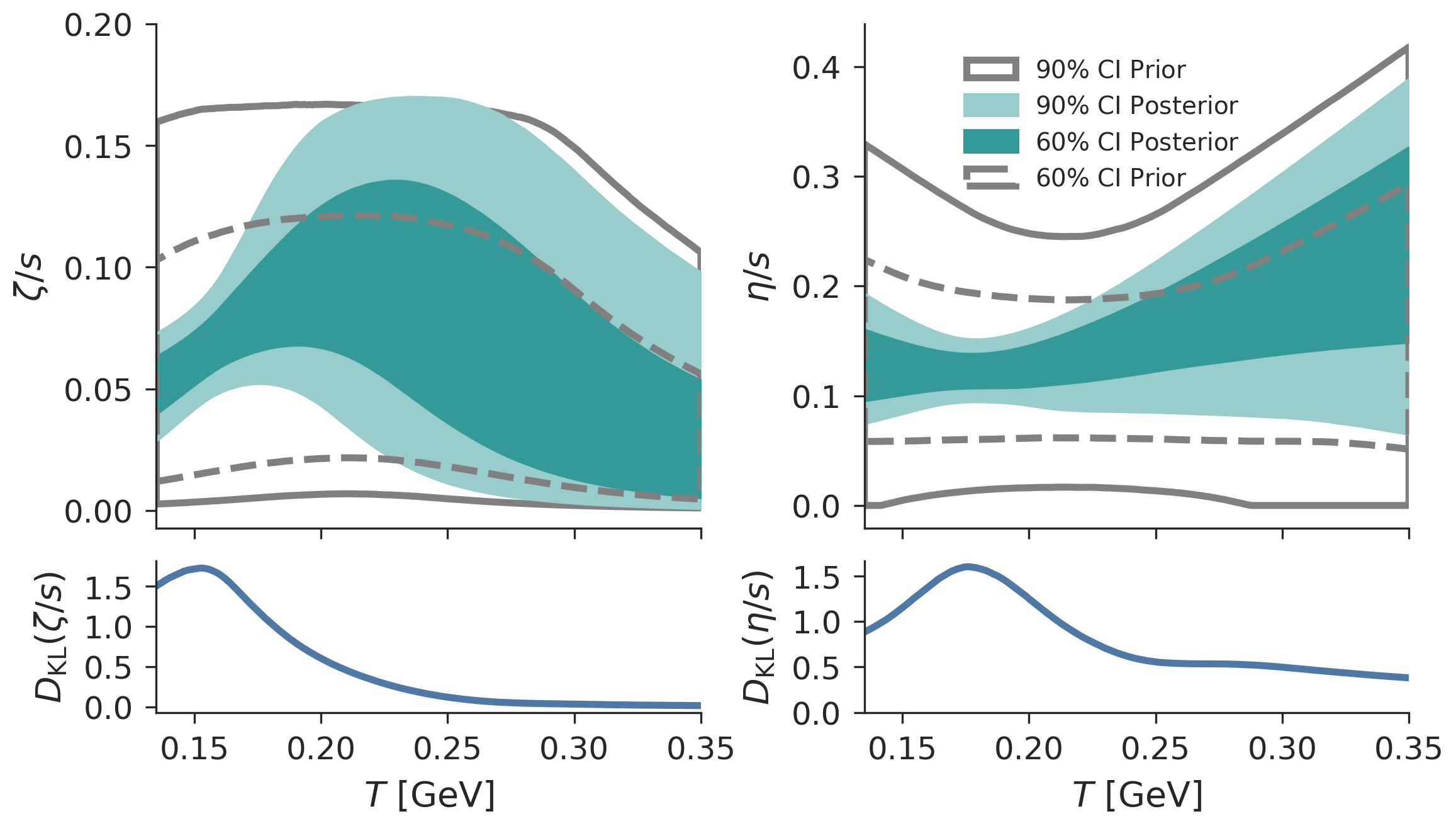}
\caption{The posteriors of specific shear (left) and bulk (right) viscosities for the mixture of Grad and Pratt-Torrieri-Bernhard models, marginalized over all other model parameters and the mixing parameter $\beta$. The mixture model has been calibrated to both Pb-Pb $\sqrts{}=2.76$ and Au-Au $\sqrts{}=0.2$ data.}
\label{fig:model_mix_0_3_shear_bulk}
\end{figure*}

Finally in Figs.~\ref{fig:model_mix_0_1_shear_bulk} and ~\ref{fig:model_mix_0_3_shear_bulk} we show the posteriors of the temperature-dependent specific shear and bulk viscosities, for the mixtures of Grad and Chapman-Enskog (purple) or Grad and Pratt-Torrieri-Bernhard models (cyan), respectively. In Fig.~\ref{fig:model_mix_0_1_shear_bulk}, we see that the Chapman-Enskog model contributes non-negligibly to the posteriors, and has the effect of reducing the overall magnitude of both the specific shear and bulk viscosities compared to the Grad model alone.
In Fig.~\ref{fig:model_mix_0_3_shear_bulk}, we see again see the influence of the strong sensitivity of the Pratt-Torrieri-Bernhard model to the bulk viscosity at low temperatures, yielding tight constraints on the specific bulk viscosity at temperatures below $200$ MeV. Additionally, we see that the negative slopes of the high-temperature specific shear viscosity are mostly ruled out by this particular mixture model.

Considering the results of both additive mixture models, our previous insights regarding the influence of the viscous correction uncertainty remain largely in tact. The propagation of model uncertainty sourced by particlization still currently limits our ability to constrain the specific shear and bulk viscosities. 
\chapter{Conclusions and Outlook}
\label{ch7}

\section{Conclusions}
\label{ch7:conclusions}

In this thesis, a program of Bayesian inference and model-building has been applied towards the study of heavy-ion collision experimental data. In particular, we have focused on comparing viscous hydrodynamic hybrid models with the data observed in high-energy Pb-Pb and Xe-Xe collisions measured at the LHC, and Au-Au collisions measured at RHIC, and found that these data are consistent with moderately small values of the specific shear and bulk viscosities at temperatures below $250$ MeV. However, the specific shear and bulk viscosities have large sensitivities to the prior specified at higher temperatures, and therefore the observables studied have little to no constraint at high temperatures. 

Additionally, we have estimated an important source of model uncertainty resulting from the particlization of the fluid into kinetic degrees of freedom. Moreover, this uncertainty was propagated in the estimates of all model parameters, including the transport coefficients, via model averaging and mixing. These methodologies will undoubtedly provide a useful framework for uncertainty quantification in future efforts in heavy-ion physics~\cite{10.1088/1361-6471/abf1df, Coleman:2019}. 

\section{Prior elicitation}
\label{ch7:priors}

It is my personal view that the community should seek \textit{less informed} priors regarding the specific shear and bulk viscosities in future endeavors. Specifically, the requirement that the viscosities obey fixed temperature-dependent parameterizations (e.g. those used in this thesis, Eqns. \ref{eq:eta_s_lin}, \ref{eq:zeta_s_cauchy}) should probably be abandoned wholesale. Such parametrizations introduce strong correlations between the values of the viscosities at different temperatures. Alternative possibilities exist, for example treating $\eta/s$ and $\zeta/s$ at on a grid of temperature as the unknowns, and performing a continuous interpolation between grid values to define $\eta/s(T)$ and $\zeta/s(T)$ when running the hydrodynamic transport. Nonparametric models given by Gaussian processes have already been applied in other contexts~\cite{Drischler:2020yad}, and are also promising candidates.

There are probably methodologies from information field theory~\cite{Ensslin:2018pno} that practitioners of Bayesian inference for heavy-ion collision could benefit from studying and implementing. Those methodologies are typically applied to spatially or temporally varying fields, however with some massaging they may be useful in building minimally-biased models of temperature-dependent quantities such as the transport coefficients, equation of state, etc...

\section{Improving pre-hydrodynamic modeling}
\label{ch7:model_bias_pre_hydro}

The pre-hydrodynamic models currently used to evolve the heavy-ion collision in phenomenological studies likely introduce biases. Moreover, it is not clear that any of the existing models can smoothly match to a non-conformal and hydrodynamized QGP. There have recently been many efforts towards more theoretically motivated descriptions of this stage, which include non-trivial scattering kernels. A systematic Bayesian study estimating and propagating this source of model uncertainty, and systematically comparing the weakly-coupled and strongly-coupled approaches is of very high interest and could address and quantify some of the largest outstanding uncertainties. 

\section{Improving particlization modeling}
\label{ch7:model_bias_particlization}

As demonstrated in this thesis, the different models used to particlize the fluid into hadronic degrees of freedom introduce biases in the estimation of model parameters, and in particular the viscosities. I am not convinced that a satisfactory particlization model exists at this time. Very little is known from first-principles regarding the physics of hadronization, and its close proximity to the fluid/gas interface makes theoretically motivated particlization models a difficult endeavor. In the absence of theoretically robust physics models, model-averaging and model-mixing provide avenues to estimate and propagate uncertainties. 

There has been much progress in the understanding and modeling of quark-gluon plasma physics; the era of quantified uncertainties is now reaching sufficient maturity to tackle the largest outstanding scientific questions regarding the creation and dynamics of QGP. I hope that the methodologies described in this thesis will be useful in such future endeavors.

\backmatter
\begin{appendices}
\chapter{Model and Statistical Validation}

\section{Validation of principal component analysis}\label{pca_valid}

\begin{figure*}[!htb]
\makebox[\textwidth][c]{\includegraphics[width=1.1\textwidth]{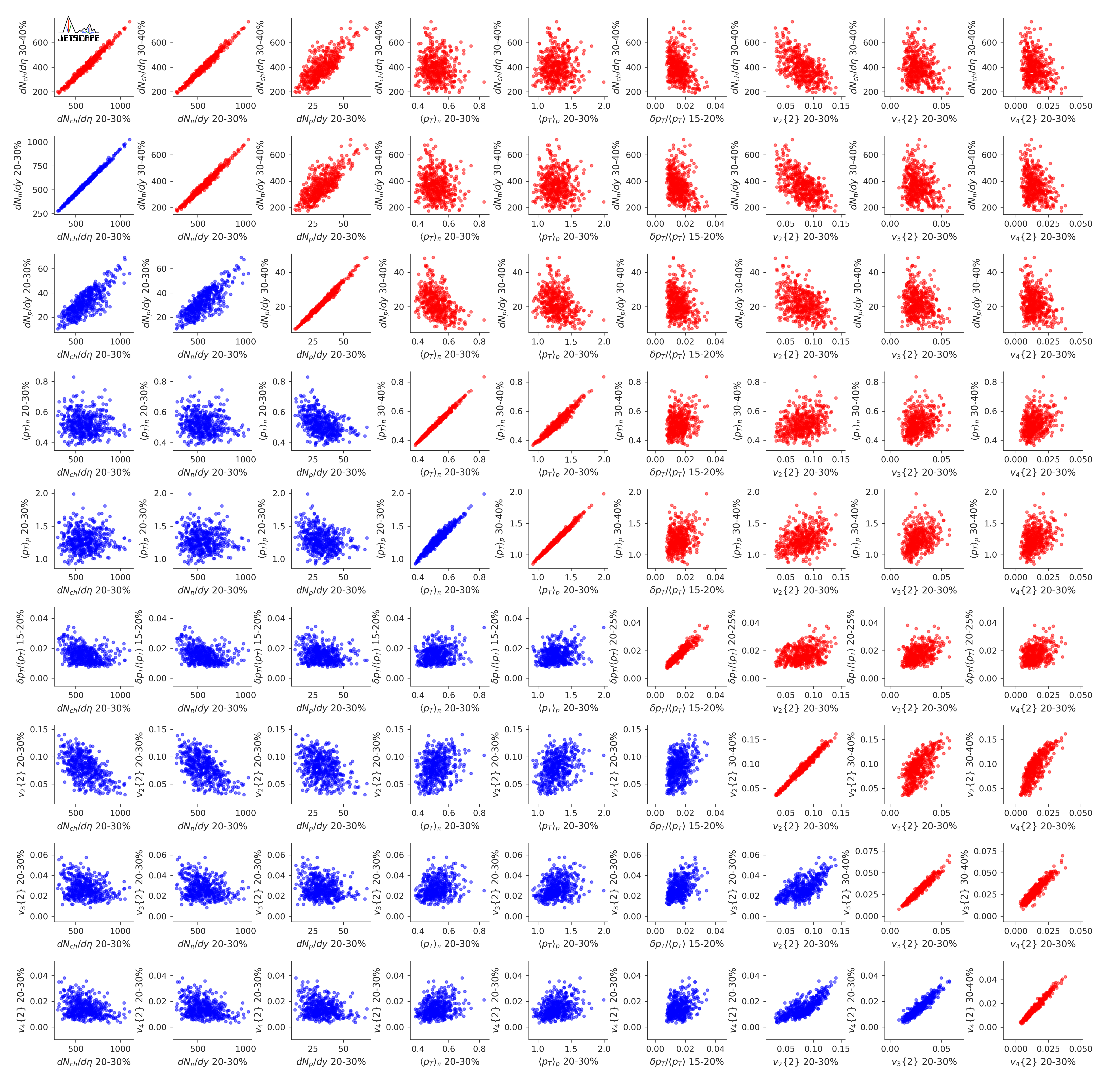}}%
\caption{Scatter plots of selected pairs of observables predicted by our model for Pb-Pb collisions at $\sqrts{} = 2.76$ TeV for the 500-point parameter design. Some pairs in the same centrality bin are shown in blue, while all pairs of different centrality bins are shown in red. Many pairs of observables have strong linear correlations, in which case they do not contain significant mutual information (knowing one is nearly sufficient). Pairs of observables which do not have strong linear correlations carry independent information about the parameters.}
\label{fig:observables_corr}
\end{figure*}

Principal component analysis acts to identify the linear correlations among pairs of observables resulting from changes in the model parameters. A figure showing the correlations among all possible pairs of observables would be far too large to plot, but we plot a subset of possible pairs in Fig.~\ref{fig:observables_corr} and make some important observations.  

\begin{figure}[!htb]
\centering
\includegraphics[trim=0 0 0 16, clip, width=0.7\linewidth]{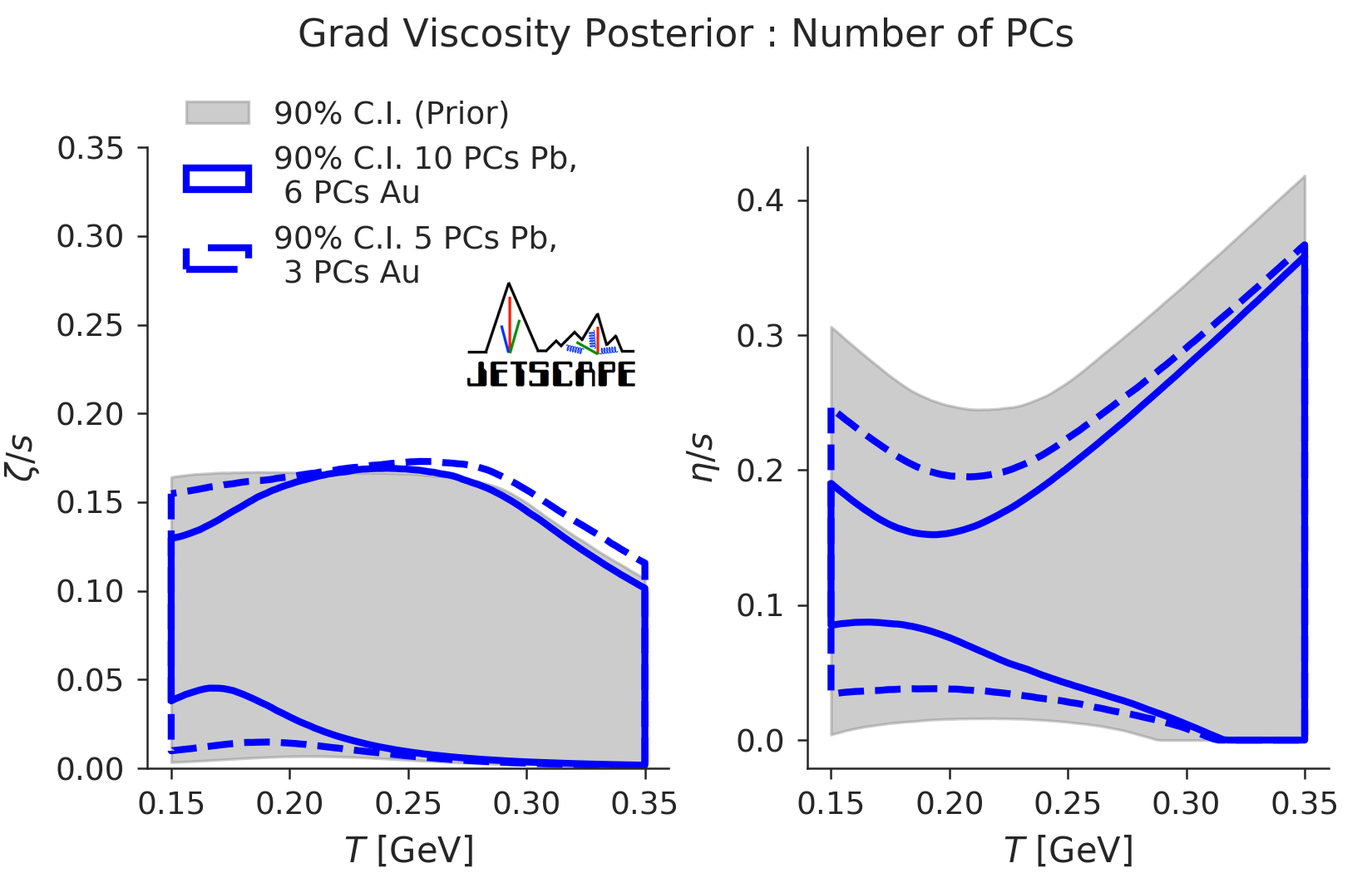}
\caption{Comparing the viscosity posteriors when we perform Bayesian parameter estimation with less principal components. The solid blue results from estimation using 10 principal components for the Pb-Pb $\sqrts{} = 2.76$ TeV emulator and $6$ principal components for the Au-Au $\sqrts{} = 0.2$ TeV emulator. The dashed blue results from $5$ principal components for Pb-Pb $\sqrts{} = 2.76$ TeV and $3$ principal components for Au-Au $\sqrts{} = 0.2$ TeV. }
\label{compare_posterior_npc}
\end{figure}

We see that certain pairs of observables have strong linear correlations: for instance the yield $dN/dy$ of pions in the $20-30$\% centrality bin and yield of charged particles $dN_{\rm ch}/d\eta$ in the $30-40$\% centrality bin. For such pairs, nearly all the information about the model parameters is contained in just one of the observables. Uncorrelated pairs contain independent information about the model parameters. No pair of observables displays a significant non-linear correlation except for the elliptic flow $v_2\{2\}$ and quadrangular flow $v_4\{2\}$, which shows a correlation $v_4\{2\} \sim (v_2\{2\})^2$. The scarcity of strong non-linear correlations suggests that ordinary principal component analysis is a suitable method for dimensionality reduction. 
To test that our model emulator used for Bayesian inference is not overfit to features from statistical noise in the hybrid model, we have examined the effect on the posterior when we reduce the number of principal components by a factor of two for each system. In this case, five principal components explain about $94$\% of the variance of Pb-Pb collisions at $\sqrts{} = 2.76$ TeV and three principal components about $91$\% of the variance of Au-Au collisions at $\sqrts{} = 200$ GeV data. The posteriors of the specific viscosities in these two cases are compared in \fig{compare_posterior_npc}. 

The uncertainty contributed by the principal components that we omit contributes to the total emulator uncertainty and the posterior of specific shear and bulk viscosities is broadened in the case with fewer principal components included. To be sure that an emulator (and the choice of the number of principal components) is not underfit or overfit, one must perform emulator validation (see Ch.~\ref{ch4:emu_valid_scatter} and Ch.~\ref{ch4:emu_validation}). An emulator which is overfit will fit the training points very well, but will perform poorly in predicting the observables for a novel testing point.

\section{Experimental covariance matrix} \label{app_exp_cov}

\begin{figure}[!htb]
  \centering
  \includegraphics[trim=0 0 0 16, clip, width=0.7\linewidth]{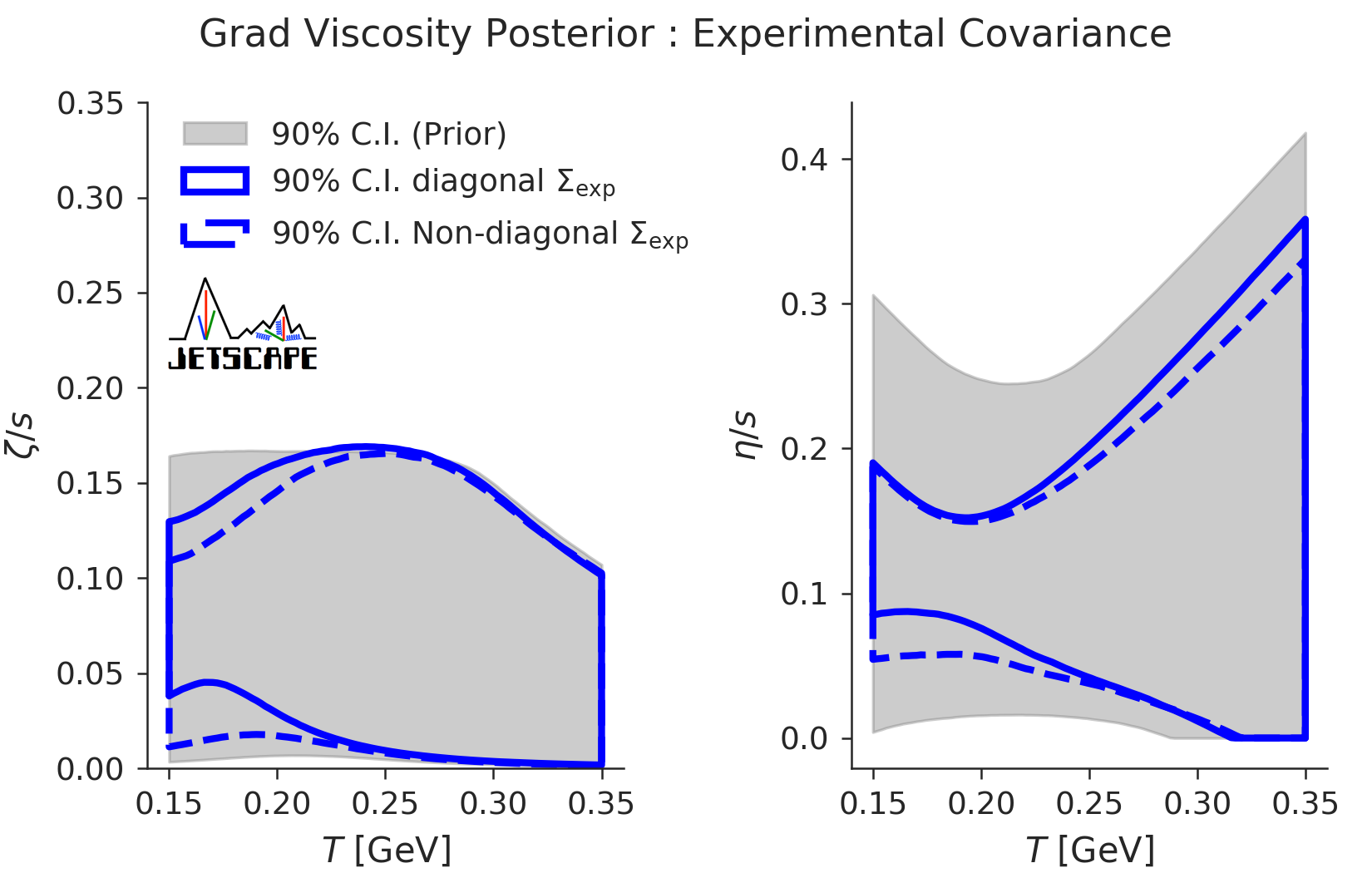}
    \caption{The change in the viscous posterior resulting from assuming a diagonal experimental covariance matrix (solid blue band) or correlated experimental covariance matrix (dashed blue band). Details regarding the magnitude of correlations in text body. }
    \label{fig:posterior_exp_cov}
\end{figure}

Currently, only the diagonal terms in the experimental covariance matrix are reported by the ALICE and STAR experiments. We have assumed a diagonal covariance matrix when performing parameter estimation. However there are undeniably nontrivial correlations in the systematic uncertainties of measured observables and centrality bins. This is important, since systematic uncertainties are generally the dominant source, larger than statistical uncertainties. We test qualitatively how these correlated uncertainties may affect our analysis. The assumed covariance matrix will affect the posterior for all model parameters, but for simplicity we quantify its effect on the posteriors of specific shear and bulk viscosities. This is shown in \fig{fig:posterior_exp_cov} for the Grad viscous correction model. 

We follow the methods used in Ref.~\cite{Bernhard:2018hnz}. In the case of the correlated experimental covariance matrix, given centrality bins $c_i$ and $c_j$ of the same observable, the experimental covariance is assumed to be
\begin{equation}
    \Sigma^{\text{exp}}_{i,j} = \rho \sigma_i \sigma_j
\end{equation}
where
\begin{equation}
    \rho = \exp\left(-(c_i - c_j)^2 / l^2\right)
\end{equation}
and $\sigma_i$ is the standard deviation of the observable in centrality bin $c_i$.
Observables are organized in groups: (i) multiplicities; (ii) mean transverse momenta; (iii) harmonic flows; and (iv) transverse momentum fluctuations. For pairs of different observables within the same `group', we take the same correlation coefficient defined above and multiply by an overall factor of $0.8$. For pairs of different observables in different groups, we assume zero correlation. The ``correlation length'' between centrality bins is assumed $l = 0.5$~\cite{Bernhard:2019bmu}. 

We see that an ansatz for the covariance matrix that includes nonzero correlations has the effect of broadening the viscous posterior, increasing the overall uncertainty. In the absence of a reported experimental covariance matrix, a more Bayesian approach would be to treat the correlation length $l$ and magnitude $\rho$ as uncertain nuisance parameters in the Bayesian parameter estimation, with priors guided by the knowledge and study of the experimental collaborations, and marginalize over them. This is an important extension that we leave for future studies.  

\section{Reducing experimental uncertainty} \label{app_exp_uncertainty}

\begin{figure}[!htb] 
\centering
\includegraphics[trim=0 0 0 16, clip, width=0.7\linewidth]{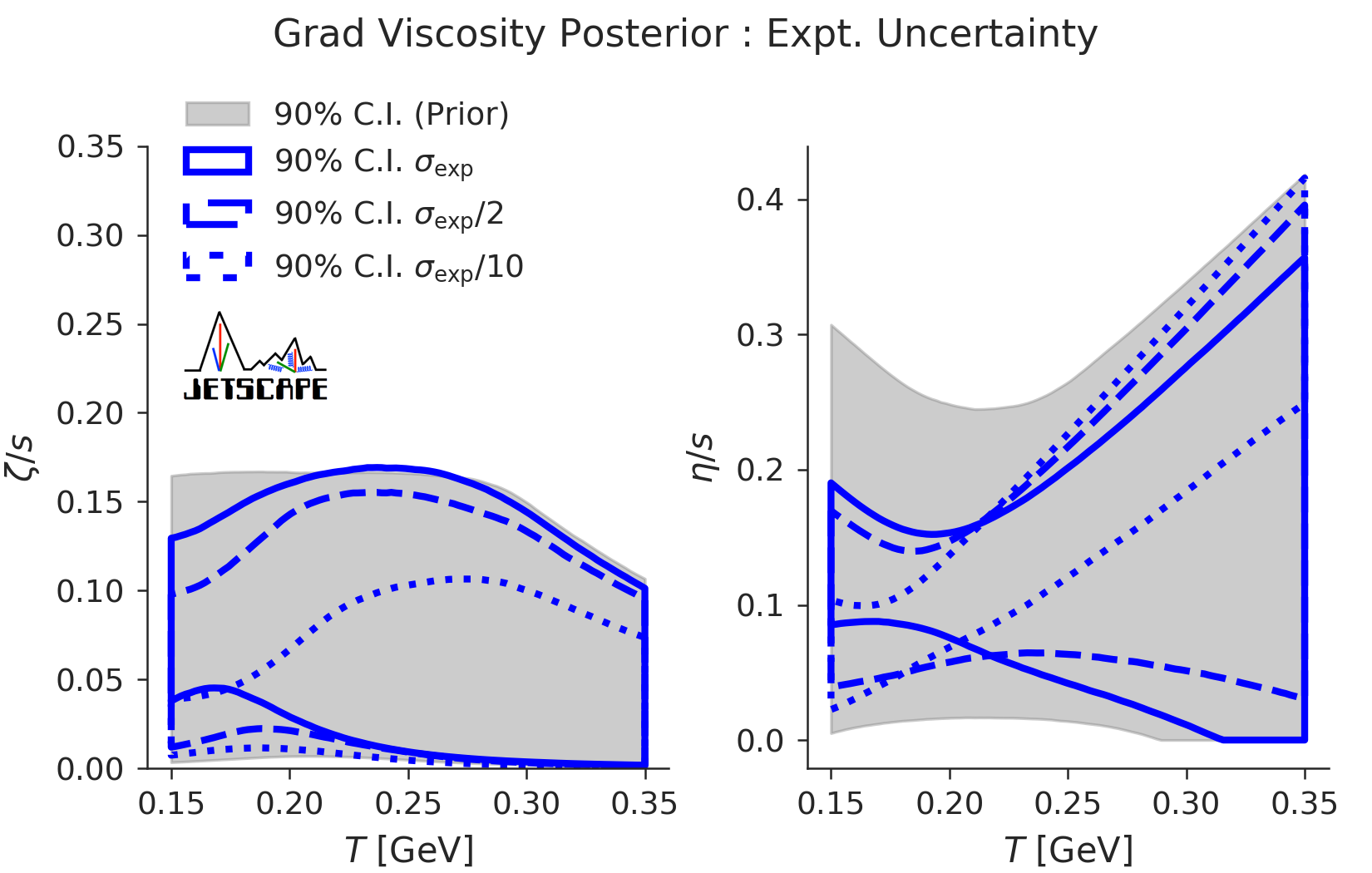}
\caption{The posterior for specific bulk (left) and shear (right) viscosities depending on whether includes the full experimental uncertainties (filled blue) or divides them by a factor of 2 (dashed blue) or 10 (dotted blue). The model emulator always contributes non-zero uncertainty. }
\label{visc_posterior_exp_uncertainty}
\end{figure}

We quantify the extent to which the experimental uncertainty contributes to the total uncertainty in our posterior for the specific shear and bulk viscosities. Besides experimental uncertainty, there is always nonzero uncertainty contributed by the use of a model emulator. We quantify this  by changing artificially the uncertainty on experimentally measured observables during the parameter estimation; the result is shown in \fig{visc_posterior_exp_uncertainty}. 

We see that significantly reducing the experimental error has the potential to qualitatively move our posterior for the specific bulk and shear viscosities. Perhaps more importantly, even if we reduce all of the experimental uncertainty by a factor of two, the credible intervals for the specific bulk and shear viscosities still remain quite large at high temperatures. This hints that in the future we should include additional observables and systems which are more sensitive to the viscosities at high temperatures.

\section{Bulk relaxation time} 
\label{app_bulk_relax}

\begin{figure}[!htb]
  \centering
    \includegraphics[trim=0 0 0 25, clip,width=4cm]{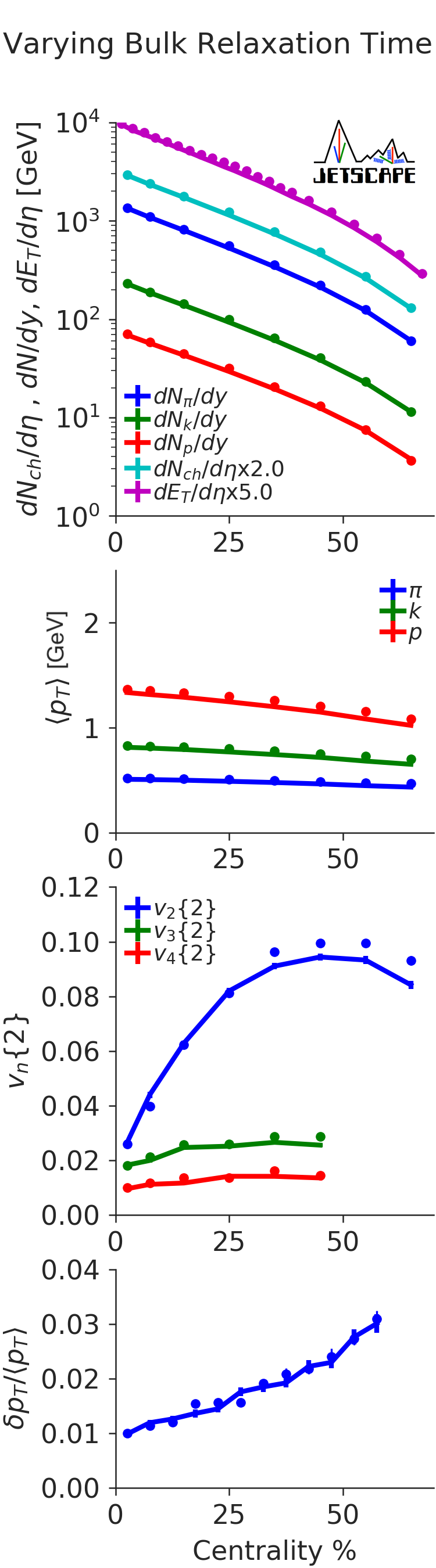}
    \caption{The solid lines are averages over five thousand Pb-Pb $\sqrts{} = 2.76$ TeV events generated with the MAP parameters for the Grad model, and the default bulk relaxation time factor $b_{\Pi} = 1/14.55$. The circles are generated with the same set of parameters except $b_{\Pi} = 2/14.55$.}
    \label{fig:obs_bulk_relax}
\end{figure}

Throughout this study, we have used a parametrization of the specific bulk viscosity given by
\begin{equation}
    \tau_{\Pi} = b_{\Pi} \frac{\zeta}{\left(\frac{1}{3} - c_s^2\right)^2 (\epsilon + p)}
\end{equation}
where $b_{\Pi} = 1/14.55$ \cite{Denicol:2014vaa}. We study how a change in $b_{\Pi}$ translates into a change in our observables using the Maximum A Posteriori parameters for the Grad viscous correction model. This is shown in \fig{fig:obs_bulk_relax}. 

Because our pre-hydrodynamic expansion model is conformal, the resulting bulk pressure at Landau matching is large and positive. The bulk pressure will relax to its Navier--Stokes value $\Pi_{\text{NS}} = -\zeta \theta$ on a time scale given by $\tau_{\Pi}$. Increasing $\tau_{\Pi}$ we find that the bulk pressure stays positive for a longer time. Therefore, comparing the two sets of events, the calculations with the larger bulk relaxation time have larger mean transverse momenta and transverse energy. In future studies, it will be important to study to what extent the bulk relaxation time affects the posterior for the specific bulk and shear viscosities. 

As the bulk relaxation time is further increased, our model also has the feature that the bulk pressure may not have time to relax to its Navier-Stokes value $\Pi_{\rm NS} = -\zeta \theta$ during the lifetime of the hydrodynamic phase. In this case, the evolution of the bulk pressure becomes less sensitive to the value of the specific bulk viscosity and is dominated by its initial conditions. Inferring the likely values of the specific bulk viscosity then becomes more challenging. 

\section{Comparison to previous studies}
\label{appendix:comparison_nature_physics}

In this section we enumerate the largest differences between the parameter estimation presented in this analysis and the analysis found in Ref.~\cite{Bernhard:2019bmu}.

\subsection{Physics models}

\paragraph{Pre-hydrodynamic free-streaming:}
Both Ref.~\cite{Bernhard:2019bmu} and this thesis used free-streaming as a pre-hydrodynamic expansion model. Different numerical implementations were used, but they were validated against each other and found to be in excellent numerical agreement (see App.~\ref{app:fs_valid}). However, in this work we have allowed the free-streaming time to be dependent on the energy of each collision. This additional feature is manifest in the parameter $\alpha$; when $\alpha$ is fixed to zero, both studies have the same physics for the pre-hydrodynamic free-streaming. 

\paragraph{Hydrodynamics: equation of state and viscosities:}

The largest differences in the hydrodynamic models include the equation of state and the parametrization of specific shear and bulk viscosities. 
The equation of state used in Ref.~\cite{Bernhard:2019bmu} was given by the HotQCD lattice result at high temperatures matched to the 2017 PDG table of hadronic resonances at low temperatures. In particular, this included a very light $\sigma$ meson with a mass of about $500$ MeV. Our study has matched the same HotQCD lattice equation of state at high temperatures to a table of hadronic resonances entering in the \SMASH{} afterburner.
In particular, we excluded the $\sigma$ meson entirely in the construction of the equation of state.
Besides the list of resonances which compose the hadron resonance gas component, Ref.~\cite{Bernhard:2019bmu} also computed the hadronic equation of state assuming relativistic Breit-Wigner resonances with nonzero width, while this study assumed all resonances on mass-shell in constructing the equation of state. 

In the parametrization of the specific shear viscosity, Ref.~\cite{Bernhard:2019bmu} included a curvature parameter for the specific shear viscosity at high temperatures, which was not included in this work. On the other hand, we varied the slope of the low-temperature specific shear viscosity, as well as the position of the ``kink'' in this work, while both of these were fixed in Bernhard's study. For the specific bulk viscosity, we allowed the parametrization to have a nonzero skewness, which was not present in Bernhard's study. 

\paragraph{Particlization, resonance width and $\sigma$ resonance:}

Ref.~\cite{Bernhard:2019bmu} fixed the particlization model to be what we have referred to as the Pratt-Torrieri-Bernhard viscous correction model (see Ch.~\ref{ch3:particlization}), while in this work we also investigated other models. 
For all viscous correction models in this work, the particles were sampled on their mass-shell, while particlization in Ref.~\cite{Bernhard:2019bmu}  sampled the particles mass from a relativistic Breit-Wigner function. 
In addition, as already mentioned, Ref.~\cite{Bernhard:2019bmu} sampled unstable $\sigma$ resonances with a mass of about $500$ MeV, which significantly increased the number of pions at low momenta once they decayed. This study excluded the $\sigma$ resonance from sampling during particlization.

\paragraph{Hadronic afterburner:}

Finally, the hadronic afterburner used in Ref.~\cite{Bernhard:2019bmu} was \URQMD{}, while we use \SMASH{}. Although these two models include somewhat different lists of resonances as well as slightly different hadronic cross-sections, we checked in \Appendix{app:smash} that \URQMD{} and \SMASH{} have excellent agreement when used with the model parameters that agree well with data. For that reason, we believe at this time that the difference in hadronic afterburners is negligible in comparison to the other differences listed above. 

\subsection{Prior distributions}

The prior used in Ref.~\cite{Bernhard:2019bmu} is nearly a subspace of the prior used in this study, with the exception of the high-temperature behavior of the specific shear viscosity. Ref.~\cite{Bernhard:2019bmu} allowed the specific shear viscosity to have a nonzero curvature, i.e. quadratic temperature dependence at high temperatures. In this study, we have not allowed such a quadratic temperature dependence in the specific shear viscosity at high temperature. 

\subsection{Experimental data}

Both Ref.~\cite{Bernhard:2019bmu} and this study have included the ALICE $p_T$-integrated, centrality-dependent data for Pb-Pb collisions at $\sqrts{} = 2.76$ TeV. However, Ref.~\cite{Bernhard:2019bmu} additionally included data for Pb-Pb collisions at $\sqrts{} = 5.02$ TeV, which are not included in this work. Instead, we have included STAR data for Au-Au collisions at $\sqrts{} = 200$ GeV, as well as ALICE data for Xe-Xe $\sqrts{} = 5.44$ TeV which were not included in Ref.~\cite{Bernhard:2019bmu}.

\section{Multistage model validation}
\label{appendix:validation}

Several of the numerical implementations of models used in this work are used for the first time; other required modifications and expansions.
For this reason we include validations of these codes against counterparts which have been used extensively in previous studies.

\subsection{Validation of second-order viscous hydrodynamics implementation}\label{sec:app_hydro_val}

In this section, we compare two different numerical implementations of the same underlying second-order relativistic hydrodynamics equations~\cite{Denicol:2012es}. The first implementation is the one used throughout this work is \music{}~\cite{Schenke:2010nt,Schenke:2010rr,Paquet:2015lta}. The second implementation is a slightly modified version of the \vishnew{} 2+1D hydrodynamics code~\cite{Song:2009gc, Shen:2014vra}, \texttt{osu-hydro}~\cite{osuhydro}, used in previous studies~\cite{Bernhard:2016tnd, Bernhard:2018hnz, Moreland:2019szz}. Both \music{} and \vishnew{} solve the same hydrodynamic equations of motion~\cite{Denicol:2012es} but with two different numerical schemes: \vishnew{} uses SHASTA~\cite{Boris:1973tjt} while \music{} uses the Kurganov-Tadmor algorithm \cite{2000JCoPh.160..241K}. Despite differences in the numerical algorithms --- amounting to approximations of spatial derivatives --- for sufficiently smooth hydrodynamic fields the two codes should agree well.

\begin{figure*}[!htb]
\noindent\makebox[\textwidth]{%
  \centering
  \begin{minipage}{0.5\textwidth}
    \includegraphics[width=\textwidth]{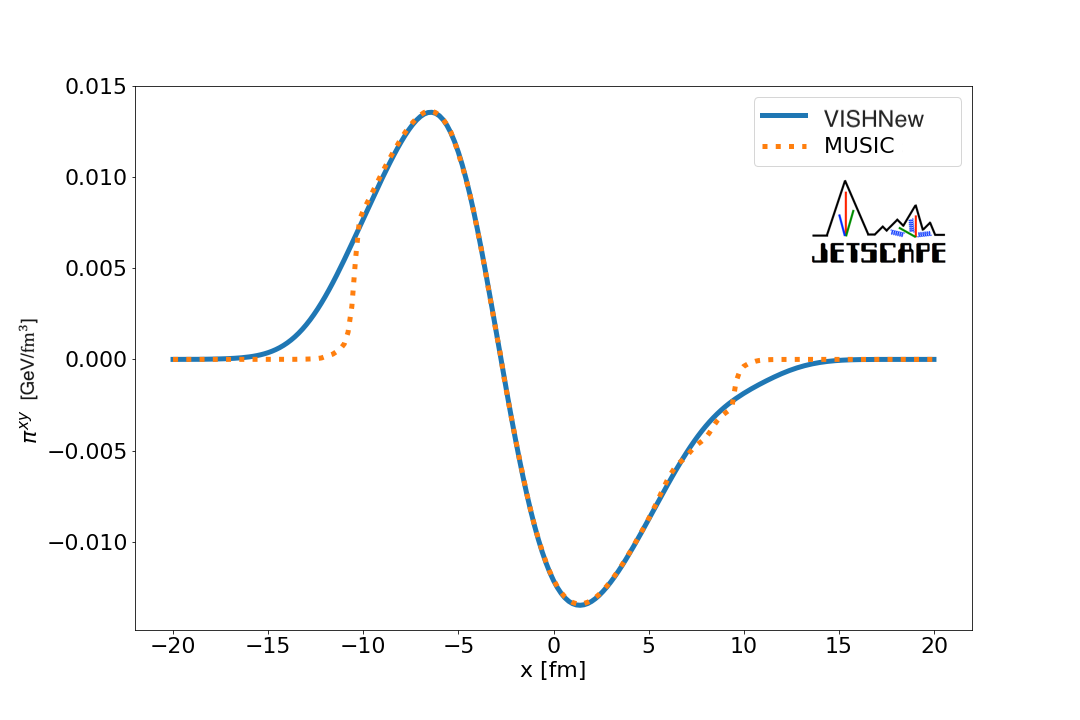}
  \end{minipage}
  \begin{minipage}{0.5\textwidth}
    \includegraphics[trim=0 0 0 16, clip, width=\textwidth]{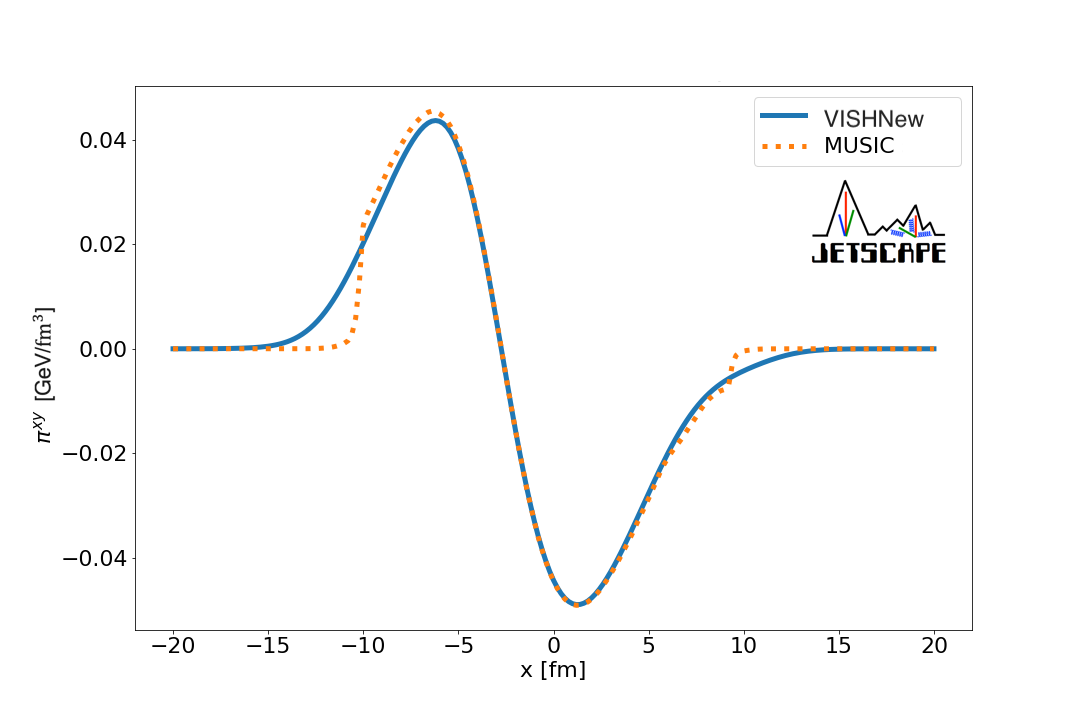}
  \end{minipage}
  }
  \caption{The results of the hydrodynamic evolution of the shear stress for a smooth initial condition, just before freeze-out, for $\eta/s = 0.08$ (left) and $\eta/s = 0.3$ (right). The \music{} regulation scheme allows larger inverse Reynolds numbers inside of the switching surface than the \vishnew{} scheme. }
  \label{shear_reg}
\end{figure*}

Besides the different numerical schemes, \vishnew{} and \music{} have different viscous current regulation schemes. The regulation scheme used in \vishnew{} is described in Ref.~\cite{Shen:2014vra} while that used in \music{} can be found in Ref.~\cite{Denicol:2018wdp, Schenke:2020mbo}. For small to moderate values of $\eta/s$ and $\zeta/s$, neither of these schemes should regulate the viscous currents close to or inside the constant energy density (or temperature) switching hypersurface. Because our hydrodynamic model will explore moderate and large values of $\eta/s$ and $\zeta/s$, it is important to compare the hydrodynamic fields. For a fixed $\eta/s = 0.08$, we run the same smooth initial conditions used for the ideal hydrodynamic comparison through free-streaming and either \music{} or \vishnew{} with zero bulk viscosity and the conformal equation of state $\epsilon = 3p$. These are shown in \fig{shear_reg}.

At late times, there are differences in the shear stress $\pi^{xy}$ near the dilute regions of the grid. These differences do not propagate into the region inside the particlization surface ($\epsilon \gtrsim 0.2$ GeV/fm$^3$).
We have also run the exact same event through viscous hydro with a fixed $\eta/s = 0.3$. The larger specific shear viscosity will incur stronger regulation.  We find that the \music{} scheme, while aggressive in low temperature regions, allows larger values of shear pressure inside the region $\epsilon > 0.2 $ GeV/fm$^3$.

\begin{figure*}[!htb]
\noindent\makebox[\textwidth]{%
  \centering
  \begin{minipage}{0.5\textwidth}
    \includegraphics[width=\textwidth]{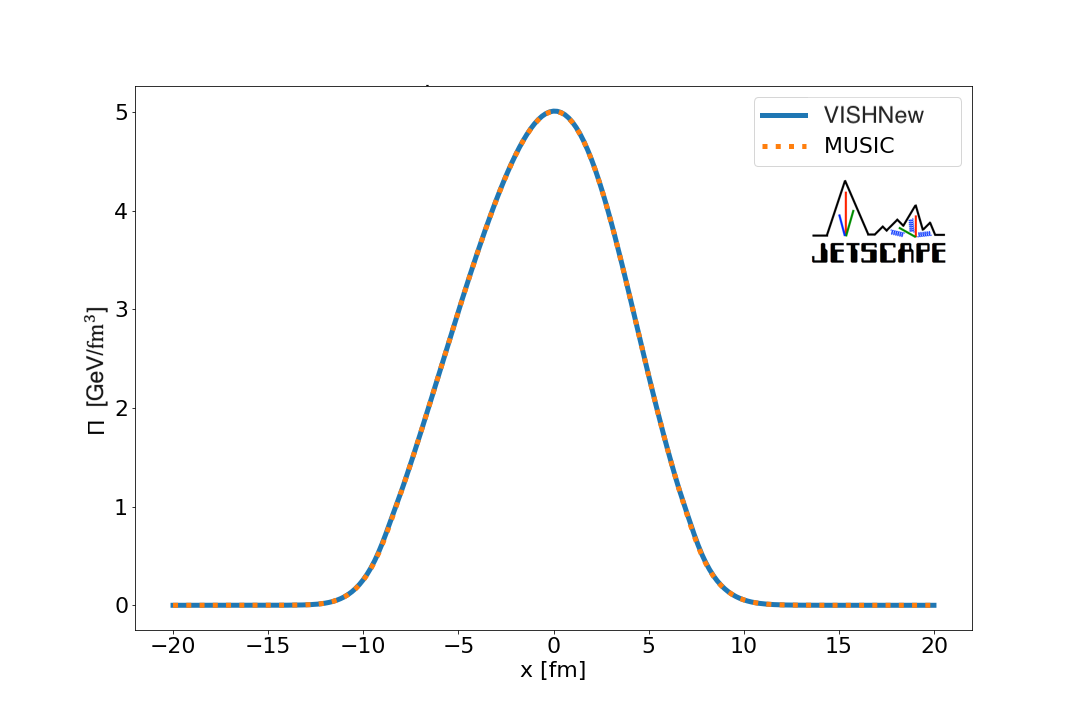}
  \end{minipage}
  \begin{minipage}{0.5\textwidth}
    \includegraphics[width=\textwidth]{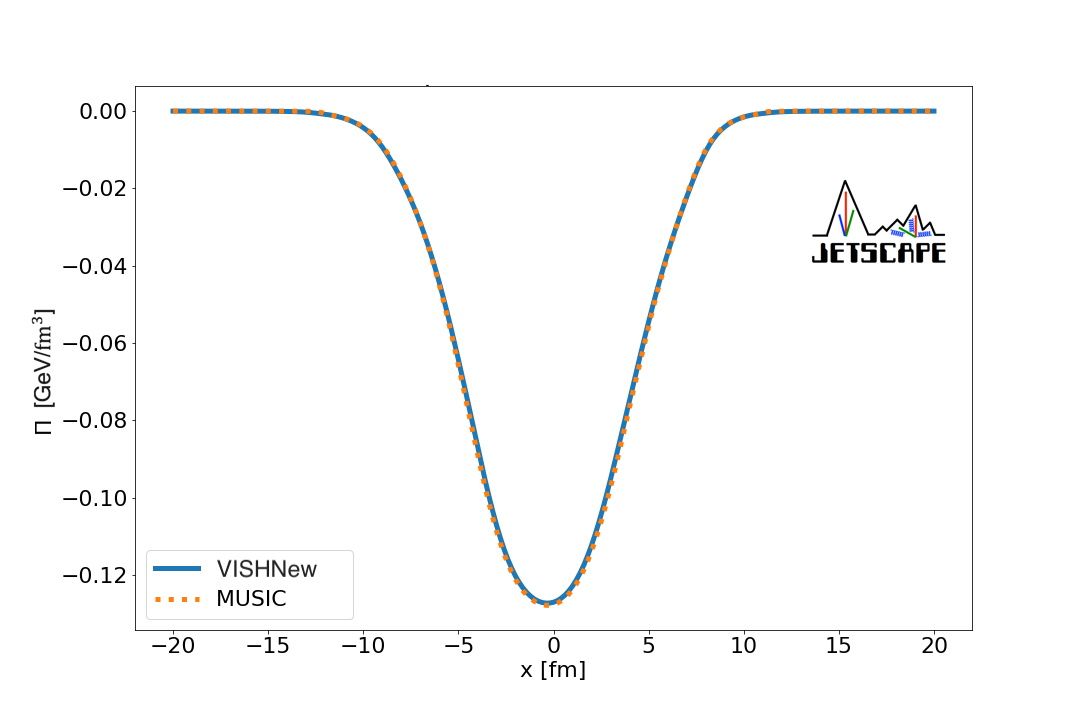}
  \end{minipage}
  }
  \caption{The initial bulk pressure (left) and bulk pressure just before freeze-out (right), resulting from hydrodynamic evolution of a smooth initial condition. The specific shear viscosity was fixed $\eta/s = 0.08$, and specific bulk viscosity $(\zeta/s)(T)$ was given by~\cite{Bernhard:2018hnz} for this test.}
  \label{bulk_final_Pi}
\end{figure*}

\begin{figure*}[!htb]
\noindent\makebox[\textwidth]{%
  \centering
  \begin{minipage}{0.5\textwidth}
    \includegraphics[width=\textwidth]{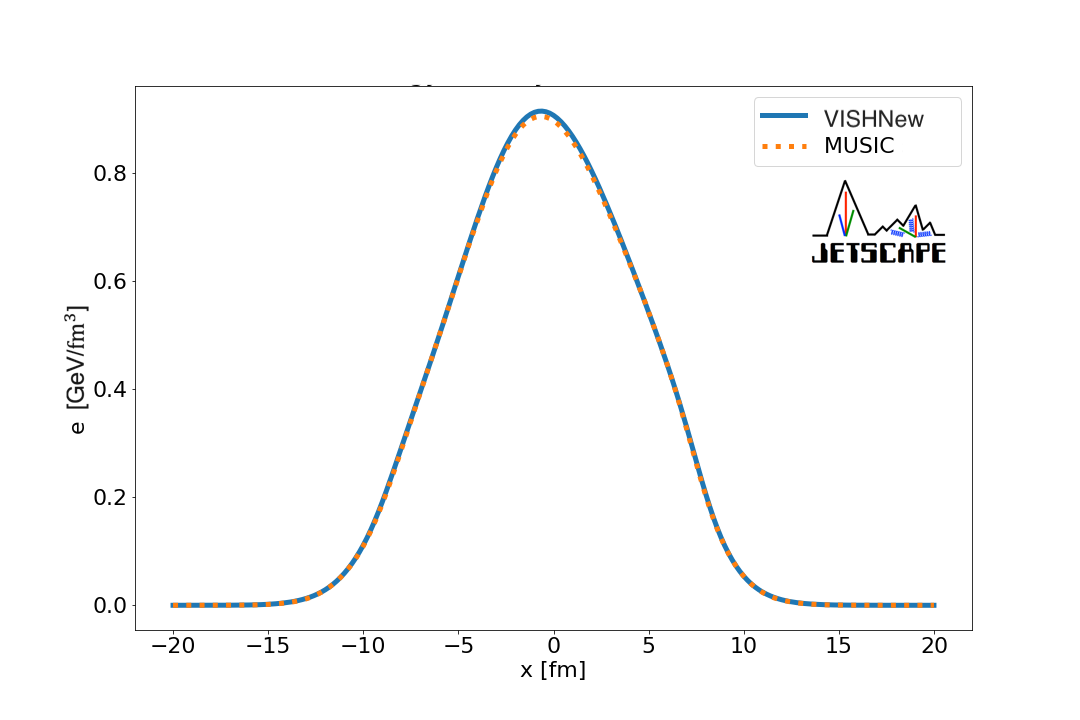}
  \end{minipage}
  \begin{minipage}{0.5\textwidth}
    \includegraphics[width=\textwidth]{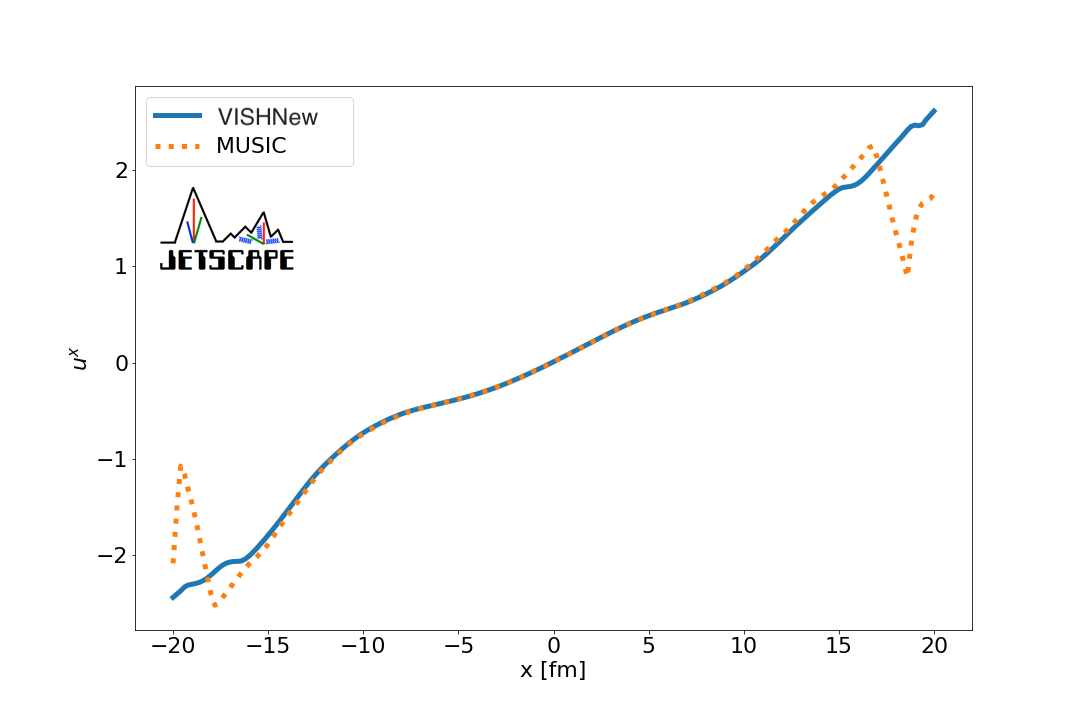}
  \end{minipage}
  }
  \caption{The energy density (left) and flow (right) after hydrodynamic evolution of a smooth initial condition. The specific shear viscosity was fixed $\eta/s = 0.08$, and specific bulk viscosity $(\zeta/s)(T)$ was given by~\cite{Bernhard:2018hnz} for this test.}
  \label{bulk_final_u}
\end{figure*}

As additional validation, we repeated the previous test with a QCD equation of state, again with a fixed specific shear viscosity $\eta/s = 0.08$ but this time with a temperature dependent specific bulk viscosity $(\zeta/s)(T)$ from Ref.~\cite{Bernhard:2018hnz}.
The bulk pressure, energy density and flow are shown in \figs{bulk_final_Pi} and \ref{bulk_final_u}. Good agreement is found between the two codes.

\begin{figure*}[!htb]
\noindent\makebox[\textwidth]{%
  \centering
  \begin{minipage}{0.5\textwidth}
    \includegraphics[width=8cm]{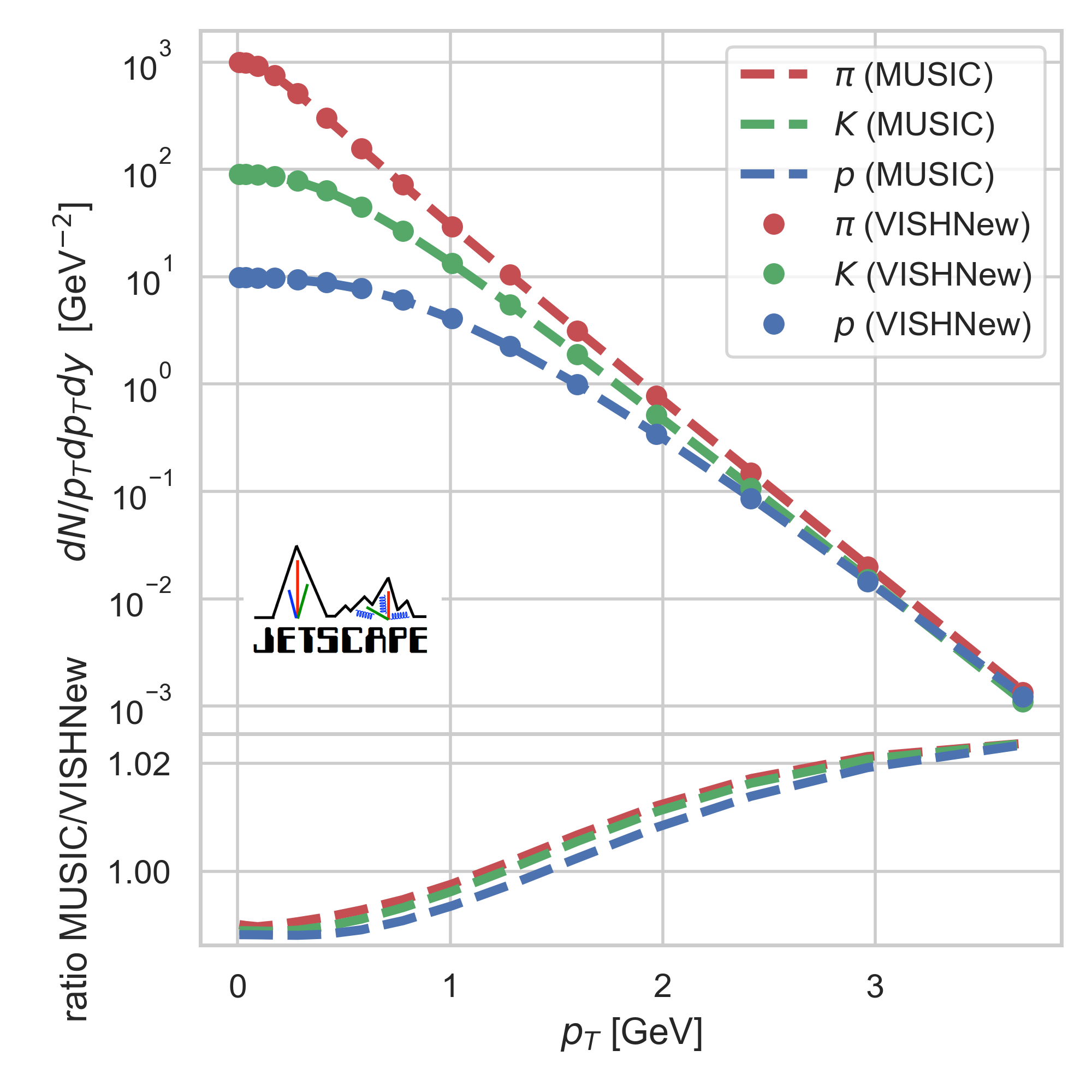}
  \end{minipage}
  \begin{minipage}{0.5\textwidth}
    \includegraphics[width=8cm]{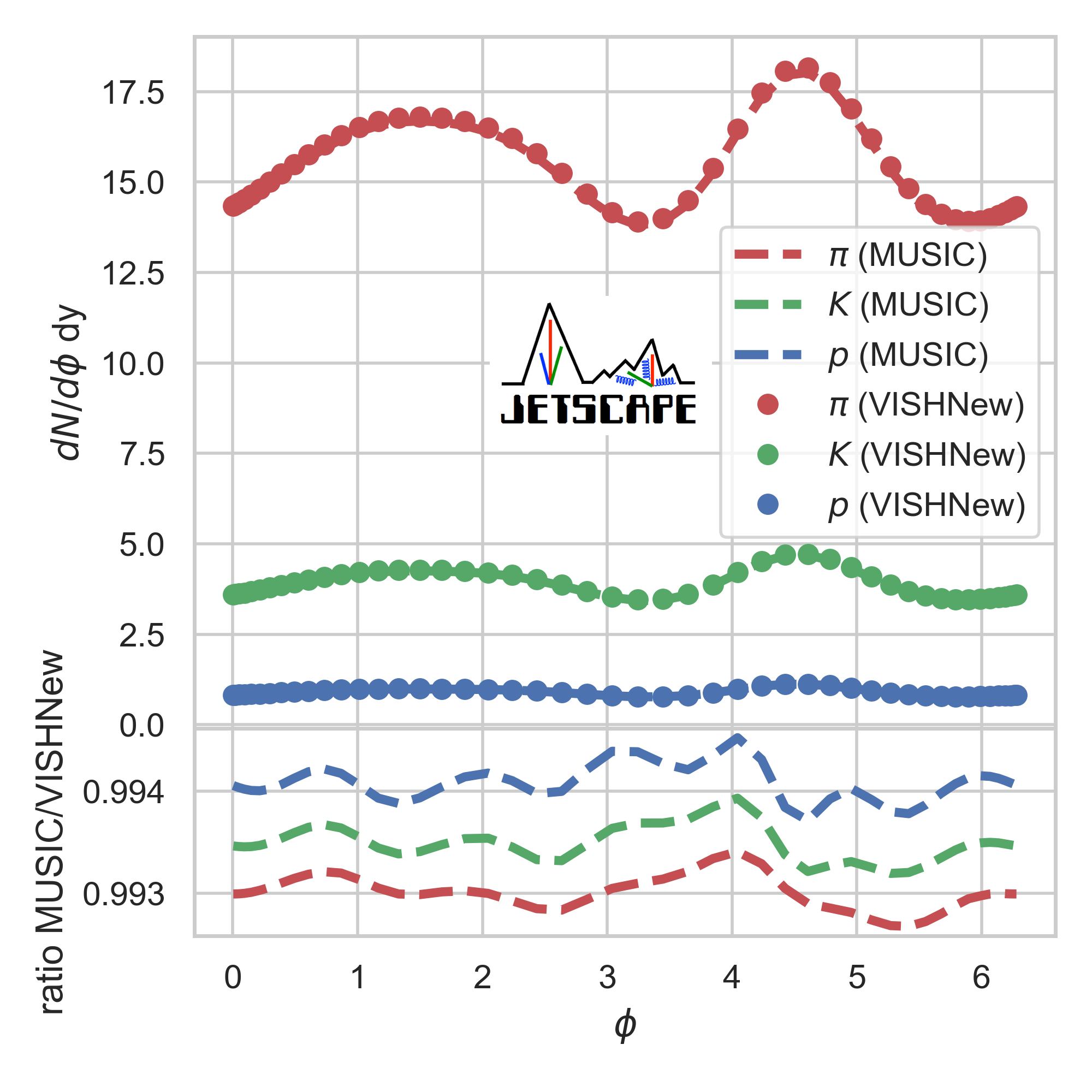}
  \end{minipage}
  }
  \caption{Comparison of the transverse momentum $p_T$ spectra (left) and azimuthal $\phi_p$ spectra (right) generated from the \music{} and \vishnew{} freezeout surfaces. The freezeout surface was generated using the events compared above, with fixed $\eta/s = 0.08$, and specific bulk viscosity $(\zeta/s)(T)$ was given by~\cite{Bernhard:2018hnz}. }
\label{spectra_ratio_pT_phi}
\end{figure*}

In order to quantify the effects of any small differences that the hydrodynamics may have on our hadronic observables, we have evaluated the smooth Cooper-Frye integral over the switching surface generated by each hydrodynamics code. The hydrodynamic event used was the same event with bulk and shear pressures for which the hydrodynamic evolution was compared above. We used \texttt{iS3D} to perform the smooth Cooper-Frye integral over each surface, including bulk and shear Grad viscous corrections, and plotted the comparisons below for pions, kaons and protons. In general, the agreement in the spectra is very good. These are shown in 
\fig{spectra_ratio_pT_phi}. These differences of about 1\% or less in the differential observables yield differences $\lesssim 1\%$ in the $p_T$ and $\phi_p$ integrated observables. 

\subsection{SMASH}\label{app:smash}

The use of \SMASH{} as an afterburner for event-by-event studies of heavy-ion collisions is still fairly new. For this reason, we have made a comparison between \URQMD{} and \SMASH{} with respect to the predicted transverse-momentum-integrated observables. We generated five thousand fluctuating initial conditions for Au-Au $\sqrts{}=0.2$\,TeV collisions with parameters fixed by the Maximum A Posteriori parameters found in \cite{Bernhard:2018hnz} except for the initial energy density normalization, which was scaled to fit the multiplicities. 

We allowed each initial condition to free-stream for the same time and then used these initial conditions for hydrodynamics in two different models:
\begin{enumerate}
    \item \SMASH{} model: We matched the HotQCD lattice equation of state to the \SMASH{} list of resonances (excluding the $\sigma$ meson). Each initial condition was propagated through viscous hydrodynamics with this equation of state, followed by particlization using the Pratt-Torrieri-Bernhard viscous correction ansatz, followed by dynamics in \SMASH{}. 
    \item \URQMD{} model: We matched the HotQCD lattice equation of state to the list of resonances which can be propagated in \URQMD{}. Each initial condition was propagated through viscous hydrodynamics with this equation of state, followed by particlization using the Pratt-Torrieri-Bernhard viscous correction ansatz, followed by dynamics in \URQMD{}. 
\end{enumerate}

\begin{figure*}[!htb]
  \centering
    \includegraphics[width=\linewidth]{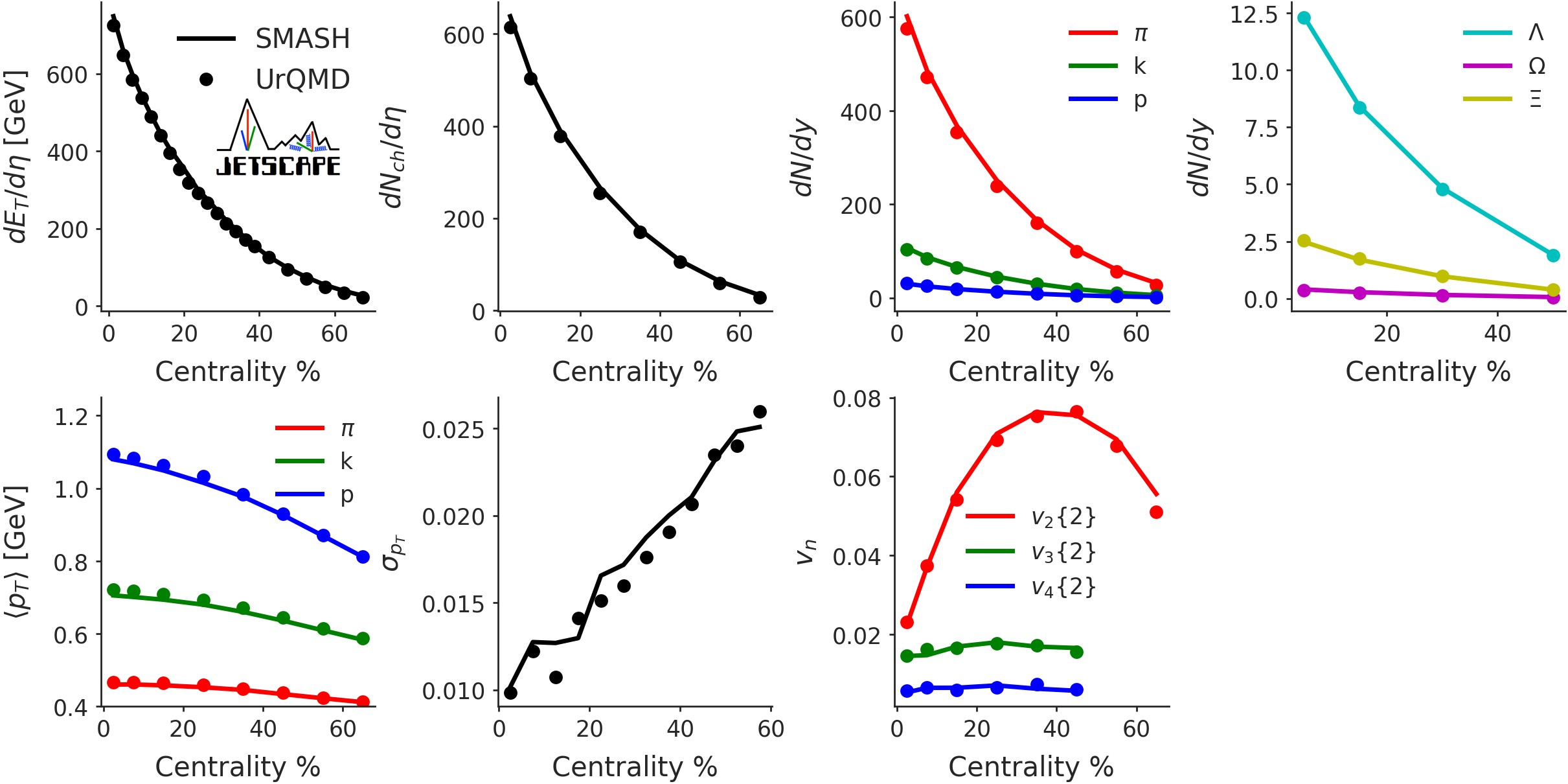}
    \caption{
    Comparison of soft hadronic observables for Au-Au $\sqrts{}=0.2$\,TeV collisions using the \SMASH{} (solid lines) or \URQMD{} (dots) afterburner.}
    \label{compare_urqmd_smash}
\end{figure*}

We compared the observables predicted by the two models, shown in \fig{compare_urqmd_smash}. 
For the observables we considered, we found very good agreement. In particular, heavier resonances have spectra that are more strongly influenced by the hadronic afterburner than lighter resonances and the agreement in the multiplicity and transverse momenta of the proton and $\Lambda$ is strong. The same level of agreement between the two models was found for the same comparison made for Pb-Pb $\sqrts{}=2.76$ TeV collisions, which are not shown. 

\begin{figure*}[!htb]
  \centering
    \includegraphics[width=0.5\linewidth]{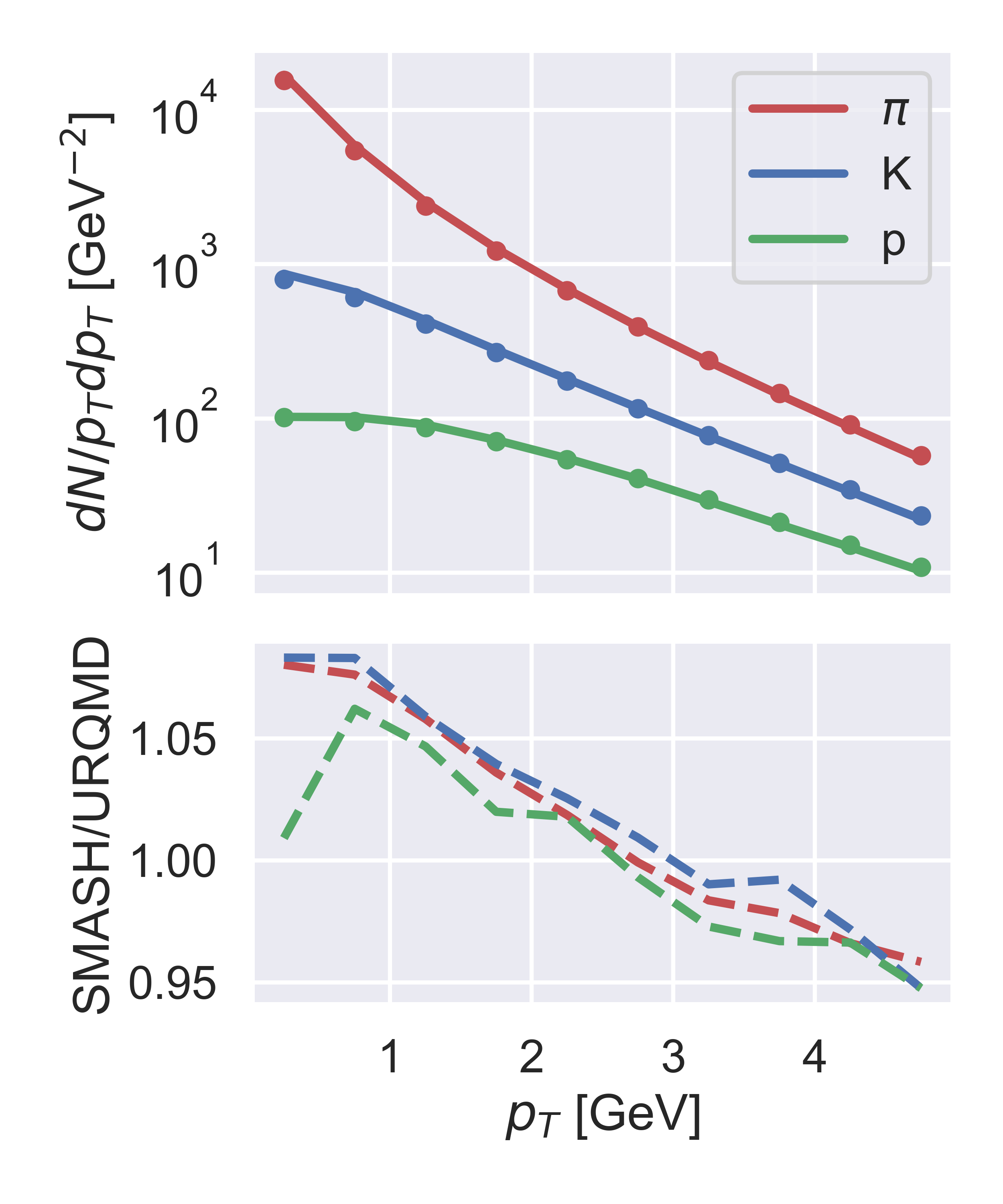}
    \caption{The top row displays a comparison of \SMASH{} afterburner (solid lines) and \URQMD{} afterburner (dots) predicted invariant $p_T$-differential spectra of pions, kaons and protons produced in $20-30\%$ centrality Pb-Pb $\sqrts{}=2.76$\,TeV collisions. The bottom row is the ratio between the two.}
    \label{compare_urqmd_smash_diff_pT}
\end{figure*}

Additionally, we checked the agreement in the differential spectra of pions, kaons and protons produced by both models described above. Roughly five-hundred Pb-Pb $\sqrts{}=2.76$ TeV events in the $20-30\%$ centrality class were generated and averaged. These comparisons are shown in Fig.~\ref{compare_urqmd_smash_diff_pT}. We see that there is very good agreement, to the level of $5\%$ for hadrons with momenta $p_T < 5$ GeV, and that the spectra of particles produced via the \SMASH{} afterburner tend to be only slightly softer.

We also ran hydrodynamics with a fixed equation of state matched to the \SMASH{} hadron resonance gas particle content and then switched at the same temperature to \URQMD{} or \SMASH{}. Because of the mismatch between the equation of state generated with the \SMASH{}  and the \URQMD{} resonance gases, there is a discrepancy at particlization in all of the thermodynamic variables. For example, at the same temperature, the energy density of the \SMASH{} resonance gas and \URQMD{}'s are different. This leads to a disagreement in observables. In particular,  observables sensitive to the normalization of energy density, such as multiplicities and the transverse energy, showed a discrepancy at the level of approximately five percent. It is easy to understand that the energy density of the \URQMD{} resonance gas is a few percent smaller than \SMASH{}'s at the same temperature because of the different species and masses of hadrons. More details can be found in \Appendix{appendix:eos}. 

Given the novelty of using \SMASH{} as an afterburner, we share for completeness the numerical parameters that we used with \SMASH{}. These parameters, shown in \Table{smash_params}, gave sufficient accuracy without unreasonable loss of speed.
\begin{table}[!htb]
\centering
\begin{tabular}{|l|l|}
\hline
Modus            & Afterburner \\ \hline
Time\_Step\_Mode & Fixed       \\ \hline
Delta\_Time      & 1.0         \\ \hline
End\_Time        & 1000.0      \\ \hline
\end{tabular}
\caption{\SMASH{} parameters used event-by-event throughout this study.}
\label{smash_params}
\end{table}

\subsection{Comparison of JETSCAPE with hic-eventgen}

\begin{figure*}[!htb]
  \centering
    \includegraphics[width=\linewidth]{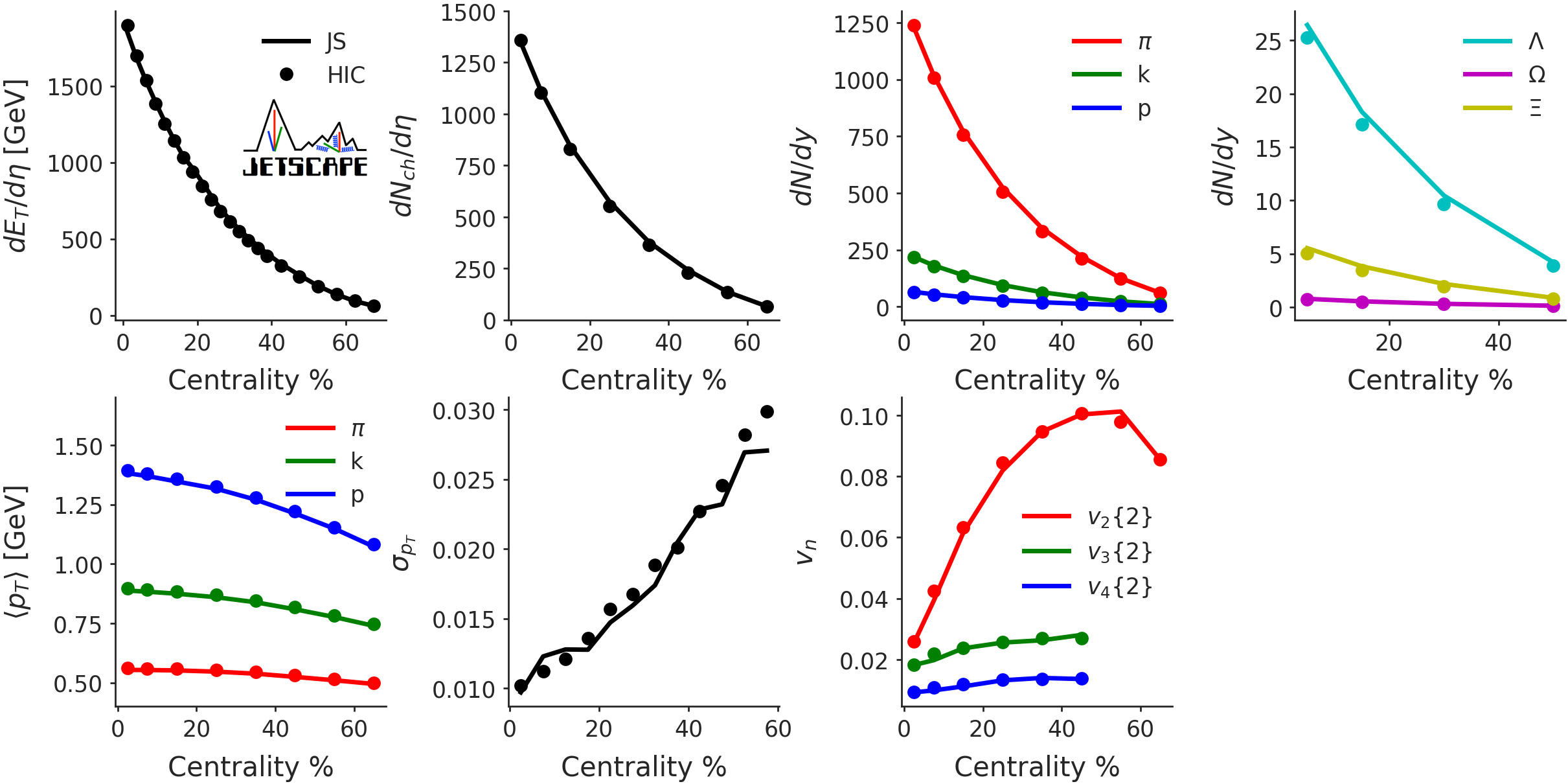}
    \caption{
    Comparison of soft hadronic observables depending on whether one uses the \texttt{JETSCAPE} event generator (solid lines) or \texttt{hic-eventgen} (dots), averaged over five-thousand Pb-Pb $\sqrts{}=2.76$ TeV events.}
    \label{compare_js_hiceventgen}
\end{figure*}

In addition to validating of all the separate model components, we also have checked that the centrality-averaged observables predicted by our \texttt{JETSCAPE} model agree very well with a version of \texttt{hic-eventgen}, the event generator used in Ref.~\cite{Bernhard:2018hnz}. This was performed by restricting our parametrizations to be the same as the Maximum A Posteriori parameters found in that study. The results of this comparison are shown in \fig{compare_js_hiceventgen}, in which we have averaged over five thousand fluctuating Pb-Pb $\sqrts{}=2.76$ TeV collision events.

In general, we find excellent agreement between the two hybrid models. For this level of agreement, the $\sigma$ meson had to be excluded from the \texttt{hic-eventgen} model; all resonances were also sampled on their mass-shell in \texttt{frzout}~\cite{frzout}, the particle sampler in \texttt{hic-eventgen}. The equation of state used during the hydrodynamic evolution was constructed to match the hadron resonance gas used in \texttt{frzout} (excluding the $\sigma$ meson). 

\subsection{The $\sigma$ meson}
\label{app:sigma_effect}

The effects of including a $\sigma$ meson resonance in our hadron resonance gas are studied using the \texttt{frzout} module~\cite{frzout}, which is designed with the option to sample the $\sigma$ resonance as a thermal resonance and perform its decay to pions. In particular, we compare three scenarios:
\begin{itemize}
    \item Excluding the $\sigma$ meson from sampling (labeled by $m\rightarrow \infty$).
    \item Sampling the $\sigma$ meson with the PDG pole mass ($\sim 500$ MeV) \cite{Tanabashi:2018oca}.
    \item Sampling the $\sigma$ meson with the mass used in \SMASH{} ($\sim 800$ MeV).
\end{itemize}

The \texttt{frzout} module was used to sample particles from a hypersurface generated by the \music{} simulation of a mid-central Pb-Pb event. The initial condition, free-streaming, and hydrodynamic transport parameters were set by the Maximum A Posteriori parameters given in \cite{Bernhard:2018hnz}. The switching temperature was 151 MeV. Sampled particles are then propagated to \URQMD{} to perform hadronic rescatterings. Note that \URQMD{} does not have a $\sigma$ meson: the effect of the  $\sigma$ meson is purely being tested at the level of the particlization, not in the afterburner. A total number of $100$ over-samples were generated to increase the statistics. The results on charged-particle multiplicity, transverse energy, and pion multiplicity and mean transverse momentum are shown in \Table{sigma_sensitivity}.

\begin{table}[!htb]
\centering
\begin{tabular}{||c|c|c|c|c||}
\hline
$\sigma$       & $dN_\text{ch}/d\eta$ & $dE_T/d\eta$ {[}GeV{]} & $dN_{\pi}/dy$ & $\langle p_T^{\pi}\rangle$ {[}GeV{]} \\ 
\hline
$m = 475$ MeV & 615       & 777                  & 569        & 0.54                                       \\ 
\hline
$m = 800$ MeV & 583       & 754                  & 534        & 0.55                                       \\ 
\hline
$m\rightarrow \infty$        & 579       & 743                  & 531        & 0.54                                       \\
\hline
\end{tabular}
\caption{Sensitivity of observables to inclusion of the $\sigma$ meson. The observables were computed using the same cuts as the ALICE experiment.}
\label{sigma_sensitivity}
\end{table}

Because the $\sigma$ meson decays into pions, we see that the pion yield can differ by 7\% for the lightest $\sigma$ resonance. For the higher mass $\sigma$, the results are close to not sampling a $\sigma$ meson. Additional differences would manifest if we included the effect of varying the $\sigma$ mass in constructing the hadron gas equation of state as this would also have an effect on the hydrodynamic evolution. As has been explained in the main text, we chose to omit the $\sigma$ from the equation of state and particlization, following \cite{Broniowski:2015oha}.

\subsection{QCD equations of state with different hadron resonance gases}
\label{appendix:eos}

The QCD equation of state used in hydrodynamic simulations of heavy ion collision matches a lattice calculation at high temperature ($T \gtrsim 120$~MeV) with a hadron resonance gas calculation at low temperature. In this work, we use the lattice calculations from Ref.~\cite{Bazavov:2014pvz}. As explained in Ref.~\cite{Bazavov:2014pvz}, the trace anomaly calculated from the lattice is used to compute the pressure by integration of
\begin{equation}
    \frac{p(T)}{T^4}=\frac{p_0(T)}{T_0^4}+\int_{T_0}^{T} \frac{d T^\prime}{T^\prime} \frac{\Theta^{\mu\mu}(T^\prime)}{T^{\prime 4}}.
\end{equation}
Energy density and entropy density then follow. The integration constant for the pressure is obtained from a hadron resonance gas calculation at $T=130$~MeV.

Reference~\cite{Bernhard:2018hnz} followed a related but modified approach. To ensure energy-momentum conservation at particlization, the lattice QCD trace anomaly is matched to a hadron resonance gas in a temperature range $[T_a,T_b]$. The trace anomaly below $T_a$ is calculated according to the hadron resonance gas. Above $T_b$, the trace anomaly is that of the lattice QCD. The trace anomaly between $T_a$ and $T_b$ is an interpolation between the resonance gas and the lattice QCD trace anomaly.
Using this new trace anomaly, which differs from that of the lattice below $T_b$, the pressure is computed by integration using $p_0(T_0=50 \textrm{MeV})$ as reference; energy density and entropy density are then calculated.

We illustrate first the differences between the lattice pressure, and the pressure obtained with the above matching. If the temperature is below the matching point $T_a$, the pressure from the lattice case is given by
\begin{equation}
    \frac{p_{L}(T)}{T^4}=\frac{p_{L,0}}{T_0^4}+\int_{T_0}^{T} \frac{d T^\prime}{T^\prime} \frac{\Theta_L^{\mu\mu}(T^\prime)}{T^{\prime 4}}.
\end{equation}
where the integration constant $p_{L,0}$ is the only input from the hadron resonance gas that enters in the definition of the pressure.

In the ``matched'' equation of state, however, the entire thermodynamics is determined by the hadron resonance gas below the lower matching temperature $T_a$:
\begin{equation}
    \frac{p_{M}(T)}{T^4}=\frac{p_{M,0}}{T_0^4}+\int_{T_0}^{T} \frac{d T^\prime}{T^\prime} \frac{\Theta_{HRG}^{\mu\mu}(T^\prime)}{T^{\prime 4}}.
    \label{eq:pressure_matched}
\end{equation}
There is no information from the lattice calculations entering in \eq{eq:pressure_matched} if $T<T_a$.
This example makes is clear that any mismatch between the trace anomaly of lattice calculations and that of the hadron resonance gas results in a difference in the equation of state. This is of course the case even if the exact same hadron resonance gas are used to fix $p_{L,0}$ --- $p_{L,0} = p_{M,0}$ --- which is arguably never the case. These uncertainties are difficult to eliminate: any mismatch between the hadron resonance gas and the lattice calculation would result in a discontinuity at particlization. Moreover, there is uncertainty in how one should interpolate the trace anomaly between $T_a$ and $T_b$; this source of uncertainty has been neglected here, but is studied in Ref.~\cite{Auvinen:2020mpc}.

\begin{figure}[!htb]
  \centering
    \includegraphics[width=0.7\linewidth]{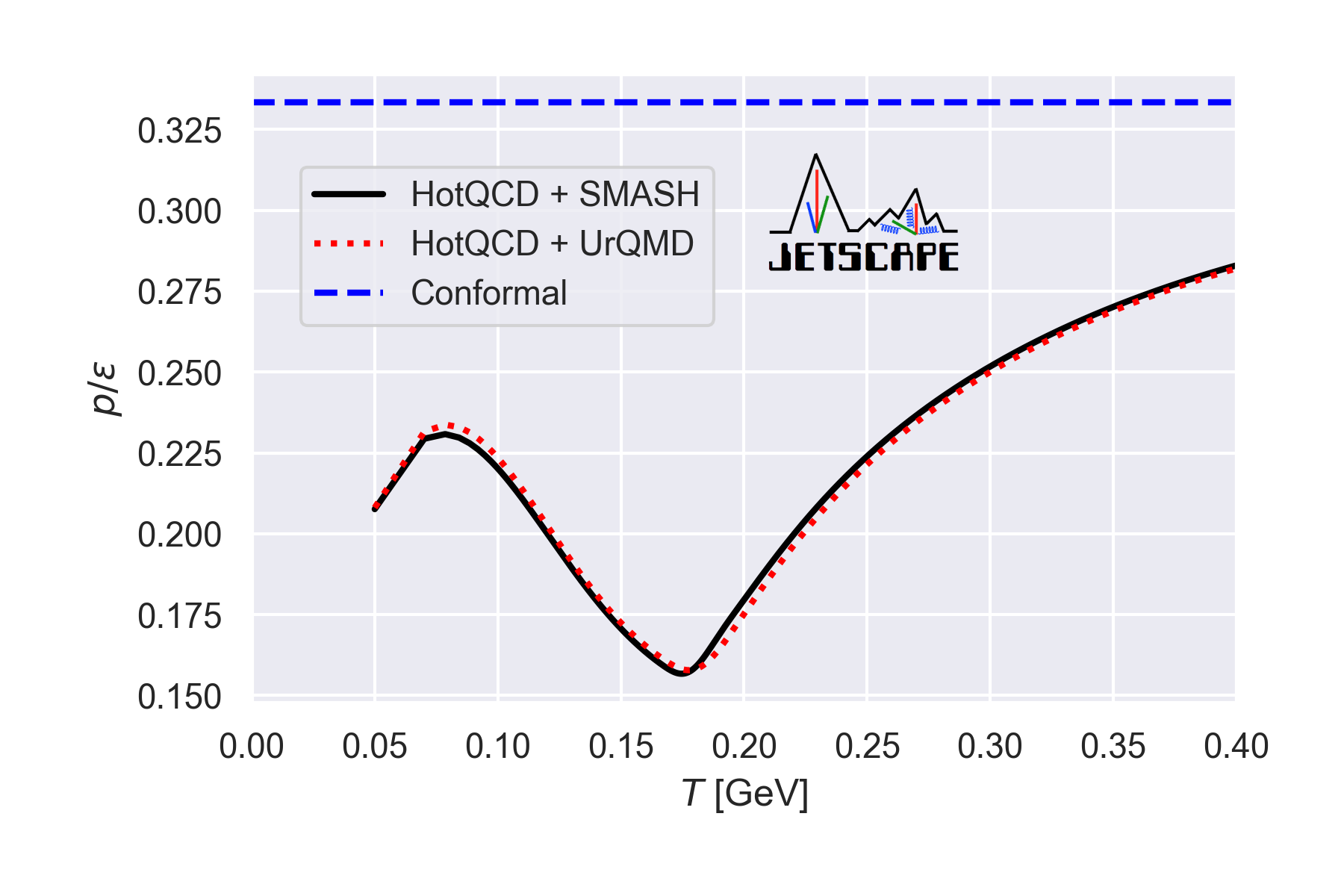}
    \caption{The equation of state used throughout this work for parameter estimation `HotQCD + \SMASH{}' is shown as well as a different equation of state that has been matched to the list of resonances propagated in \URQMD{}. The conformal equation of state is included as a visual reference.}
    \label{fig:eos_comp}
\end{figure}

\begin{figure}[!htb]
  \centering
    \includegraphics[width=0.7\linewidth]{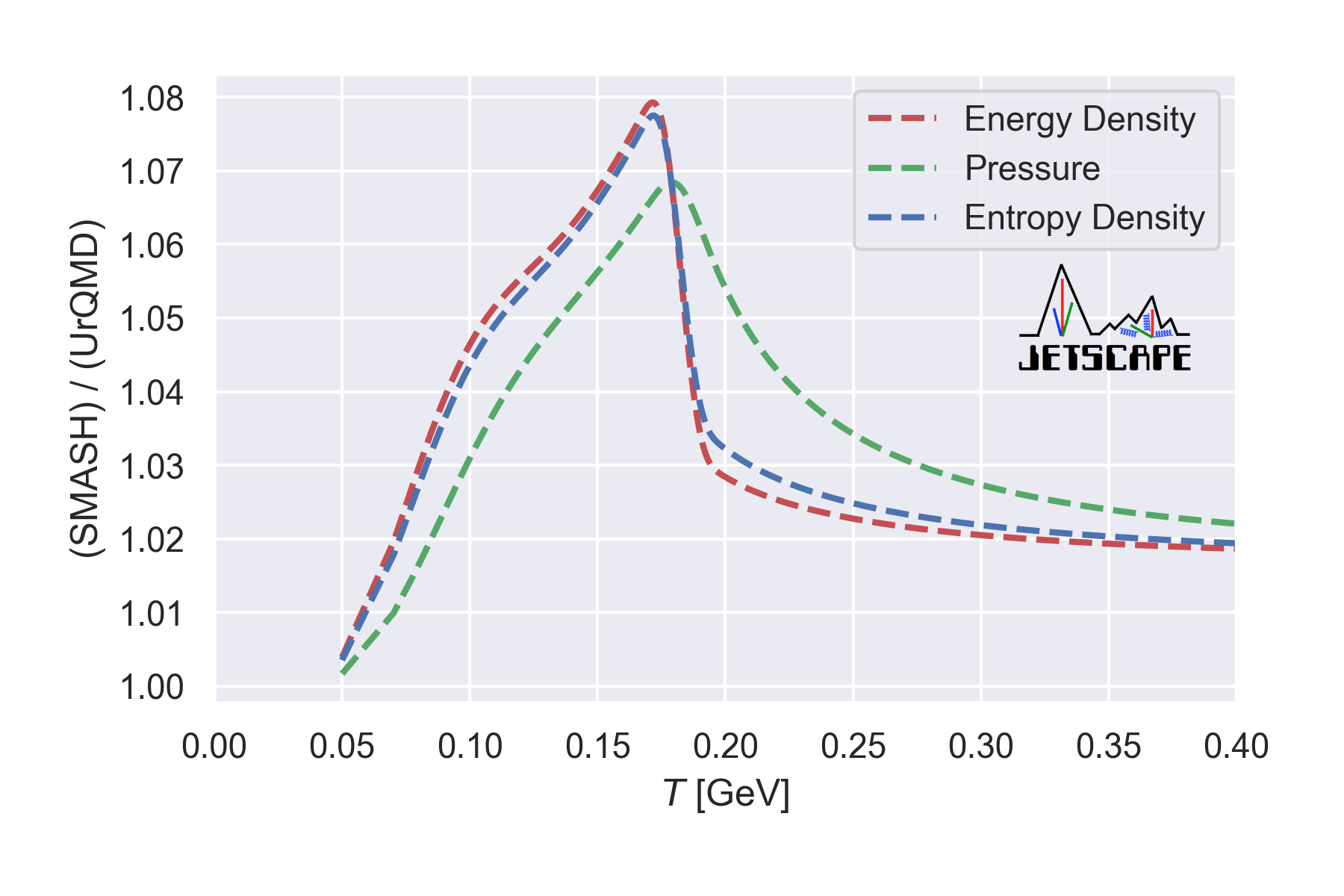}
    \caption{Ratio of \SMASH{} / \URQMD{} at the same temperature for three thermodynamic quantities: energy density (red), equilibrium pressure (green), and entropy density (blue). Each equation of state is constructed by matching the $l$QCD equation of state to a hadron resonance gas matching the list/masses of particles for each code. We see that the disagreement is largest near the region of the switching temperature.}
    \label{fig:eos_ratio}
\end{figure}

Evidently, even with the same matching procedure between the hadron resonance gas and the lattice calculation, the exact content of the hadron resonance gas is important. In the present case, we are interested in two configurations: one used the particle content from \SMASH{}, while the other uses \URQMD{}'s. Both are matched to HotQCD's lattice calculation as described above. The equation of state with the \SMASH{} hadron resonance gas is the one that has been used to perform parameter estimation in this study. We compare the two equations of state in \figs{fig:eos_comp} and~\ref{fig:eos_ratio}. The differences between the two equations of state amount to up to 8\%.

In \Appendix{app:smash}, we compared the predictions of two hybrid models, one model using \SMASH{} as afterburner and the other using \URQMD{}. To obtain such a level of agreement in the observables, it was necessary to use, in the hydrodynamics, equations of state that matched consistently the chosen hadronic transport afterburner. This is consistent with what we see in \fig{fig:eos_comp}: inside the window of particlization temperature, the differences between the equations of state can be larger than $\sim$5\%, and can undeniably produce noticeably different hadronic observables.

We note that this work uses a fixed equation of state which does not parametrize any potential theoretical uncertainties. See Ref.~\cite{Auvinen:2020mpc} for a recent study which includes uncertainty in the lattice-matched equation of state.


\subsection{Validation of free-streaming}
\label{app:fs_valid}
\begin{figure}[!htb]
\centering
\includegraphics[width=0.6\linewidth]{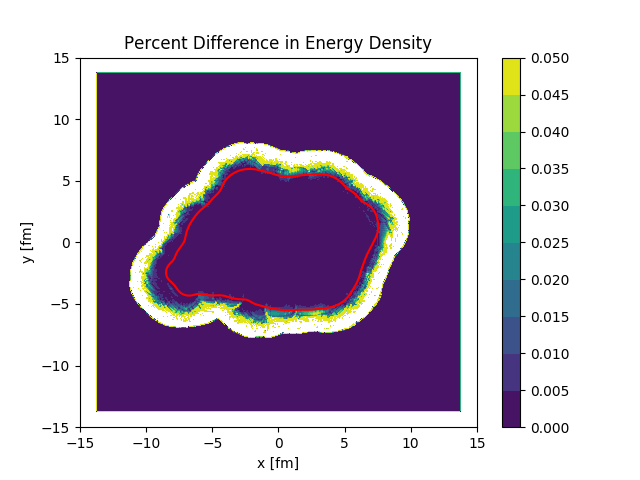}
\caption{The percent difference in energy density between OSU and Duke free-streaming codes. The red contour traces cells for which energy density is $0.1$ GeV/fm$^3$.}
\label{fs_percent_diff_e}
\end{figure}

\begin{figure*}[!htb]
\noindent\makebox[\linewidth]{%
  \centering
  \begin{minipage}{0.5\textwidth}
    \includegraphics[width=\textwidth]{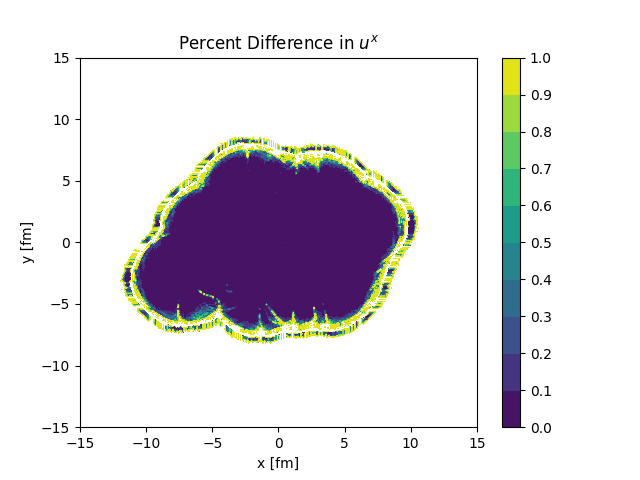}
  \end{minipage}
  \begin{minipage}{0.5\textwidth}
    \includegraphics[width=\textwidth]{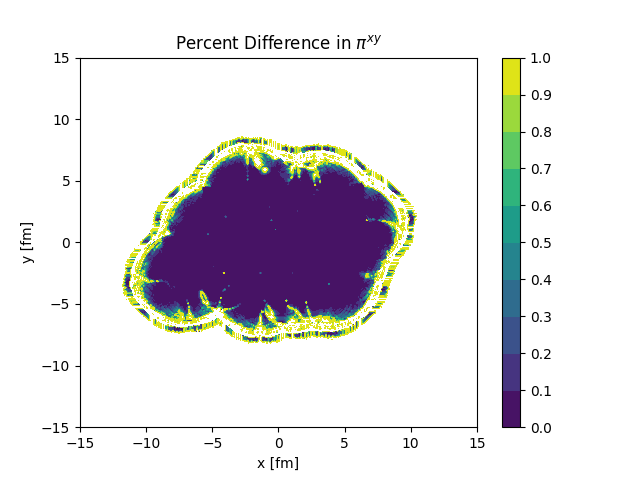}
  \end{minipage}
  }
  \caption{Percent difference in transverse flow $u^{x}$ (left) and shear stress $\pi^{xy}$ (right) between OSU and Duke free-streaming.}
  \label{fs_percent_diff_pi}
\end{figure*}

We will refer to the free-streaming code in JETSCAPE as the `OSU' code, and refer to the code used in J. Bernhard's study\footnote{\url{https://github.com/Duke-QCD/free-stream.git}} as the `Duke' code. The OSU code generalizes the free-streaming formalism to include nontrivial longitudinal motion, but in this study it was used in its boost-invariant mode. In this case the two codes are based on the same formalism and (up to numerical implementation differences) should give the same results. As a test, we ran both the Duke and OSU free-streaming codes for the same initial energy density profile and free-streaming time $\tau_s$ and compare the results below. We chose a non-central Pb+Pb event at 2.76 TeV with a nucleon width $w = 0.5$fm. The grid step $dx$ was determined by taking $dx = 0.15 * w$. The free-streaming time was set to $\tau_s = 1.16$\,fm/$c$. Below we compare the energy density, flow velocity and a component of the shear stress tensor. Any significant differences only occur in the very dilute regions, where it becomes harder to find a numerical solution of the eigenvalue problem.
These are shown in Figs.~\ref{fs_percent_diff_e} through \ref{fs_percent_diff_pi}.

\end{appendices}

\bibliography{references}
\bibliographystyle{elsarticle-num}


\end{document}